\documentclass{article}
\usepackage[active]{srcltx}
\usepackage{amsmath}
\usepackage{bbm}
\usepackage{amsthm}
\usepackage{amsfonts}
\usepackage{amssymb}
\usepackage{amsbsy}
\usepackage{amscd}
\usepackage{cases}
\usepackage{a4wide}
\usepackage{bm}
\usepackage{latexsym}
\usepackage{color}
\usepackage{dsfont}
\usepackage{indentfirst}
\usepackage[latin1]{inputenc}
\usepackage{mathrsfs}
\usepackage{yfonts}

\def\qed{\hfill $\blacksquare$}

\newtheorem{definition}{\underline{Definition}}[section]
\newtheorem{proposition}[definition]{Proposition}
\newtheorem{theorem}[definition]{Theorem}
\newtheorem{lema}[definition]{Lemma}

\newtheorem{remark}[definition]{Remark}
\newtheorem*{statement1*}{($\mathscr{S}_{1}$)}
\newtheorem*{statement2*}{($\mathscr{S}_{2}$)}
\newtheorem*{statement3*}{($\mathscr{S}_{3}$)}
\newtheorem*{statement4*}{($\mathscr{S}_{4}$)}
\newtheorem*{assumption1*}{($\mathscr{A}_{\mathrm{p}}$)}
\newtheorem*{assumption2*}{($\mathscr{A}_{\mathrm{r}}$)}

\numberwithin{equation}{section}

\DeclareMathAlphabet{\mathpzc}{OT1}{pzc}{m}{it}
\DeclareMathAlphabet{\mathpzcb}{OT1}{pzc}{m}{b}

\begin{document}
\begin{center}
{\Large{\textbf{A rigorous proof of the Bohr-van Leeuwen theorem\\ in the semiclassical limit.}}}

\medskip

\today

\end{center}

\begin{center}
\small{Baptiste Savoie\footnote{Dublin Institute for Advanced Studies,
School of Theoretical Physics, 10 Burlington Road, Dublin 4, Ireland; e-mail: baptiste.savoie@gmail.com .}}

\end{center}

\begin{abstract}
The original formulation of \textit{the Bohr-van Leeuwen (BvL) theorem} states that, in a uniform magnetic field and in thermal equilibrium, the magnetization of an electron gas in the classical \textit{Drude-Lorentz model} vanishes identically. This stems from classical statistics which assign the canonical momenta all values ranging from $-\infty$ to $\infty$ what makes the free energy density magnetic-field-independent. When considering a classical (Maxwell-Boltzmann) interacting electron gas, it is usually admitted that the BvL theorem holds upon condition that the potentials modeling the
interactions are particle-velocities-independent and do not cause the system to rotate after turning on the magnetic field. From a rigorous viewpoint, when treating large macroscopic systems one expects the BvL theorem to hold provided the thermodynamic limit of the free energy density exists (and the equivalence of ensemble holds). This requires suitable assumptions on the many-body interactions potential and on the possible external potentials to prevent the system from collapsing or flying apart. Starting from quantum statistical mechanics, the purpose of this article is to give, within the linear-response theory, a proof of the BvL theorem in the semiclassical limit when considering a dilute electron gas in the canonical conditions subjected to a class of translational invariant external potentials.
\end{abstract}

\noindent
PACS-2010 number: 75.20.-g, 51.60.+a, 75.20.Ck, 75.30.Cr

\medskip

\noindent
MSC-2010 number: 81Q10, 82B10, 82B21, 82D05, 82D20, 82D40

\medskip


\medskip

\tableofcontents

\medskip

\section{Introduction \& the main result.}

\subsection{An historical review.}

To highlight the significant role the BvL theorem played in the understanding of the origins of magnetism, J.H.  van Vleck characterized, in his Nobel lecture   \cite{VV2'} in 1977, the works of N. Bohr at the basis of the BvL theorem as '\textit{perhaps the most deflationary publication of all time in Physics}'. Through a brief historical review, let us list the main works which gave rise to the BvL theorem.\\
\indent In 1905, P. Langevin published in \cite{Lan1,Lan2} his 'microscopic theory' of magnetism to account for dia- and paramagnetism phenomena observed in ions/molecules gases. His theory is based on classical statistical mechanics and electrodynamics. Considering that matter is formed by electrons in stable periodic motion (the mutual actions between electrons assure the mechanical stability), he supposed two \textit{ad hoc} assumptions: the molecules contain at least one closed electron orbit with a fixed magnetic moment out of any external field, and, the different orbits in each molecule have such a moment and orientations that their resultant moment may vanish or not. From those assumptions, Langevin calculated the mean variation of the magnetic moment of electrons moving in closed orbits under the influence of an external constant magnetic field. This led to the so-called \textit{Langevin formula} in
\cite[pp. 89]{Lan2} for the diamagnetic susceptibility of electrons. Also, he recovered the analytic expression of the Curie's empirical law for paramagnetic molecules gases  in \cite[pp. 119]{Lan2}. We mention that
Langevin's theory was the trigger of a series of articles discussing his assumptions/results, and in that sense, it was at the root of the BvL theorem.\\
\indent In all likelihood, N. Bohr was the first one to point out that a classical free electron gas  in thermal equilibrium can display no magnetic effects. The results derived in his PhD dissertation from 1911 are the basis for the BvL theorem. Among other things, Bohr was interested in the influence of a magnetic field on the motion of free electrons in metals within the Drude-Lorentz model (see Sec. \ref{fgfdc} for the assumptions of the model). His conclusion can be stated as follows: \textit{'a piece of metal in electric and thermal equilibrium will not possess any magnetic properties whatever due to the presence of free electrons'}, see \cite[pp. 380]{Niels}. To come to such a conclusion, he showed that whenever a state of equilibrium exists, the presence of magnetic forces does not affect the classical statistical distribution of the electrons. Since the electron velocities in any arbitrary volume element are equally distributed in all directions, then no magnetic effects can arise from such a volume
element.\\ 
\indent On the 1910's, a series of works came in response to Langevin's theory of magnetism. Essentially, the purpose consisted in computing within the framework of classical statistical mechanics the thermal average resultant magnetic moment (and then the thermal average magnetic susceptibility via the linear response theory) of various sorts of molecules taking into account specific kinds of collisions. The case of free electrons in metals has been considered as well. For a review, we refer to \cite[Sec. X]{VL}. It seems that H.-A. Lorentz derived a result backing up the Bohr conclusions.\\
\indent In 1921, H.J. van Leeuwen published the article \cite{VL} that will be taken later on as the seminal paper for the BvL theorem.
In response to Langevin's theory, \cite[Sec. II-VII]{VL} are devoted to magnetism of molecules gases in thermal equilibrium. Considering molecules carrying either a current or charges placed in an external constant magnetic field, she computed the thermal average magnetization (i.e. resultant magnetic moment per unit-volume) and susceptibility within the framework of classical statistical mechanics. 
The results heavily depend on the type of collisions considered, and also on the permissible values of the canonical momenta associated with the degrees of freedom of a single molecule in a given configuration. In some cases, she recovered some of the Langevin results (temperature independence of the diamagnetic susceptibility, Curie's law for the paramagnetic susceptibility), for some others the magnetization vanishes. An attempt to find sufficient conditions leading to a vanishing magnetization for arbitrary molecules is given in
\cite[Sec. IX]{VL}. \cite[Sec. VIII]{VL} deals with the magnetic response of free electrons in metals within the Drude-Lorentz (DL) model (see Sec. \ref{fgfdc} for the assumptions). Essentially, van Leeuwen derived a result in the direction of Bohr conclusions. In the presence of a constant magnetic field, she showed that, whenever a state of equilibrium exists, the thermal average magnetization of the electron gas in the DL model vanishes. Additionally, if the collisions between electrons are not neglected, then it may happen that the system of electrons rotates with a constant angular velocity (as an equilibrium state) after turning on the magnetic field. In that case, the magnetization does not vanish, and corresponds to induced current (as a result of the Larmor precession theorem).\\
\indent The apparent contradiction between the Langevin and Bohr-van Leeuwen results is one of the main factors which led to the emergence of quantum mechanics in the 1920's. Quantum mechanics will remove this paradox: the Langevin assumptions (stationary of the electron orbits and permanence of magnetic moments) are of a quantum nature. Furthermore, it will provide the framework to explain the origins of dia- and paramagnetism phenomena observed in solids and molecules/ions gases. Below, we list the most important works within the framework of quantum mechanics.\\
\indent In 1927, W. Pauli investigated in \cite{Paul} the induced magnetization arising from the coupling between the spin magnetic moment and an external constant magnetic field. Considering a free electron gas obeying Fermi-Dirac statistics, he found that the spin contribution to the magnetic susceptibility in the weak-field limit and in the low-temperature regime is purely paramagnetic and temperature-independent to first-order. The formula is known as \textit{Pauli susceptibility formula}.\\
\indent In 1929, J.H. van Vleck revisited in \cite{VV2,VV1} Langevin's theory of magnetism within the framework of quantum mechanics. For a brief summary of his works, we refer to
\cite[Sec. 1.1]{Sa2}.\\
\indent In 1930, L. Landau investigated in \cite{Lan} the induced magnetization arising from the 'helical motion' of electrons induced by the Zeeman Hamiltonian in an external constant magnetic field. Considering a free electron gas confined in a box, and obeying Fermi-Dirac statistics while disregarding the spin, he found that the bulk value (i.e. independent of the boundary effects) of the orbital susceptibility in the weak-field limit and in the low-temperature regime is to first-order purely diamagnetic and temperature-independent.  The formula, known as \textit{Landau susceptibility formula}, is exactly one-third the Pauli susceptibility formula in absolute value. Thus, diamagnetism of free electrons gas arises from the quantization of the radii of the helical paths, and it is a weak phenomenon. Besides, Landau pointed out the existence of another contribution whose magnitude oscillates with the magnetic field: the so-called \textit{de Haas-van Alphen effect} discovered experimentally the same year. We mention that the Landau's article generated a huge amount of papers during about 40 years in which the influence of the walls of the box on the bulk value of the orbital susceptibility is discussed. Indeed, the contribution coming from the boundary effects is expected to be strongly paramagnetic and it has been treated in \cite{Lan} by a semiclassical argument. The corrections to the bulk value have been computed in low/high-temperatures regime for various shapes of boxes, and various confining potentials modeling the boundary of the box. For a review, we refer to \cite{Tho,ABN}.

\subsection{The BvL theorem, a comprehensive overview.}
\label{fgfdc}

\subsubsection{The original statement.}

From \cite{VL}, the BvL theorem can be stated as follows:
\begin{statement1*} In a constant magnetic field and in thermal equilibrium, the magnetization of an electron gas in the classical Drude-Lorentz (DL) model is identically zero.
\end{statement1*}
The DL model, originally introduced to investigate the properties of metals, is based on classical kinetic theory of gases, see e.g. \cite[Sec. 1.2]{Wil}. Within this model, the valence electrons are treated as an ideal gas of free particles surrounding the ion cores. The latter occupy a small volume of the metal. The electrons are assimilated to solid spheres and the ion cores are assimilated to heavy rigid spheres in thermal equilibrium which may vibrate when collisions occur. Only the collisions with ion cores are considered, and the collisions are instantaneous and elastic. During collisions, the electrons transfer energy and momentum. Between successive collisions, the electrons have \textit{random thermal motions} and the interactions electron-electron are neglected. The electrons are assumed to achieve thermal equilibrium with their surroundings only through collisions with ion cores, and their velocity distribution  in thermal equilibrium follows the Maxwell-Boltzmann distribution. Let us turn to the 'proof' of statement $\mathrm{(\mathscr{S}_{1})}$ in \cite[Sec. VIII]{VL}. Consider a $3$-dimensional electron gas in the DL model. The metal has a large extent (in the DL model, it often has an infinite extent). The electrons are subjected to a constant magnetic field $\bold{B}$. We choose it parallel to the third direction of $\mathbb{R}^{3}$, i.e. $\mathbf{B}:= B \mathbf{e}_{3}$ with $B>0$ and we use the symmetric gauge $\mathbf{A}(\mathbf{x}):= \frac{1}{2} \mathbf{B} \times \mathbf{x}$ s.t. $\mathbf{B} = \nabla \times \mathbf{A}(\mathbf{x})$. We assume that the system has achieved thermal equilibrium and that an equilibrium state exists. Let $\Omega$ be an arbitrary macroscopic element volume of the metal containing an assembly of $N$ electrons. To lighten the derivation, we assimilate the electrons to point-particles and the presence of fixed scatters is disregarded. From the fundamental law of dynamics, an electron moves in the presence of the magnetic field along a circular helix whose axis is parallel to the magnetic field. Considering the projection of the helical orbits on a plane orthogonal to the field, the projected orbits are circular. The magnetic moment of the $j$-th particle projected along $\bold{e}_{3}$ reads as:
\begin{equation}
\label{mu3}
\mu_{j,3} = \boldsymbol{\mu}_{j} \cdot\mathbf{e}_{3} := - \frac{e}{2m c} \mathbf{L}_{j} \cdot \mathbf{e}_{3}  = \frac{e}{2 c} (x_{j,2} v_{j,1}-x_{j,1} v_{j,2}) = - \frac{1}{B} \frac{e}{c}  \bold{v}_{j} \cdot \bold{A}(\bold{x}_{j}),\quad j \in \{1,\ldots,N\},
\end{equation}
where $\mathbf{L}_{j}:= \mathbf{x}_{j} \times (m \mathbf{v}_{j})$ is the gauge-invariant (or kinetic) angular momentum
of the $j$-th particle. Here and hereafter, $e$, $m$ and $c$ denote respectively the elementary charge, the electron rest mass and the speed of light in vacuum. $\mathbf{x}_{j}=(x_{j,1},x_{j,2},x_{j,3})$ and $\mathbf{v}_{j}=(v_{j,1},v_{j,2},v_{j,3})$ are respectively the position and velocity vectors of the $j$-th particle. The Lagrangian of the system is defined as:
\begin{equation}
\label{Lagr}
\mathcal{L}\left(\{\mathbf{x}_{j}\},\{\mathbf{v}_{j}\};B\right) := \sum_{j=1}^{N} \left(\frac{1}{2} m \bold{v}_{j}^{2} + \boldsymbol{\mu}_{j} \cdot \bold{B}\right) = \sum_{j=1}^{N} \left(\frac{1}{2} m \mathbf{v}_{j}^{2} - \frac{e}{c} \mathbf{v}_{j}\cdot \mathbf{A}(\mathbf{x}_{j})\right).
\end{equation}
The classical Hamiltonian is obtained by performing a Legendre-transform of the Lagrangian:
\begin{gather}
\label{Hamilt}
\mathcal{H}\left(\{\mathbf{x}_{j}\},\{\mathbf{p}_{j}\};B\right) := \sum_{j=1}^{N} \mathbf{p}_{j} \cdot \mathbf{v}_{j} - \mathcal{L}\left(\{\mathbf{x}_{j}\},\{\mathbf{v}_{j}\};B\right) = \sum_{j=1}^{N}  \frac{1}{2m}\left(\mathbf{p}_{j} + \frac{e}{c}\mathbf{A}(\mathbf{x}_{j})\right)^{2}, \\
\textrm{with:}\quad p_{j,l} = \mathbf{p}_{j} \cdot \mathbf{e}_{l} := \frac{\partial \mathcal{L}}{\partial v_{j,l}}\left(\{\mathbf{x}_{j}\},\{\mathbf{v}_{j}\};B\right) = m v_{j,l} - \frac{e}{c} A_{l}(\mathbf{x}_{j}),\quad l=1,2,3,\nonumber
\end{gather}
where $p_{j,l}$ stands for the canonical momentum of the $j$-th particle projected along the $l$-th direction.
The thermal average magnetization (along the third direction) in $\Omega$ with volume $\mathrm{V}(\Omega)$ reads as:
\begin{equation}
\label{M3beta}
\frac{1}{\mathrm{V}(\Omega)} \frac{\displaystyle{\int_{\Omega} \mathrm{d}\mathbf{x}_{1} \dotsb \int_{\Omega} \mathrm{d}\mathbf{x}_{N} \int_{\mathbb{R}^{3}} \mathrm{d}\mathbf{p}_{1} \dotsb \int_{\mathbb{R}^{3}}\mathrm{d}\mathbf{p}_{N}\, \sum_{j=1}^{N} \mu_{j,3}(\{\mathbf{x}_{j}\},\{\mathbf{v}_{j}\})\, \mathrm{e}^{-\beta \mathcal{H}(\{\mathbf{x}_{j}\},\{\mathbf{p}_{j}\};B)}}}{\displaystyle{\int_{\Omega} \mathrm{d}\mathbf{x}_{1} \dotsb \int_{\Omega}  \mathrm{d}\mathbf{x}_{N} \int_{\mathbb{R}^{3}} \mathrm{d}\mathbf{p}_{1} \dotsb \int_{\mathbb{R}^{3}} \mathrm{d}\mathbf{p}_{N}\,\mathrm{e}^{-\beta \mathcal{H}(\{\mathbf{x}_{j}\},\{\mathbf{p}_{j}\};B)}}},
\end{equation}
where $\beta := (k_{B}T)^{-1}$ is the 'inverse' temperature and $k_{B}$ denotes the Boltzmann constant.
The integral over each one of the components of the canonical momenta $\bold{p}_{j}$ runs from $-\infty$ to $\infty$ in accordance with classical statistics. Inserting \eqref{mu3} into \eqref{M3beta} and remarking that $v_{j,l}= \partial_{p_{j,l}} \mathcal{H}(\{\mathbf{x}_{j}\},\{\mathbf{p}_{j}\};B)$, then the expression in the numerator of \eqref{M3beta} consists of a sum of $2N$ terms whose a generical term is:
\begin{equation*}
\frac{e}{2c \beta} \int_{\Omega^{N}} \mathrm{d}\mathbf{x}_{1} \dotsb \mathrm{d}\mathbf{x}_{N} \int_{\mathbb{R}^{3N}} \mathrm{d}\mathbf{p}_{1} \dotsb  \mathrm{d}\mathbf{p}_{N}\, x_{j,l_{1}} \partial_{p_{j,l_{2}}} \mathrm{e}^{-\beta \mathcal{H}(\{\mathbf{x}_{j}\},\{\mathbf{p}_{j}\};B)},\quad (l_{1},l_{2}) \in \{1,2\}^{2},\, l_{1} \neq l_{2}.
\end{equation*}
In view of \eqref{Hamilt}, performing the integrations over the $p_{j,l}$s gives a null value. Therefore, the thermal average magnetization in \eqref{M3beta} vanishes identically. Such a derivation is independent of the choice of the macroscopic element volume $\Omega$.\\
\indent What about when considering the 'true' DL model? In that case, the configuration space $\Omega$ has to be slightly modified since the distances between pairs of particles and between particles and scatters remain larger than a certain constant (both are assimilated to solid spheres). Also, the Hamiltonian in \eqref{Hamilt} has to be supplemented with the condition that the particles undergo an elastic collision when colliding with the ion cores. The rest of the derivation remains unchanged. Besides, if one initially considers  the electron gas in the whole of the metal, then one has to assume that the gas is confined in a container $\Lambda$ with reflecting walls in accordance with classical kinetic theory. Such a confinement can be modeled by a potential energy $U_{\mathrm{c}}$ satisfying $U_{\mathrm{c}}(\bold{x})= 0$ if $\bold{x} \in \Lambda$, $U_{\mathrm{c}}(\bold{x})=\infty$ otherwise. Elastic and specular reflections are assured, and then no kinetic energy is lost. It follows that the permissible values for the canonical momenta still run from $-\infty$ to $\infty$, and the thermal average magnetization vanishes identically. As pointed out in \cite[Sec. 26]{VV2'}, such a result is not sensitive to the shape of the container, and to the smoothness of the boundary.\\
\indent Turning to the interpretation of the BvL theorem, we reproduce the text in \cite{Mol}: \textit{'The orbits of a cloud of electrons in space give no net current density in bulk, but build a surface current orthogonal to the magnetic field. Only the outer electrons feed this diamagnetic field, but since their ratio to the total number vanishes for larger and larger clouds, in thermodynamic limit no magnetization results. If the motion of several electrons is confined in a box, a current also develops along the boundary due to electrons that bounce on it. This current density is opposite and exactly cancels the surface current due to electrons that do not hit the walls. Thus paramagnetic and diamagnetic terms compensate and again, no magnetization survives'}. See also \cite[Sec. 26]{VV2'}.

\subsubsection{The modern formulations: Assessing the assumptions.}

So far, we have focused on the original formulation of the BvL theorem. We now turn to the modern formulations usually dealing with the classical (Maxwell-Boltzmann) electron gas apart from the DL model.
The BvL theorem is thought of as a basic result and it can be found in any textbook working with magnetic phenomenon. Some state it only for the classical ideal electron gas, see e.g. \cite[pp. 256]{Mor}. The statement that is often taken as a reference is the following one, see \cite[Sec. 1.6]{Matt}:
\begin{statement2*}
At any finite temperature, and in all finite applied electric or magnetic fields, the net magnetization of a collection of classical electrons in thermal equilibrium vanishes identically.
\end{statement2*}
\noindent The statements encountered in literature are generally much deeper. For instance
\cite[Sec. 4.3]{Pei}:
\begin{statement3*}
In classical mechanics, there can be no magnetization.
\end{statement3*}
\noindent Of the same type, we can cite: \textit{In classical statistics, there are no macroscopic magnetic properties of matter}, see \cite[Sec. 52]{Land}, or \textit{the phenomenon of diamagnetism does not exist in classical physics}, see \cite[p. 168]{Hu}. Written in this way, statements of type $\mathrm{(\mathscr{S}_{3})}$ can lead to some misunderstanding. Indeed, it is known that the BvL theorem breaks down if the system of classical charged particles uniformly rotates (as an equilibrium state) after turning on the magnetic field. A textbook model leading to such a situation and involving the Larmor precession theorem is discussed in
\cite[Sec. 34.5]{FLS}. Such classical systems exhibit a non-zero induced magnetization, and therefore diamagnetism occurs. See \cite{OZ} for related discussions.
The proof given in literature for statements $\mathrm{(\mathscr{S}_{2})}$, $\mathrm{(\mathscr{S}_{3})}$ and $\mathrm{(\mathscr{S}_{3})}$-like is standard, and is nothing but a variant of the Van Leeuwen's derivation. We shall reproduce the arguments in the next paragraph. In \cite[Sec. 4.3]{Pei} and \cite[Sec. 1.6]{Matt}, a classical non-relativistic interacting electron gas is considered. The many-body interactions and possible interactions with external electric fields are modeled by an arbitrary potential energy. However, no assumptions are made regarding the potential energy. In view of the above counterexample (uniformly rotating systems), the question of validity of the BvL theorem for classical interacting systems should be addressed. The aim of the two following paragraphs is to give necessary and/or sufficient conditions on the potential energies assuring the BvL theorem to hold. We successively treat the case of finite systems and infinite systems (thermodynamic behavior).

\paragraph{Case of finite systems.} \label{para1} Consider a $3$-dimensional assembly of $N$ classical electrons (assimilated to point-particles) confined in a container $\Lambda$, say a cube centered at the origin, with reflecting walls. We assume that the electrons interact with each other, and also that each electron interacts with an external electric field modeling the underlying medium. The confinement is modeled by $U_{\mathrm{c}}$ defined as previously, and we denote by $U_{\mathrm{int}}$ and $U_{\mathrm{el}}$ the many-body interactions and electric potential energy respectively. All the involved potential energies are assumed to be independent of the particle-velocities. Besides, the system is plunged into a constant magnetic field $\bold{B}$. We choose it as $\mathbf{B}:= B \mathbf{e}_{3}$, with $B>0$ and we use the symmetric gauge $\mathbf{A}(\mathbf{x}):= \frac{1}{2} \mathbf{B} \times \mathbf{x}$. We suppose that the interactions do not cause the system to rotate after turning on the magnetic field, and that the degrees of freedom are only translational. We also assume that the system is in thermal equilibrium with a heat bath. Under such conditions, a stationary state (in equilibrium) occurs. The Lagrangian and Hamiltonian in \eqref{Lagr} and \eqref{Hamilt} have respectively to be replaced by:
\begin{gather}
\label{secLan}
\tilde{\mathcal{L}}(\{\mathbf{x}_{j}\},\{\mathbf{v}_{j}\};B) := \sum_{j=1}^{N}  \left(\frac{1}{2} m \mathbf{v}_{j}^{2}  + \boldsymbol{\mu}_{j} \cdot \bold{B} - U_{\mathrm{el}}(\mathbf{x}_{j}) - U_{\mathrm{c}}(\mathbf{x}_{j})\right) - U_{\mathrm{int}}(\bold{x}_{1},\ldots,\bold{x}_{N}),\\
\label{secHam}
\tilde{\mathcal{H}}(\{\mathbf{x}_{j}\},\{\mathbf{p}_{j}\};B) :=  \sum_{j=1}^{N} \left(\frac{1}{2m}\left(\mathbf{p}_{j} + \frac{e}{c}\mathbf{A}(\mathbf{x}_{j})\right)^{2} + U_{\mathrm{el}}(\mathbf{x}_{j}) + U_{\mathrm{c}}(\mathbf{x}_{j})\right) + U_{\mathrm{int}}(\bold{x}_{1},\ldots,\bold{x}_{N}).
\end{gather}
We now give the derivation at a formal level, the convergence issues will be discussed afterwards.  In the canonical ensemble of classical statistical mechanics, the free energy density is given by:
\begin{equation*}
\mathcal{F}(\beta,N,\mathrm{V}(\Lambda);B) := -\frac{1}{\mathrm{V}(\Lambda)} \frac{1}{\beta} \ln \left(\mathcal{Z}(\beta,N,\mathrm{V}(\Lambda);B)\right),
\end{equation*}
where $\mathcal{Z}(\beta,N,\mathrm{V}(\Lambda);B)$ denotes the classical canonical partition function defined as:
\begin{equation}
\label{parf}
\mathcal{Z}(\beta,N,\mathrm{V}(\Lambda);B) := \frac{1}{h^{3N} N!} \int_{\Lambda} \mathrm{d}\mathbf{x}_{1} \dotsb \int_{\Lambda} \mathrm{d}\mathbf{x}_{N} \int_{\mathbb{R}^{3}} \mathrm{d}\mathbf{p}_{1} \dotsb \int_{\mathbb{R}^{3}} \mathrm{d}\mathbf{p}_{N}\,\mathrm{e}^{-\beta \tilde{\mathcal{H}}(\{\mathbf{x}_{j}\},\{\mathbf{p}_{j}\};B)}.
\end{equation}
We added the correction factor $h^{-3N} (N!)^{-1}$, where $h$ is the Planck's constant, to make the classical partition function dimensionless (while making it agree with the quantum behavior in the high-temperature limit). The integral over each one of the components of the $\bold{p}_{j}$s runs from $-\infty$ to $\infty$ in accordance with classical statistics. The canonical magnetization is defined as:
\begin{equation}
\label{Rex}
\mathcal{M}(\beta,N,\mathrm{V}(\Lambda);B) := - \frac{\partial \mathcal{F}}{\partial B}(\beta,N,\mathrm{V}(\Lambda);B) = \frac{1}{\mathrm{V}(\Lambda)} \frac{1}{\beta} \frac{\partial}{\partial B}\ln \left(\mathcal{Z}(\beta,N,\mathrm{V}(\Lambda);B)\right).
\end{equation}
To make the connection with the derivation given below statement $\mathrm{(\mathscr{S}_{1})}$, the quantity in \eqref{M3beta} is nothing but \eqref{Rex}. Indeed, the quantity in \eqref{mu3} can be rewritten as $- (\partial_{B} \mathcal{H})(\{\mathbf{x}_{j}\},\{\mathbf{p}_{j}\};B)$. Next, remark that the classical partition function is invariant under a transformation from canonical momentum $p_{j,l}$ to kinetic momentum $v_{j,l}$ (the Jacobian of the transformation is equal to $1$). Ergo, the vector potential disappears from the Boltzmann factor, and \eqref{parf} can be rewritten as:
\begin{equation}
\label{rewriled}
\mathcal{Z}(\beta,N,\mathrm{V}(\Lambda);B) = \mathcal{Z}(\beta,N,\mathrm{V}(\Lambda);B=0) = \lambda_{\beta}^{-3N} \mathcal{Z}_{\mathrm{conf}}(\beta,N,\mathrm{V}(\Lambda)),
\end{equation}
where $\lambda_{\beta}$ denotes the thermal de Broglie wavelength and $\mathcal{Z}_{\mathrm{conf}}$ the so-called \textit{configuration integral}:
\begin{equation}
\label{Zconf}
\mathcal{Z}_{\mathrm{conf}}(\beta,N,\mathrm{V}(\Lambda)) := \frac{1}{N!} \int_{\Lambda^{N}} \mathrm{d}\mathbf{x}_{1} \dotsb \mathrm{d}\mathbf{x}_{N}\, \mathrm{e}^{-\beta \sum_{j=1}^{N}  U_{\mathrm{el}}(\mathbf{x}_{j})} \mathrm{e}^{-\beta U_{\mathrm{int}}(\bold{x}_{1},\ldots,\bold{x}_{N})}.
\end{equation}
Here, we got rid of $\sum_{j=1}^{N} U_{\mathrm{c}}(\bold{x}_{j})$ following the convention. The free energy density then reads as:
\begin{equation}
\label{sfree2}
\mathcal{F}(\beta,N,\mathrm{V}(\Lambda);B) = \mathcal{F}(\beta,N,\mathrm{V}(\Lambda);B=0) = -\frac{1}{\mathrm{V}(\Lambda)} \frac{1}{\beta} \ln \left(\lambda_{\beta}^{-3N} \mathcal{Z}_{\mathrm{conf}}(\beta,N,\mathrm{V}(\Lambda))\right).
\end{equation}
Provided that \eqref{Zconf} exists (in the Lebesgue-sense), the free energy density in \eqref{sfree2} is well-defined and $B$-independent. Due to \eqref{Rex}, the canonical magnetization vanishes identically. Note that, using the same method as above, it is argued in \cite{Matt} that a potential deriving from the Coulomb force can be considered. But for such a potential, \eqref{Zconf} is not defined. For the existence of \eqref{Zconf}, $U_{\mathrm{int}}$ and $U_{\mathrm{el}}$ have to be bounded from below.

\paragraph{Case of infinite systems (thermodynamic behavior).} The derivation in Sec. \ref{para1} applies to \textit{finite systems} in which boundary effects usually play a significant role. Turning to large macroscopic systems, the thermodynamic description is obtained in statistical mechanics by considering the bulk limit or thermodynamic limit (TL), i.e. taking the limit of an infinitely large system with a finite particle density: $N \rightarrow \infty$, $\mathrm{V}(\Lambda) \rightarrow \infty$ while $\rho=N \mathrm{V}(\Lambda)^{-1}>0$ is held fixed. In this limit, the surface effects disappear and we are left with the \textit{bulk properties}. Whenever the thermodynamic limit exists and depends \textit{only} on the \textit{intensive quantities}, then the system has the extensive property which means that the thermodynamic quantities are asymptotically proportional to the system size. 
We stress the point that the existence of the TL depends on the nature of the interactions involved. Indeed, if the interaction potentials decrease fast enough such that the interactions for a particle mainly come from the first neighbors, then increasing $N$ while keeping the density fixed has 'almost' no effect on the bulk, and physical properties are 'almost' independent of $N$. However, this is not true anymore if the potential for a particle is dominated by the influence of far away particles. Such a situation may occur when long-range interactions are involved (by long-range, it is generally understood that the interaction potentials behave at long distance like $\mathcal{O}(\vert \bold{x}\vert^{-\eta})$, $0<\eta < 3$). Note that systems with long-range interactions are known to exhibit peculiar behaviors: they can be spatially inhomogeneous, the TL may not exist and the additivity of energy is usually broken.
Getting back to the thermodynamic description, proving mathematically the existence of the limit $\Lambda \uparrow \mathbb{R}^{d}$ in some sense (typically, by considering a sequence of convex and bounded domains whose the surface areas do not increase too rapidly compared to their volume) while $N\mathrm{V}(\Lambda)^{-1}$ is held fixed, is however not sufficient to establish a complete connection with thermodynamics. In addition, one has to show the consistency of the thermodynamic quantities defined by means of the various ensembles (i.e., the equivalence problem), and also the thermodynamic stability. Both are interrelated. The thermodynamic stability is a result of the convexity of the free energy density w.r.t. the density of particle $\rho$. We refer to \cite[Sec. 4]{Mun} for further details.\\
\indent When considering large macroscopic systems, it follows from the foregoing that the result of the BvL theorem holds whenever the TL of \eqref{sfree2} exists and the equivalence of ensemble holds. Let us give some generical assumptions leading to this situation. We shall distinguish three cases.
\begin{itemize}
\item[$\mathrm{(i)}$] \textit{$U_{\mathrm{int}}\neq0$ and $U_{\mathrm{el}}=0$.}
\end{itemize}
On the basis of physical considerations, the two following assumptions are required: $\mathrm{(h1)}$ $U_{\mathrm{int}}$ is symmetric in the $N$ variables $\bold{x}_{j}$ (identical particles assumption); $\mathrm{(h2)}$ $U_{\mathrm{int}}$ is invariant under translations. We also suppose: $\mathrm{(h3)}$ The configuration integral in \eqref{Zconf} exists as a Lebesgue-integral. Clearly, assumptions $\mathrm{(h1)}$-$\mathrm{(h3)}$ are not sufficient to assure the existence of the TL. On the one hand, the attraction forces could be so strong that the system collapses into a bounded region of $\mathbb{R}^{3}$ as the number of particles increases. This occurs when the TL of \eqref{sfree2} diverges to $-\infty$. On the other hand, the repulsion forces could decrease so little with increasing separation that the TL of \eqref{sfree2} diverges to $\infty$. To avoid the first situation, $U_{\mathrm{int}}$ has to satisfy the stability condition $\mathrm{(C1)}$, i.e. there exists $c\geq 0$ s.t. $\forall N \geq 1$ and $\forall \bold{x}_{j} \in \mathbb{R}^{3}$, $U_{\mathrm{int}}(\bold{x}_{1},\ldots,\bold{x}_{N}) \geq - Nc$. To avoid the second situation, $U_{\mathrm{int}}$ has to satisfy the temperedness (or 'weak-tempering') condition $\mathrm{(C2)}$. If $U_{\mathrm{int}}$ is a pair-interaction potential, i.e. of type $U_{\mathrm{int}}(\bold{x}_{1},\ldots,\bold{x}_{N}) = \frac{1}{2} \sum_{i < j} \Phi(\bold{x}_{i} - \bold{x}_{j})$ with $\Phi$ Lebesgue-mesurable with values in $\mathbb{R}\cup \{\infty\}$, then $\mathrm{(C2)}$ amounts to the condition that the pair-potential $\Phi$ is a \textit{short-range potential}: there exists $\lambda>3$, $R_{0} > 0$ and $C \geq 0$ s.t. $\Phi(\bold{x}) \leq C \vert \bold{x} \vert^{-\lambda}$ for $\vert \bold{x}\vert \geq R_{0}$. We refer to \cite[Sec. 3.1]{Ru} for a precise definition of $\mathrm{(C2)}$ for general many-body interaction potentials, and to \cite[Sec. 3.2]{Ru} for a series of criteria assuring $\mathrm{(C1)}$. We also refer to \cite[Sec. 4]{Gav} for further discussions on these conditions. Under assumptions $\mathrm{(h1)}$-$\mathrm{(h3)}$ and $\mathrm{(C1)}$-$\mathrm{(C2)}$, it is proven in \cite[Sec. 3.3]{Ru} that the TL of \eqref{sfree2} exists when the limit $\Lambda \uparrow \mathbb{R}^{3}$ is taken in the Fisher-sense, see \cite[Def. 2.1.2]{Ru}. Note that the factor $1/N!$ in \eqref{Zconf} is absolutely necessary to obtain such a result. Moreover, the equivalence between ensembles holds (and ergo,  the thermodynamic stability follows). In the presence of hard cores (i.e. when $U_{\mathrm{int}}$ takes the value $\infty$ for the excluded configurations), the TL of \eqref{sfree2} is a convex function w.r.t. $\rho$ if $\rho$ is less than a critical value (the so-called 'close-packing' density).
\begin{itemize}
\item[$\mathrm{(ii)}$] \textit{$U_{\mathrm{int}}=0$ and $U_{\mathrm{el}} \neq 0$.}
\end{itemize}
In that case, \eqref{sfree2} can be decomposed into two contributions:
\begin{equation}
\label{sfree3}
\mathcal{F}(\beta,N,\mathrm{V}(\Lambda); B=0) = -\frac{1}{\beta \mathrm{V}(\Lambda)} \ln\left(\frac{1}{N!} \left(\mathrm{V}(\Lambda) \lambda_{\beta}^{-3}\right)^{N}\right) + \mathcal{F}_{\mathrm{el}}(\beta,N,\mathrm{V}(\Lambda)),
\end{equation}
where the first term in the r.h.s. is nothing but the free energy density of the ideal electron gas and $\mathcal{F}_{\mathrm{el}}$ the contribution arising from the external potential defined as:
\begin{equation}
\label{spfel}
\mathcal{F}_{\mathrm{el}}(\beta,N,\mathrm{V}(\Lambda)) := - \frac{N}{\beta \mathrm{V}(\Lambda)} \ln\left(\frac{1}{\mathrm{V}(\Lambda)} \int_{\Lambda} \mathrm{d}\bold{x}\, \mathrm{e}^{-\beta U_{\mathrm{el}}(\bold{x})}\right).
\end{equation}
In view of \eqref{spfel}, whenever $\lim_{\Lambda \uparrow \mathbb{R}^{3}} \mathrm{V}(\Lambda)^{-1} \int_{\Lambda} \mathrm{d}\bold{x}\, \mathrm{e}^{-\beta U_{\mathrm{el}}(\bold{x})}$ exists and is non-zero, the TL of \eqref{sfree3} exists. This requires $U_{\mathrm{el}}$ to be bounded from below otherwise the system collapses. Note that if $U_{\mathrm{el}} \rightarrow \infty$ when $\vert \bold{x} \vert \uparrow \infty$ then the TL diverges to $\infty$ (the system flies apart). Now, let us turn to the generical situations. There are two: either the external potential is homogeneous or weak-inhomogeneous. The first situation corresponds to translational invariant potentials. Since the case of $U_{\mathrm{el}}=cste$ is obvious, then consider that $U_{\mathrm{el}}$ is periodic, say over an infinite regular lattice in $\mathbb{R}^{3}$ with unit-cell $\Gamma$. This models perfect crystalline solids. Whenever $\mathrm{e}^{-\beta U_{\mathrm{el}}}$ is Lebesgue-integrable over $\Gamma$, $\lim_{\Lambda \uparrow \mathbb{R}^{3}} \mathrm{V}(\Lambda)^{-1} \int_{\Lambda} \mathrm{d}\bold{x}\, \mathrm{e}^{-\beta U_{\mathrm{el}}(\bold{x})} = \mathrm{V}(\Gamma)^{-1} \int_{\Gamma} \mathrm{d}\bold{x}\, \mathrm{e}^{-\beta U_{\mathrm{el}}(\bold{x})}$ and thus the TL of \eqref{sfree3} exists. The second situation corresponds to slowly varying potentials. A widespread model is as follows. Consider a sequence of domains $\{\Lambda_{L}\}_{L \geq 1}$ obtained by an isotropic dilatation of a bounded subset $\Lambda_{1} \subset \mathbb{R}^{3}$, i.e. $\Lambda=\Lambda_{L} := \{\bold{x} : \bold{x}=L \bold{y},\, \bold{y} \in \Lambda_{1}\}$. On $\Lambda_{L}$, set $U_{\mathrm{el},L}(\bold{x}) := U_{\mathrm{el}}(L^{-1}\bold{x})$ with $U_{\mathrm{el}}$ initially defined on $\Lambda_{1}$. When $L$ is large enough, then the potential is slowly varying on $\Lambda_{L}$. Let $\{N_{L}\}_{L\geq 1}$ be any sequence of integers s.t. $N_{L} \mathrm{V}(\Lambda_{L})^{-1} = N_{1} \mathrm{V}(\Lambda_{1})^{-1} = \rho>0$. Consider now the sequence $\{\mathcal{F}(\beta,N_{L}, \mathrm{V}(\Lambda_{L}))\}_{L \geq 1}$ with $\mathcal{F}(\beta,N_{L}, \mathrm{V}(\Lambda_{L}))$ defined as in \eqref{sfree3}. Whenever $\mathrm{e}^{-\beta U_{\mathrm{el}}}$ is Lebesgue-integrable over $\Lambda_{1}$, then $\lim_{L \uparrow \infty} \mathrm{V}(\Lambda_{L})^{-1} \int_{\Lambda_{L}} \mathrm{d}\bold{x}\, \mathrm{e}^{-\beta U_{\mathrm{el},L}(\bold{x})} = \mathrm{V}(\Lambda_{1})^{-1} \int_{\Lambda_{1}} \mathrm{d}\bold{x}\, \mathrm{e}^{-\beta U_{\mathrm{el}}(\bold{x})}$ and the sequence $\{\mathcal{F}(\beta,N_{L}, \mathrm{V}(\Lambda_{L}))\}_{L \geq 1}$ converges. Therefore, the TL of \eqref{sfree3} exists. We mention that there exists another approach to treat weak-inhomogeneous systems. Suppose that the external potential is s.t. the confining container $\Lambda$ can be divide into subregions small enough to allow the potential energy to be 'almost' constant in them, and in the same time, large enough so that the subregions are macroscopic and statistically independent from each other. All the parts have to be in relative thermal equilibrium, i.e. having a unique temperature. This situation is usually obtained by considering \textit{macroscopic fields}. We will get back on it in the next paragraph. In view of the latter situation, one would be tempted to use the TL description on each subregion since they are 'almost' homogeneous, and then to derive a barometric formula giving the value of the TL for each macroscopic position. However, this method fails since the subregions are not individual closed systems. The grand-canonical ensemble lends itself better to this kind of description.

\begin{itemize}
\item[$(iii)$] \textit{$U_{\mathrm{int}},U_{\mathrm{el}} \neq 0$.}
\end{itemize}
This situation is considered in \cite{GP,GS}. Let us start with \cite{GP}. $U_{\mathrm{int}}$ is a pair-interaction potential of type $\sum_{i < j} \Phi(\bold{x}_{i} - \bold{x}_{j})$ where $\Phi = q + \gamma^{3} K(\gamma \cdot\,)$, with $\gamma$ a positive parameter. $q$ is a hard-core potential (then the electrons are assimilated to spherical hard-cores) of short-range type obeying: $q(\bold{x}) = q(-\bold{x})$ $\forall \bold{x} \in \mathbb{R}^{3}$, $q(\bold{x}) = \infty$ for $\vert \bold{x} \vert < r_{0}$ and $\vert q(\bold{x})\vert < C \vert \bold{x}\vert^{-3 -\epsilon}$ for $\vert \bold{x}\vert \geq r_{0}$. Here $C,r_{0},\epsilon$ are positive constants. $K$ obeys the assumptions of \cite[Thm. 1]{Dob}; its range is proportional to $\gamma^{-1}$. Note that $q$ and $K$ both satisfy the stability and temperedness conditions $\mathrm{(C1)}$-$\mathrm{(C2)}$. $U_{\mathrm{el}}$ is given by $\psi(\gamma \cdot\,)$ where $\psi$ is periodic over an infinite cubic lattice, uniformly bounded and Riemann-integrable over any bounded region of $\mathbb{R}^{3}$. Gates \textit{et al.} were interested in the van der Waals limit of the bulk free energy density consisting in taking the limit $\gamma \rightarrow 0$ in the TL. Although the above assumptions are chosen to address the van der Waals limit existence question, the authors mention that whenever $U_{\mathrm{el}}$ is periodic and bounded from below, the TL of the free energy density exists and the equivalence between the canonical and grand-canonical ensembles holds. They refer to \cite[Thm. 1]{Dob} and mention that the proof can be readapted to include such external potentials.\\
Subsequently, we turn to \cite{GS}. $U_{\mathrm{int}}$ is typically a many-body interactions potential satisfying the conditions of stability and temperedness $\mathrm{(C1)}$-$\mathrm{(C2)}$. The $U_{\mathrm{el}}$s considered are typically step potentials and uniform limits of sequences of steps potentials. Let us go further into details. A step potential $\phi_{r}$ of $r$ steps is defined as $\phi_{r}(\bold{x})= \sum_{i=1}^{r} a_{i} \chi_{\Lambda_{i}}(\bold{x})$ on a bounded subset $\Lambda \subset \mathbb{R}^{3}$ where $a_{i} \in \mathbb{R}$, $\{\Lambda_{i}\}_{i=1}^{r}$ is a partition of $\Lambda$ into $r$ disjoint subregions s.t. $\cup_{i=1}^{r} \Lambda_{i}=\Lambda$, and $\chi_{\Lambda_{i}}$ is the characteristic function of $\Lambda_{i}$. It is assumed that each $\Lambda_{i}$ has a connected interior, $\mathrm{V}(\Lambda_{i}) \mathrm{V}(\Lambda)^{-1} = \omega_{i}$ with $\omega_{i} \in \mathbb{R}$ and $\mathrm{V}(\Lambda_{i}) \rightarrow \infty$ in the Fisher-sense. Consider a sequence of such step potentials denoted by $\{(\Lambda_{\alpha}, \phi_{r,\alpha})\}_{\alpha}$, where each $\phi_{r,\alpha}$ is paired with one member of an expanding sequence of domains $\{\Lambda_{\alpha}\}_{\alpha}$. The sequence $\{(\Lambda_{\alpha}, \phi_{r,\alpha})\}_{\alpha}$ is obtained from an initial pair $(\Lambda_{0},\phi_{r,0})$ ($\phi_{r,0}$ is a step function defined on a bounded domain $\Lambda_{0}$) by means of an isotropic expansion, that is $\Lambda_{\alpha} :=\{\bold{x}: \bold{x}= \alpha^{\frac{1}{3}} \bold{y},\, \bold{y} \in \Lambda_{0}\}$ and $\phi_{r,\alpha}(\bold{x}) = \phi_{r,0}(\alpha^{-\frac{1}{3}} \bold{x})$. Then $\mathrm{V}(\Lambda_{\alpha}) = \alpha \mathrm{V}(\Lambda_{0})$ and $\mathrm{V}(\Lambda_{\alpha}) \rightarrow \infty$ in the Fisher-sense as $\alpha \rightarrow \infty$. Let $\{N(\Lambda_{\alpha})\}_{\alpha}$ be any sequence of positive integers s.t. $N(\Lambda_{\alpha}) \mathrm{V}(\Lambda_{\alpha})^{-1} \rightarrow \rho_{0} = N(\Lambda_{0}) \mathrm{V}(\Lambda_{0})^{-1}$ as $\mathrm{V}(\Lambda_{\alpha}) \rightarrow \infty$. Define the sequence of free energy density $\{\mathcal{F}(\beta,N(\Lambda_{\alpha}),\mathrm{V}(\Lambda_{\alpha}))\}_{\alpha}$ with $\mathcal{F}(\beta,N(\Lambda_{\alpha}),\mathrm{V}(\Lambda_{\alpha}))$ as in \eqref{sfree2} (but with $\phi_{r,\alpha}$ instead of $U_{\mathrm{el}}$ in \eqref{Zconf}). Here is the first result in
\cite[Thm. 2.1]{GS}. Provided $\mathrm{U}_{\mathrm{int}}$ satisfies in addition the asymptotically additive condition in \cite[Def. 2.1 $(ii)$]{GS}, then the sequence $\{\mathcal{F}(\beta,N(\Lambda_{\alpha}),\mathrm{V}(\Lambda_{\alpha}))\}_{\alpha}$ converges when $\alpha \rightarrow \infty$. The same result holds if one considers instead a sequence $\{(\Lambda_{\alpha}, \phi_{\alpha})\}_{\alpha}$ constructed as above, where $\phi_{\alpha}$ is a uniform limit of a sequence of step potentials. Moreover, under the same assumptions a general expression for the TL is derived in \cite[Coro. 5.1]{GS} from which the equivalence between the canonical and grand-canonical ensembles is proven.\\
Finally, we refer to \cite{MP} for the treatment of macroscopic fields in the grand-canonical ensemble.

\subsubsection{Questioning the validity for classical interacting systems.}

All the above discussions lean on the assumption that the involved interaction potential energies are independent of the particle velocities. This feature allowed us to get rid of the magnetic field via a change of variables in the classical translational partition function. In \cite{Ess2,EssF}, Ess\'en \textit{et al.} have recently revisited classical diamagnetism, and discussed the validity of the BvL theorem. In particular, they consider a classical charged particles system described by the so-called \textit{Darwin Lagrangian} in \cite[Eq. (13)]{Ess2} (instead of \eqref{secLan}) which takes into account the internal magnetic fields generated by the moving charged particles of the system itself. Although there is no close formula for the Darwin Hamiltonian, an expression for the second-order Hamiltonian can be found in \cite[Eq. (1)]{Ess}. Note that the internal magnetic potential vectors depend on the particle velocities. From formal arguments, Ess\'en \textit{et al.} show that the BvL theorem breaks down whenever one takes into account, through the Darwin magnetic interactions, the internal magnetic fields produced by the moving particles. In this condition, classical charged particles systems do exhibit diamagnetism.

\subsection{The motivations of the paper.}

In the light of Sec. \ref{fgfdc},  handling the presence of an external magnetic field acting on a classical charged particles system in thermal equilibrium does not cause any difficulty in the framework of classical statistical mechanics since the magnetic field simply disappears from the translational partition function after a change of variables. 
When dealing with large macroscopic systems, the proof of the BvL theorem therefore reduces to proving the existence of the thermodynamic limit for the zero-field free energy density. As pointed out, this requires suitable assumptions on the potential energies considered. An alternative proof of the BvL theorem would be to use the framework of quantum statistical mechanics.  Provided the thermodynamic limit of the magnetization exists, say for a confined electron gas obeying Fermi-Dirac statistics placed in a weak uniform magnetic field and in the canonical situation, one expects to recover the result of the BvL theorem (at least, within the linear-response theory) by taking the semiclassical limit $\hbar \rightarrow 0$ (here, $\hbar$ is seen as a parameter). Due to the presence of the magnetic field, proving the thermodynamic limit of the magnetization and magnetic susceptibility is mathematically a tricky problem to address. Indeed, singular terms appear in the thermodynamic limit arising from the linear growth of the magnetic vector potential. An interesting problem to tackle would consist in recovering the result of the BvL theorem in the semiclassical limit when considering an external electric potential obeying for instance assumptions of same type as those in \cite{GP}. This paper takes place in this direction.\\
\indent The system we consider is typically a 3-dimensional confined non-relativistic electron gas, obeying Fermi-Dirac statistics and placed in a non-zero external uniform magnetic field. The particles are subjected to external electric fields modeled by a class of translational invariant potentials. After thermal equilibrium is achieved, we assume that the gas is sufficiently diluted so that the many-body interactions can be neglected. The dynamics of a single-particle is described by the usual Pauli Hamiltonian. Theorem \ref{THMF} is our main result. Under suitable assumptions on the external potential, see assumptions $(\mathscr{A}_{\mathrm{p}})$-$(\mathscr{A}_{\mathrm{r}})$ pp. \pageref{deftheta}, we derive an asymptotic expansion for the bulk zero-field canonical magnetic susceptibility in the semiclassical limit. The asymptotic holds for any 'inverse' temperature $\beta>0$. The leading term is quadratic in the 'Planck's constant', independent of the external potential, and moreover, coincides with the leading term of the asymptotic expansion in the high-temperature regime. Since the bulk zero-field canonical magnetic susceptibility vanishes identically when performing the limit $\hbar \rightarrow 0$, the result of the BvL theorem is recovered within the linear-response theory. A series of remarks placed below Theorem \ref{THMF} discuss the assumptions, the results and the link with some well-known related results in literature.\\
\indent In Mathematical-Physics literature, a few works deal with the magnetic response of an electron gas in the semiclassical limit, see e.g. \cite{Fou1,Fou3,CR}. In \cite{CR}, Combescure \textit{et al.} investigated the semiclassical limit of the bulk grand-canonical orbital magnetization and susceptibility in weak magnetic field and in various regimes of temperature. To make the connection between our present work and theirs, we shall summarize their main results. The system considered is a $n$-dimensional non-relativistic electron gas obeying Fermi-Dirac statistics while neglecting the spin, and placed in a magnetic field. The particles interacts with a confining potential obeying the growth condition: $V(\bold{x}) \geq c_{0}(1 + \vert \bold{x}\vert^{2})^{\frac{s}{2}}$ for some $s,c_{0}>0$. The vector potential and electric potential are assumed to be smooth functions whose the derivatives satisfy suitable assumptions. When the temperature is large compared to $\hbar$, typically in the regimes $\beta \leq \hbar^{\epsilon - \frac{2}{3}}$ and $\hbar^{1-\epsilon}\leq T \leq \hbar^{\frac{2}{3}-\epsilon}$ for some $\epsilon>0$, they prove that the magnetization and orbital susceptibility in weak magnetic field admit a complete asymptotic expansion in powers of $\hbar$. By performing successively the limits $\beta \uparrow \infty$ and $\hbar \downarrow 0$ in the leading term of the asymptotic expansion derived in the regime $\beta \leq \hbar^{\epsilon - \frac{2}{3}}$, they recover the Landau orbital susceptibility formula for the 2-dimensional free electron gas. Turning to the regime of temperatures of the same order as $\hbar$ ('mesoscopic regime'), typically $\hbar \beta \in [\sigma_{0},\sigma_{1}]$ with $\sigma_{1}>\sigma_{0}>0$, they introduce a 'smeared out' magnetization and orbital susceptibility (different from the 'true' quantities). They prove that these two quantities in weak magnetic field can be split into two parts: 'an average part' having a complete asymptotic expansion in $\hbar$, and an 'oscillating part' in $\hbar$. The oscillating part is identified with the contribution of the periodic orbits of the classical motion, and it is a generalization of the de Haas-van Alphen oscillations.\\
\indent For completeness' sake, we mention that the rigorous study of orbital magnetism and more generally of diamagnetism, have been the subject of numerous works. We list below the main ones among them. The first rigorous proof of the Landau susceptibility formula for the free electron gas came as late as 1975, due to Angelescu \textit{et al.} in \cite{ABN}. Then in 1990, Helffer \textit{et al.} developed for the first time in \cite{HeSj} a rigorous theory based on the Peierls substitution and considered the connection with the diamagnetism of Bloch electrons and the de Haas-van Alphen effect.  These and many more results were reviewed in 1991 by G. Nenciu in \cite{Nen}. In 2012, Briet \textit{et al.} gave for the first time in \cite{BCS2} a rigorous justification of the Landau-Peierls approximation for the bulk zero-field orbital susceptibility of Bloch electron gases. Recently, the present author revisited the atomic orbital magnetism in \cite{Sa2} and gave a rigorous derivation of the diamagnetic Larmor contribution and the 'complete' orbital Van Vleck paramagnetic contribution in the tight-binding approximation.

\subsection{The setting and the main result.}
\label{mainre}

Consider a 3-dimensional quantum gas composed of a large number of non-relativistic identical and indistinguishable particles, with charge $q \neq 0$, mass $m=1$ and spin $\frac{1}{2}$, obeying Fermi-Dirac statistics. The Fermi gas is confined in a box, and the system is placed in an external non-zero constant magnetic field together with an external periodic potential. This latter may model the electric potential created by an ideal lattice of fixed ions in crystalline ordered solids. The effects arising from the spin-orbit coupling are disregarded. Furthermore, the interactions between particles are neglected (strongly diluted gas assumption) and the gas is at thermal equilibrium.\\
\indent Let us make our assumptions more precise. The gas is confined in a cubic box $\Lambda_{L}:= (-\frac{L}{2},\frac{L}{2})^{3}$ with $L>0$, centered at the origin of coordinates. We denote its Lebesgue-measure by $\vert \Lambda_{L}\vert$. We consider a uniform magnetic field $\mathbf{B} := B \mathbf{e}_{3}$, $\mathbf{e}_{3}:=(0,0,1)$ parallel to the third direction of the canonical basis of $\mathbb{R}^{3}$. We choose the symmetric gauge, i.e. the magnetic vector potential is defined  by $\mathbf{A}(\mathbf{x}) := \frac{1}{2} \mathbf{B} \times \mathbf{x} = B\mathbf{a}(\mathbf{x})$, $\mathbf{a}(\mathbf{x}) := \frac{1}{2}(-x_{2},x_{1},0)$ so that $\mathbf{B} = \nabla \times \bold{A}(\mathbf{x})$ and $\nabla \cdot \mathbf{A}(\mathbf{x}) = 0$. In the following, we denote by $b := \frac{q}{c}B \in \mathbb{R}$ the cyclotron frequency. The electric potential $V:\mathbb{R}^{3} \rightarrow \mathbb{R}$ (hereafter, we use the same notation to denote the electric potential energy) satisfies:
\begin{assumption1*}
$V$ is periodic with respect to the $\mathbb{Z}^{3}$-lattice.
\end{assumption1*}
\begin{assumption2*}
$V$ is globally H\"older-continuous with exponent $\theta \in (0,1]$, i.e. there exists a real $C>0$ s.t.
\begin{equation}
\label{deftheta}
\sup_{\substack{(\mathbf{x},\mathbf{y}) \in \mathbb{R}^{6} \\ \mathbf{x} \neq \mathbf{y}}} \frac{\left\vert V(\mathbf{x}) - V(\mathbf{y})\right\vert}{\vert \mathbf{x} - \mathbf{y}\vert^{\theta}} \leq C < \infty.
\end{equation}
\end{assumption2*}
The choice of the assumptions $(\mathscr{A}_{\mathrm{p}})$ and $(\mathscr{A}_{\mathrm{r}})$ are discussed below Theorem 1.1, see Remark \ref{discuss}.
Hereafter, we denote by $\Omega= (-\frac{1}{2},\frac{1}{2})^{3}$ the unit-cell centered at the origin of coordinates (which corresponds to the Wigner-Seitz cell of the $\mathbb{Z}^{3}$-lattice), and we denote by $\vert \Omega\vert$ its Lebesgue-measure.\\

\indent Let us first introduce the Hamiltonian determining the dynamics of a particle of spin $\frac{1}{2}$. Denote $\mathfrak{h}_{L}:= L^{2}(\Lambda_{L})$. The underlying Hilbert-space is the space of spinors $\mathfrak{h}_{L} \oplus \mathfrak{h}_{L} \equiv L^{2}(\Lambda_{L};\mathbb{C}^{2})$. In the absence of relativistic corrections, the Pauli operator takes into account the energy interaction $- \boldsymbol{\mu}_{S} \cdot \mathbf{B}$ (the so-called Stern-Gerlach term) between the spin magnetic moment $\boldsymbol{\mu}_{S} :=  g \kappa \mathbf{S}$ and the magnetic field $\mathbf{B}$. Here $\kappa := \frac{q}{2c}$ and $g$ is the Land\'e $g$-factor. Without loss of generality, we assume that $g \leq 2$ in the following (note that $g=2$ in the case of the electrons).
Then define on $\mathcal{C}_{0}^{\infty} (\Lambda_{L})\oplus \mathcal{C}_{0}^{\infty} (\Lambda_{L})$ the family of Pauli-like operators:
\begin{equation}
\label{HL}
\mathpzc{H}_{\mathpzc{h},L}(b) := \frac{1}{2} (- i \mathpzc{h} \nabla - b \mathbf{a})^{2} \mathpzc{I}_{\mathrm{d}} - \frac{g}{4} \mathpzc{h} b \sigma_{3} + V \mathpzc{I}_{\mathrm{d}} = \begin{pmatrix}
  H_{\mathpzc{h},L}^{-}(b) & 0 \\
  0 & H_{\mathpzc{h},L}^{+}(b)
 \end{pmatrix} =H_{\mathpzc{h},L}^{-}(b) \oplus H_{\mathpzc{h},L}^{+}(b),
\end{equation}
where $\sigma_{3} := \bigl(\begin{smallmatrix}1 & 0\\ 0& -1 \end{smallmatrix} \bigr)$ stands for the 'third' Pauli matrix, $\mathpzc{I}_{\mathrm{d}}$ the identity operator on $\mathfrak{h}_{L}\oplus \mathfrak{h}_{L}$, and:
\begin{equation}
\label{HL0}
H_{\mathpzc{h},L}^{\mp}(b) := \frac{1}{2} (-i \mathpzc{h} \mathpzc \nabla - b\mathbf{a})^{2} \mp \frac{g}{4} \mathpzc{h} b + V, \quad \mathpzc{h}>0,\, b  \in \mathbb{R}.
\end{equation}
For any $\mathpzc{h} >0$ and $b\in\mathbb{R}$, \eqref{HL} extends to a family of self-adjoint and semi-bounded operators for any $L \in (0,\infty)$, denoted again by $\mathpzc{H}_{\mathpzc{h},L}(b)$, with domain $\mathrm{D}(\mathpzc{H}_{\mathpzc{h},L}(b)) = D(H_{\mathpzc{h},L}^{-}(b)) \oplus D(H_{\mathpzc{h},L}^{+}(b))$. Under our conditions, $D(H_{\mathpzc{h},L}^{\mathpzc{s}}(b)) := \mathcal{H}_{0}^{1}(\Lambda_{L}) \cap \mathcal{H}^{2}(\Lambda_{L})$, $\mathpzc{s} \in \{-,+\}$. This definition corresponds to choose Dirichlet boundary conditions on $\partial\Lambda_{L}$. Moreover, since the inclusion $\mathcal{H}_{0}^{1}(\Lambda_{L}) \hookrightarrow L^{2}(\Lambda_{L})$ is compact, then $H_{\mathpzc{h},L}^{\mathpzc{s}}(b)$ has a purely discrete spectrum with an accumulation point at infinity. The same holds true for $\mathpzc{H}_{\mathpzc{h},L}(b)$ since $\sigma(\mathpzc{H}_{\mathpzc{h},L}(b)) = \sigma(H_{\mathpzc{h},L}^{-}(b)) \cup \sigma(H_{\mathpzc{h},L}^{+}(b))$. Hereafter, we denote  by $\{\lambda_{\mathpzc{h},L}^{\mathpzc{s}, (j)}(b)\}_{j \geq 1}$, the set of eigenvalues of $H_{\mathpzc{h},L}^{\mathpzc{s}}(b)$  counting multiplicities and in increasing order.\\
\indent When $\Lambda_{L}$ fills the whole space, define on $\mathcal{C}_{0}^{\infty}(\mathbb{R}^{3})\oplus \mathcal{C}_{0}^{\infty}(\mathbb{R}^{3})$ the Pauli-like operator:
\begin{equation}
\label{Hinfini}
\mathpzc{H}_{\mathpzc{h}}(b) := \begin{pmatrix}
  H_{\mathpzc{h}}^{-}(b) & 0 \\
  0 & H_{\mathpzc{h}}^{+}(b)
 \end{pmatrix} = H_{\mathpzc{h}}^{-}(b) \oplus H_{\mathpzc{h}}^{+}(b), \quad
\end{equation}
where:
\begin{equation}
\label{Hinfini1}
H_{\mathpzc{h}}^{\mp}(b) := \frac{1}{2} (-i \mathpzc{h} \mathpzc \nabla - b\mathbf{a})^{2} \mp \frac{g}{4} \mathpzc{h} b + V, \quad \mathpzc{h}>0,\, b  \in \mathbb{R}.
\end{equation}
By \cite[Thm. B.13.4]{Si1}, $\forall \mathpzc{h} >0$ and $\forall b \in \mathbb R$, \eqref{Hinfini} is essentially self-adjoint and its self-adjoint extension, denoted again by $\mathpzc{H}_{\mathpzc{h}}(b)$, is bounded from below. Due to the assumption $(\mathscr{A}_{\mathrm{p}})$, it only has essential spectrum. Moreover, the following inequality holds $\forall L>0$, $\forall b \in \mathbb{R}$ and $\forall \mathpzc{h}>0$:
\begin{equation}
\label{infspectrum}
-\Vert V \Vert_{\infty} \leq \mathpzc{h} \frac{\vert b \vert}{2}\left(1 - \frac{g}{2}\right) - \Vert V \Vert_{\infty} \leq \mathpzc{E}_{\mathpzc{h}}(b) \leq \inf \sigma\left(\mathpzc{H}_{\mathpzc{h},L}(b)\right),\quad \mathpzc{E}_{\mathpzc{h}}(b) := \inf \sigma\left(\mathpzc{H}_{\mathpzc{h}}(b)\right),
\end{equation}
$\Vert V \Vert_{\infty} := \sup_{\mathbf{x} \in (-\frac{1}{2},\frac{1}{2}]} \vert V(\mathbf{x})\vert$. The lower bound for $\mathpzc{E}_{\mathpzc{h}}(b)$ stems from the variational principle, the upper bound from the min-max principle, see \cite[Sec. XIII.1]{RS4}. Remind that we assumed $g \leq 2$.\\
\indent Next, we introduce the Hamiltonian determining the dynamics of $N$ confined Fermions of spin $\frac{1}{2}$ with identical spin projection $m_{\frac{1}{2}} = \mathpzc{s}\frac{1}{2}$, $\mathpzc{s}\in\{-,+\}$. Denoting by $\bigwedge$ the antisymmetric tensor product, let $\mathfrak{h}_{L}^{N,\textswab{a}} := \bigwedge_{1}^{N} \mathfrak{h}_{L}$ be the antisymmetric $N$-particles Hilbert space. It is a subspace of $\mathfrak{h}_{L}^{N} := \bigotimes_{1}^{N} \mathfrak{h}_{L} \cong L^{2}(\Lambda_{L}^{N})$, and consists of those functions odd under interchange of coordinates.
In view of \eqref{HL0}, for any $L>0$ and any integer $N \geq 2$, define on $\bigwedge_{1}^{N} \mathrm{D}(H_{\mathpzc{h},L}^{\mathpzc{s}}(b)) \subset \mathfrak{h}_{L}^{N, \textswab{a}}$ the family of operators:
\begin{equation}
\label{HamiN}
H_{\mathpzc{h},L}^{\mathpzc{s}\,(N)}(b) := H_{\mathpzc{h},L}^{\mathpzc{s}}(b) \otimes I_{\mathrm{d}} \otimes \dotsb \otimes I_{\mathrm{d}} + \dotsb + I_{\mathrm{d}} \otimes \dotsb \otimes I_{\mathrm{d}} \otimes H_{\mathpzc{h},L}^{\mathpzc{s}}(b),\quad \mathpzc{h}>0,\, b \in \mathbb{R},
\end{equation}
where the r.h.s. of \eqref{HamiN} consists of '$N$ terms' and $I_{\mathrm{d}}$ denotes the identity operator on $\mathfrak{h}_{L}$. By \cite[Thm. VIII.33]{RS1}, $\forall \mathpzc{h}>0$ and $\forall b \in \mathbb{R}$, \eqref{HamiN}  extends to a family of self-adjoint and semi-bounded operators $\forall L\in (0,\infty)$, denoted again by $H_{\mathpzc{h},L}^{\mathpzc{s}\,(N)}(b)$. We denote  by $\mathrm{D}(H_{\mathpzc{h},L}^{\mathpzc{s}\,(N)}(b))$ its domain. \\
\indent Then, we turn to the second quantization of $H_{\mathpzc{h},L}^{\mathpzc{s}}(b)$. Let $\mathfrak{F}_{L}(\mathfrak{h}_{L}) := \mathbb{C} \oplus \bigoplus_{N \geq 1} \mathfrak{h}_{L}^{N}$ be the Fermi Fock space. Here, $\oplus$ denotes the direct sum in the sense of Hilbert spaces, i.e. the completion of the algebraic direct sum. Let $\mathfrak{F}_{L}^{\textswab{a}}(\mathfrak{h}_{L}) := \mathbb{C} \oplus \bigoplus_{N \geq 1} \mathfrak{h}_{L}^{N,\textswab{a}}$ be the antisymmetric Fermi Fock space.  From \eqref{HamiN} and by setting $H_{\mathpzc{h},L}^{\mathpzc{s}\,(1)}(b):= H_{\mathpzc{h},L}^{\mathpzc{s}}(b)$, introduce in $\mathfrak{F}_{L}^{\textswab{a}}(\mathfrak{h}_{L})$ the family of operators:
\begin{equation}
\label{secqu}
\mathrm{d}\Gamma\left(H_{\mathpzc{h},L}^{\mathpzc{s}}(b)\right) := 0 \oplus \bigoplus_{N \geq 1} H_{\mathpzc{h},L}^{\mathpzc{s}\,(N)}(b),\quad \mathpzc{h}>0,\,b\in\mathbb{R}.
\end{equation}
From \cite[pp. 302]{RS1}, $\forall \mathpzc{h}>0$, $\forall b \in \mathbb{R}$ and $\forall L \in (0,\infty)$, \eqref{secqu} is essentially self-adjoint on the domain $\{ \Psi = (\Psi^{(0)},\Psi^{(1)},\ldots) \in \mathfrak{F}_{L}^{\textswab{a}}(\mathfrak{h}_{L})\,:\, \Psi^{(n)}=0\,\textrm{for $n$ large enough},\, \textrm{and}\, \Psi^{(n)} \in  \mathrm{D}(H_{\mathpzc{h},L}^{\mathpzc{s}\,(n)}(b))\,\textrm{for each $n$}\}$.
Its self-adjoint closure, denoted again by $\mathrm{d}\Gamma(H_{\mathpzc{h},L}^{\mathpzc{s}}(b))$, is the second quantization of $H_{\mathpzc{h},L}^{\mathpzc{s}}(b)$.
Also, define in $\mathfrak{F}_{L}^{\textswab{a}}(\mathfrak{h}_{L})$ the number operator as the second quantization of the identity $I_{\mathrm{d}}$ on $\mathfrak{h}_{L}$:
\begin{equation}
\label{nbrepar}
\mathrm{d}\Gamma\left( I_{\mathrm{d}}\right) := 0 \oplus \bigoplus_{N \geq 1} N,
\end{equation}
acting in any $\mathfrak{h}_{L}^{N,\textswab{a}}$ as the multiplication with $N$. Its self-adjoint closure is denoted again by $\mathrm{d}\Gamma( I_{\mathrm{d}})$.
On the same way, introduce also the second quantization of $\mathpzc{H}_{\mathpzc{h},L}(b)= H_{\mathpzc{h},L}^{-}(b)\oplus H_{\mathpzc{h},L}^{+}(b)$ and $\mathpzc{I}_{\mathrm{d}}=I_{\mathrm{d}} \oplus I_{\mathrm{d}}$ in $\mathfrak{F}_{L}^{\textswab{a}}(\mathfrak{h}_{L}\oplus \mathfrak{h}_{L})$ denoted respectively by $\mathrm{d}\Gamma(\mathpzc{H}_{\mathpzc{h},L}(b))$ and $\mathrm{d}\Gamma(\mathpzc{I}_{d})$.\\

\indent Let us now define some quantities related to the confined Fermi gas of spin $\frac{1}{2}$ introduced above within the framework of quantum statistical mechanics. In the grand-canonical ensemble, let $(\beta,z,\vert \Lambda_{L}\vert)$ be the fixed external parameters. Here $\beta := (k_{B}T)^{-1} > 0$ is the 'inverse' temperature ($k_{B}$ is the Boltzmann constant) and $z:= \mathrm{e}^{\beta \mu} >0$ ($\mu \in \mathbb{R}$ is the chemical potential) is the fugacity. When dealing with the canonical ensemble, $(\beta,\rho,\vert \Lambda_{L}\vert)$ are the fixed external parameters, with $\rho>0$ the density of particles. The number of particles is related to the density by $N_{L} = \rho \vert \Lambda_{L}\vert$.
For any $L>0$, $\beta>0$, $z>0$, $b \in \mathbb{R}$ and $\mathpzc{h}>0$, the grand-canonical partition function reads as, see e.g. \cite{BR2}:
\begin{equation}
\label{parttl}
\Xi_{\mathpzc{h},L}(\beta,z,b) := \mathrm{Tr}_{\mathfrak{F}_{L}^{\textswab{a}}(\mathfrak{h}_{L}\oplus\mathfrak{h}_{L})}\left\{\mathrm{e}^{-\beta \left[\mathrm{d} \Gamma\left(\mathpzc{H}_{\mathpzc{h},L}(b)\right) - \mu \mathrm{d}\Gamma
\left(\mathpzc{I}_{\mathrm{d}}\right)\right]}\right\}.
\end{equation}
From \eqref{secqu} and \eqref{nbrepar}, introduce under the conditions of \eqref{parttl}:
\begin{equation}
\label{GCparf}
\Xi_{\mathpzc{h},L}^{\mathpzc{s}}(\beta,z,b) := \mathrm{Tr}_{\mathfrak{F}_{L}^{\textswab{a}}(\mathfrak{h}_{L})}\left\{\mathrm{e}^{-\beta \left[\mathrm{d} \Gamma\left(H_{\mathpzc{h},L}^{\mathpzc{s}}(b)\right) - \mu \mathrm{d}\Gamma
\left(I_{\mathrm{d}}\right)\right]}\right\},\quad \mathpzc{s} \in \{-,+\}.
\end{equation}
By the exponential law for Fock spaces, see e.g. \cite[Sec. 5.6]{Der}, \eqref{parttl} can be rewritten as:
\begin{equation}
\label{partifunc}
\Xi_{\mathpzc{h},L}(\beta,z,b)  = \prod_{\mathpzc{s} \in \{-,+\}} \Xi_{\mathpzc{h},L}^{\mathpzc{s}}(\beta,z,b) = \prod_{\mathpzc{s} \in \{-,+\}} \prod_{j=1}^{\infty} \left( 1 + z \mathrm{e}^{-\beta \lambda_{\mathpzc{h},L}^{\mathpzc{s},(j)}(b)}\right).
\end{equation}
The product over the index $\mathpzc{s}$ in \eqref{partifunc} reflects the fact that, in the grand-canonical formalism, the Fermi gas of spin $\frac{1}{2}$ is treated as two independent subsystems of Fermions with identical spin projection $m_{\frac{1}{2}}$, but with opposite sign: $m_{\frac{1}{2}}=-\frac{1}{2}$ and $m_{\frac{1}{2}}=\frac{1}{2}$. Ergo, the grand-canonical quantities associated to the Fermi gas are obtained by superposing the contributions coming from each one of the subsystems treated separately. From now on, we define the grand-canonical quantities only for the Fermi gas of spin $\frac{1}{2}$ with identical spin projection $m_{\frac{1}{2}} = \mathpzc{s}\frac{1}{2}$, $\mathpzc{s}\in\{-,+\}$.\\
For any $L>0$, $\beta>0$, $z>0$, $b \in \mathbb{R}$ and $\mathpzc{h}>0$, the finite-volume grand-canonical pressure and density of particles are respectively defined by, see e.g. \cite{BR2,ABN}:
\begin{gather}
\label{pressurefv}
P_{\mathpzc{h},L}^{\mathpzc{s}}(\beta,z,b) :=  \frac{1}{\beta \vert \Lambda_{L}\vert} \ln\left(\Xi_{\mathpzc{h},L}^{\mathpzc{s}}(\beta,z,b)\right) = \frac{1}{\beta \vert \Lambda_{L}\vert}  \sum_{j=1}^{\infty} \ln\left(1 + z \mathrm{e}^{-\beta \lambda_{\mathpzc{h},L}^{\mathpzc{s},(j)}(b)}\right),\\
\label{densityfv}
\rho_{\mathpzc{h},L}^{\mathpzc{s}}(\beta,z,b) := \beta z \frac{\partial P_{\mathpzc{h},L}^{\mathpzc{s}}}{\partial z}(\beta,z,b) = \frac{1}{\vert \Lambda_{L}\vert}
\sum_{j=1}^{\infty} \frac{z \mathrm{e}^{-\beta \lambda_{\mathpzc{h},L}^{\mathpzc{s},(j)}(b)}}{1 + z\mathrm{e}^{-\beta \lambda_{\mathpzc{h},L}^{\mathpzc{s},(j)}(b)}}.
\end{gather}
Since the semigroups $\{\mathrm{e}^{-\beta H_{\mathpzc{h},L}^{\mathpzc{s}}(b)},\, \beta>0\}$, $\mathpzc{s} \in \{-,+\}$ are trace-class $\forall L>0$, $\forall \mathpzc{h}>0$ and $\forall b \in \mathbb{R}$, see
\cite[Eq. (2.12)]{BCS1}, then \eqref{partifunc} is well-defined by \cite[Prop. 5.2.22]{BR2}. Therefore, the series in \eqref{pressurefv} and \eqref{densityfv} are absolutely convergent. Under the same conditions, $\mu \mapsto P_{\mathpzc{h},L}^{\mathpzc{s}}(\beta,\mathrm{e}^{\beta \mu},b)$ is a convex function on $\mathbb{R}$ and $b \mapsto P_{\mathpzc{h},L}^{\mathpzc{s}}(\beta,z,b)$ is an even function on $\mathbb{R}$. Moreover, from \cite[Thm. 1.1 (i)]{BCS1}, $\forall L>0$, $\forall \beta>0$, $\forall b \in \mathbb{R}$ and $\forall \mathpzc{h}>0$, $P_{\mathpzc{h},L}^{\mathpzc{s}}(\beta,\cdot\,,b)$ admits an analytic extension to the complex domain $\mathcal{D}(E_{\mathpzc{h}}^{\mathpzc{s}}(b)):= \mathbb{C} \setminus (-\infty, - \mathrm{e}^{\beta E_{\mathpzc{h}}^{\mathpzc{s}}(b)}]$ with $E_{\mathpzc{h}}^{\mathpzc{s}}(b) :=  \inf \sigma(H_{\mathpzc{h}}^{\mathpzc{s}}(b))$. We denote by $\widehat{P}_{\mathpzc{h},L}^{\mathpzc{s}}(\beta,\cdot\,,b)$ the analytic continuation of $P_{\mathpzc{h},L}^{\mathpzc{s}}(\beta,\cdot\,,b)$ to $\mathcal{D}(E_{\mathpzc{h}}^{\mathpzc{s}}(b))$. Furthermore, $\forall L>0$, $\forall \beta>0$ and $\forall \mathpzc{h}>0$,  $P_{\mathpzc{h},L}^{\mathpzc{s}}(\beta,\cdot\,,\cdot\,)$ is jointly real analytic in $(z,b) \in (0,\infty) \times \mathbb{R}$. This allows us to define the finite-volume grand-canonical magnetization and magnetic susceptibility respectively as the first and second derivative of the pressure w.r.t. the intensity $B$ of the magnetic field, see e.g. \cite[Prop. 2]{ABN}:
\begin{gather*}
\mathcal{M}_{\mathpzc{h},L}^{\mathpzc{s}\,\mathrm{(GC)}}(\beta,z,b) := \left(\frac{q}{c}\right) \frac{\partial P_{\mathpzc{h},L}^{\mathpzc{s}}}{\partial b}(\beta,z,b),\\
\mathcal{X}_{\mathpzc{h},L}^{\mathpzc{s}\,\mathrm{(GC)}}(\beta,z,b) := \left(\frac{q}{c}\right)^{2} \frac{\partial^{2} P_{\mathpzc{h},L}^{\mathpzc{s}}}{\partial b^{2}}(\beta,z,b).
\end{gather*}
Note that, for each one of the quantities defined above, the contributions corresponding to $\mathpzc{s}=+$ and $\mathpzc{s}=-$ are identical when the magnetic field vanishes, see \eqref{HL0}.\\
\indent When $\Lambda_{L}$ fills the whole space (i.e. in the limit $L \uparrow \infty$), the thermodynamic limits of the four grand-canonical quantities defined above generically exist, see e.g. \cite[Thms. 1.1 \& 1.2]{BS} and \cite[Sec. 3.1]{BCS2}. Denoting $\forall \beta>0$, $\forall z >0$, $\forall b \in \mathbb{R}$ and $\forall \mathpzc{h} >0$ the bulk grand-canonical pressure by:
\begin{equation}
\label{bulkpres}
P_{\mathpzc{h}}^{\mathpzc{s}}(\beta,z,b) :=  \lim_{L \uparrow \infty} P_{\mathpzc{h},L}^{\mathpzc{s}}(\beta,z,b),
\end{equation}
then under the same conditions, we have the following point-wise convergences:
\begin{gather}
\label{densitylim}
\rho_{\mathpzc{h}}^{\mathpzc{s}}(\beta,z,b) := \beta z \frac{\partial P_{\mathpzc{h}}^{\mathpzc{s}}}{\partial z}(\beta,z,b) = \lim_{L \uparrow \infty} \beta z  \frac{\partial P_{\mathpzc{h},L}^{\mathpzc{s}}}{\partial z}(\beta,z,b),\\
\label{magzlim}
\mathcal{M}_{\mathpzc{h}}^{\mathpzc{s}\,\mathrm{(GC)}}(\beta,z,b) := \left(\frac{q}{c}\right) \frac{\partial P_{\mathpzc{h}}^{\mathpzc{s}}}{\partial b}(\beta,z,b) = \lim_{L \uparrow \infty} \left(\frac{q}{c}\right) \frac{\partial P_{\mathpzc{h},L}^{\mathpzc{s}}}{\partial b}(\beta,z,b),\\
\label{susceptilim}
\mathcal{X}_{\mathpzc{h}}^{\mathpzc{s}\,\mathrm{(GC)}}(\beta,z,b) := \left(\frac{q}{c}\right)^{2} \frac{\partial^{2} P_{\mathpzc{h}}^{\mathpzc{s}}}{\partial b^{2}}(\beta,z,b) = \lim_{L \uparrow \infty} \left(\frac{q}{c}\right)^{2} \frac{\partial^{2} P_{\mathpzc{h},L}^{\mathpzc{s}}}{\partial b^{2}}(\beta,z,b),
\end{gather}
and the convergences are compact w.r.t. $(\beta,z,b) \in (0,\infty)\times(0,\infty)\times \mathbb{R}$. Hence, the limit $L \uparrow \infty$ commutes with the first derivative (resp. the first two derivatives) of the pressure w.r.t. the fugacity $z$ (resp. w.r.t. the cyclotron frequency $b$). Note that under the conditions of \eqref{bulkpres}, $\mu \mapsto P_{\mathpzc{h}}^{\mathpzc{s}}(\beta,\mathrm{e}^{\beta \mu},b)$ is convex on $\mathbb{R}$ as point-wise limit of a sequence of convex functions and $b \mapsto P_{\mathpzc{h}}^{\mathpzc{s}}(\beta,z,b)$ is even on $\mathbb{R}$ as point-wise limit of a sequence of even functions. Moreover, from \cite[Thm. 1.1 (ii)]{BCS1}, $\forall \beta>0$, $\forall b\in \mathbb{R}$ and $\forall \mathpzc{h} >0$, $P_{\mathpzc{h}}^{\mathpzc{s}}(\beta,\cdot\,,b)$ admits an analytic extension to the domain $\mathcal{D}(E_{\mathpzc{h}}^{\mathpzc{s}}(b))$. Further, $\forall \beta >0$ and $\forall \mathpzc{h} >0$, $P_{\mathpzc{h}}^{\mathpzc{s}}(\beta,\cdot\,,\cdot\,)$ is jointly smooth in $(z,b) \in (0,\infty) \times \mathbb{R}$, see \cite{Sa}.\\

\indent Getting back to the Fermi gas of spin $\frac{1}{2}$, the bulk grand-canonical quantities are obtained by adding the two contributions corresponding to $\mathpzc{s}=-$ and $\mathpzc{s}=+$. Hereafter, we denote by $P_{\mathpzc{h}}$, $\rho_{\mathpzc{h}}$, $\mathcal{M}_{\mathpzc{h}}^{\mathrm{(GC)}}$ and $\mathcal{X}_{\mathpzc{h}}^{\mathrm{(GC)}}$  respectively the bulk grand-canonical pressure, density, magnetization and magnetic susceptibility of the Fermi gas.\\
\indent Next, we switch to the canonical conditions (still within the grand-canonical ensemble) and we assume that the density of particles of the Fermi gas of spin $\frac{1}{2}$ $\rho>0$ is a fixed external parameter. \\
Denote $\forall \beta>0$, $\forall \rho^{\mathpzc{s}}>0$, $\forall b \in \mathbb{R}$ and $\forall\mathpzc{h}>0$ by $\overline{z}_{\mathpzc{h}}^{\mathpzc{s}}(\beta,\rho^{\mathpzc{s}},b) \in (0,\infty)$ the unique solution of:
\begin{equation}
\label{knock}
\rho_{\mathpzc{h}}^{\mathpzc{s}}(\beta,z,b) = \rho^{\mathpzc{s}} >0,\quad \mathpzc{s} \in \{-,+\}.
\end{equation}
The inversion of the relation between the bulk density and the fugacity $z$ relies on the fact that $\forall \beta>0$, $\forall b \in \mathbb{R}$ and $\forall \mathpzc{h}>0$, $z \mapsto \rho_{\mathpzc{h}}^{\mathpzc{s}}(\beta,z,b)$ is a strictly increasing function on $(0,\infty)$ and actually defines a $\mathcal{C}^{\infty}$-diffeomorphism of this interval into itself, see \cite{Sa,BCS2}. Since $\rho_{\mathpzc{h}}^{-}(\beta,z,0)=\rho_{\mathpzc{h}}^{+}(\beta,z,0)$ when the magnetic field vanishes, then $\overline{z}_{\mathpzc{h}}^{-}(\beta,\rho^{\mathpzc{s}},0) =\overline{z}_{\mathpzc{h}}^{+}(\beta,\rho^{\mathpzc{s}},0)$. By setting $\rho = \rho^{-} + \rho^{+}>0$ and $\rho^{-}=\rho^{+}$, we hereafter denote $\forall \beta>0$ and $\forall\mathpzc{h}>0$ by $\overline{z}_{\mathpzc{h}}(\beta,\frac{\rho}{2},0)>0$ the unique solution of:
\begin{equation}
\label{invdensi}
\rho = \rho_{\mathpzc{h}}(\beta,z,0),\quad  \rho_{\mathpzc{h}}(\beta,z,0) := \rho_{\mathpzc{h}}^{-}(\beta,z,0) + \rho_{\mathpzc{h}}^{+}(\beta,z,0)=2 \rho_{\mathpzc{h}}^{\mathpzc{s}}(\beta,z,0).
\end{equation}
From \eqref{magzlim} and \eqref{susceptilim}, the bulk grand-canonical magnetization and magnetic susceptibility of the Fermi gas of spin $\frac{1}{2}$ at fixed density $\rho>0$ are respectively defined $\forall \beta>0$, $\forall b \in \mathbb{R}$ and $\forall \mathpzc{h}>0$ as:
\begin{gather}
\label{magnfi2}
\mathcal{M}_{\mathpzc{h}}^{\mathrm{(GC)}}\left(\beta,\rho,b\right) :=  \mathcal{M}_{\mathpzc{h}}^{-\,\mathrm{(GC)}}\left(\beta,\rho^{-},b\right) + \mathcal{M}_{\mathpzc{h}}^{+\,\mathrm{(GC)}}\left(\beta,\rho^{+},b\right),\\
\label{susfix2}
\mathcal{X}_{\mathpzc{h}}^{\mathrm{(GC)}}\left(\beta,\rho,b\right) :=  \mathcal{X}_{\mathpzc{h}}^{-\,\mathrm{(GC)}}\left(\beta,\rho^{-},b\right) + \mathcal{X}_{\mathpzc{h}}^{+\,\mathrm{(GC)}}\left(\beta,\rho^{+},b\right),
\end{gather}
where we set $\rho = \rho^{-}+\rho^{+}$, $\rho^{\mathpzc{s}}>0$ and with:
\begin{gather}
\label{magnfi23}
\mathcal{M}_{\mathpzc{h}}^{\mathpzc{s}\,\mathrm{(GC)}}\left(\beta,\rho^{\mathpzc{s}},b\right) :=  \mathcal{M}_{\mathpzc{h}}^{\mathpzc{s}\,\mathrm{(GC)}}\left(\beta, \overline{z}_{\mathpzc{h}}^{\mathpzc{s}}(\beta,\rho^{\mathpzc{s}},b) ,b\right),\quad \mathpzc{s} \in \{-,+\},\\
\label{susfix23}
\mathcal{X}_{\mathpzc{h}}^{\mathpzc{s}\,\mathrm{(GC)}}\left(\beta,\rho^{\mathpzc{s}},b\right) :=  \mathcal{X}_{\mathpzc{h}}^{\mathpzc{s} \,\mathrm{(GC)}}\left(\beta,\overline{z}_{\mathpzc{h}}^{\mathpzc{s}}(\beta,\rho^{\mathpzc{s}},b) ,b\right).
\end{gather}
Since $b \mapsto P_{\mathpzc{h}}^{\mathpzc{s}}(\beta,z,b)$ is an even function on $\mathbb{R}$, then when the magnetic field vanishes:
\begin{equation}
\label{magzdens0}
\mathcal{M}_{\mathpzc{h}}^{\mathrm{(GC)}}\left(\beta,\rho,0\right) = 0;
\end{equation}
and when $\rho^{-}=\rho^{+}$ in addition, \eqref{susfix2} becomes:
\begin{equation}
\label{susfixdens}
\mathcal{X}_{\mathpzc{h}}^{\mathrm{(GC)}}\left(\beta,\rho,0\right) =  \mathcal{X}_{\mathpzc{h}}^{\mathrm{(GC)}}\left(\beta,\overline{z}_{\mathpzc{h}}\left(\beta,\frac{\rho}{2},0\right) ,0\right) = 2 \mathcal{X}_{\mathpzc{h}}^{\mathpzc{s}\,\mathrm{(GC)}}\left(\beta,\overline{z}_{\mathpzc{h}}\left(\beta, \frac{\rho}{2},0\right) ,0\right).
\end{equation}

The above procedure is the usual one which allows one to mimic the canonical conditions within the grand-canonical ensemble, see e.g. \cite{ABN}. We emphasize that, in thermodynamic limit the grand-canonical quantities at fixed density $\rho$ may not coincide with the corresponding quantities defined in the canonical ensemble for the corresponding density $\rho$.\\

Before stating our main result involving \eqref{susfixdens}, we go back to the 'true' canonical ensemble. For any $L>0$, $\beta>0$, $b \in \mathbb{R}$ and $\mathpzc{h}>0$ the canonical partition function of the confined Fermi gas with identical spin projection $m_{\frac{1}{2}}= \mathpzc{s}\frac{1}{2}$ at fixed density $\rho^{\mathpzc{s}}>0$ is defined as, see e.g. \cite{Ru}:
\begin{equation}
\label{ZZZs}
\mathcal{Z}_{\mathpzc{h},L}^{\mathpzc{s}}(\beta,\rho^{\mathpzc{s}},b) := \mathrm{Tr}_{\mathfrak{h}_{L}^{N_{L}^{\mathpzc{s}},\textswab{a}}}\left\{\mathrm{e}^{-\beta H_{\mathpzc{h},L}^{\mathpzc{s}\,(N_{L}^{\mathpzc{s}})}(b)}\right\},\quad N_{L}^{\mathpzc{s}}:=\rho^{\mathpzc{s}} \vert \Lambda_{L}\vert,\, \mathpzc{s} \in \{-,+\}.
\end{equation}
It is related to the grand-canonical partition function in \eqref{GCparf} via, see e.g. \cite{Hu}:
\begin{equation*}
\Xi_{\mathpzc{h},L}^{\mathpzc{s}}(\beta,z,b) = \sum_{n=0}^{\infty} z^{n} \mathrm{Tr}_{\mathfrak{h}_{L}^{n,\textswab{a}}}\left\{\mathrm{e}^{-\beta H_{\mathpzc{h},L}^{\mathpzc{s}\,(n)}(b)}\right\}.
\end{equation*}
Furthermore, under the conditions of \eqref{ZZZs} the canonical partition function can be rewritten from the finite-volume grand-canonical pressure in \eqref{pressurefv} as follows, see e.g. \cite[Eq. (2.27)]{C1}:
\begin{equation}
\label{goore}
\mathcal{Z}_{\mathpzc{h},L}^{\mathpzc{s}}(\beta,\rho^{\mathpzc{s}},b) = \frac{1}{2i \pi} \int_{\mathscr{C}} \mathrm{d}\zeta \, \frac{1}{\zeta} \left( \frac{\exp\left(\frac{\beta}{\rho^{\mathpzc{s}}} \widehat{P}_{\mathpzc{h},L}^{\mathpzc{s}}(\beta,\zeta,b)\right)}{\zeta}\right)^{N_{L}^{\mathpzc{s}}},
\end{equation}
where $\mathscr{C}$ is any positively oriented simple closed contour surrounding the origin and avoiding the cut $(-\infty, -\mathrm{e}^{\beta E_{\mathpzc{h}}^{\mathpzc{s}}(b)}]$, and $\widehat{P}_{\mathpzc{h},L}^{\mathpzc{s}}(\beta,\cdot\,,b)$ is the analytic continuation of  $P_{\mathpzc{h},L}^{\mathpzc{s}}(\beta,\cdot\,,b)$ to $\mathcal{D}(E_{\mathpzc{h}}^{\mathpzc{s}}(b))$. 
Note that the canonical partition function associated to the Fermi gas of spin $\frac{1}{2}$ reads as, see e.g. \cite{Ru,Hu}:
\begin{equation}
\label{partifuncC}
\mathcal{Z}_{\mathpzc{h},L}(\beta,\rho,b) := \mathrm{Tr}_{\bigwedge_{1}^{N_{L}}(\mathfrak{h}_{L} \oplus \mathfrak{h}_{L})}\left\{\mathrm{e}^{-\beta \mathrm{d}\Gamma(\mathpzc{H}_{\mathpzc{h},L}(b))}\right\},\quad N_{L} := \rho \vert \Lambda_{L}\vert,
\end{equation}
where $\bigwedge_{1}^{N_{L}}(\mathfrak{h}_{L} \oplus \mathfrak{h}_{L})$ is the $N_{L}$-fold  antisymmetric tensor product of $\mathfrak{h}_{L} \oplus \mathfrak{h}_{L}$. Since $\bigwedge_{1}^{N_{L}} (\mathfrak{h}_{L}\oplus \mathfrak{h}_{L}) \cong \oplus_{N_{L}^{+}=0}^{N_{L}} \mathfrak{h}_{L}^{N_{L}^{+},\textswab{a}} \otimes \mathfrak{h}_{L}^{N_{L} - N_{L}^{+},\textswab{a}}$ with $N_{L}^{+} + N_{L}^{-} = N_{L}$, then \eqref{partifuncC} can be rewritten in terms of \eqref{ZZZs}.\\
For any $L>0$, $\beta>0$, $b \in \mathbb{R}$ and $\mathpzc{h}>0$ the free energy density of the confined Fermi gas with identical spin projection $m_{\frac{1}{2}}=\mathpzc{s}\frac{1}{2}$ at fixed density $\rho^{\mathpzc{s}}>0$ is defined as, see e.g. \cite{Ru,Hu}:
\begin{equation}
\label{Helm}
\mathcal{F}_{\mathpzc{h},L}^{\mathpzc{s}}(\beta,\rho^{\mathpzc{s}},b) := -\frac{1}{\beta \vert \Lambda_{L} \vert} \ln\left( \mathcal{Z}_{\mathpzc{h},L}^{\mathpzc{s}}(\beta,\rho^{\mathpzc{s}},b)\right),\quad \mathpzc{s} \in \{-,+\}.
\end{equation}
From \cite[Coro. 3.9]{BCS1}, $\forall L>0$, $\forall \beta >0$, $\forall \rho^{\mathpzc{s}}>0$ and $\forall \mathpzc{h}>0$, $b \mapsto \mathcal{F}_{\mathpzc{h},L}^{\mathpzc{s}}(\beta,\rho^{\mathpzc{s}},b)$ is an even and real analytic function on $\mathbb{R}$. This latter feature allows us to define the finite-volume canonical magnetization and magnetic susceptibility respectively as the first and second derivative of the free energy density w.r.t. the intensity $B$ of the magnetic field,
see e.g. \cite[Eq. (3.17)]{BCS1}:
\begin{gather}
\mathcal{M}_{\mathpzc{h},L}^{\mathpzc{s}\,\mathrm{(C)}}(\beta,\rho^{\mathpzc{s}},b) := -\left(\frac{q}{c}\right) \frac{\partial \mathcal{F}_{\mathpzc{h},L}^{\mathpzc{s}}}{\partial b}(\beta,\rho^{\mathpzc{s}},b),\nonumber \\
\label{gfdse4}
\mathcal{X}_{\mathpzc{h},L}^{\mathpzc{s}\,\mathrm{(C)}}(\beta,\rho^{\mathpzc{s}},b) := -\left(\frac{q}{c}\right)^{2} \frac{\partial^{2} \mathcal{F}_{\mathpzc{h},L}^{\mathpzc{s}}}{\partial b^{2}}(\beta,\rho^{\mathpzc{s}},b).
\end{gather}

We are now ready to state the main result of this paper:

\begin{theorem}
\label{THMF}
Suppose that the assumptions $(\mathscr{A}_{\mathrm{p}})$ and $(\mathscr{A}_{\mathrm{r}})$ hold.\\
$\mathrm{(i)}$. For any $\beta >0$ and $\rho^{\mathpzc{s}}>0$, $\mathpzc{s} \in \{-,+\}$ there exists $0< \hat{\mathpzc{h}}\leq 1$ s.t. $\forall \mathpzc{h} \in (0,\hat{\mathpzc{h}}]$, the canonical zero-field magnetic susceptibility of the Fermi gas of spin $\frac{1}{2}$ with identical spin projection $m_{\frac{1}{2}}= \mathpzc{s}\frac{1}{2}$ admits the thermodynamic limit. Moreover, for such $\mathpzc{h}$s the thermodynamic limit coincides with the bulk grand-canonical zero-field magnetic susceptibility defined in \eqref{susfix23} (with $b=0$) for the corresponding density, i.e.
\begin{equation}
\label{limca2}
\mathcal{X}_{\mathpzc{h}}^{\mathpzc{s}\,\mathrm{(C)}}\left(\beta,\rho^{\mathpzc{s}},0\right) := \lim_{L \uparrow \infty} \mathcal{X}_{\mathpzc{h},L}^{\mathpzc{s}\,\mathrm{(C)}}\left(\beta,\rho^{\mathpzc{s}},0\right) = \mathcal{X}_{\mathpzc{h}}^{\mathpzc{s}\,\mathrm{(GC)}}\left(\beta,\rho^{\mathpzc{s}},0\right).
\end{equation}
$\mathrm{(ii)}$. For any $\beta>0$, $\rho>0$ with $\rho = \rho^{-}+\rho^{+}$ and $\rho^{\mathpzc{s}}>0$, and $0< \mathpzc{h} \leq \hat{\mathpzc{h}}$ denote:
\begin{equation}
\label{limca}
\mathcal{X}_{\mathpzc{h}}^{\mathrm{(C)}}\left(\beta,\rho,0\right) := \mathcal{X}_{\mathpzc{h}}^{-\,\mathrm{(C)}}\left(\beta,\rho^{-},0\right) + \mathcal{X}_{\mathpzc{h}}^{+\,\mathrm{(C)}}\left(\beta,\rho^{+},0\right) = \mathcal{X}_{\mathpzc{h}}^{\mathrm{(GC)}}\left(\beta, \rho,0\right).
\end{equation}
By setting $\rho^{-} = \rho^{+}$, \eqref{limca} admits the following asymptotic expansion in the semiclassical limit:
\begin{equation}
\label{semisusdens}
\mathcal{X}_{\mathpzc{h}}^{\mathrm{(C)}}(\beta,\rho,0) = \mathscr{X}_{\mathpzc{h}}^{\mathrm{(orbit)}}(\beta,\rho) + \mathscr{X}_{\mathpzc{h}}^{(\mathrm{spin})}(\beta,\rho) +  o(\mathpzc{h}^{2})\quad \textrm{when $\mathpzc{h} \downarrow 0$},
\end{equation}
and the leading terms read as:
\begin{gather}
\label{susorbital}
\mathscr{X}_{\mathpzc{h}}^{\mathrm{(orbit)}}(\beta,\rho) := - \frac{1}{3} \left(\frac{q}{2c}\right)^{2} \rho \beta \mathpzc{h}^{2},\\
\label{susspin}
\mathscr{X}_{\mathpzc{h}}^{(\mathrm{spin})}(\beta,\rho) := \frac{g^{2}}{4} \left(\frac{q}{2c}\right)^{2} \rho \beta \mathpzc{h}^{2}.
\end{gather}
\end{theorem}

\begin{remark}[\textbf{Comments on the assumptions $(\mathscr{A}_{\mathrm{p}})$ and $(\mathscr{A}_{\mathrm{r}})$}]
\label{discuss}
Let us first discuss $(\mathscr{A}_{\mathrm{p}})$. Except for particular systems, proving the existence of the thermodynamic limit requires some invariance properties for the infinite-volume Hamiltonian. In our conditions, \eqref{Hinfini} is invariant under the magnetic translations of the lattice, see e.g. \cite[Sec. V]{Nen}. In \cite{BS}, the existence of the thermodynamic limit of the derivatives w.r.t. $b$ and $z$ of the grand-canonical pressure is proven for singular $\mathbb{R}^{3}$-ergodic random field potentials, see \cite[Sec. 1.3]{BS}. This kind of potentials covers the periodic and almost-periodic case. Note that such invariance assumptions are also required when proving the existence of the integrated density of states, see e.g. \cite{HLMW}. Besides, we emphasize that our analysis does not imply any restriction on the Bravais-lattice provided that it is non-degenerate. Considering the $\mathbb{Z}^{3}$-lattice simply allows us to dodge a number of technical difficulties arising from the shape of the Wigner-Seitz cell. Subsequently, we turn to $(\mathscr{A}_{\mathrm{r}})$. If $\theta=1$, \eqref{deftheta} corresponds to the global Lipschitz-continuity. This implies that $V$ is differentiable almost everywhere on $\mathbb{R}^{3}$ (i.e. outside a set of Lebesgue-measure zero) with essentially bounded first derivatives. Note that $\theta=1$ covers the case of continuously differentiable functions on $\mathbb{R}^{3}$ with globally bounded derivative. If $0<\theta < 1$, $V$ may be nowhere differentiable (think of the Weierstrass function). Finally, we mention that $(\mathscr{A}_{\mathrm{p}})$ and $(\mathscr{A}_{\mathrm{r}})$ together imply that $V$ is bounded and continuous on $\mathbb{R}^{3}$.
\end{remark}

\begin{remark}[\textbf{Equivalence of ensembles}]
\label{enssembl}
Theorem \ref{THMF} $\mathrm{(i)}$ reflects the equivalence between the grand-canonical and canonical ensembles. In fact, we prove in Proposition \ref{weakequi} that $\forall \beta>0$, $\forall b \in \mathbb{R}$ and for suitable values of $\rho^{\mathpzc{s}}$, $\lim_{L \uparrow \infty} (\partial_{b}^{l} \mathcal{F}_{\mathpzc{h},L}^{\mathpzc{s}})(\beta,\rho^{\mathpzc{s}},b)$, $l=0,1,2$ exists and identifies with $(\partial_{b}^{l}(P_{\mathpzc{h}}^{\mathpzc{s}})^{*})(\beta,\rho^{\mathpzc{s}},b)$, where $(P_{\mathpzc{h}}^{\mathpzc{s}})^{*}$ is the Legendre-transform of the bulk pressure defined in \eqref{Tleg}. While $\mathcal{M}_{\mathpzc{h}}^{\mathpzc{s}\,(C)}(\beta,\rho^{\mathpzc{s}},b) := \lim_{L \uparrow \infty} \mathcal{M}_{\mathpzc{h},L}^{\mathpzc{s}\,\mathrm{(C)}}(\beta,\rho^{\mathpzc{s}},b) =  - (\frac{q}{c}) (\partial_{b} (P_{\mathpzc{h}}^{\mathpzc{s}})^{*})(\beta,\rho^{\mathpzc{s}},b)$ coincides with $\mathcal{M}_{\mathpzc{h}}^{\mathpzc{s}\,\mathrm{(GC)}}(\beta,\rho^{\mathpzc{s}},b)$, $\mathcal{X}_{\mathpzc{h}}^{\mathpzc{s}\,(C)}(\beta,\rho^{\mathpzc{s}},b) = - (\frac{q}{c})^{2} (\partial_{b}^{2} (P_{\mathpzc{h}}^{\mathpzc{s}})^{*})(\beta,\rho^{\mathpzc{s}},b)$ coincides with $\mathcal{X}_{\mathpzc{h}}^{\mathpzc{s}\,\mathrm{(GC)}}(\beta,\rho^{\mathpzc{s}},b)$ only when the magnetic field vanishes, see Lemma \ref{derPstar}. Otherwise, there is an additional term which generically does not cancel, see the proof of Lemma \ref{derPstar} along with Remark \ref{canorem}. We stress the point that these equivalence properties are established for a certain regime of the density $\rho$. The restriction on $\rho$ can be removed in zero-magnetic field provided that $\mathpzc{h}$ is small enough. In particular, we prove that $\forall \beta>0$, $\forall \rho^{\mathpzc{s}}>0$ and for $\mathpzc{h}$ sufficiently small, $\mathcal{M}_{\mathpzc{h}}^{\mathpzc{s}\,(C)}(\beta,\rho^{\mathpzc{s}},0) = \mathcal{M}_{\mathpzc{h}}^{\mathpzc{s}\,(GC)}(\beta,\rho^{\mathpzc{s}},0) = 0$.
\end{remark}

\begin{remark}[\textbf{Removing the condition on the $\rho^{\mathpzc{s}}$s in Theorem \ref{THMF} $\mathrm{(ii)}$}] Theorem \ref{THMF} $\mathrm{(ii)}$ is stated with the condition $\rho^{-} = \rho^{+}$ in \eqref{limca}. This restriction can be removed since we work in zero-magnetic field. The asymptotic expansion in \eqref{semisusdens} with leading terms \eqref{susorbital}-\eqref{susspin} still hold true when $\rho^{-} \neq \rho^{+}$ obeying $\rho^{-}+\rho^{+} = \rho$. Imposing such a restriction on the $\rho^{\mathpzc{s}}$s allows us to deal with only one fugacity (see \eqref{susfixdens}) instead of two (see \eqref{susfix2}-\eqref{susfix23}) when proving \eqref{semisusdens}.
\end{remark}

\begin{remark}[\textbf{Connection with the BvL theorem within the linear-response theory}]
For any $L>0$, $\beta>0$, $\rho>0$, $b \in \mathbb{R}$ and $\mathpzc{h}>0$ define the finite-volume grand-canonical magnetization at fixed density $\mathcal{M}_{\mathpzc{h},L}^{\mathrm{(GC)}}(\beta,\rho,b)$ as in \eqref{magnfi2}-\eqref{magnfi23} but with $\overline{z}_{\mathpzc{h},L}^{\mathpzc{s}}(\beta,\rho^{s},b)$, $\mathpzc{s} \in \{-,+\}$ denoting the unique solution of the equation $\rho_{\mathpzc{h},L}^{\mathpzc{s}}(\beta,z,b) = \rho^{\mathpzc{s}}$, see \eqref{densityfv}. In \cite{Sa}, it is proven that, under the same conditions, $b \mapsto \mathcal{M}_{\mathpzc{h},L}^{\mathrm{(GC)}}(\beta,\rho,b)$ is a real analytic function on $\mathbb{R}$. Therefore, it admits in a neighborhood of $b=0$ a convergent Taylor series expansion in powers of $b$. In Physics literature, see e.g.
\cite[Sec. 52]{Land}, it is common to truncate the series to the first-order as an approximation:
\begin{equation}
\label{LRT}
m_{\mathpzc{h},L}^{\mathrm{LRT}}(\beta,\rho,b) := \mathcal{M}_{\mathpzc{h},L}^{\mathrm{(GC)}}(\beta,\rho,0) + \mathcal{X}_{\mathpzc{h},L}^{\mathrm{(GC)}}(\beta,\rho,0) b = \mathcal{X}_{\mathpzc{h},L}^{\mathrm{(GC)}}(\beta,\rho,0) b,
\end{equation}
where we used that $b \mapsto P_{\mathpzc{h},L}^{\mathpzc{s}}(\beta,z,b)$ is an even function. \eqref{LRT} is often referred to as the magnetization formula in the linear-response theory. From Theorem \ref{THMF} $\mathrm{(i)}$, for $\mathpzc{h}$ sufficiently small:
\begin{equation*}
m_{\mathpzc{h}}^{\mathrm{LRT}}(\beta,\rho,b) := \lim_{L \uparrow \infty} m_{\mathpzc{h},L}^{\mathrm{LRT}}(\beta,\rho,b) = \mathcal{X}_{\mathpzc{h}}^{\mathrm{(C)}}(\beta,\rho,0) b.
\end{equation*}
From Theorem \ref{THMF} $\mathrm{(ii)}$, one recovers the result of the BvL theorem within the linear-response theory by performing the semiclassical limit $\mathpzc{h} \downarrow 0$. Note that the same result still holds if one defines the 'LRT'-magnetization from the canonical ensemble, see also Remark \ref{enssembl}.
\end{remark}

\begin{remark}[\textbf{Bloch electrons in the semiclassical limit: free electrons behavior}]\label{there}
The\\ leading term in \eqref{semisusdens} is made up of two terms: a diamagnetic contribution in \eqref{susorbital} arising from the particle 'motions' induced by the Zeeman Hamiltonian, and a paramagnetic contribution in \eqref{susspin} arising from the particle spin through the Stern-Gerlach term. Such a decomposition holds since the spin-orbit coupling has been disregarded. We emphasize that the external potential is not involved in both contributions. This goes in the direction of a result stated by Kohn \textit{et al.} in \cite{KoL}: for weak potentials, the correction to the bulk zero-field magnetic susceptibility is second-order in the potential.
In the case of electrons, the gas is globally paramagnetic, and we recover the relation:
\begin{equation*}
\mathscr{X}_{\mathpzc{h}}^{\mathrm{(orbit)}}(\beta,\rho) = - \frac{1}{3} \mathscr{X}_{\mathpzc{h}}^{(\mathrm{spin})}(\beta,\rho).
\end{equation*}
Therefore, the Bloch electron gas in the semiclassical limit behaves to first-order like the free electron gas.
Concerning the remainder term in \eqref{semisusdens}, we prove that it behaves like $\mathcal{O}(\mathpzc{h}^{2+\theta(1-\alpha)})$, with $\theta \in (0,1]$ appearing in \eqref{deftheta} and $0<\alpha<1$. Its properties are discussed in Sec. \ref{openpb}.
\end{remark}

\begin{remark}[\textbf{Recovering the well-known results in the high-temperature regime}] In Physics literature, the diamagnetic contribution in \eqref{susorbital} corresponds to the high-temperature Landau susceptibility of free electrons, and the paramagnetic contribution in \eqref{susspin} to the Curie susceptibility of free spins. Both are usually derived from Maxwell-Boltzmann statistics justified in the high-temperature regime or low-particle density regime, see e.g. \cite[Eq. (5-2.1)]{Mor} along with \cite[Eq. (5-3.11)]{Mor}. This means that, in the semiclassical limit, the Fermi-Dirac distribution can be approximated to first-order with the (classical) Maxwell-Boltzmann distribution.\\
We point out that, from our analysis, we can derive the asymptotic behavior of the bulk zero-field magnetic susceptibility at fixed density $\rho>0$ in the regime of high-temperature:
\begin{equation*}
\mathcal{X}_{\mathpzc{h}}^{\mathrm{(C)}}(\beta,\rho,0) = \mathscr{X}_{\mathpzc{h}}^{\mathrm{(orbit)}}(\beta,\rho) + \mathscr{X}_{\mathpzc{h}}^{(\mathrm{spin})}(\beta,\rho) +  o(\beta)\quad \textrm{when $\beta \downarrow 0$},
\end{equation*}
which follows from \eqref{semisusdens} by setting $\beta=1$, then by replacing $\mathpzc{h}$ with $(\sqrt{\beta}\mathpzc{h})$.
\end{remark}

\subsection{A brief outline of the proof of Theorem \ref{THMF}--The contents.}

In Sec. \ref{appr}, we start with a technical result involving the operator in \eqref{tildHb}. It is obtained from \eqref{Hinfini}-\eqref{Hinfini1} by moving the semiclassical parameter $\mathpzc{h}$ into the argument of the potential via the unitary transformation \eqref{trnsfounit}.  On this way, the potential in \eqref{tildHb} is $\mathpzc{h}\mathbb{Z}^{3}$-periodic. Then for $\mathpzc{h}$ small enough, it is  slowly varying with period $\mathpzc{h}^{-1}$. From this feature, we develop a geometric perturbation theory to write down an approximation for the resolvent operator in \eqref{restildH}. To achieve that, we chop up for $\mathpzc{h}<1$ sufficiently small the dilated unit cell $\Omega_{\mathpzc{h}}:= (-\frac{1}{2\mathpzc{h}},\frac{1}{2 \mathpzc{h}})^{3}$ in disjoint closed cubic boxes centered at points in $\mathpzc{h}^{-\alpha} \mathbb{Z}^{3}$ and with side length $\mathpzc{h}^{-\alpha}$,  $0<\alpha<1$. Since the potential is slowly varying, the key-idea consists in approximating to first-order the resolvent in \eqref{restildH} by the operator in \eqref{Sepsi} which is a sum of terms, each locally approximating $(\tilde{H}_{\mathpzc{h}} - \xi)^{-1}$ (i.e. kind of 'local' resolvent with an operator having a constant potential, see \eqref{HconstV}). By the use of cutoff functions, each 'local' resolvent is localized on a small number of cubic boxes. The key-identity is \eqref{identtoot}, and it is the starting-point of our analysis. We prove that the operator norm $\Vert (\tilde{H}_{\mathpzc{h}} - \xi)^{-1} - \mathscr{R}_{\mathpzc{h}}(\xi) \Vert$ behaves like $\mathcal{O}(\mathpzc{h}^{\theta(1-\alpha)})$, $\theta \in (0,1]$ in \eqref{deftheta}. We emphasize that the H\"older continuity of $V$ is used when estimating the operator norm of \eqref{Teps}, see \eqref{esfgv3} and \eqref{slot}-\eqref{accroissfi}. We end Sec. \ref{appr} by giving a series of estimates on kernels and norms we use throughout the paper.\\
\indent Sec. \ref{sec3t} is devoted to the proof of Theorem \ref{THMF}. In Sec. \ref{invfug}, we focus on a preliminary result: Proposition \ref{lemrho} in which we write down an asymptotic expansion in the semiclassical limit of the fugacity which is the unique solution  of the equation \eqref{invdensi}. The starting-point in the proof of \eqref{semifug} is the formula \eqref{vrairho} for the bulk grand-canonical zero-field density of particles derived in \cite{BS}. By using the unitary transformation in \eqref{trnsfounit}, the formula in \eqref{vrairho} can be rewritten as \eqref{rhovr} involving the resolvent $(\tilde{H}_{\mathpzc{h}} - \xi)^{-1}$. Next, we use the results of Sec. \ref{appr} to derive an asymptotic expansion in the semiclassical limit. From the identity in \eqref{identtoot} and due to \eqref{esfgv2}-\eqref{esfgv3}, we expect the contribution obtained by replacing $(\tilde{H}_{\mathpzc{h}} - \xi)^{-1}$ in \eqref{rhovr} with $\mathscr{R}_{\mathpzc{h}}(\xi)$ to give rise to the leading term in the expansion. This is in fact the case, see Lemmas \ref{isola0}-\ref{isola1} whose the proofs lie in Sec. \ref{append21}. After rewriting this leading term to isolate the main $\mathpzc{h}$-dependent contribution, we arrive at Proposition \ref{asymrho}. To conclude the proof of \eqref{semifug}, it remains to solve a fixed-point equation \eqref{pzcG}.\\
\indent In Sec. \ref{THMi}, we prove Theorem \ref{THMF} $\mathrm{(i)}$. The proof relies on Proposition \ref{weakequi} along with Proposition \ref{invfug}. In Proposition \ref{weakequi}, we show that the thermodynamic limit of the free energy density of the Fermi gas of spin $\frac{1}{2}$ with identical spin projection $m_{\frac{1}{2}}= \mathpzc{s}\frac{1}{2}$ at fixed density $\rho^{\mathpzc{s}}>0$ and its two derivatives w.r.t. $b$ at $b=0$ identify respectively with the Legendre transform of the bulk grand-canonical pressure of the Fermi gas of spin $\frac{1}{2}$ with identical spin projection $m_{\frac{1}{2}}= \mathpzc{s}\frac{1}{2}$ in \eqref{Tleg} and its two derivatives w.r.t. $b$ at $b=0$. We emphasize that Proposition \ref{weakequi} only holds for a certain regime of density. But from the asymptotic expansion in \eqref{semifug}, this condition on the density is fulfilled provided that $\mathpzc{h}$ is sufficiently small. The proof of $\mathrm{(i)}$ finally follows from the identifications in Lemma \ref{derPstar}. The proof of Proposition \ref{weakequi} lies in Sec. \ref{append25} and it is based on the Darwin-Fowler method in \cite[Sec. 9.1]{Hu}. Such a method has already been used in \cite{C1} to treat the free spin-0 Bose gas. When applying it to the Fermi gas, a limiting condition on the density is required. However, this is enough for the application we have in mind.\\
\indent In Sec. \ref{sec33}, we prove Theorem \ref{THMF} $\mathrm{(ii)}$. From Theorem \ref{THMF} $\mathrm{(i)}$ together with \eqref{susfixdens}, the only thing we have to do is derive an asymptotic expansion in the semiclassical limit of the bulk grand-canonical zero-field magnetic susceptibility (orbital and spin contributions). This is contained in Proposition \ref{thmsus}. The proof of $\mathrm{(ii)}$ directly follows from Proposition \ref{thmsus} along with Proposition \ref{invfug}. The starting-point in the proof of \eqref{semisuo} (resp. \eqref{semisup}) is the formula \eqref{suso} (resp. \eqref{susp}) for the bulk grand-canonical zero-field orbital (resp. spin) susceptibility obtained from \cite{BS}. We emphasize that such a two-terms decomposition holds since the spin-orbit coupling is disregarded. By using the unitary transformation in \eqref{trnsfounit}, the formulas in \eqref{suso}-\eqref{susp} can be rewritten respectively as \eqref{suso2}-\eqref{susp2} involving the resolvent $(\tilde{H}_{\mathpzc{h}} - \xi)^{-1}$. In Sec. \ref{subsec32}, we use the results of Sec. \ref{appr} to derive a first asymptotic expansion in the semiclassical limit, see Proposition \ref{gropro2}. From the identity in \eqref{identtoot} and due to \eqref{esfgv2}-\eqref{esfgv3}, we expect the contribution obtained by replacing each resolvent $(\tilde{H}_{\mathpzc{h}} - \xi)^{-1}$ (even the ones in \eqref{tildW1}-\eqref{tildW2} in the kernels sense) with $\mathscr{R}_{\mathpzc{h}}(\xi)$ to give rise to the leading term in the expansion. This is in fact the case, see Lemmas \ref{isola11}-\ref{isola13} whose the proofs lie in Sec. \ref{append23}. The rest of the proof consists in isolating the main $\mathpzc{h}$-dependent contribution from \eqref{suso5}-\eqref{susp5}. The procedure is a bit technical and it is contained in Sec. \ref{subsec33}.

\subsection{Concluding remarks and a few open problems.}
\label{openpb}

As mentioned in Remark \ref{there}, we show in the proof of Theorem \ref{THMF} $\mathrm{(ii)}$ that the remainder in the asymptotic expansion \eqref{semisusdens} behaves like $\mathcal{O}(\mathpzc{h}^{2 + \theta(1-\alpha)})$, with $\theta\in (0,1]$ in \eqref{deftheta} and $\alpha \in (0,1)$ an arbitrary parameter coming from the geometric perturbation theory in Sec. \ref{appr}. Note that $2+\theta(1-\alpha)$ approaches the value 2 when $\theta$ approaches 0 independently of $\alpha$ (or when $\theta=1$ and $\alpha$ approaches 1). We emphasize that our method allows us to write down an explicit expression for the remainder. However, such an expression contains a large number of terms depending on the cutting of the unit-cell that we use in the geometric perturbation theory (hence the $\alpha$-dependence). In particular, the remainder term involves the operator in \eqref{Teps} which appears in the identity \eqref{identtoot}. \eqref{deftheta} plays a crucial role when estimating the operator norm of \eqref{Teps}, see \eqref{slot}-\eqref{accroissfi}. If the potential $V$ belongs to $\mathcal{C}^{p}(\mathbb{R}^{3};\mathbb{R})$, $p\geq 1$ and has globally bounded derivatives, then one can use the Taylor's formula with integral remainder to rewrite the difference of potentials in \eqref{Teps}. In that way, the remainder of the asymptotic expansion in \eqref{semisusdens} will involve the derivatives of $V$, and its behavior will be of order $\mathcal{O}(\mathpzc{h}^{3-\alpha})$. We expect the optimal behavior to be $\mathcal{O}(\mathpzc{h}^{3})$, and we believe that a more refined geometric perturbation might allow us to remove the $\alpha$-dependence.\\
\indent In Physics literature, Jennings \textit{et al.} were interested in \cite{JB} in the expansion in powers of Planck's constant of the bulk zero-field orbital susceptibility for a 3-dimensional non-interacting electron gas obeying either Maxwell-Boltzmann statistics in the high-temperature limit, or Fermi-Dirac statistics in the low-temperature limit. Considering a confining smooth potential barrier $U$ of arbitrary shape (but with bounded high-order derivatives), the two-lowest-order terms in the expansion in powers of Planck's constant for the Maxwell-Boltzmann electron gas read as:
\begin{equation}
\label{avoir}
-\frac{1}{3} \left(\frac{q}{2c}\right)^{2} \rho \beta \hbar^{2} + \frac{1}{3} \left(\frac{q}{2c}\right)^{2} \frac{1}{60} \frac{\displaystyle{\int \mathrm{d}\bold{x}\, \mathrm{e}^{-\beta U(\bold{x})} \left(\frac{\partial^{2} U}{\partial x_{1}^{2}}(\bold{x}) + \frac{\partial^{2} U}{\partial x_{2}^{2}}(\bold{x})\right)}}{\displaystyle{\int \mathrm{d}\bold{x}\, \mathrm{e}^{-\beta U(\bold{x})}}} \rho \beta^{3} \hbar^{4}.
\end{equation}
If $U$ is periodic, the integrations are over the Wigner-Seitz cell. Note that the first term corresponds to \eqref{susorbital}. The method they use consists in expanding the Maxwell-Boltzmann factor involved in the partition function in powers of $\hbar$ via the Wigner-Kirkwood expansion, see \cite{WiKi}. We mention that D. Bivin showed in \cite{Bi} that \eqref{avoir} represents the first two terms in the expansion for the bulk zero-field orbital susceptibility in the high-temperature limit for a certain type of potentials.\\
\indent Finding a 'wider' class of potentials than the one covered by assumptions $(\mathscr{A}_{\mathrm{p}})$-$(\mathscr{A}_{\mathrm{r}})$ for which the conclusion of the BvL theorem holds true, and deriving the second lowest-order term in the expansion \eqref{semisusdens} which holds for any 'inverse' temperature $\beta>0$, are both challenging problems.



\section{An approximation of the resolvent via a geometric perturbation theory.}
\label{appr}

In the whole of this section, we suppose that assumptions $(\mathscr{A}_{\mathrm{p}})$ and $(\mathscr{A}_{\mathrm{r}})$ hold.\\
Here, we are interested in the self-adjoint realization in $L^{2}(\mathbb{R}^{3};\mathbb{C}^{2})$ of the family of operators:
\begin{equation}
\label{tildHb}
\tilde{\mathpzc{H}}_{\mathpzc{h}} := \begin{pmatrix} \tilde{H}_{\mathpzc{h}} & 0 \\ 0 & \tilde{H}_{\mathpzc{h}} \end{pmatrix} = \tilde{H}_{\mathpzc{h}} \oplus \tilde{H}_{\mathpzc{h}}, \quad \tilde{H}_{\mathpzc{h}} := \frac{1}{2}(-i \nabla)^{2} + V (\mathpzc{h} \cdot\,),\quad \mathpzc{h}>0,
\end{equation}
defined originally on $\mathcal{C}_{0}^{\infty}(\mathbb{R}^{3};\mathbb{C}^{2})$. Since the period of the dilated potential is $\mathpzc{h}^{-1}$ (keep in mind that $V$ is chosen $\mathbb{Z}^{3}$-periodic), then the potential is slowly varying if $\mathpzc{h}$ is sufficiently small.\\
Define on $L^{2}(\mathbb{R}^{3};\mathbb{C}^{2})$ the family $\{\mathpzc{U}_{\mathpzc{h}},\, \mathpzc{h}>0\}$ of unitary operators by:
\begin{equation}
\label{trnsfounit}
(\mathpzc{U}_{\mathpzc{h}} \Psi)(\mathbf{x}) := \mathpzc{h}^{\frac{3}{2}} \Psi(\mathpzc{h} \mathbf{x}), \quad \mathbf{x} \in \mathbb{R}^{3},\,\Psi \in L^{2}(\mathbb{R}^{3};\mathbb{C}^{2}).
\end{equation}
Then, under the transformation \eqref{trnsfounit}, the operator $\tilde{\mathpzc{H}}_{\mathpzc{h}}$ in \eqref{tildHb} and $\mathpzc{H}_{\mathpzc{h}}(0)$ in \eqref{Hinfini} are unitarily equivalent. Using the shorthand notation $\mathpzc{H}_{\mathpzc{h}} = \mathpzc{H}_{\mathpzc{h}}(0)$, one has for any $\mathpzc{h}>0$:
\begin{equation}
\label{equivalu}
\mathpzc{U}_{\mathpzc{h}} \mathpzc{H}_{\mathpzc{h}} \mathpzc{U}^{-1}_{\mathpzc{h}} = \tilde{\mathpzc{H}}_{\mathpzc{h}}.
\end{equation}

The aim of this section consists in using a geometric perturbation theory to obtain a series of approximations for the resolvent operator $(\tilde{\mathpzc{H}}_{\mathpzc{h}} - \xi)^{-1}$, $\xi \in \varrho(\tilde{\mathpzc{H}}_{\mathpzc{h}})$ on $L^{2}(\mathbb{R}^{3};\mathbb{C}^{2})$:
\begin{equation}
\label{restildH}
\left(\tilde{\mathpzc{H}}_{\mathpzc{h}}-\xi\right)^{-1} :=  \begin{pmatrix} (\tilde{H}_{\mathpzc{h}} - \xi)^{-1} & 0 \\ 0 & (\tilde{H}_{\mathpzc{h}} - \xi)^{-1}\end{pmatrix} = \left(\tilde{H}_{\mathpzc{h}}-\xi\right)^{-1} \oplus \left(\tilde{H}_{\mathpzc{h}} - \xi\right)^{-1}.
\end{equation}

Let us turn to the geometric perturbation theory, for further applications see \cite{CFFH,CN3,Sa2}.\\
For any $\mathpzc{h}>0$, let $\Omega_{\mathpzc{h}} := (-\frac{1}{2 \mathpzc{h}},\frac{1}{2\mathpzc{h}})^{3}$ be the dilated unit cell centered at the origin of coordinates. For any $0 < \alpha < 1$ and $\mathpzc{h} \in (0,1]$, we cover $\Omega_{\mathpzc{h}}$ with disjoint closed cubic boxes parallel to the coordinate axis, centered at points in $\mathpzc{h}^{- \alpha} \mathbb{Z}^{3}$, and with side length $\mathpzc{h}^{-\alpha}$. Denote by $\mathscr{E}$ the set of centers of those cubes which have common points with $\Omega_{\mathpzc{h}}$. Note that $\mathrm{Card}(\mathscr{E}) = \mathcal{O}(\mathpzc{h}^{3 \alpha - 3})$ for $\mathpzc{h}$ small enough. Denote by $\mathrm{C}(\boldsymbol{\gamma}, r)$ the cube centered at $\boldsymbol{\gamma} \in \mathscr{E}$ with side length $r \geq \mathpzc{h}^{-\alpha}$. In the following, for $0<\alpha <1$ kept fixed, by $\mathpzc{h}$ sufficiently small, we mean:
\begin{equation}
\label{epsilon0}
0<\mathpzc{h} \leq \mathpzc{h}_{0}\quad \textrm{with} \quad \mathpzc{h}_{0}=\mathpzc{h}_{0}(\alpha) \leq 1 \quad \textrm{s.t.}\quad  \mathrm{C}\left(0, 69 \mathpzc{h}_{0}^{-\alpha}\right) \subsetneq \Omega_{\mathpzc{h}_{0}}.
\end{equation}

\indent Let us introduce some well-chosen families of smooth cutoff functions.\\
Let $\{\tau_{\mathpzc{h},\boldsymbol{\gamma}}\}_{\boldsymbol{\gamma} \in \mathscr{E}}$, $\mathpzc{h} \in (0,\mathpzc{h}_{0}]$ be a partition of unity of $\Omega_{\mathpzc{h}}$ satisfying:
\begin{gather}
\label{partiden0}
\mathrm{Supp}\left(\tau_{\mathpzc{h},\boldsymbol{\gamma}}\right) \subset \mathrm{C}\left(\boldsymbol{\gamma},2\mathpzc{h}^{-\alpha}\right),\quad  0 \leq \tau_{\mathpzc{h},\boldsymbol{\gamma}} \leq 1;\\
\label{partiden}
\sum_{\boldsymbol{\gamma} \in \mathscr{E}} \tau_{\mathpzc{h},\boldsymbol{\gamma}}(\mathbf{x}) = 1,\quad \forall \mathbf{x} \in \Omega_{\mathpzc{h}}.
\end{gather}
Moreover, there exists a constant $C>0$ s.t.
\begin{equation*}
\forall \mathpzc{h} \in (0,\mathpzc{h}_{0}],\, \forall \boldsymbol{\gamma} \in \mathscr{E},\quad \left\Vert D^{s} \tau_{\mathpzc{h},\boldsymbol{\gamma}} \right\Vert_{\infty} \leq C \mathpzc{h}^{\vert s \vert \alpha},\quad \forall\, \vert s \vert \leq 2,\, s \in \mathbb{N}^{3}.
\end{equation*}
Also let $\{\hat{\tau}_{\mathpzc{h},\boldsymbol{\gamma}}\}_{\boldsymbol{\gamma} \in \mathscr{E}}$ and $\{\hat{\hat{\tau}}_{\mathpzc{h},\boldsymbol{\gamma}}\}_{\boldsymbol{\gamma} \in \mathscr{E}}$,  $\mathpzc{h} \in (0,\mathpzc{h}_{0}]$ satisfying:
\begin{gather}
\mathrm{Supp}\left(\hat{\tau}_{\mathpzc{h},\boldsymbol{\gamma}}\right) \subset \mathrm{C}\left(\boldsymbol{\gamma},4\mathpzc{h}^{-\alpha}\right), \quad
\hat{\tau}_{\mathpzc{h},\boldsymbol{\gamma}}(\mathbf{x}) = 1\,\,\textrm{if $\mathbf{x} \in \mathrm{C}\left(\boldsymbol{\gamma},3 \mathpzc{h}^{-\alpha}\right)$},\quad 0 \leq \hat{\tau}_{\mathpzc{h},\boldsymbol{\gamma}} \leq 1;\nonumber\\
\label{supp2hat}
\mathrm{Supp}\left(\hat{\hat{\tau}}_{\mathpzc{h},\boldsymbol{\gamma}}\right) \subset \mathrm{C}\left(\boldsymbol{\gamma},6\mathpzc{h}^{-\alpha}\right), \quad
\hat{\hat{\tau}}_{\mathpzc{h},\boldsymbol{\gamma}}(\mathbf{x}) = 1\,\,\textrm{if $\mathbf{x} \in \mathrm{C}\left(\boldsymbol{\gamma},5 \mathpzc{h}^{-\alpha}\right)$}, \quad 0 \leq \hat{\hat{\tau}}_{\mathpzc{h},\boldsymbol{\gamma}} \leq 1.
\end{gather}
Moreover, there exists another constant $C>0$ s.t.
\begin{equation}
\label{norminf}
\forall \mathpzc{h} \in (0,\mathpzc{h}_{0}],\, \forall \boldsymbol{\gamma} \in \mathscr{E},\quad \max \left\{
\left\Vert D^{s} \hat{\tau}_{\mathpzc{h},\boldsymbol{\gamma}} \right\Vert_{\infty}, \left\Vert D^{s} \hat{\hat{\tau}}_{\mathpzc{h},\boldsymbol{\gamma}} \right\Vert_{\infty}\right\} \leq C \mathpzc{h}^{\vert s \vert \alpha},\quad \forall \vert s\vert \leq 2,\, s \in \mathbb{N}^{3}.
\end{equation}
From the above definitions, we have the following identities:
\begin{align}
\label{hatggep}
\forall \mathpzc{h} \in (0,\mathpzc{h}_{0}],\, \forall \boldsymbol{\gamma} \in \mathscr{E},\quad \hat{\tau}_{\mathpzc{h},\boldsymbol{\gamma}} \tau_{\mathpzc{h},\boldsymbol{\gamma}} &= \tau_{\mathpzc{h},\boldsymbol{\gamma}},\\
\label{hatggep2}
\hat{\hat{\tau}}_{\mathpzc{h},\boldsymbol{\gamma}} \hat{\tau}_{\mathpzc{h},\boldsymbol{\gamma}} &= \hat{\tau}_{\mathpzc{h},\boldsymbol{\gamma}};
\end{align}
and moreover, under the same conditions, there exists another constant $C>0$ s.t.
$\forall 1 \leq \vert s\vert\leq 2$:
\begin{equation}
\label{support}
\max\left\{\mathrm{dist}\left(\mathrm{Supp}\left(D^{s} \hat{\tau}_{\mathpzc{h},\boldsymbol{\gamma}}\right), \mathrm{Supp}\left(\tau_{\mathpzc{h},\boldsymbol{\gamma}}\right)\right),
\mathrm{dist}\left(\mathrm{Supp}\left(D^{s} \hat{\hat{\tau}}_{\mathpzc{h},\boldsymbol{\gamma}}\right), \mathrm{Supp}\left(\tau_{\mathpzc{h},\boldsymbol{\gamma}}\right)\right)\right\} \geq C \mathpzc{h}^{-\alpha}.
\end{equation}
\indent Next, introduce a series of operators which will be used to approximate, for $\mathpzc{h}$ small enough, the resolvent operator $(\tilde{\mathpzc{H}}_{\mathpzc{h}} - \xi)^{-1}$ on $L^{2}(\mathbb{R}^{3};\mathbb{C}^{2})$, with $\xi \in \varrho(\tilde{\mathpzc{H}}_{\mathpzc{h}})$ away from the interval $[\inf \sigma(\tilde{\mathpzc{H}}_{\mathpzc{h}}),\infty)$. Since \eqref{equivalu} leads to $\varrho(\tilde{\mathpzc{H}}_{\mathpzc{h}}) = \varrho(\mathpzc{H}_{\mathpzc{h}})$, then we can restrict to $\xi \in \mathbb{C} \setminus [-\Vert V \Vert_{\infty},\infty)$ due to \eqref{infspectrum}. Moreover, in view of \eqref{restildH}, it is enough to restrict our analysis to $(\tilde{H}_{\mathpzc{h}} - \xi)^{-1}$ on $L^{2}(\mathbb{R}^{3})$.\\
\indent Introduce the self-adjoint realization in $L^{2}(\mathbb{R}^{3})$ of the family of reference operators:
\begin{equation}
\label{locham}
\tilde{H}_{\mathpzc{h},\boldsymbol{\gamma}}^{\mathrm{(ref)}} := \frac{1}{2}(-i \nabla)^{2} + V(\mathpzc{h} \cdot\,) \hat{\hat{\tau}}_{\mathpzc{h},\boldsymbol{\gamma}}  + \left(1 - \hat{\hat{\tau}}_{\mathpzc{h},\boldsymbol{\gamma}}\right) V\left(\mathpzc{h} \mathpzc{h}^{-\alpha} \boldsymbol{\gamma}\right),\quad  \mathpzc{h} \in (0,\mathpzc{h}_{0}],\, \boldsymbol{\gamma} \in \mathscr{E},
\end{equation}
defined originally on $\mathcal{C}_{0}^{\infty}(\mathbb{R}^{3})$. Define $\forall \mathpzc{h} \in (0,\mathpzc{h}_{0}]$ and $\forall \xi \in \mathbb{C} \setminus [-\Vert V \Vert_{\infty},\infty)$ on $L^{2}(\mathbb{R}^{3})$:
\begin{equation}
\label{chapS}
\mathcal{R}_{\mathpzc{h}}(\xi) := \sum_{\boldsymbol{\gamma} \in \mathscr{E}} \hat{\tau}_{\mathpzc{h},\boldsymbol{\gamma}} \left(\tilde{H}_{\mathpzc{h},\boldsymbol{\gamma}}^{\mathrm{(ref)}} - \xi\right)^{-1} \tau_{\mathpzc{h},\boldsymbol{\gamma}}.
\end{equation}
Note an important thing. Due to the property \eqref{hatggep2}, then $\forall \mathpzc{h} \in (0,\mathpzc{h}_{0}]$ and $\forall \boldsymbol{\gamma} \in \mathscr{E}$:
\begin{equation*}
\left(\tilde{H}_{\mathpzc{h}} - \xi\right) \hat{\tau}_{\mathpzc{h},\boldsymbol{\gamma}} = \left(\tilde{H}_{\mathpzc{h},\boldsymbol{\gamma}}^{\mathrm{(ref)}} - \xi\right) \hat{\tau}_{\mathpzc{h},\boldsymbol{\gamma}}.
\end{equation*}
Since $\mathrm{Ran}(\mathcal{R}_{\mathpzc{h}}(\xi)) \subset \mathrm{Dom}(\tilde{H}_{\mathpzc{h}})$ by standard arguments, then (below $[\cdot\,,\cdot\,]$ denotes the commutator):
\begin{equation*}
\left(\tilde{H}_{\mathpzc{h}} - \xi\right) \mathcal{R}_{\mathpzc{h}}(\xi) = \sum_{\boldsymbol{\gamma} \in \mathscr{E}} \hat{\tau}_{\mathpzc{h},\boldsymbol{\gamma}} \tau_{\mathpzc{h},\boldsymbol{\gamma}} + \sum_{\boldsymbol{\gamma} \in \mathscr{E}} \left[\tilde{H}_{\mathpzc{h},\boldsymbol{\gamma}}^{\mathrm{(ref)}}, \hat{\tau}_{\mathpzc{h},\boldsymbol{\gamma}}\right]
\left(\tilde{H}_{\mathpzc{h},\boldsymbol{\gamma}}^{\mathrm{(ref)}} - \xi\right)^{-1} \tau_{\mathpzc{h},\boldsymbol{\gamma}} = \mathbbm{1} + \mathcal{W}_{\mathpzc{h}}(\xi),
\end{equation*}
where we used \eqref{hatggep} then \eqref{partiden} in the second equality, and $\forall \mathpzc{h} \in (0,\mathpzc{h}_{0}]$ and $\forall \xi \in \mathbb{C}\setminus [-\Vert V \Vert_{\infty},\infty)$:
\begin{equation}
\label{hatTep}
\mathcal{W}_{\mathpzc{h}}(\xi) := \sum_{\boldsymbol{\gamma} \in \mathscr{E}} \left\{-\frac{1}{2}\left(\Delta \hat{\tau}_{\mathpzc{h},\boldsymbol{\gamma}}\right) - \left(\nabla \hat{\tau}_{\mathpzc{h},\boldsymbol{\gamma}}\right) \cdot \nabla \right\} \left(\tilde{H}_{\mathpzc{h},\boldsymbol{\gamma}}^{\mathrm{(ref)}} - \xi\right)^{-1} \tau_{\mathpzc{h},\boldsymbol{\gamma}}.
\end{equation}
Since $\mathcal{W}_{\mathpzc{h}}(\xi)$ is bounded on $L^{2}(\mathbb{R}^{3})$, see \cite[Lem. 5.1]{BS}, this means in the bounded operators sense:
\begin{equation}
\label{approxres}
\left(\tilde{H}_{\mathpzc{h}} - \xi\right)^{-1} = \mathcal{R}_{\mathpzc{h}}(\xi) - \left(\tilde{H}_{\mathpzc{h}} - \xi\right)^{-1} \mathcal{W}_{\mathpzc{h}}(\xi).
\end{equation}
Below, we show that the operator norm  $\Vert(\tilde{H}_{\mathpzc{h}} - \xi)^{-1} - \mathcal{R}_{\mathpzc{h}}(\xi)\Vert$ grows slower than any power of $\mathpzc{h}$.\\
\indent Afterwards, let us continue with another approximation for the operator $\mathcal{R}_{\mathpzc{h}}(\xi)$ in \eqref{chapS}. Introduce the self-adjoint realization in $L^{2}(\mathbb{R}^{3})$ of the family of operators with constant potential:
\begin{equation}
\label{HconstV}
\tilde{H}_{\mathpzc{h},\boldsymbol{\gamma}}^{\mathrm{(cste)}} := \frac{1}{2}(-i \nabla)^{2}  + V(\mathpzc{h}^{1-\alpha} \boldsymbol{\gamma}),\quad \mathpzc{h} \in (0,\mathpzc{h}_{0}],\, \boldsymbol{\gamma} \in \mathscr{E},
\end{equation}
defined originally on $\mathcal{C}_{0}^{\infty}(\mathbb{R}^{3})$. Define $\forall \mathpzc{h} \in (0,\mathpzc{h}_{0}]$ and $\forall \xi \in \mathbb{C}\setminus [-\Vert V \Vert_{\infty},\infty)$ on $L^{2}(\mathbb{R}^{3})$:
\begin{equation}
\label{Sepsi}
\mathscr{R}_{\mathpzc{h}}(\xi) := \sum_{\boldsymbol{\gamma} \in \mathscr{E}} \hat{\tau}_{\mathpzc{h},\boldsymbol{\gamma}} \left(\tilde{H}_{\mathpzc{h},\boldsymbol{\gamma}}^{\mathrm{(cste)}} - \xi\right)^{-1} \tau_{\mathpzc{h},\boldsymbol{\gamma}}.
\end{equation}
By making use of the second resolvent equation, $\mathcal{R}_{\mathpzc{h}}(\xi)$ in \eqref{chapS} can be rewritten as:
\begin{equation}
\label{idd}
\mathcal{R}_{\mathpzc{h}}(\xi) =  \mathscr{R}_{\mathpzc{h}}(\xi) + \mathscr{W}_{\mathpzc{h}}(\xi),
\end{equation}
where $\forall \mathpzc{h} \in (0,\mathpzc{h}_{0}]$ and $\forall \xi \in \mathbb{C} \setminus [-\Vert V \Vert_{\infty},\infty)$:
\begin{equation}
\label{Teps}
\mathscr{W}_{\mathpzc{h}}(\xi) := \sum_{\boldsymbol{\gamma} \in \mathscr{E}} \hat{\tau}_{\mathpzc{h},\boldsymbol{\gamma}} \left(\tilde{H}_{\mathpzc{h},\boldsymbol{\gamma}}^{\mathrm{(ref)}} - \xi\right)^{-1}\left\{\hat{\hat{\tau}}_{\mathpzc{h},\boldsymbol{\gamma}} \left(V(\mathpzc{h}^{1-\alpha} \boldsymbol{\gamma}) - V(\mathpzc{h}\cdot\,)\right)\right\} \left(\tilde{H}_{\mathpzc{h},\boldsymbol{\gamma}}^{\mathrm{(cste)}} - \xi\right)^{-1} \tau_{\mathpzc{h},\boldsymbol{\gamma}}.
\end{equation}
Below, we prove that the operator norm of $\mathscr{W}_{\mathpzc{h}}(\xi)$ behaves like $\mathcal{O}(\mathpzc{h}^{\theta(1-\alpha)})$, $\theta \in (0,1]$,  see \eqref{esfgv3}.\\
Gathering \eqref{approxres} and \eqref{idd} together, then we get in the bounded operators sense on $L^{2}(\mathbb{R}^{3})$:
\begin{equation}
\label{identtoot}
\left(\tilde{H}_{\mathpzc{h}} - \xi\right)^{-1} = \mathscr{R}_{\mathpzc{h}}(\xi) + \mathscr{W}_{\mathpzc{h}}(\xi) - \left(\tilde{H}_{\mathpzc{h}} - \xi\right)^{-1} \mathcal{W}_{\mathpzc{h}}(\xi).\\
\end{equation}

We end this paragraph by giving a series of estimates we will use throughout this paper.\\
Let us recall that from \cite[Thm. B.7.2]{Si1}, for any $\xi \in \varrho(\tilde{H}_{\mathpzc{h}})$ the resolvent operator $(\tilde{H}_{\mathpzc{h}}-\xi)^{-1}$ is an integral operator with integral kernel $(\tilde{H}_{\mathpzc{h}}-\xi)^{-1}(\cdot\,,\cdot\,)$ jointly continuous on $\mathbb{R}^{6}\setminus D$, $D:= \{(\mathbf{x},\mathbf{y}) \in \mathbb{R}^{6}: \mathbf{x} = \mathbf{y}\}$ standing for the diagonal. The same holds true for the integral kernel $(H_{\mathpzc{h}}-\xi)^{-1}(\cdot\,,\cdot\,)$. Below, for any $\zeta \in \mathbb{C}$ and real number $\ell >0$, we use the shorthand notation:
\begin{equation}
\label{short}
\ell_{\zeta} := \ell (1 + \vert \zeta\vert)^{-1}.
\end{equation}

\begin{lema}
\label{lem0}
For every $\eta>0$, there exists a constant $\vartheta = \vartheta(\eta) >0$ and a polynomial $p(\cdot\,)$ s.t. $\forall \mathpzc{h}>0$, $\forall \xi \in \mathbb{C}$ satisfying $\mathrm{dist}(\xi, \sigma(H_{\mathpzc{h}})) \geq \eta$ (or $\mathrm{dist}(\xi, \sigma(\tilde{H}_{\mathpzc{h}})) \geq \eta$) and $\forall(\mathbf{x},\mathbf{y}) \in \mathbb{R}^{6}\setminus D$:
\begin{align}
\label{trures}
\max\left\{\left\vert \left(H_{\mathpzc{h}} - \xi\right)^{-1}(\mathbf{x},\mathbf{y})\right\vert, \left\vert \left(\tilde{H}_{\mathpzc{h}} - \xi\right)^{-1}(\mathbf{x},\mathbf{y})\right\vert\right\} &\leq p(\vert \xi\vert) \frac{\mathrm{e}^{- \vartheta_{\xi} \vert \mathbf{x} - \mathbf{y}\vert}}{\vert \mathbf{x} - \mathbf{y}\vert},\\
\label{trueder}
\max\left\{\left\vert \nabla_{\mathbf{x}} \left(H_{\mathpzc{h}} - \xi\right)^{-1}(\mathbf{x},\mathbf{y})\right\vert, \left\vert \nabla_{\mathbf{x}} \left(\tilde{H}_{\mathpzc{h}} - \xi\right)^{-1}(\mathbf{x},\mathbf{y})\right\vert\right\} &\leq p(\vert \xi\vert) \frac{\mathrm{e}^{- \vartheta_{\xi} \vert \mathbf{x} - \mathbf{y}\vert}}{\vert \mathbf{x} - \mathbf{y}\vert^{2}}.
\end{align}
\end{lema}

\noindent \textbf{Proof}. \eqref{trures} and \eqref{trueder} follow from \cite[Thm. B.7.2]{Si1} and \cite[Lem. 2.4]{BS} respectively. \qed \\

In view of the definitions \eqref{locham} and \eqref{HconstV}, we straightforwardly get:

\begin{lema}
\label{lem01}
Let $0<\alpha<1$ and $\mathpzc{h}_{0}=\mathpzc{h}_{0}(\alpha) \leq 1$ as in \eqref{epsilon0}. For every $\eta>0$, there exists a constant $\vartheta = \vartheta(\eta) >0$ and a polynomial $p(\cdot\,)$ s.t. $\forall \mathpzc{h} \in (0,\mathpzc{h}_{0}]$, $\forall \boldsymbol{\gamma} \in \mathscr{E}$, $\forall \xi \in \mathbb{C}$ satisfying $\mathrm{dist}(\xi, [-\Vert V \Vert_{\infty},\infty)) \geq \eta$ and $\forall(\mathbf{x},\mathbf{y}) \in \mathbb{R}^{6}\setminus D$:
\begin{align}
\label{otyp}
\max\left\{\left\vert \left(\tilde{H}_{\mathpzc{h},\boldsymbol{\gamma}}^{\mathrm{(ref)}} - \xi\right)^{-1}(\mathbf{x},\mathbf{y})\right\vert, \left\vert \left(\tilde{H}_{\mathpzc{h},\boldsymbol{\gamma}}^{\mathrm{(cste)}} - \xi\right)^{-1}(\mathbf{x},\mathbf{y})\right\vert\right\} &\leq p(\vert \xi\vert) \frac{\mathrm{e}^{- \vartheta_{\xi} \vert \mathbf{x} - \mathbf{y}\vert}}{\vert \mathbf{x} - \mathbf{y}\vert},\\
\label{otypder}
\max\left\{\left\vert \nabla_{\mathbf{x}} \left(\tilde{H}_{\mathpzc{h},\boldsymbol{\gamma}}^{\mathrm{(ref)}} - \xi\right)^{-1}(\mathbf{x},\mathbf{y})\right\vert, \left\vert \nabla_{\mathbf{x}} \left(\tilde{H}_{\mathpzc{h},\boldsymbol{\gamma}}^{\mathrm{(cste)}} - \xi\right)^{-1}(\mathbf{x},\mathbf{y})\right\vert\right\} &\leq p(\vert \xi\vert) \frac{\mathrm{e}^{- \vartheta_{\xi} \vert \mathbf{x} - \mathbf{y}\vert}}{\vert \mathbf{x} - \mathbf{y}\vert^{2}}.
\end{align}
\end{lema}

\begin{remark} Since the potential in \eqref{HconstV} is nothing but a constant, then the Green function of the operator $\tilde{H}_{\mathpzc{h},\boldsymbol{\gamma}}^{\mathrm{(cste)}}$ is explicitly known. It reads on $\mathbb{R}^{6}\setminus D$ as, see e.g. \cite[Sec. 7.4]{Tes}:
\begin{equation}
\label{Greenf}
\left(\tilde{H}_{\mathpzc{h},\boldsymbol{\gamma}}^{\mathrm{(cste)}} - \xi\right)^{-1}(\mathbf{x},\mathbf{y}) = \frac{1}{2\pi} \frac{\mathrm{e}^{- \varsigma_{\mathpzc{h},\boldsymbol{\gamma}}(\xi) \vert \mathbf{x} - \mathbf{y}\vert}}{\vert \mathbf{x} - \mathbf{y}\vert}, \quad \varsigma_{\mathpzc{h},\boldsymbol{\gamma}}(\xi) := \sqrt{-2(\xi - V(\mathpzc{h}^{1-\alpha} \boldsymbol{\gamma}))}.
\end{equation}
\end{remark}

Finally, we have (the $\theta$ appearing in the below lemma corresponds to the $\theta$ in \eqref{deftheta}):

\begin{lema}
\label{lem1}
Let $0<\alpha<1$ and $\mathpzc{h}_{0}=\mathpzc{h}_{0}(\alpha) \leq 1$ as in \eqref{epsilon0}. For every $\eta>0$, there exists a constant $\vartheta=\vartheta(\eta)>0$ and a polynomial $p(\cdot\,)$ s.t.  $\forall\mathpzc{h} \in (0,\mathpzc{h}_{0}]$, $\forall \theta \in (0,1]$ and $\forall \xi \in \mathbb{C}$ satisfying $\mathrm{dist}(\xi,[-\Vert V\Vert_{\infty},\infty)) \geq \eta$:
\begin{align}
\label{kerSep}
\forall(\mathbf{x},\mathbf{y}) \in \mathbb{R}^{6}\setminus D,\quad \max\left\{\left\vert (\mathcal{R}_{\mathpzc{h}}(\xi))(\mathbf{x},\mathbf{y})\right\vert, \left\vert (\mathscr{R}_{\mathpzc{h}}(\xi))(\mathbf{x},\mathbf{y})\right\vert\right\} &\leq p(\vert \xi \vert) \frac{\mathrm{e}^{- \vartheta_{\xi}\vert \mathbf{x} - \mathbf{y}\vert}}{\vert \mathbf{x} - \mathbf{y}\vert},\\
\label{kerSepder}
\max\left\{\left\vert \nabla_{\mathbf{x}} (\mathcal{R}_{\mathpzc{h}}(\xi))(\mathbf{x},\mathbf{y})\right\vert, \left\vert \nabla_{\mathbf{x}} (\mathscr{R}_{\mathpzc{h}}(\xi))(\mathbf{x},\mathbf{y})\right\vert\right\} &\leq p(\vert \xi \vert) \frac{\mathrm{e}^{- \vartheta_{\xi}\vert \mathbf{x} - \mathbf{y}\vert}}{\vert \mathbf{x} - \mathbf{y}\vert^{2}};
\end{align}
\begin{align}
\label{kerT}
\forall(\mathbf{x},\mathbf{y}) \in \mathbb{R}^{6},\quad \left\vert \left(\mathcal{W}_{\mathpzc{h}}(\xi)\right)(\mathbf{x},\mathbf{y})\right\vert &\leq p(\vert \xi \vert) \mathrm{e}^{- \vartheta_{\xi} \mathpzc{h}^{- \alpha}} \mathrm{e}^{- \vartheta_{\xi}\vert \mathbf{x} - \mathbf{y}\vert},\\
\label{kerresT2}
\left\vert \left(\mathscr{W}_{\mathpzc{h}}(\xi)\right)(\mathbf{x},\mathbf{y})\right\vert &\leq p(\vert \xi \vert) \mathpzc{h}^{\theta(1 - \alpha)} \mathrm{e}^{- \vartheta_{\xi}\vert \mathbf{x} - \mathbf{y}\vert}.
\end{align}
\end{lema}

\noindent \textbf{Proof.} Let $0<\alpha<1$ and $\eta>0$ be fixed. In view of \eqref{chapS} and \eqref{Sepsi}, \eqref{kerSep} follows from \eqref{otyp}. Here, we used the estimate \eqref{norminf} uniform in $\boldsymbol{\gamma} \in \mathscr{E}$, along with the fact that:
\begin{equation}
\label{sumim}
\forall \mathbf{x} \in \mathbb{R}^{3},\quad \sum_{\boldsymbol{\gamma} \in \mathscr{E}} \tau_{\mathpzc{h},\boldsymbol{\gamma}}(\mathbf{x}) \leq C,
\end{equation}
for some $\mathpzc{h}$-independent constant $C\geq 1$. Indeed, the sum is zero if $\bold{x} \in \mathbb{R}^{3}\setminus (\cup_{\boldsymbol{\gamma} \in \mathscr{E}} \mathrm{Supp}(\tau_{\mathpzc{h},\boldsymbol{\gamma}}))$ and equals $1$ if $\bold{x} \in \Omega_{\mathpzc{h}}$, see \eqref{partiden0}-\eqref{partiden}. If $\bold{x} \in (\cup_{\boldsymbol{\gamma} \in \mathscr{E}} \mathrm{Supp}(\tau_{\mathpzc{h},\boldsymbol{\gamma}}))\setminus \Omega_{\mathpzc{h}}$, it is bounded since only a finite number of cutoff functions overlap. Next, we turn to \eqref{kerSepder}. $\forall \mathpzc{h} \in (0,\mathpzc{h}_{0}]$, we have on $\mathbb{R}^{6}\setminus D$:
\begin{multline*}
\nabla_{\mathbf{x}}(\mathcal{R}_{\mathpzc{h}}(\xi))(\mathbf{x},\mathbf{y}) = \\\sum_{\boldsymbol{\gamma} \in \mathscr{E}} \left\{\left(\nabla \hat{\tau}_{\mathpzc{h},\boldsymbol{\gamma}}\right)(\mathbf{x}) \left(\tilde{H}_{\mathpzc{h},\boldsymbol{\gamma}}^{\mathrm{(ref)}} - \xi\right)^{-1}(\mathbf{x},\mathbf{y}) + \hat{\tau}_{\mathpzc{h},\boldsymbol{\gamma}}(\mathbf{x}) \nabla_{\mathbf{x}}\left(\tilde{H}_{\mathpzc{h},\boldsymbol{\gamma}}^{\mathrm{(ref)}} - \xi\right)^{-1}(\mathbf{x},\mathbf{y})\right\}\tau_{\mathpzc{h},\boldsymbol{\gamma}}(\mathbf{y}).
\end{multline*}
Then \eqref{kerSepder} follows from \eqref{otyp}-\eqref{otypder} combined with \eqref{keyes} below, \eqref{norminf} and \eqref{sumim}. In view of \eqref{Sepsi}, the same holds true for $\vert\nabla_{\mathbf{x}}(\mathscr{R}_{\mathpzc{h}}(\xi))(\cdot\,,\cdot\,)\vert$. Next, we turn to \eqref{kerT}. From \eqref{support} along with \eqref{otyp}-\eqref{otypder}, there exists a constant $\vartheta>0$ and a polynomial $p(\cdot\,)$ s.t. $\forall\mathpzc{h} \in (0,\mathpzc{h}_{0}]$ and $\forall\boldsymbol{\gamma}\in \mathscr{E}$:
\begin{multline}
\label{bigl}
\forall(\mathbf{x},\mathbf{y})\in \mathbb{R}^{6},\quad \max\left\{\left\vert \left(\Delta \hat{\tau}_{\mathpzc{h},\boldsymbol{\gamma}}\right)(\mathbf{x}) \left(\tilde{H}_{\mathpzc{h},\boldsymbol{\gamma}}^{\mathrm{(ref)}} - \xi\right)^{-1}(\mathbf{x},\mathbf{y}) \tau_{\mathpzc{h},\boldsymbol{\gamma}}(\mathbf{y})\right\vert, \right.\\
\left. \left\vert \left(\nabla \hat{\tau}_{\mathpzc{h},\boldsymbol{\gamma}}\right)(\mathbf{x}) \cdot \nabla_{\mathbf{x}}\left(\tilde{H}_{\mathpzc{h},\boldsymbol{\gamma}}^{\mathrm{(ref)}} - \xi\right)^{-1}(\mathbf{x},\mathbf{y}) \tau_{\mathpzc{h},\boldsymbol{\gamma}}(\mathbf{y})\right\vert\right\} \leq p(\vert \xi\vert) \mathrm{e}^{- \vartheta_{\xi} \mathpzc{h}^{- \alpha}} \mathrm{e}^{- \vartheta_{\xi} \vert \mathbf{x} - \mathbf{y}\vert}.
\end{multline}
In view of \eqref{hatTep}, we obtain under the conditions of \eqref{bigl}:
\begin{equation*}
\forall(\mathbf{x},\mathbf{y}) \in \mathbb{R}^{6},\quad \left\vert \left(\mathcal{W}_{\mathpzc{h}}(\xi)\right)(\mathbf{x},\mathbf{y})\right\vert \leq 2 p(\vert \xi \vert) \mathpzc{h}^{-\alpha(3\alpha^{-1} - 3)} \mathrm{e}^{- \vartheta_{\xi} \mathpzc{h}^{- \alpha}} \mathrm{e}^{- \vartheta_{\xi}\vert \mathbf{x} - \mathbf{y}\vert},
\end{equation*}
where we used that $\mathrm{Card}(\mathscr{E}) = \mathcal{O}(\mathpzc{h}^{3\alpha - 3})$ to get rid of the sum. \eqref{kerT} follows from the estimate:
\begin{equation}
\label{keyes}
\forall \mu>0,\, \forall \nu>0, \quad t^{\nu} \mathrm{e}^{-\mu t} \leq \left(\frac{2 \nu}{\mathrm{e} \mu}\right)^{\nu} \mathrm{e}^{-\frac{\mu}{2} t},\quad t \geq 0.
\end{equation}
Finally, let us prove \eqref{kerresT2}. For any $\mathpzc{h} \in (0,\mathpzc{h}_{0}]$, the kernel of $\mathscr{W}_{\mathpzc{h}}(\xi)$ in \eqref{Teps} reads on $\mathbb{R}^{6}$ as:
\begin{equation}
\label{slot}
\begin{split}
\left(\mathscr{W}_{\mathpzc{h}}(\xi)\right)(\mathbf{x},\mathbf{y}) = &\sum_{\boldsymbol{\gamma} \in \mathscr{E}} \int_{\mathbb{R}^{3}} \mathrm{d}\mathbf{z}\, \hat{\tau}_{\mathpzc{h},\boldsymbol{\gamma}}(\mathbf{x}) \left(\tilde{H}_{\mathpzc{h},\boldsymbol{\gamma}}^{\mathrm{(ref)}} - \xi\right)^{-1}(\mathbf{x},\mathbf{z}) \\ &\times \left\{\hat{\hat{\tau}}_{\mathpzc{h},\boldsymbol{\gamma}}(\mathbf{z}) \left(V(\mathpzc{h}^{1-\alpha} \boldsymbol{\gamma}) - V(\mathpzc{h}\mathbf{z})\right)\right\} \left(\tilde{H}_{\mathpzc{h},\boldsymbol{\gamma}}^{\mathrm{(cste)}} - \xi\right)^{-1}(\mathbf{z},\mathbf{y}) \tau_{\mathpzc{h},\boldsymbol{\gamma}}(\mathbf{y}).
\end{split}
\end{equation}
From \eqref{deftheta} followed by \eqref{supp2hat}, there exist two constants $C, C^{'}>0$ independent of $\mathpzc{h}$ and $\boldsymbol{\gamma}$ s.t.
\begin{equation}
\label{accroissfi}
\forall\mathbf{z} \in \mathrm{Supp}\left(\hat{\hat{\tau}}_{\mathpzc{h},\boldsymbol{\gamma}}\right),\quad \left\vert V(\mathpzc{h}^{1-\alpha} \boldsymbol{\gamma}) - V(\mathpzc{h} \mathbf{z})\right\vert \leq C \mathpzc{h}^{\theta}  \left\vert \mathpzc{h}^{-\alpha} \boldsymbol{\gamma} - \mathbf{z} \right\vert^{\theta} \leq C^{'} \mathpzc{h}^{\theta(1-\alpha)}.
\end{equation}
It remains to use \eqref{norminf}, then \eqref{otyp} together with \cite[Lem. A.2]{BS}, and finally \eqref{sumim}.  \qed

\begin{remark}
From Lemmas \ref{lem01}-\ref{lem1} along with the Shur-Holmgren criterion, $\forall \mathpzc{h} \in (0,\mathpzc{h}_{0}]$:
\begin{gather}
\label{esfgv}
\max\left\{\left\Vert (H_{\mathpzc{h}} - \xi)^{-1} \right\Vert, \left\Vert (\tilde{H}_{\mathpzc{h}} - \xi)^{-1} \right\Vert, \left\Vert \mathcal{R}_{\mathpzc{h}}(\xi)\right\Vert, \left\Vert \nabla \mathcal{R}_{\mathpzc{h}}(\xi) \right\Vert, \left\Vert \mathscr{R}_{\mathpzc{h}}(\xi)\right\Vert, \left\Vert \nabla \mathscr{R}_{\mathpzc{h}}(\xi) \right\Vert\right\} \leq p(\vert \xi \vert),\\
\label{esfgv2}
\left\Vert \mathcal{W}_{\mathpzc{h}}(\xi)\right\Vert \leq p(\vert \xi\vert) \mathrm{e}^{- \vartheta_{\xi} \mathpzc{h}^{- \alpha}},\\
\label{esfgv3}
\Vert \mathscr{W}_{\mathpzc{h}}(\xi)\Vert \leq p(\vert \xi\vert) \mathpzc{h}^{\theta(1- \alpha)},
\end{gather}
for another constant $\vartheta>0$ and polynomial $p(\cdot\,)$ both independent of $\mathpzc{h} \in (0,\mathpzc{h}_{0}]$.
\end{remark}

\begin{remark}
\label{HSn}
Let $(\mathfrak{I}_{2}(L^{2}(\mathbb{R}^{3})),\Vert \cdot\,\Vert_{\mathfrak{I}_{2}})$ be the Banach space of Hilbert-Schmidt operators on $L^{2}(\mathbb{R}^{3})$. Under the conditions of Lemma \ref{lem01}, we have $\forall \mathpzc{h} \in (0,\mathpzc{h}_{0}]$ and $\forall \boldsymbol{\gamma} \in \mathscr{E}$:
\begin{gather}
\label{I2n1}
\max\left\{\left\Vert \hat{\tau}_{\mathpzc{h},\boldsymbol{\gamma}} \left(\tilde{H}_{\mathpzc{h},\boldsymbol{\gamma}}^{(\wp)} - \xi\right)^{-1}\right\Vert_{\mathfrak{I}_{2}}, \left\Vert \left(\tilde{H}_{\mathpzc{h},\boldsymbol{\gamma}}^{(\wp)} - \xi\right)^{-1} \tau_{\mathpzc{h},\boldsymbol{\gamma}} \right\Vert_{\mathfrak{I}_{2}}\right\} \leq p(\vert \xi\vert) \mathpzc{h}^{-\frac{3}{2}\alpha},\quad \wp = \mathrm{ref}\,\, \textrm{or}\,\, \mathrm{cste},\\
\max\left\{\left\Vert (\Delta \hat{\tau}_{\mathpzc{h},\boldsymbol{\gamma}}) \left(\tilde{H}_{\mathpzc{h},\boldsymbol{\gamma}}^{(\mathrm{ref})} - \xi\right)^{-1}\tau_{\mathpzc{h},\boldsymbol{\gamma}}\right\Vert_{\mathfrak{I}_{2}}, \left\Vert (\nabla \hat{\tau}_{\mathpzc{h},\boldsymbol{\gamma}}) \cdot \nabla \left(\tilde{H}_{\mathpzc{h},\boldsymbol{\gamma}}^{(\mathrm{ref})} - \xi\right)^{-1}\tau_{\mathpzc{h},\boldsymbol{\gamma}}\right\Vert_{\mathfrak{I}_{2}}\right\} \leq p(\vert \xi\vert) \mathrm{e}^{-\vartheta_{\xi} \mathpzc{h}^{-\alpha}},\nonumber
\end{gather}
for some constant $\vartheta>0$ and polynomial $p(\cdot\,)$ both independent of $\mathpzc{h},\boldsymbol{\gamma}$.
\end{remark}

\section{Proof of Theorem \ref{THMF}.}
\label{sec3t}


\subsection{A preliminary result.}
\label{invfug}

The main result of this paragraph gives an asymptotic expansion in the semiclassical limit for the unique solution of the equation \eqref{invdensi}:

\begin{proposition}
\label{lemrho}
Suppose that the assumptions $(\mathscr{A}_{\mathrm{p}})$ and $(\mathscr{A}_{\mathrm{r}})$ hold.\\
For any $0<\alpha<1$, $0 < \theta \leq 1$, $\beta >0$ and $\rho>0$, it holds:
\begin{equation}
\label{semifug}
\overline{z}_{\mathpzc{h}}\left(\beta,\frac{\rho}{2},0\right) =  \frac{1}{2} \rho\vert \Omega \vert (2\pi \beta)^{\frac{3}{2}} \left(\int_{\Omega} \mathrm{d}\mathbf{x}\, \mathrm{e}^{-\beta V(\mathbf{x})}\right)^{-1} \mathpzc{h}^{3} + \mathcal{O}\left(\mathpzc{h}^{3+ \theta(1-\alpha)}\right).
\end{equation}
\end{proposition}

The rest of this section is devoted to the proof of Proposition \ref{lemrho}.\\
We start by writing down a suitable expression for the bulk density of the Fermi gas of spin $\frac{1}{2}$.\\ In the grand-canonical situation, let $\beta:= (k_{B}T)^{-1} >0$ and $z := \mathrm{e}^{\beta \mu} >0$. In view of \eqref{infspectrum}, let $\mathscr{C}_{\beta}$ be the counter-clockwise oriented simple contour around the interval $[-\Vert V \Vert_{\infty}, \infty)$ defined as:
\begin{gather}
\label{Gamma}
\mathscr{C}_{\beta} := \left\{ \Re \xi \in [\delta,\infty),\, \Im\xi = \pm \frac{\pi}{2\beta}\right\} \cup \left\{ \Re \xi = \delta,\, \Im\xi \in \left[-\frac{\pi}{2\beta},\frac{\pi}{2\beta}\right]\right\},\\
\label{delta}
\delta := -\Vert V\Vert_{\infty} - 1.
\end{gather}
The closed subset surrounding by $\mathscr{C}_{\beta}$ is a strict subset of $\mathfrak{D} := \{\zeta \in \mathbb{C}: \Im \zeta \in (-\frac{\pi}{\beta},\frac{\pi}{\beta})\}$, the holomorphic domain of the Fermi-Dirac distribution function  $\mathfrak{f}_{FD}(\beta,z;\xi) := z \mathrm{e}^{-\beta \xi}(1 + z \mathrm{e}^{-\beta \xi})^{-1}$. Note that $\mathfrak{f}_{FD}(\beta,z;\cdot\,)$ admits an exponential decay on $\mathscr{C}_{\beta}$, i.e. there exists a constant $c>0$ s.t.
\begin{equation}
\label{estimo}
\forall \beta>0,\,\forall\xi\in\mathscr{C}_{\beta},\quad  \left\vert \mathfrak{f}_{FD}(\beta,z;\xi)\right\vert \leq c z \mathrm{e}^{-\beta \Re \xi}.
\end{equation}
The function $\mathfrak{f}(\beta,z;\xi) := \ln(1 + z \mathrm{e}^{-\beta \xi})$ satisfying $(\partial_{\xi} \mathfrak{f})(\beta,z;\xi) = - \beta \mathfrak{f}_{FD}(\beta,z;\xi)$ also obeys \eqref{estimo}.\\
\indent The bulk density reads $\forall \beta>0$, $\forall z>0$, $\forall b \in \mathbb{R}$ and $\forall\mathpzc{h}>0$ as, see e.g. \cite[Eq. (2.2)]{BCS2}:
\begin{equation}
\label{rhobgf}
\rho_{\mathpzc{h}}(\beta,z,b) = \frac{1}{\vert \Omega \vert} \frac{i}{2\pi} \mathrm{Tr}_{L^{2}(\mathbb{R}^{3};\mathbb{C}^{2})}\left\{\chi_{\Omega}\mathpzc{I}_{\mathrm{d}} \left(\int_{\mathscr{C}_{\beta}} \mathrm{d}\xi\, \mathfrak{f}_{FD}(\beta,z;\xi) (\mathpzc{H}_{\mathpzc{h}}(b) - \xi)^{-1}\right)\chi_{\Omega}\mathpzc{I}_{\mathrm{d}}\right\},
\end{equation}
where $\chi_{\Omega}\mathpzc{I}_{\mathrm{d}} = \chi_{\Omega} \oplus \chi_{\Omega}$ denotes the multiplication operator by the indicator function of the unit cell $\Omega$ on $L^{2}(\mathbb{R}^{3};\mathbb{C}^{2})$, and $(\mathpzc{H}_{\mathpzc{h}}(b) - \xi)^{-1}$ the resolvent operator of $\mathpzc{H}_{\mathpzc{h}}(b)$ in \eqref{Hinfini}:
\begin{equation*}
(\mathpzc{H}_{\mathpzc{h}}(b) - \xi)^{-1} := \begin{pmatrix} (H_{\mathpzc{h}}^{-}(b) - \xi)^{-1} & 0 \\ 0 & (H_{\mathpzc{h}}^{+}(b) - \xi)^{-1}\end{pmatrix} = (H_{\mathpzc{h}}^{-}(b) - \xi)^{-1} \oplus (H_{\mathpzc{h}}^{+}(b) - \xi)^{-1},
\end{equation*}
with $H_{\mathpzc{h}}^{\mp}(b)$ defined in \eqref{Hinfini1}. Under the conditions of \eqref{rhobgf}, we have:
\begin{gather*}
\rho_{\mathpzc{h}}(\beta,z,b) = \rho_{\mathpzc{h}}^{-}(\beta,z,b) + \rho_{\mathpzc{h}}^{+}(\beta,z,b),\\
\rho_{\mathpzc{h}}^{\mathpzc{s}}(\beta,z,b) := \frac{1}{\vert \Omega \vert} \frac{i}{2\pi} \mathrm{Tr}_{L^{2}(\mathbb{R}^{3})}\left\{\chi_{\Omega}\left(\int_{\mathscr{C}_{\beta}} \mathrm{d}\xi\, \mathfrak{f}_{FD}(\beta,z;\xi) (H_{\mathpzc{h}}^{\mathpzc{s}}(b) - \xi)^{-1}\right)\chi_{\Omega}\right\},\quad \mathpzc{s} \in \{-,+\}.
\end{gather*}
When the magnetic field vanishes, we have $H_{\mathpzc{h}}^{+}(0)= H_{\mathpzc{h}}^{-}(0)=:H_{\mathpzc{h}}$ and then:
\begin{equation}
\label{vrairho}
\rho_{\mathpzc{h}}(\beta,z,0) =2 \rho_{\mathpzc{h}}^{\mathpzc{s}}(\beta,z,0) = \frac{2}{\vert \Omega \vert} \frac{i}{2\pi} \mathrm{Tr}_{L^{2}(\mathbb{R}^{3})}\left\{\chi_{\Omega}\left(\int_{\mathscr{C}_{\beta}} \mathrm{d}\xi\, \mathfrak{f}_{FD}(\beta,z;\xi) (H_{\mathpzc{h}} - \xi)^{-1}\right)\chi_{\Omega}\right\}.
\end{equation}
Performing an integration by parts w.r.t. the $\xi$-variable, the r.h.s. of \eqref{vrairho} can be rewritten as:
\begin{equation}
\label{vrairho2}
\rho_{\mathpzc{h}}(\beta,z,0) = \frac{2}{\beta \vert \Omega \vert} \frac{i}{2\pi} \int_{\mathscr{C}_{\beta}} \mathrm{d}\xi\, \mathfrak{f}(\beta,z;\xi) \int_{\Omega} \mathrm{d}\mathbf{x}\, (H_{\mathpzc{h}} - \xi)^{-2}(\mathbf{x},\mathbf{x}).
\end{equation}
Here, we used that $(H_{\mathpzc{h}} - \xi)^{-2}$ is locally trace-class on $L^{2}(\mathbb{R}^{3})$ since $(H_{\mathpzc{h}} - \xi)^{-1}$ is locally Hilbert-Schmidt on $L^{2}(\mathbb{R}^{3})$.  Note that the integral kernel $(H_{\mathpzc{h}} - \xi)^{-2}(\cdot\,,\cdot\,)$ is jointly continuous on $\mathbb{R}^{6}$ and its diagonal part is uniformly bounded by some polynomial in $\vert \xi\vert$, see the estimate in \eqref{trures} combined with \cite[Lem. A.1 \& A.2]{BS}. Therefore, \eqref{vrairho2} is well-defined due to  \eqref{estimo}.\\
\indent Next, let us rewrite the r.h.s. of \eqref{vrairho} in a more convenient way. In Sec. \ref{appr}, we gave  an approximation of the resolvent operator $(\tilde{\mathpzc{H}}_{\mathpzc{h}} - \xi)^{-1}$ via a geometric perturbation theory. Remind that under the transformation \eqref{trnsfounit}, the operators $\tilde{\mathpzc{H}}_{\mathpzc{h}}$ and $\mathpzc{H}_{\mathpzc{h}}:= \mathpzc{H}_{\mathpzc{h}}(b=0)$ are unitarily equivalent. Since:
\begin{align}
\label{equivun}
\mathpzc{U}_{\mathpzc{h}} \left(\mathpzc{H}_{\mathpzc{h}}-\xi\right)^{-1} \mathpzc{U}_{\mathpzc{h}}^{-1} &= \left(\tilde{\mathpzc{H}}_{\mathpzc{h}} -\xi\right)^{-1},\\
\label{equivun1}
\mathpzc{U}_{\mathpzc{h}} \chi_{\Omega}\mathpzc{I}_{\mathrm{d}} \mathpzc{U}_{\mathpzc{h}}^{-1} &= \chi_{\Omega_{\mathpzc{h}}} \mathpzc{I}_{\mathrm{d}},
\end{align}
where $\Omega_{\mathpzc{h}} := (-\frac{1}{2\mathpzc{h}},\frac{1}{2\mathpzc{h}})^{3}$ denotes the dilated unit cube centered at the origin of coordinates, then
under the conditions of \eqref{vrairho}, the bulk zero-field density of particles can be rewritten as:
\begin{equation}
\label{rhovr}
\rho_{\mathpzc{h}}(\beta,z,0) = \frac{2}{\vert \Omega \vert} \frac{i}{2\pi} \mathrm{Tr}_{L^{2}(\mathbb{R}^{3})} \left\{\chi_{\Omega_{\mathpzc{h}}} \left(\int_{\mathscr{C}_{\beta}} \mathrm{d}\xi\, \mathfrak{f}_{FD}(\beta,z;\xi) \left(\tilde{H}_{\mathpzc{h}} - \xi\right)^{-1}\right)\chi_{\Omega_{\mathpzc{h}}}\right\}.
\end{equation}

The continuation of the proof consists in using the approximation \eqref{identtoot} in order to isolate the main $\mathpzc{h}$-dependent contribution from \eqref{rhovr}. Due to \eqref{identtoot} along with \eqref{esfgv2}-\eqref{esfgv3}, we expect the contribution involving only the operator $\mathscr{R}_{\mathpzc{h}}(\xi)$ defined in \eqref{Sepsi} to give rise to the leading term in the asymptotic expansion. This is in fact the case, and furthermore:

\begin{proposition}
\label{asymrho}
For any $0 < \alpha < 1$, $0<\theta \leq 1$, $\beta >0$ and $z>0$, it holds:
\begin{equation}
\label{semidens}
\rho_{\mathpzc{h}}(\beta,z,0) = \frac{2}{(2\pi \mathpzc{h})^{3} \vert \Omega \vert} \int_{\Omega} \mathrm{d}\mathbf{x} \int_{\mathbb{R}^{3}} \mathrm{d}\mathbf{k}\, \frac{z \mathrm{e}^{-\beta\left(\frac{1}{2} \mathbf{k}^{2} + V(\mathbf{x})\right)}}{1+ z\mathrm{e}^{-\beta\left(\frac{1}{2}\mathbf{k}^{2} + V(\mathbf{x})\right)}} + \mathcal{O}\left(z \mathpzc{h}^{-3 + \theta(1-\alpha)}\right).
\end{equation}
\end{proposition}

The proof of Proposition \ref{asymrho} is based on the two following lemmas whose proof lie in Sec. \ref{append21}:

\begin{lema}
\label{isola0}
Let $0 < \alpha < 1$ and $\mathpzc{h}_{0}= \mathpzc{h}_{0}(\alpha)\leq 1$ as in \eqref{epsilon0}. Let $\mathscr{C}_{\beta}$, $\beta>0$ as in \eqref{Gamma}.
Then:\\
$\mathrm{(i)}$. $\forall N >0$ there exists a polynomial $p(\cdot\,)$ s.t. $\forall\mathpzc{h} \in (0,\mathpzc{h}_{0}]$ and $\forall \xi \in \mathscr{C}_{\beta}$:
\begin{equation}
\label{rftgh}
\left\vert \mathrm{Tr}_{L^{2}(\mathbb{R}^{3})}\left\{\chi_{\Omega_{\mathpzc{h}}}\left(\tilde{H}_{\mathpzc{h}} - \xi\right)^{-1}\mathcal{W}_{\mathpzc{h}}(\xi)\chi_{\Omega_{\mathpzc{h}}}\right\}\right\vert \leq p(\vert \xi \vert) \mathpzc{h}^{N}.
\end{equation}
$\mathrm{(ii)}$. There exists another polynomial $p(\cdot\,)$ s.t. $\forall \theta \in (0,1]$, $\forall\mathpzc{h} \in (0,\mathpzc{h}_{0}]$ and $\forall \xi \in \mathscr{C}_{\beta}$:
\begin{equation}
\label{rftgh2}
\left\vert \mathrm{Tr}_{L^{2}(\mathbb{R}^{3})}\left\{\chi_{\Omega_{\mathpzc{h}}} \mathscr{W}_{\mathpzc{h}}(\xi)\chi_{\Omega_{\mathpzc{h}}}\right\}\right\vert \leq p(\vert \xi\vert) \mathpzc{h}^{-3 + \theta(1-\alpha)}.
\end{equation}
\end{lema}

\begin{lema}
\label{isola1}
For any $0 < \alpha <1$, $0<\theta \leq 1$, $\beta >0$ and $z>0$, it holds:
\begin{multline}
\label{rftgh3}
\frac{2}{\vert \Omega\vert} \frac{i}{2\pi} \mathrm{Tr}_{L^{2}(\mathbb{R}^{3})}\left\{\chi_{\Omega_{\mathpzc{h}}} \left(\int_{\mathscr{C}_{\beta}} \mathrm{d}\xi\, \mathfrak{f}_{FD}(\beta,z;\xi) \mathscr{R}_{\mathpzc{h}}(\xi)\right)\chi_{\Omega_{\mathpzc{h}}}\right\} = \\
\frac{2}{(2\pi \mathpzc{h})^{3} \vert \Omega\vert} \int_{\Omega} \mathrm{d}\mathbf{x} \int_{\mathbb{R}^{3}} \mathrm{d}\mathbf{k}\, \frac{z \mathrm{e}^{-\beta\left(\frac{1}{2}\mathbf{k}^{2} + V(\mathbf{x})\right)}}{1+ z \mathrm{e}^{-\beta\left(\frac{1}{2}\mathbf{k}^{2} + V(\mathbf{x})\right)}} + \mathcal{O}\left(z\mathpzc{h}^{-3 + \theta(1-\alpha)}\right).
\end{multline}
\end{lema}

\noindent \textbf{Proof of Proposition \ref{asymrho}}. 
From \eqref{rhovr} along with \eqref{identtoot}, then $\forall \beta>0$, $\forall z>0$ and $\forall \mathpzc{h} \in (0,\mathpzc{h}_{0}]$:
\begin{multline}
\label{secote}
\rho_{\mathpzc{h}}(\beta,z,0) = \frac{2}{\vert \Omega\vert} \frac{i}{2\pi} \mathrm{Tr}_{L^{2}(\mathbb{R}^{3})}\left\{\chi_{\Omega_{\mathpzc{h}}} \left(\int_{\mathscr{C}_{\beta}} \mathrm{d}\xi\, \mathfrak{f}_{FD}(\beta,z;\xi) \mathscr{R}_{\mathpzc{h}}(\xi)\right)\chi_{\Omega_{\mathpzc{h}}}\right\} \\
+ \frac{2}{\vert \Omega\vert} \frac{i}{2\pi} \int_{\mathscr{C}_{\beta}} \mathrm{d}\xi\, \mathfrak{f}_{FD}(\beta,z;\xi) \mathrm{Tr}_{L^{2}(\mathbb{R}^{3})}\left\{\chi_{\Omega_{\mathpzc{h}}} \left[\mathscr{W}_{\mathpzc{h}}(\xi) - \left(\tilde{H}_{\mathpzc{h}} - \xi\right)^{-1}\mathcal{W}_{\mathpzc{h}}(\xi)\right]\chi_{\Omega_{\mathpzc{h}}}\right\}.
\end{multline}
Now, \eqref{semidens} follows from Lemma \ref{isola1}, together with \eqref{rftgh}-\eqref{rftgh2} combined with \eqref{estimo}. \qed \\

We are now ready for:\\
\noindent \textit{\textbf{Proof of Proposition \ref{lemrho}}}. For any $\beta>0$ and $z \geq 0$, define
$\forall \mathpzc{h}>0$ the following quantity:
\begin{equation}
\label{rhdfgy}
R_{\mathpzc{h}}(\beta,z) := \beta (\partial_{z} \widehat{P}_{\mathpzc{h}})(\beta,z,0),
\end{equation}
where $\widehat{P}_{\mathpzc{h}}(\beta,\cdot\,,b)$ denotes the analytic continuation of the bulk grand-canonical pressure of the Fermi gas of spin $\frac{1}{2}$ $P_{\mathpzc{h}}(\beta,\cdot\,,b)$ to the domain $\mathbb{C}\setminus (-\infty,-\mathrm{e}^{\beta \mathpzc{E}_{\mathpzc{h}}(b)}]$ with $\mathpzc{E}_{\mathpzc{h}}(b) := \inf \sigma(\mathpzc{H}_{\mathpzc{h}}(b))$, see
\cite[Thm. 1.1 (ii)]{BS}. Thus, $z \mapsto R_{\mathpzc{h}}(\beta,z)$ is continuous. Besides, one can prove that $R_{\mathpzc{h}}(\beta,\cdot\,)$ never vanishes on $[0,\infty)$ and it is a strictly decreasing function. These features are used in the following.\\
\indent From \eqref{semidens} together with \eqref{rhdfgy}, introduce $\forall \beta>0$, $\forall z \geq 0$ and $\forall \mathpzc{h}>0$:
\begin{equation}
\label{lacalA}
\Theta_{\mathpzc{h}}(\beta,z) := R_{\mathpzc{h}}(\beta,z)
- \frac{2}{\vert \Omega \vert (2\pi \mathpzc{h})^{3}} \int_{\Omega} \mathrm{d}\mathbf{x}\int_{\mathbb{R}^{3}} \mathrm{d}\mathbf{k}\, \frac{ \mathrm{e}^{- \beta\left(\frac{1}{2} \mathbf{k}^{2} + V(\mathbf{x})\right)}}{1 + z \mathrm{e}^{- \beta\left(\frac{1}{2} \mathbf{k}^{2} + V(\mathbf{x})\right)}}.
\end{equation}
Let $\rho>0$ be fixed. For any $\mathpzc{h}>0$, let $\overline{z}_{\mathpzc{h}}(\beta) = \overline{z}_{\mathpzc{h}}(\beta,\frac{\rho}{2},0)>0$ be the unique solution of the equation $\rho=\rho_{\mathpzc{h}}(\beta,z,0) = z R_{\mathpzc{h}}(\beta,z)$. Under the conditions of Proposition \ref{lemrho}, by \eqref{lacalA} we have:
\begin{equation}
\label{xcxc}
\forall \mathpzc{h}>0,\quad \overline{z}_{\mathpzc{h}}(\beta) =  \mathcal{G}_{\mathpzc{h}}\left(\beta,\overline{z}_{\mathpzc{h}}(\beta)\right),
\end{equation}
with for any $u \in \mathbb{R}_{+}$:
\begin{equation}
\label{pzcG}
\mathcal{G}_{\mathpzc{h}}(\beta,u) := \frac{\rho}{R_{\mathpzc{h}}(\beta,u)} = \frac{1}{2} \frac{\rho \vert \Omega \vert (2\pi \mathpzc{h})^{3}}{\displaystyle{\int_{\Omega} \mathrm{d}\mathbf{x}\int_{\mathbb{R}^{3}} \mathrm{d}\mathbf{k}\, \frac{\mathrm{e}^{- \beta\left(\frac{1}{2} \mathbf{k}^{2} + V(\mathbf{x})\right)}}{1 + u\mathrm{e}^{- \beta\left(\frac{1}{2} \mathbf{k}^{2} + V(\mathbf{x})\right)}} + \frac{\vert \Omega \vert (2\pi \mathpzc{h})^{3}}{2} \Theta_{\mathpzc{h}}(\beta,u)}}.
\end{equation}
Therefore, $\overline{z}_{\mathpzc{h}}(\beta)$ obeys a fixed-point equation.
Let us continue by giving a series of estimates.
Let $0<\alpha <1$, $0<\theta \leq 1$. From \eqref{semidens}, there exists a constant $c=c(\beta)>0$ s.t.
\begin{equation}
\label{calAasymptot}
\forall \mathpzc{h} \in (0,\mathpzc{h}_{0}],\,\forall z \geq 0,\quad \vert \Theta_{\mathpzc{h}}(\beta,z)\vert \leq c \mathpzc{h}^{-3 + \theta(1-\alpha)}.
\end{equation}
In view of \eqref{xcxc}-\eqref{pzcG} together with \eqref{calAasymptot}, we then expect $\overline{z}_{\mathpzc{h}}(\beta)$ to behave like $\mathcal{O}(\mathpzc{h}^{3})$. Next, we need to estimate $(\partial_{z}\Theta_{\mathpzc{h}})(\beta,\cdot\,)$. In Sect. \ref{append21}, we give an explicit expression of $\Theta_{\mathpzc{h}}(\beta,\cdot\,)$. It is in fact the sum of the second term in the r.h.s. of \eqref{secote} and the second term in the r.h.s. of \eqref{groPh} but with the function $\xi \mapsto \mathrm{e}^{-\beta \xi}(1+z \mathrm{e}^{-\beta \xi})^{-1} = z^{-1} \mathfrak{f}_{FD}(\beta,z;\xi)$ instead of $\xi \mapsto \mathfrak{f}_{FD}(\beta,z;\xi)$. From these expressions, we can prove that there exists another constant $c=c(\beta)>0$ s.t.
\begin{equation}
\label{calAasymptot2}
\forall \mathpzc{h} \in (0,\mathpzc{h}_{0}],\,\forall z \geq 0,\quad \vert (\partial_{z} \Theta_{\mathpzc{h}})(\beta,z)\vert \leq c \mathpzc{h}^{-3 + \theta(1-\alpha)}.
\end{equation}
From \eqref{pzcG} and \eqref{calAasymptot}-\eqref{calAasymptot2}, then for every $\kappa>0$, there exists a constant $c_{\kappa}(\beta,\rho)>0$ s.t.
\begin{equation*}
\textrm{$\forall \mathpzc{h} \in (0,\mathpzc{h}_{0}]$ and small enough},\quad \sup_{u \in [0,\kappa]} \vert (\partial_{u} \mathcal{G}_{\mathpzc{h}})(\beta,u)\vert \leq c_{\kappa}(\beta,\rho) \mathpzc{h}^{3}.
\end{equation*}
Therefore, $\sup_{u \in [0,\kappa]} \vert (\partial_{u} \mathcal{G}_{\mathpzc{h}})(\beta,u)\vert < 1$  for $\mathpzc{h}$ sufficiently small. Hence, for such $\mathpzc{h}$'s, $\mathcal{G}_{\mathpzc{h}}(\beta,\cdot\,)$ is a contraction, and then $\overline{z}_{\mathpzc{h}}(\beta)$ must belong to the interval $[0,\kappa]$. It remains to use the iteration procedure provided by the Banach fixed-point theorem to obtain the leading term in \eqref{semifug}. \qed

\subsection{Proof of Theorem \ref{THMF} $\mathrm{(i)}$.}
\label{THMi}

Let us introduce the Legendre-transform of the bulk grand-canonical pressure in \eqref{bulkpres} for the Fermi gas of spin $\frac{1}{2}$ with identical spin projection $m_{s}=\mathpzc{s}\frac{1}{2}$. It is defined $\forall \beta>0$, $\forall \rho^{\mathpzc{s}}>0$, $\forall b \in \mathbb{R}$ and $\forall \mathpzc{h}>0$ as:
\begin{equation}
\label{Tleg}
\begin{split}
(P_{\mathpzc{h}}^{\mathpzc{s}})^{*}(\beta,\rho^{\mathpzc{s}},b) :&= \sup_{\mu \in \mathbb{R}} \left(\rho^{\mathpzc{s}}\mu - P_{\mathpzc{h}}^{\mathpzc{s}}\left(\beta,\mathrm{e}^{\beta \mu},b\right)\right)
\\&= \frac{\rho^{\mathpzc{s}}}{\beta} \log\left(\overline{z}_{\mathpzc{h}}^{\mathpzc{s}}(\beta,\rho^{\mathpzc{s}},b)\right) - P_{\mathpzc{h}}^{\mathpzc{s}}\left(\beta, \overline{z}_{\mathpzc{h}}^{\mathpzc{s}}(\beta,\rho^{\mathpzc{s}},b),b\right),\quad \mathpzc{s} \in \{-,+\},
\end{split}
\end{equation}
where $\overline{z}_{\mathpzc{h}}^{\mathpzc{s}}(\beta,\rho^{\mathpzc{s}},b)>0$ denotes the unique solution of the equation $\rho_{\mathpzc{h}}^{\mathpzc{s}}(\beta,z,b) = \rho^{\mathpzc{s}}$. The quantity in \eqref{Tleg} is well-defined since $\mathbb{R} \owns \mu \mapsto P_{\mathpzc{h}}^{\mathpzc{s}}(\beta,\mathrm{e}^{\beta \mu},b)$ is a convex function, see below \eqref{susceptilim}.\\
\indent We start with the following lemma whose proof lies in the Appendix, see Sec. \ref{append25}:

\begin{lema}
\label{derPstar}
For any $\beta>0$, $\rho^{\mathpzc{s}}>0$, $\mathpzc{h}>0$, $\mathbb{R} \owns b \mapsto (P_{\mathpzc{h}}^{\mathpzc{s}})^{*}(\beta,\rho^{\mathpzc{s}},b)$ is a $\mathcal{C}^{\infty}$-function. Moreover:
\begin{gather}
\label{truidma}
-\left(\frac{q}{c}\right) \frac{\partial (P_{\mathpzc{h}}^{\mathpzc{s}})^{*}}{\partial b}(\beta,\rho^{\mathpzc{s}},b) = \mathcal{M}_{\mathpzc{h}}^{\mathpzc{s}\,\mathrm{(GC)}}(\beta,\rho^{\mathpzc{s}},b), \quad b \in \mathbb{R},\, \mathpzc{s} \in \{-,+\},\\
\label{truid}
-\left(\frac{q}{c}\right)^{2} \frac{\partial^{2} (P_{\mathpzc{h}}^{\mathpzc{s}})^{*}}{\partial b^{2}}(\beta,\rho^{\mathpzc{s}},0) =  \mathcal{X}_{\mathpzc{h}}^{\mathpzc{s}\,\mathrm{(GC)}}(\beta,\rho^{\mathpzc{s}},0).
\end{gather}
\end{lema}

\begin{remark} We emphasize that an identity of type \eqref{truid} generically does not hold in non-vanishing magnetic field. For further details, see Remark \ref{canorem}.
\end{remark}

Subsequently, we switch to the 'true' canonical ensemble. In \eqref{Helm}, we introduced the finite-volume free energy density $\mathcal{F}_{\mathpzc{h},L}^{\mathpzc{s}}$ of the Fermi gas of spin $\frac{1}{2}$ with identical spin projection $m_{s}=\mathpzc{s}\frac{1}{2}$ at fixed density $\rho^{\mathpzc{s}}>0$. Remind that $E_{\mathpzc{h}}^{\mathpzc{s}}(b) := \inf \sigma(H_{\mathpzc{h}}^{\mathpzc{s}}(b))$, $\mathpzc{s} \in \{-,+\}$. The proof of Theorem \ref{THMF} $\mathrm{(i)}$ is based on the following:

\begin{proposition}
\label{weakequi}
For any $\beta>0$, $b \in \mathbb{R}$, $\mathpzc{h}>0$ and
$\rho^{\mathpzc{s}} \in (0, \rho_{\mathpzc{h}}^{\mathpzc{s}}(\beta,\mathrm{e}^{\beta E_{\mathpzc{h}}^{\mathpzc{s}}(b)},b))$, it holds:
\begin{gather}
\label{enlib}
\mathcal{F}_{\mathpzc{h}}^{\mathpzc{s}}(\beta,\rho^{\mathpzc{s}},b):= \lim_{L \uparrow \infty} \mathcal{F}_{\mathpzc{h},L}^{\mathpzc{s}}(\beta,\rho^{\mathpzc{s}},b) = (P_{\mathpzc{h}}^{\mathpzc{s}})^{*}(\beta,\rho^{\mathpzc{s}},b),\\
\label{magzcano}
\mathcal{M}_{\mathpzc{h}}^{\mathpzc{s}\,\mathrm{(C)}}(\beta,\rho^{\mathpzc{s}},b) := - \lim_{L \uparrow \infty} \left(\frac{q}{c}\right) \frac{\partial \mathcal{F}_{\mathpzc{h},L}^{\mathpzc{s}}}{\partial b}(\beta,\rho^{\mathpzc{s}},b) = - \left(\frac{q}{c}\right) \frac{\partial (P_{\mathpzc{h}}^{\mathpzc{s}})^{*}}{\partial b}(\beta,\rho^{\mathpzc{s}},b),\\
\label{succano}
\mathcal{X}_{\mathpzc{h}}^{\mathpzc{s}\,\mathrm{(C)}}(\beta,\rho^{\mathpzc{s}},b) := - \lim_{L \uparrow \infty} \left(\frac{q}{c}\right)^{2} \frac{\partial^{2} \mathcal{F}_{\mathpzc{h},L}^{\mathpzc{s}}}{\partial b^{2}}(\beta,\rho^{\mathpzc{s}},b) = - \left(\frac{q}{c}\right)^{2} \frac{\partial^{2} (P_{\mathpzc{h}}^{\mathpzc{s}})^{*}}{\partial b^{2}}(\beta,\rho^{\mathpzc{s}},b).
\end{gather}
\end{proposition}

The proof of Proposition \ref{weakequi} lies in Appendix, see Sec. \ref{append25}. We are now ready for:\\
\noindent \textit{\textbf{Proof of Theorem \ref{THMF} $\mathrm{(i)}$}}. Let $\beta>0$ and $\rho^{\mathpzc{s}}>0$ be fixed. In view of \eqref{succano} along with \eqref{truid}, the only thing we have to prove is that for $\mathpzc{h}>0$ sufficiently small:
\begin{equation*}
\rho^{\mathpzc{s}} = \rho_{\mathpzc{h}}^{\mathpzc{s}}\left(\beta, \overline{z}_{\mathpzc{h}}^{\mathpzc{s}}(\beta,\rho^{\mathpzc{s}},0),0\right) < \rho_{\mathpzc{h}}^{\mathpzc{s}}\left(\beta,\mathrm{e}^{\beta E_{\mathpzc{h}}^{\mathpzc{s}}(0)},0\right).
\end{equation*}
To do so, we need an asymptotic expansion of $\overline{z}_{\mathpzc{h}}^{\mathpzc{s}}(\beta,\rho^{\mathpzc{s}},0)$ for $\mathpzc{h}>0$ sufficiently small. By mimicking the proof of \eqref{semifug}, it suffices to replace the factor $\rho/2$ by $\rho^{\mathpzc{s}}$ in the leading term of the asymptotic expansion in \eqref{semifug}. Due to \eqref{infspectrum}, then there exists a $\hat{\mathpzc{h}} = \hat{\mathpzc{h}}(\beta,\rho^{\mathpzc{s}},\Vert V\Vert_{\infty}) > 0$ s.t.
\begin{equation*}
\forall  \mathpzc{h} \in (0, \hat{\mathpzc{h}}],\quad \overline{z}_{\mathpzc{h}}^{\mathpzc{s}}\left(\beta,\rho^{\mathpzc{s}},0\right) \leq \frac{\mathrm{e}^{-\beta \Vert V \Vert_{\infty}}}{2} < \mathrm{e}^{-\beta \Vert V \Vert_{\infty}} \leq \mathrm{e}^{\beta E_{\mathpzc{h}}^{\mathpzc{s}}(0)}.
\end{equation*}
It remains to use that $\forall\mathpzc{h}>0$, $z \mapsto \rho_{\mathpzc{h}}^{\mathpzc{s}}(\beta,z,0)$ is a strictly increasing function on $(0,\infty)$. \qed


\subsection{Proof of Theorem \ref{THMF} $\mathrm{(ii)}$.}
\label{sec33}

The main result of this paragraph gives an asymptotic expansion in the semiclassical limit for the bulk zero-field magnetic susceptibility under the grand-canonical conditions.\\
From now on, we drop the superscript '$\mathrm{(GC)}$' when dealing with the bulk magnetic susceptibility.\\
\indent Let us introduce the (complete) Fermi-Dirac function $f_{\nu}: (-1,\infty) \rightarrow \mathbb{R}$ with $\nu> 0$. Denoting by $\Gamma(\cdot\,)$ the usual Euler Gamma function, it is defined as, see e.g. \cite{McDS}:
\begin{equation}
\label{funcFermi}
\forall u > -1,\quad f_{\nu}(u) := \frac{1}{\Gamma(\nu)}\int_{0}^{\infty} \mathrm{d}t\, \frac{u}{\mathrm{e}^{t}+u} t^{\nu-1}.
\end{equation}

\begin{proposition}
\label{thmsus}
Suppose that the assumptions $(\mathscr{A}_{\mathrm{p}})$ and $(\mathscr{A}_{\mathrm{r}})$ hold.\\
For any $0<\alpha <1$, $0<\theta \leq 1$, $\beta >0$ and $z>0$, we have the asymptotic expansion:
\begin{equation*}
\mathcal{X}_{\mathpzc{h}}(\beta,z, 0) =
\tilde{\mathscr{X}}_{\mathpzc{h}}^{\mathrm{(orbit)}}(\beta,z, 0) + \tilde{\mathscr{X}}_{\mathpzc{h}}^{\mathrm{(spin)}}(\beta,z, 0)  +
\mathcal{O}\left(z \mathpzc{h}^{-1+\theta(1-\alpha)}\right),
\end{equation*}
with:
\begin{gather}
\label{semisuo}
\tilde{\mathscr{X}}_{\mathpzc{h}}^{\mathrm{(orbit)}}(\beta,z, 0) := -\left(\frac{q}{c}\right)^{2} \frac{1}{\sqrt{\beta} \vert \Omega\vert} \frac{1}{\mathpzc{h}} \frac{1}{6(2\pi)^{\frac{3}{2}}} \int_{\Omega} \mathrm{d}\mathbf{x}\, f_{\frac{1}{2}}\left(z \mathrm{e}^{-\beta V(\mathbf{x})}\right),\\
\label{semisup}
\tilde{\mathscr{X}}_{\mathpzc{h}}^{\mathrm{(spin)}}(\beta,z, 0) := \left(\frac{q}{c}\right)^{2} \frac{1}{\sqrt{\beta} \vert \Omega\vert} \frac{1}{\mathpzc{h}} \frac{g^{2}}{4}  \frac{1}{2(2\pi)^{\frac{3}{2}}} \int_{\Omega} \mathrm{d}\mathbf{x}\, f_{\frac{1}{2}}\left(z \mathrm{e}^{-\beta V(\mathbf{x})}\right).
\end{gather}
\end{proposition}

From Proposition \ref{thmsus} together with Proposition \ref{lemrho}, we can turn to:\\
\noindent \textbf{Proof of Theorem \ref{THMF} $\mathrm{(ii)}$.} Let $0<\alpha<1$ and $0<\theta\leq 1$ be fixed. For any $\beta>0$ and $\rho>0$, let $\overline{z}_{\mathpzc{h}}(\beta):= \overline{z}_{\mathpzc{h}}(\beta,\frac{\rho}{2},0) > 0$ be the unique solution of the equation $\rho_{\mathpzc{h}}(\beta,z,0) = \rho$.\\ From \eqref{susfixdens} and by using the results of Proposition \ref{thmsus}, one has:
\begin{gather*}
\mathcal{X}_{\mathpzc{h}}\left(\beta,\rho,0\right) = \tilde{\mathscr{X}}_{\mathpzc{h}}^{\mathrm{(orbit)}}\left(\beta,\rho, 0\right) + \tilde{\mathscr{X}}_{\mathpzc{h}}^{\mathrm{(spin)}}\left(\beta,\rho, 0\right)  +
\mathcal{O}\left(\overline{z}_{\mathpzc{h}}(\beta) \mathpzc{h}^{-1+\theta(1-\alpha)}\right),\\
\tilde{\mathscr{X}}_{\mathpzc{h}}^{\mathrm{(\wp)}}\left(\beta,\rho, 0\right) := \tilde{\mathscr{X}}_{\mathpzc{h}}^{\mathrm{(\wp)}}\left(\beta,\overline{z}_{\mathpzc{h}}(\beta), 0\right),\quad \textrm{$\wp = \mathrm{orbit}$ or $\mathrm{spin}$}.
\end{gather*}
It remains to use the asymptotic in \eqref{semifug} along with the following one derived from \eqref{funcFermi}:
\begin{equation*}
f_{\frac{1}{2}}(u) = u - \frac{1}{\sqrt{2}} u^{2} + \mathcal{O}\left(u^{3}\right). \tag*{\qed}
\end{equation*}

The rest of this section is devoted to the proof of Proposition \ref{thmsus}. It requires three steps.

\subsubsection{A formula for the bulk grand-canonical zero-field magnetic susceptibility.}

In the grand-canonical situation, let $\beta>0$ and $z >0$. Let $\mathscr{C}_{\beta}$ be the contour around the interval $[-\Vert V \Vert_{\infty},\infty)$ defined in \eqref{Gamma}. The closed subset surrounding by $\mathscr{C}_{\beta}$ is a strict subset of $\mathfrak{D}:= \{\zeta \in \mathbb{C}: \Im \zeta \in (-\frac{\pi}{\beta},\frac{\pi}{\beta})\}$, the holomorphic domain  of $\xi \mapsto \mathfrak{f}(\beta,z;\xi) := \ln(1 + z \mathrm{e}^{-\beta \xi})$. Note that $\mathfrak{f}(\beta,z;\cdot\,)$ admits an exponential decay on $\mathscr{C}_{\beta}$, i.e. there exists a constant $c>0$ s.t.
\begin{equation}
\label{estimo2}
\forall \beta>0,\,\forall \xi \in \mathscr{C}_{\beta},\quad \vert \mathfrak{f}(\beta,z;\xi)\vert \leq c z \mathrm{e}^{-\beta \Re \xi}.
\end{equation}
For any $\mathpzc{h}>0$ and $\xi \in \varrho(\mathpzc{H}_{\mathpzc{h}})$, introduce on $L^{2}(\mathbb{R}^{3};\mathbb{C}^{2})$ the operators $\mathpzc{T}_{\mathpzc{h},j}(\xi)$, with $j=1,2$:
\begin{equation}
\label{bfTj}
\mathpzc{T}_{\mathpzc{h},j}(\xi) := \begin{pmatrix} T_{\mathpzc{h},j}(\xi) & 0 \\ 0 & T_{\mathpzc{h},j}(\xi) \end{pmatrix} = T_{\mathpzc{h},j}(\xi) \oplus T_{\mathpzc{h},j}(\xi),\quad j=1,2,
\end{equation}
where the operators $T_{\mathpzc{h},j}(\xi)$ on $L^{2}(\mathbb{R}^{3})$ are generated via their kernel respectively defined as:
\begin{align*}
\forall (\mathbf{x},\mathbf{y}) \in \mathbb{R}^{6}\setminus D,\quad T_{\mathpzc{h},1}(\mathbf{x},\mathbf{y};\xi) &:= \mathbf{a}(\mathbf{x}-\mathbf{y}) \cdot (i\mathpzc{h}\nabla_{\mathbf{x}}) (H_{\mathpzc{h}} - \xi)^{-1}(\mathbf{x},\mathbf{y}),\\
T_{\mathpzc{h},2}(\mathbf{x},\mathbf{y};\xi) &:= \frac{1}{2} \mathbf{a}^{2}(\mathbf{x}-\mathbf{y})(H_{\mathpzc{h}} - \xi)^{-1}(\mathbf{x},\mathbf{y}).
\end{align*}
Above, we used the notation $\mathpzc{H}_{\mathpzc{h}} = \mathpzc{H}_{\mathpzc{h}}(0)$, and $H_{\mathpzc{h}} = H_{\mathpzc{h}}^{+}(0) = H_{\mathpzc{h}}^{-}(0)$ in \eqref{Hinfini1}.\\
The bulk zero-field magnetic susceptibility reads $\forall \beta>0$, $\forall z>0$ and $\forall \mathpzc{h}>0$ as:
\begin{gather}
\label{sus}
\mathcal{X}_{\mathpzc{h}}(\beta,z,0) = \mathcal{X}_{\mathpzc{h}}^{\mathrm{(orbit)}}(\beta,z,0) + \mathcal{X}_{\mathpzc{h}}^{(\mathrm{spin})}(\beta,z,0),\\
\label{suso}
\begin{split}
\mathcal{X}_{\mathpzc{h}}^{\mathrm{(orbit)}}(\beta,z,0) := &\left(\frac{q}{c}\right)^{2}\frac{2}{\beta \vert \Omega\vert} \frac{i}{2\pi} \int_{\mathscr{C}_{\beta}} \mathrm{d}\xi\, \mathfrak{f}(\beta,z;\xi) \\
&\times \mathrm{Tr}_{L^{2}(\mathbb{R}^{3};\mathbb{C}^{2})}\left\{\chi_{\Omega}\mathpzc{I}_{\mathrm{d}} \left(\mathpzc{H}_{\mathpzc{h}} - \xi\right)^{-1}\left[\mathpzc{T}_{\mathpzc{h},1}(\xi)\mathpzc{T}_{\mathpzc{h},1}(\xi) - \mathpzc{T}_{\mathpzc{h},2}(\xi)\right]\chi_{\Omega}\mathpzc{I}_{\mathrm{d}}\right\},
\end{split} \\
\label{susp}
\mathcal{X}_{\mathpzc{h}}^{(\mathrm{spin})}(\beta,z,0) := \left(\frac{q}{c}\right)^{2} \left(\frac{g \mathpzc{h}}{4}\right)^{2} \frac{2}{\beta \vert \Omega\vert}  \frac{i}{2\pi} \int_{\mathscr{C}_{\beta}} \mathrm{d}\xi\, \mathfrak{f}(\beta,z;\xi) \mathrm{Tr}_{L^{2}(\mathbb{R}^{3};\mathbb{C}^{2})}\left\{\chi_{\Omega}\mathpzc{I}_{\mathrm{d}} \left(\mathpzc{H}_{\mathpzc{h}} - \xi\right)^{-3} \chi_{\Omega}\mathpzc{I}_{\mathrm{d}}\right\},
\end{gather}
where $\chi_{\Omega}\mathpzc{I}_{\mathrm{d}} = \chi_{\Omega} \oplus \chi_{\Omega}$ denotes the multiplication operator by the indicator function of the unit cell $\Omega$ on $L^{2}(\mathbb{R}^{3};\mathbb{C}^{2})$ and $(\mathpzc{H}_{\mathpzc{h}}-\xi)^{-1}$ the resolvent operator of $\mathpzc{H}_{\mathpzc{h}}=\mathpzc{H}_{\mathpzc{h}}(0)$ in \eqref{Hinfini}:
\begin{equation}
\label{trres}
(\mathpzc{H}_{\mathpzc{h}} - \xi)^{-1} := \begin{pmatrix} (H_{\mathpzc{h}}^{-} - \xi)^{-1} & 0 \\ 0 & (H_{\mathpzc{h}}^{+} - \xi)^{-1}\end{pmatrix} = (H_{\mathpzc{h}}^{-} - \xi)^{-1} \oplus (H_{\mathpzc{h}}^{+} - \xi)^{-1}.
\end{equation}
The derivation of the orbital contribution in \eqref{suso} is the main purpose of \cite{BS}, and it is based on the so-called \textit{gauge invariant magnetic perturbation theory}. The same method can be applied when taking into account the Stern-Gerlach term in the Hamiltonian (i.e. the interaction between the spin and the magnetic field), and allows us to derive \eqref{susp} when the magnetic field vanishes.
Actually, we can prove the following formula: for any $\beta>0$, $z>0$, $b\in \mathbb{R}$ and $\mathpzc{h}>0$,
\begin{equation*}
\mathcal{X}_{\mathpzc{h}}(\beta,z,b) = \mathcal{X}_{\mathpzc{h}}^{-}(\beta,z,b) + \mathcal{X}_{\mathpzc{h}}^{+}(\beta,z,b),
\end{equation*}
\begin{multline*}
\mathcal{X}_{\mathpzc{h}}^{-}(\beta,z,b) := \left(\frac{q}{c}\right)^{2}\frac{2}{\beta \vert \Omega\vert} \frac{i}{2\pi} \int_{\mathscr{C}_{\beta}} \mathrm{d}\xi\, \mathfrak{f}(\beta,z;\xi)
\mathrm{Tr}_{L^{2}(\mathbb{R}^{3})}\{\chi_{\Omega} R_{\mathpzc{h}}^{-}(b;\xi) [T_{\mathpzc{h},1}^{-}(b;\xi)T_{\mathpzc{h},1}^{-}(b;\xi) - T_{\mathpzc{h},2}^{-}(b;\xi) + \\
- \frac{g\mathpzc{h}}{4} T_{\mathpzc{h},1}^{-}(b;\xi) R_{\mathpzc{h}}^{-}(b;\xi) - \frac{g\mathpzc{h}}{4} R_{\mathpzc{h}}^{-}(b;\xi) T_{\mathpzc{h},1}^{-}(b;\xi) + \frac{(g\mathpzc{h})^{2}}{16} R_{\mathpzc{h}}^{-}(b;\xi)R_{\mathpzc{h}}^{-}(b;\xi)]\chi_{\Omega}\},
\end{multline*}
\begin{multline*}
\mathcal{X}_{\mathpzc{h}}^{+}(\beta,z,b) := \left(\frac{q}{c}\right)^{2}\frac{2}{\beta \vert \Omega\vert} \frac{i}{2\pi} \int_{\mathscr{C}_{\beta}} \mathrm{d}\xi\, \mathfrak{f}(\beta,z;\xi)
\mathrm{Tr}_{L^{2}(\mathbb{R}^{3})}\{\chi_{\Omega} R_{\mathpzc{h}}^{+}(b;\xi) [T_{\mathpzc{h},1}^{+}(b;\xi)T_{\mathpzc{h},1}^{+}(b;\xi) - T_{\mathpzc{h},2}^{+}(b;\xi) + \\
+ \frac{g\mathpzc{h}}{4} T_{\mathpzc{h},1}^{+}(b;\xi) R_{\mathpzc{h}}^{+}(b;\xi) + \frac{g\mathpzc{h}}{4} R_{\mathpzc{h}}^{+}(b;\xi) T_{\mathpzc{h},1}^{+}(b;\xi) + \frac{(g\mathpzc{h})^{2}}{16} R_{\mathpzc{h}}^{+}(b;\xi)R_{\mathpzc{h}}^{+}(b;\xi)]\chi_{\Omega}\},
\end{multline*}
where we set $R_{\mathpzc{h}}^{\mathpzc{s}}(b;\xi):= (H_{\mathpzc{h}}^{\mathpzc{s}}(b) - \xi)^{-1}$ and where the operators $T_{\mathpzc{h},j}^{\mathpzc{s}}(b;\xi)$ on $L^{2}(\mathbb{R}^{3})$ are generated via their kernel respectively defined as:
\begin{align*}
\forall (\mathbf{x},\mathbf{y}) \in \mathbb{R}^{6}\setminus D,\quad T_{\mathpzc{h},1}^{\mathpzc{s}}(\mathbf{x},\mathbf{y};b,\xi) &:= \mathbf{a}(\mathbf{x}-\mathbf{y}) \cdot (i\mathpzc{h}\nabla_{\mathbf{x}}+ b \mathbf{a}(\mathbf{x})) (H_{\mathpzc{h}}^{\mathpzc{s}}(b) - \xi)^{-1}(\mathbf{x},\mathbf{y}),\\
T_{\mathpzc{h},2}^{\mathpzc{s}}(\mathbf{x},\mathbf{y};b,\xi) &:= \frac{1}{2} \mathbf{a}^{2}(\mathbf{x}-\mathbf{y})(H_{\mathpzc{h}}^{\mathpzc{s}}(b) - \xi)^{-1}(\mathbf{x},\mathbf{y}).
\end{align*}
We point out that the spin contribution to the bulk magnetic susceptibility is made up of three kinds of terms in non-vanishing magnetic field, and two of them involve the coupling between the linear part of the Zeeman Hamiltonian and the Stern-Gerlach term. Note that the decomposition into an orbital and a spin contribution is made possible since the spin-orbit coupling has been disregarded. Getting back to \eqref{sus}, from \eqref{trres} and \eqref{bfTj}, we have under the same conditions:
\begin{gather*}
\begin{split}
\mathcal{X}_{\mathpzc{h}}^{\mathrm{(orbit)}}(\beta,z,0) = &\left(\frac{q}{c}\right)^{2}\frac{4}{\beta \vert \Omega\vert} \frac{i}{2\pi} \int_{\mathscr{C}_{\beta}} \mathrm{d}\xi\, \mathfrak{f}(\beta,z;\xi) \\
&\times \mathrm{Tr}_{L^{2}(\mathbb{R}^{3})}\left\{\chi_{\Omega}\left(H_{\mathpzc{h}} - \xi\right)^{-1}\left[T_{\mathpzc{h},1}(\xi)T_{\mathpzc{h},1}(\xi) - T_{\mathpzc{h},2}(\xi)\right]\chi_{\Omega}\right\},
\end{split} \\
\mathcal{X}_{\mathpzc{h}}^{(\mathrm{spin})}(\beta,z,0) = \left(\frac{q}{c}\right)^{2} \frac{(g \mathpzc{h})^{2}}{4} \frac{1}{\beta \vert \Omega\vert}  \frac{i}{2\pi} \int_{\mathscr{C}_{\beta}} \mathrm{d}\xi\, \mathfrak{f}(\beta,z;\xi) \mathrm{Tr}_{L^{2}(\mathbb{R}^{3})}\left\{\chi_{\Omega}\left(H_{\mathpzc{h}} - \xi\right)^{-3} \chi_{\Omega}\right\}.
\end{gather*}
\indent Next, let us rewrite the contributions in \eqref{suso}-\eqref{susp} in a more convenient way, i.e. involving the resolvent operator $(\tilde{\mathpzc{H}}_{\mathpzc{h}} - \xi)^{-1}$. In Sec. \ref{appr}, we gave an approximation of $(\tilde{\mathpzc{H}}_{\mathpzc{h}} - \xi)^{-1}$. For the need, introduce on $L^{2}(\mathbb{R}^{3};\mathbb{C}^{2})$ for any $\mathpzc{h}>0$ and $\xi \in \varrho(\tilde{\mathpzc{H}}_{\mathpzc{h}})$, the operators $\tilde{\mathpzc{T}}_{\mathpzc{h},j}(\xi)$, with $j=1,2$:
\begin{equation*}
\tilde{\mathpzc{T}}_{\mathpzc{h},j}(\xi) := \begin{pmatrix} \tilde{T}_{\mathpzc{h},j}(\xi) & 0 \\ 0 & \tilde{T}_{\mathpzc{h},j}(\xi)\end{pmatrix} = \tilde{T}_{\mathpzc{h},j}(\xi) \oplus \tilde{T}_{\mathpzc{h},j}(\xi),\quad j=1,2,
\end{equation*}
where the operators $\tilde{T}_{\mathpzc{h},j}(\xi)$ on $L^{2}(\mathbb{R}^{3})$ are generated via their kernel respectively defined as:
\begin{align}
\label{tildW1}
\forall (\mathbf{x},\mathbf{y}) \in \mathbb{R}^{6}\setminus D,\quad \tilde{T}_{\mathpzc{h},1}(\mathbf{x},\mathbf{y};\xi) &:= \mathbf{a}(\mathbf{x}-\mathbf{y})\cdot (i\nabla_{\mathbf{x}}) (\tilde{H}_{\mathpzc{h}} - \xi)^{-1}(\mathbf{x},\mathbf{y}),\\
\label{tildW2}
\tilde{T}_{\mathpzc{h},2}(\mathbf{x},\mathbf{y};\xi) &:= \frac{1}{2} \mathbf{a}^{2}(\mathbf{x}-\mathbf{y})(\tilde{H}_{\mathpzc{h}} - \xi)^{-1}(\mathbf{x},\mathbf{y}).
\end{align}
Since $\vert \mathbf{a}(\mathbf{x}-\mathbf{y})\vert \leq \vert \mathbf{x}-\mathbf{y}\vert$, then under the conditions of Lemma \ref{lem0} (see the notation in \eqref{short}):
\begin{equation}
\label{estTh}
\forall (\mathbf{x},\mathbf{y}) \in \mathbb{R}^{6}\setminus D,\quad \max\left\{\left\vert \tilde{T}_{\mathpzc{h},j}(\mathbf{x},\mathbf{y};\xi)\right\vert, \left\vert T_{\mathpzc{h},j}(\mathbf{x},\mathbf{y};\xi)\right \vert\right\} \leq p(\vert \xi\vert) \frac{\mathrm{e}^{- \vartheta_{\xi} \vert \mathbf{x} - \mathbf{y}\vert}}{\vert \mathbf{x} - \mathbf{y}\vert},
\end{equation}
for some constant $\vartheta>0$ and polynomial $p(\cdot\,)$. \eqref{estTh} follows from \eqref{trures}-\eqref{trueder} and \eqref{keyes}. From the unitary transformation in \eqref{trnsfounit}, $\mathpzc{T}_{\mathpzc{h},j}(\xi)$ and $\tilde{\mathpzc{T}}_{\mathpzc{h},j}(\xi)$ are related to each other through:
\begin{equation}
\label{equivun2}
\mathpzc{U}_{\mathpzc{h}} \mathpzc{T}_{\mathpzc{h},j}(\xi) \mathpzc{U}_{\mathpzc{h}}^{-1} = \mathpzc{h}^{j} \tilde{\mathpzc{T}}_{\mathpzc{h},j}(\xi),\quad j=1,2,
\end{equation}
where we used \eqref{equivun}, which leads in the kernels sense to:
\begin{equation*}
\forall(\mathbf{x},\mathbf{y}) \in \mathbb{R}^{6} \setminus D,\quad  \left(\tilde{H}_{\mathpzc{h}} - \xi \right)^{-1}(\mathbf{x},\mathbf{y}) = \mathpzc{h}^{3} \left(H_{\mathpzc{h}} - \xi \right)^{-1}\left(\mathpzc{h}\mathbf{x},\mathpzc{h}\mathbf{y}\right),\quad \mathpzc{h}>0.
\end{equation*}
From \eqref{suso}-\eqref{susp} together with \eqref{equivun}, \eqref{equivun1} and \eqref{equivun2}, the bulk grand-canonical zero-field orbital and spin susceptibilities can be rewritten respectively as:
\begin{gather}
\label{suso2}
\begin{split}
\mathcal{X}_{\mathpzc{h}}^{\mathrm{(orbit)}}(\beta,z,0) = &\left(\frac{q}{c}\right)^{2} 4 \mathpzc{h}^{2} \frac{1}{\beta \vert \Omega\vert} \frac{i}{2\pi} \int_{\mathscr{C}_{\beta}} \mathrm{d}\xi\, \mathfrak{f}(\beta,z;\xi) \\
&\times \mathrm{Tr}_{L^{2}(\mathbb{R}^{3})}\left\{\chi_{\Omega_{\mathpzc{h}}}\left(\tilde{H}_{\mathpzc{h}} - \xi\right)^{-1}\left[\tilde{T}_{\mathpzc{h},1}(\xi)\tilde{T}_{\mathpzc{h},1}(\xi) - \tilde{T}_{\mathpzc{h},2}(\xi)\right]\chi_{\Omega_{\mathpzc{h}}}\right\},
\end{split} \\
\label{susp2}
\mathcal{X}_{\mathpzc{h}}^{(\mathrm{spin})}(\beta,z,0) = \left(\frac{q}{c}\right)^{2} \frac{(g \mathpzc{h})^{2}}{4} \frac{1}{\beta \vert \Omega\vert}  \frac{i}{2\pi} \int_{\mathscr{C}_{\beta}} \mathrm{d}\xi\, \mathfrak{f}(\beta,z;\xi) \mathrm{Tr}_{L^{2}(\mathbb{R}^{3})}\left\{\chi_{\Omega_{\mathpzc{h}}}\left(\tilde{H}_{\mathpzc{h}} - \xi\right)^{-3} \chi_{\Omega_{\mathpzc{h}}}\right\}.
\end{gather}
Note that $\forall \mathpzc{h}>0$, the operators $(\tilde{H}_{\mathpzc{h}} - \xi)^{-3}$, $(\tilde{H}_{\mathpzc{h}} - \xi)^{-1} \tilde{T}_{\mathpzc{h},j}(\xi)$ and $(\tilde{H}_{\mathpzc{h}} - \xi)^{-1} (\tilde{T}_{\mathpzc{h},j}(\xi))^{2}$ are locally trace-class on $L^{2}(\mathbb{R}^{3})$ with trace-norm  bounded above by some polynomial in $\vert \xi\vert$ (by unitary equivalence, the same holds true for the corresponding operators without the tilde). This results from the fact that $(\tilde{H}_{\mathpzc{h}} - \xi)^{-1}$ and  $\tilde{T}_{\mathpzc{h},j}(\xi)$ are both locally Hilbert-Schmidt, see the estimates in \eqref{trures} and \eqref{estTh}. Due to \eqref{estimo2}, \eqref{suso2}-\eqref{susp2} (and then \eqref{suso}-\eqref{susp}) are therefore well-defined.

\subsubsection{Isolating the main $\mathpzc{h}$-dependent contribution.}
\label{subsec32}

Here we use the same strategy as the one leading to Proposition \ref{asymrho}, i.e. we use the approximation \eqref{identtoot} in order to isolate the main $\mathpzc{h}$-dependent contributions from \eqref{suso2}-\eqref{susp2}. As it was the case for the bulk zero-field density of particles, we expect the contributions involving only the operator $\mathscr{R}_{\mathpzc{h}}(\xi)$ (i.e. obtained from \eqref{suso2}-\eqref{susp2} by replacing $(\tilde{H}_{\mathpzc{h}}-\xi)^{-1}$ with $\mathscr{R}_{\mathpzc{h}}(\xi)$, including the ones in \eqref{tildW1}-\eqref{tildW2}) to give rise to the leading terms in the asymptotic expansion.\\ We will see below that this is in fact the case.\\

Introduce $\forall 0<\alpha<1$, $\forall \mathpzc{h} \in (0,\mathpzc{h}_{0}]$ and $\forall \boldsymbol{\gamma} \in \mathscr{E}$, the operator $\tilde{T}_{\mathpzc{h},\boldsymbol{\gamma};2}^{\mathrm{(cste)}}(\xi)$ on $L^{2}(\mathbb{R}^{3})$ generated via its kernel which reads as (remind that  $\tilde{H}_{\mathpzc{h},\boldsymbol{\gamma}}^{\mathrm{(cste)}}$ is defined in \eqref{HconstV}):
\begin{equation}
\label{Tgam2hat}
\forall(\mathbf{x},\mathbf{y}) \in \mathbb{R}^{6}\setminus D,\quad \tilde{T}_{\mathpzc{h},\boldsymbol{\gamma};2}^{\mathrm{(cste)}}(\mathbf{x},\mathbf{y};\xi) := \frac{1}{2} \mathbf{a}^{2}(\mathbf{x}-\mathbf{y}) \left(\tilde{H}_{\mathpzc{h},\boldsymbol{\gamma}}^{\mathrm{(cste)}} - \xi\right)^{-1}(\mathbf{x},\mathbf{y}).
\end{equation}

The following proposition identifies a first main $\mathpzc{h}$-dependent contribution to the bulk grand-canonical zero-field orbital and spin susceptibilities in \eqref{suso2}-\eqref{susp2} respectively:

\begin{proposition}
\label{gropro2}
For any $0<\alpha<1$, $0<\theta\leq 1$, $\beta >0$ and $z >0$, it holds:
\begin{multline}
\label{suso5}
\mathcal{X}_{\mathpzc{h}}^{\mathrm{(orbit)}}(\beta,z,0) =
-\left(\frac{q}{c}\right)^{2} 4 \mathpzc{h}^{2} \frac{1}{\beta \vert \Omega\vert} \frac{i}{2\pi}  \int_{\mathscr{C}_{\beta}} \mathrm{d}\xi\, \mathfrak{f}(\beta,z;\xi) \\
\times \mathrm{Tr}_{L^{2}(\mathbb{R}^{3})}\left\{\chi_{\Omega_{\mathpzc{h}}}\left(\sum_{\boldsymbol{\gamma}_{1} \in \mathscr{E}} \left(\tilde{H}_{\mathpzc{h},\boldsymbol{\gamma}_{1}}^{\mathrm{(cste)}} - \xi\right)^{-1} \tau_{\mathpzc{h},\boldsymbol{\gamma}_{1}} \sum_{\boldsymbol{\gamma}_{2} \in \mathscr{E}} \tilde{T}_{\mathpzc{h},\boldsymbol{\gamma}_{2};2}^{\mathrm{(cste)}}(\xi) \tau_{\mathpzc{h},\boldsymbol{\gamma}_{2}}\right)\chi_{\Omega_{\mathpzc{h}}}\right\}  + \mathcal{O}\left(z \mathpzc{h}^{-1 + \theta(1-\alpha)}\right),
\end{multline}
\begin{multline}
\label{susp5}
\mathcal{X}_{\mathpzc{h}}^{(\mathrm{spin})}(\beta,z,0) =
\left(\frac{q}{c}\right)^{2} \frac{(g\mathpzc{h})^{2}}{4} \frac{1}{\beta \vert \Omega\vert} \frac{i}{2\pi}  \int_{\mathscr{C}_{\beta}} \mathrm{d}\xi\, \mathfrak{f}(\beta,z;\xi)\\
\times \mathrm{Tr}_{L^{2}(\mathbb{R}^{3})}\left\{\chi_{\Omega_{\mathpzc{h}}} \left(\prod_{l=1}^{3}\sum_{\boldsymbol{\gamma}_{l} \in \mathscr{E}} \left(\tilde{H}_{\mathpzc{h},\boldsymbol{\gamma}_{l}}^{\mathrm{(cste)}} - \xi\right)^{-1} \tau_{\mathpzc{h},\boldsymbol{\gamma}_{l}}\right)\chi_{\Omega_{\mathpzc{h}}}\right\}  + \mathcal{O}\left(z \mathpzc{h}^{-1 + \theta(1-\alpha)}\right).
\end{multline}
\end{proposition}

\begin{remark} At this point, we expect the main $\mathpzc{h}$-dependent contributions coming from \eqref{suso5} and \eqref{susp5} to behave like $\mathcal{O}(\mathpzc{h}^{-1})$. Let us give the main arguments for the orbital case. Since the kernel in \eqref{Tgam2hat} obeys an estimate of type \eqref{otyp} (use that $\vert \bold{a}(\bold{x}-\bold{y})\vert \leq \vert \bold{x} - \bold{y}\vert$ followed by \eqref{keyes}), then the Hilbert-Schmidt norm of $\tilde{T}_{\mathpzc{h},\boldsymbol{\gamma};2}^{\mathrm{(cste)}}(\xi) \tau_{\mathpzc{h},\boldsymbol{\gamma}}$ obeys an estimate of type \eqref{I2n1}. Moreover, when keeping $\boldsymbol{\gamma}_{1}$ fixed, only a finite number of $\boldsymbol{\gamma}_{2}$'s
have an overlapping support. This means that the double sum in \eqref{suso5} only contains $cste \times \mathpzc{h}^{3\alpha-3}$ non-zero terms. Therefore, the trace-norm obeys:
\begin{equation*}
\left\Vert \sum_{\boldsymbol{\gamma}_{1} \in \mathscr{E}} \left(\tilde{H}_{\mathpzc{h},\boldsymbol{\gamma}_{1}}^{\mathrm{(cste)}} - \xi\right)^{-1} \tau_{\mathpzc{h},\boldsymbol{\gamma}_{1}} \sum_{\boldsymbol{\gamma}_{2} \in \mathscr{E}} \tilde{T}_{\mathpzc{h},\boldsymbol{\gamma}_{2};2}^{\mathrm{(cste)}}(\xi) \tau_{\mathpzc{h},\boldsymbol{\gamma}_{2}}\right\Vert_{\mathfrak{I}_{1}} \leq p(\vert \xi\vert) \mathpzc{h}^{-3}, \end{equation*}
for some $\mathpzc{h}$-independent polynomial $p(\cdot\,)$. From \eqref{I2n1} and the foregoing, one proves similarly:
\begin{equation*}
\left\Vert \prod_{l=1}^{3}\sum_{\boldsymbol{\gamma}_{l} \in \mathscr{E}} \left(\tilde{H}_{\mathpzc{h},\boldsymbol{\gamma}_{l}}^{\mathrm{(cste)}} - \xi\right)^{-1} \tau_{\mathpzc{h},\boldsymbol{\gamma}_{l}}\right\Vert_{\mathfrak{I}_{1}} \leq p(\vert \xi\vert) \mathpzc{h}^{-3}. \end{equation*}
\end{remark}

Before turning to the proof of Proposition \ref{gropro2}, let us introduce some new operators on $L^{2}(\mathbb{R}^{3})$. Let $0<\alpha<1$. In view of \eqref{approxres} and \eqref{idd}, define $\forall \mathpzc{h} \in (0,\mathpzc{h}_{0}]$ the operators $\mathcal{T}_{\mathpzc{h},j}(\xi)$ and $\mathscr{T}_{\mathpzc{h},j}(\xi)$, $j=1,2$ on $L^{2}(\mathbb{R}^{3})$ generated via their kernel respectively defined as:
\begin{align}
\forall (\mathbf{x},\mathbf{y}) \in \mathbb{R}^{6}\setminus D,\quad \mathcal{T}_{\mathpzc{h},1}(\mathbf{x},\mathbf{y};\xi) &:= \mathbf{a}(\mathbf{x}-\mathbf{y}) \cdot (i\nabla_{\mathbf{x}}) \left(\mathcal{R}_{\mathpzc{h}}(\xi))(\mathbf{x},\mathbf{y}\right), \nonumber\\
\mathcal{T}_{\mathpzc{h},2}(\mathbf{x},\mathbf{y};\xi) &:= \frac{1}{2} \mathbf{a}^{2}(\mathbf{x}-\mathbf{y})\left(\mathcal{R}_{\mathpzc{h}}(\xi)\right)(\mathbf{x},\mathbf{y}), \nonumber \\
\label{Wep1}
\mathscr{T}_{\mathpzc{h},1}(\mathbf{x},\mathbf{y};\xi) &:= \mathbf{a}(\mathbf{x}-\mathbf{y}) \cdot (i\nabla_{\mathbf{x}}) \left(\mathscr{R}_{\mathpzc{h}}(\xi)\right)(\mathbf{x},\mathbf{y}),\\
\label{Wep2}
\mathscr{T}_{\mathpzc{h},2}(\mathbf{x},\mathbf{y};\xi) &:= \frac{1}{2} \mathbf{a}^{2}(\mathbf{x}-\mathbf{y})\left(\mathscr{R}_{\mathpzc{h}}(\xi)\right)(\mathbf{x},\mathbf{y}).
\end{align}
Since $\vert \mathbf{a}(\mathbf{x}-\mathbf{y})\vert \leq \vert \mathbf{x}-\mathbf{y}\vert$ and by using \eqref{kerSep}-\eqref{kerSepder}, then under the conditions of Lemma \ref{lem1}:
\begin{equation}
\label{estTh2}
\forall(\mathbf{x},\mathbf{y})\in \mathbb{R}^{6}\setminus D,\quad \max\left\{\left\vert \mathcal{T}_{\mathpzc{h},j}(\mathbf{x},\mathbf{y};\xi)\right\vert, \left\vert \mathscr{T}_{\mathpzc{h},j}(\mathbf{x},\mathbf{y};\xi)\right\vert\right\} \leq p(\vert \xi\vert) \frac{\mathrm{e}^{- \vartheta_{\xi} \vert \mathbf{x} - \mathbf{y}\vert}}{\vert \mathbf{x} - \mathbf{y}\vert},\quad j=1,2,
\end{equation}
for some constant $\vartheta>0$ and polynomial $p(\cdot\,)$ both $\mathpzc{h}$-independent. Here $\vartheta_{\xi}:= \vartheta (1+\vert \xi\vert)^{-1}$. Note that, due to the estimate in \eqref{estTh2}, the operators $\mathcal{T}_{\mathpzc{h},j}(\xi)$ and $\mathscr{T}_{\mathpzc{h},j}(\xi)$, $j=1,2$ are locally Hilbert-Schmidt with H-S norms bounded above by some $\mathpzc{h}$-independent polynomial in $\vert \xi\vert$.\\

The proof of Proposition \ref{gropro2} is based on the three following lemmas whose proofs lie in Sect. \ref{append23}. Remind that $\forall 0<\alpha<1$, $\mathpzc{h}_{0}= \mathpzc{h}_{0}(\alpha)\leq 1$ is defined via \eqref{epsilon0}, and the contour $\mathscr{C}_{\beta}$  in \eqref{Gamma}.

\begin{lema}
\label{isola11}
$\forall \alpha\in (0,1)$, $\forall N >0$, there exists a polynomial $p(\cdot\,)$ s.t.  $\forall \mathpzc{h} \in (0,\mathpzc{h}_{0}]$ and $\forall \xi \in \mathscr{C}_{\beta}$:
\begin{equation}
\label{trace01}
\left\vert \mathrm{Tr}_{L^{2}(\mathbb{R}^{3})}\left\{\chi_{\Omega_{\mathpzc{h}}} \left(\tilde{H}_{\mathpzc{h}} - \xi\right)^{-3}\chi_{\Omega_{\mathpzc{h}}}\right\}
-  \mathrm{Tr}_{L^{2}(\mathbb{R}^{3})}\left\{\chi_{\Omega_{\mathpzc{h}}} \left(\mathcal{R}_{\mathpzc{h}}(\xi)\right)^{3} \chi_{\Omega_{\mathpzc{h}}}\right\}\right\vert \leq p(\vert \xi\vert) \mathpzc{h}^{N}.
\end{equation}
\begin{multline}
\label{trace1}
\biggl\vert \mathrm{Tr}_{L^{2}(\mathbb{R}^{3})}\left\{\chi_{\Omega_{\mathpzc{h}}} \left(\tilde{H}_{\mathpzc{h}} - \xi\right)^{-1}\left[\tilde{T}_{\mathpzc{h},1}(\xi)\tilde{T}_{\mathpzc{h},1}(\xi) - \tilde{T}_{\mathpzc{h},2}(\xi)\right]\chi_{\Omega_{\mathpzc{h}}}\right\}\\
-  \mathrm{Tr}_{L^{2}(\mathbb{R}^{3})}\left\{\chi_{\Omega_{\mathpzc{h}}} \mathcal{R}_{\mathpzc{h}}(\xi)\left[\mathcal{T}_{\mathpzc{h},1}(\xi)\mathcal{T}_{\mathpzc{h},1}
(\xi) - \mathcal{T}_{\mathpzc{h},2}(\xi)\right]\chi_{\Omega_{\mathpzc{h}}}\right\}\biggr \vert \leq p(\vert \xi\vert) \mathpzc{h}^{N}.
\end{multline}
\end{lema}

\begin{lema}
\label{isola12}
$\forall 0<\alpha<1$ there exists a polynomial $p(\cdot\,)$ s.t. $\forall \theta \in (0,1]$,
$\forall \mathpzc{h} \in (0,\mathpzc{h}_{0}]$ and $\forall \xi \in \mathscr{C}_{\beta}$:
\begin{equation}
\label{trace02}
\left\vert \mathrm{Tr}_{L^{2}(\mathbb{R}^{3})}\left\{\chi_{\Omega_{\mathpzc{h}}} \left(\mathcal{R}_{\mathpzc{h}}(\xi)\right)^{3}\chi_{\Omega_{\mathpzc{h}}}\right\}
-  \mathrm{Tr}_{L^{2}(\mathbb{R}^{3})}\left\{\chi_{\Omega_{\mathpzc{h}}} \left(\mathscr{R}_{\mathpzc{h}}(\xi)\right)^{3}\chi_{\Omega_{\mathpzc{h}}}\right\}\right\vert \leq p(\vert \xi\vert) \mathpzc{h}^{-3 + \theta(1-\alpha)}.
\end{equation}
\begin{multline}
\label{trace2}
\left\vert \mathrm{Tr}_{L^{2}(\mathbb{R}^{3})}\left\{\chi_{\Omega_{\mathpzc{h}}} \mathcal{R}_{\mathpzc{h}}(\xi) \left[\mathcal{T}_{\mathpzc{h},1}(\xi)\mathcal{T}_{\mathpzc{h},1}(\xi) - \mathcal{T}_{\mathpzc{h},2}(\xi)\right]\chi_{\Omega_{\mathpzc{h}}}\right\} \right.\\
\left.-  \mathrm{Tr}_{L^{2}(\mathbb{R}^{3})}\left\{\chi_{\Omega_{\mathpzc{h}}} \mathscr{R}_{\mathpzc{h}}(\xi)\left[\mathscr{T}_{\mathpzc{h},1}(\xi)\mathscr{T}_{\mathpzc{h},1}
(\xi) - \mathscr{T}_{\mathpzc{h},2}(\xi)\right]\chi_{\Omega_{\mathpzc{h}}}\right\}\right\vert \leq p(\vert \xi\vert) \mathpzc{h}^{-3 + \theta(1-\alpha)}.
\end{multline}
\end{lema}

\begin{lema}
\label{isola13}
$\forall \alpha \in (0,1)$, $\forall N >0$ there exists a polynomial $p(\cdot\,)$ s.t.  $\forall \mathpzc{h} \in (0,\mathpzc{h}_{0}]$ and $\forall \xi \in \mathscr{C}_{\beta}$:\\
\begin{gather}
\label{moust1}
\left\vert \mathrm{Tr}_{L^{2}(\mathbb{R}^{3})} \left\{ \chi_{\Omega_{\mathpzc{h}}} \left(\mathscr{R}_{\mathpzc{h}}(\xi)\right)^{3}\chi_{\Omega_{\mathpzc{h}}}\right\} -
\mathrm{Tr}_{L^{2}(\mathbb{R}^{3})} \left\{ \chi_{\Omega_{\mathpzc{h}}} \left(\prod_{l=1}^{3} \sum_{\boldsymbol{\gamma}_{l} \in \mathscr{E}} \left(\tilde{H}_{\mathpzc{h},\boldsymbol{\gamma}_{l}}^{\mathrm{(cste)}} - \xi\right)^{-1} \tau_{\mathpzc{h},\boldsymbol{\gamma}_{l}}\right)\chi_{\Omega_{\mathpzc{h}}}\right\}\right\vert \leq p(\vert \xi\vert) \mathpzc{h}^{N},\\
\label{moust}
\left\vert \mathrm{Tr}_{L^{2}(\mathbb{R}^{3})} \left\{\chi_{\Omega_{\mathpzc{h}}} \mathscr{R}_{\mathpzc{h}}(\xi)\mathscr{T}_{\mathpzc{h},1}(\xi) \mathscr{T}_{\mathpzc{h},1}(\xi)\chi_{\Omega_{\mathpzc{h}}}\right\}\right\vert \leq p(\vert \xi\vert) \mathpzc{h}^{N}.
\end{gather}
\begin{multline}
\label{etenc}
\Biggl\vert \mathrm{Tr}_{L^{2}(\mathbb{R}^{3})} \left\{ \chi_{\Omega_{\mathpzc{h}}} \mathscr{R}_{\mathpzc{h}}(\xi) \mathscr{T}_{\mathpzc{h},2}(\xi) \chi_{\Omega_{\mathpzc{h}}}\right\} - \\
\mathrm{Tr}_{L^{2}(\mathbb{R}^{3})} \left\{ \chi_{\Omega_{\mathpzc{h}}} \left(\sum_{\boldsymbol{\gamma}_{1} \in \mathscr{E}} \left(\tilde{H}_{\mathpzc{h},\boldsymbol{\gamma}_{1}}^{\mathrm{(cste)}} - \xi\right)^{-1} \tau_{\mathpzc{h},\boldsymbol{\gamma}_{1}} \sum_{\boldsymbol{\gamma}_{2} \in \mathscr{E}} \tilde{T}_{\mathpzc{h},\boldsymbol{\gamma}_{2};2}^{\mathrm{(cste)}}(\xi) \tau_{\mathpzc{h},\boldsymbol{\gamma}_{2}}\right)\chi_{\Omega_{\mathpzc{h}}}\right\}\Biggr\vert \leq p(\vert \xi\vert) \mathpzc{h}^{N}.
\end{multline}
\end{lema}

\begin{remark} All the estimates given in Lemmas \ref{isola11}-\ref{isola12}-\ref{isola13} still hold true when removing the characteristic functions $\chi_{\Omega_{\mathpzc{h}}}$ from the traces, for further details see Remark \ref{bite}.
\end{remark}

\noindent \textbf{Proof of Proposition \ref{gropro2}.} Let $\alpha \in (0,1)$ and $\theta \in (0,1]$. From \eqref{suso2} (resp. \eqref{susp2}) and by using \eqref{trace1}-\eqref{trace2}-\eqref{moust} (resp. \eqref{trace01}-\eqref{trace02}) followed by \eqref{estimo2}, then $\forall \beta>0$, $\forall z>0$ and $\forall \mathpzc{h} \in (0,\mathpzc{h}_{0}]$:
\begin{gather*}
\mathcal{X}_{\mathpzc{h}}(\beta,z,0) = \check{\mathcal{X}}_{\mathpzc{h}}^{\mathrm{(orbit)}}(\beta,z,0) + \check{\mathcal{X}}_{\mathpzc{h}}^{\mathrm{(spin)}}(\beta,z,0) + \mathcal{O}\left(z \mathpzc{h}^{-1 + \theta(1-\alpha)}\right),\\
\check{\mathcal{X}}_{\mathpzc{h}}^{\mathrm{(orbit)}}(\beta,z,0) := - \left(\frac{q}{c}\right)^{2}  \frac{4 \mathpzc{h}^{2}}{\beta \vert \Omega\vert} \frac{i}{2\pi} \int_{\mathscr{C}_{\beta}} \mathrm{d}\xi\, \mathfrak{f}(\beta,z;\xi) \mathrm{Tr}_{L^{2}(\mathbb{R}^{3})}\left\{\chi_{\Omega_{\mathpzc{h}}}\mathscr{R}_{\mathpzc{h}}(\xi) \mathscr{T}_{\mathpzc{h},2}(\xi)\chi_{\Omega_{\mathpzc{h}}}\right\},\\
\check{\mathcal{X}}_{\mathpzc{h}}^{(\mathrm{spin})}(\beta,z,0) := \left(\frac{q}{c}\right)^{2} \frac{(g \mathpzc{h})^{2}}{4} \frac{1}{\beta \vert \Omega\vert}  \frac{i}{2\pi} \int_{\mathscr{C}_{\beta}} \mathrm{d}\xi\, \mathfrak{f}(\beta,z;\xi) \mathrm{Tr}_{L^{2}(\mathbb{R}^{3})}\left\{\chi_{\Omega_{\mathpzc{h}}}\left(\mathscr{R}_{\mathpzc{h}}(\xi) \right)^{-3} \chi_{\Omega_{\mathpzc{h}}}\right\}.
\end{gather*}
To obtain \eqref{suso5} (resp. \eqref{susp5}), it remains to use \eqref{etenc} (resp. \eqref{moust1}) and \eqref{estimo2} again. \qed

\subsubsection{Isolating the main $\mathpzc{h}$-dependent contribution - Continuation and end.}
\label{subsec33}

From Proposition \ref{gropro2}, we can now turn to the end of the proof of Proposition \ref{thmsus}. We split the proof into two parts dealing respectively with the bulk zero-field orbital and spin susceptibility.

\begin{itemize}
\item \textbf{Proof of \eqref{semisuo}}.
\end{itemize}

Set $\varkappa_{\mathrm{o}} := (\frac{q}{c})^{2}\frac{4}{\vert \Omega\vert}$. In view of \eqref{suso5}, denote  $\forall 0<\alpha<1$, $\forall \beta>0$, $\forall z>0$ and $\forall\mathpzc{h} \in (0,\mathpzc{h}_{0}]$:
\begin{multline}
\label{prlaref}
\tilde{\mathcal{X}}_{\mathpzc{h}}^{\mathrm{(orbit)}}(\beta,z) := - \mathpzc{h}^{2} \frac{\varkappa_{\mathrm{o}}}{\beta} \frac{i}{2\pi} \int_{\mathscr{C}_{\beta}} \mathrm{d}\xi\, \mathfrak{f}(\beta,z;\xi) \\
\times \int_{\Omega_{\mathpzc{h}}} \mathrm{d}\mathbf{x} \int_{\mathbb{R}^{3}} \mathrm{d}\mathbf{z}\, \sum_{\boldsymbol{\gamma}_{1} \in \mathscr{E}} \left(\tilde{H}_{\mathpzc{h},\boldsymbol{\gamma}_{1}}^{\mathrm{(cste)}} - \xi\right)^{-1}(\mathbf{x},\mathbf{z}) \tau_{\mathpzc{h},\boldsymbol{\gamma}_{1}}(\mathbf{z})  \sum_{\boldsymbol{\gamma}_{2} \in \mathscr{E}} \tilde{T}_{\mathpzc{h},\boldsymbol{\gamma}_{2};2}^{\mathrm{(cste)}}(\mathbf{z},\mathbf{x};\xi) \tau_{\mathpzc{h},\boldsymbol{\gamma}_{2}}(\mathbf{x}).
\end{multline}
Splitting the integral w.r.t. $\mathbf{z}$ into two parts, introduce under the conditions of \eqref{prlaref}:
\begin{multline*}
\tilde{\mathcal{X}}_{\mathpzc{h},r_{1}}^{\mathrm{(orbit)}}(\beta,z) := \tilde{\mathcal{X}}_{\mathpzc{h}}^{\mathrm{(orbit)}}(\beta,z) + \mathpzc{h}^{2} \frac{\varkappa_{\mathrm{o}}}{\beta} \frac{i}{2\pi} \int_{\mathscr{C}_{\beta}} \mathrm{d}\xi\, \mathfrak{f}(\beta,z;\xi) \\
\times \int_{\Omega_{\mathpzc{h}}} \mathrm{d}\mathbf{x} \int_{\Omega_{\mathpzc{h}}} \mathrm{d}\bold{z} \sum_{\boldsymbol{\gamma}_{1} \in \mathscr{E}} \left(\tilde{H}_{\mathpzc{h},\boldsymbol{\gamma}_{1}}^{\mathrm{(cste)}} - \xi\right)^{-1}(\mathbf{x},\mathbf{z}) \tau_{\mathpzc{h},\boldsymbol{\gamma}_{1}}(\mathbf{z})  \sum_{\boldsymbol{\gamma}_{2} \in \mathscr{E}} \tilde{T}_{\mathpzc{h},\boldsymbol{\gamma}_{2};2}^{\mathrm{(cste)}}(\mathbf{z},\mathbf{x};\xi) \tau_{\mathpzc{h},\boldsymbol{\gamma}_{2}}(\mathbf{x}).
\end{multline*}
Remind that the above kernels are explicitly known, see \eqref{Greenf} (and \eqref{Tgam2hat}). Replacing the two above kernels with their explicit expression, then using the obvious identity which holds on $\mathbb{R}^{6}$:
\begin{equation}
\label{astuc}
\mathrm{e}^{- \varsigma_{\mathpzc{h},\boldsymbol{\gamma}}(\xi) \vert \mathbf{x} - \mathbf{y}\vert} = \mathrm{e}^{- \sqrt{-2\left(\xi - V(\mathpzc{h}\mathbf{x})\right)} \vert \mathbf{x} - \mathbf{y}\vert} + \left(\mathrm{e}^{-\varsigma_{\mathpzc{h},\boldsymbol{\gamma}}(\xi) \vert \mathbf{x} - \mathbf{y}\vert} - \mathrm{e}^{-\sqrt{-2\left(\xi - V(\mathpzc{h}\mathbf{x})\right)} \vert \mathbf{x} - \mathbf{y}\vert}\right),
\end{equation}
let us introduce under the conditions of \eqref{prlaref}:
\begin{multline*}
\tilde{\mathcal{X}}_{\mathpzc{h},r_{2}}^{\mathrm{(orbit)}}(\beta,z) := \tilde{\mathcal{X}}_{\mathpzc{h}}^{\mathrm{(orbit)}}(\beta,z) - \tilde{\mathcal{X}}_{\mathpzc{h},r_{1}}^{\mathrm{(orbit)}}(\beta,z) \\
+ \mathpzc{h}^{2} \frac{\varkappa_{\mathrm{o}}}{2 (2\pi)^{2}} \frac{1}{\beta} \frac{i}{2\pi} \int_{\mathscr{C}_{\beta}} \mathrm{d}\xi\, \mathfrak{f}(\beta,z;\xi)
\int_{\Omega_{\mathpzc{h}}} \mathrm{d}\mathbf{x} \int_{\Omega_{\mathpzc{h}}} \mathrm{d}\bold{z} \frac{\mathrm{e}^{- 2\sqrt{-2\left(\xi - V(\mathpzc{h} \mathbf{x})\right)} \vert \mathbf{x} - \mathbf{z}\vert}}{\vert \mathbf{x} - \mathbf{z}\vert^{2}} \bold{a}^{2}(\mathbf{x}-\mathbf{z}).
\end{multline*}
Finally, define under the conditions of \eqref{prlaref}:
\begin{equation*}
\tilde{\mathcal{X}}_{\mathpzc{h},r_{3}}^{\mathrm{(orbit)}}(\beta,z) := \mathpzc{h}^{2} \frac{\varkappa_{\mathrm{o}}}{8 \pi^{2}} \frac{1}{\beta} \frac{i}{2\pi} \int_{\mathscr{C}_{\beta}} \mathrm{d}\xi\, \mathfrak{f}(\beta,z;\xi)
\int_{\Omega_{\mathpzc{h}}} \mathrm{d}\mathbf{x} \int_{\mathbb{R}^{3}\setminus \Omega_{\mathpzc{h}}} \mathrm{d}\bold{z} \frac{\mathrm{e}^{- 2\sqrt{-2\left(\xi - V(\mathpzc{h}\mathbf{x})\right)} \vert \mathbf{x} - \mathbf{z}\vert}}{\vert \mathbf{x} - \mathbf{z}\vert^{2}} \bold{a}^{2}(\mathbf{x}-\mathbf{z}).
\end{equation*}
From the foregoing, one arrives $\forall 0<\alpha<1$, $\forall \beta>0$, $\forall z>0$ and $\forall\mathpzc{h} \in (0,\mathpzc{h}_{0}]$ at the decomposition:
\begin{gather}
\tilde{\mathcal{X}}_{\mathpzc{h}}^{\mathrm{(orbit)}}(\beta,z) = \tilde{\mathscr{X}}_{\mathpzc{h}}^{\mathrm{(orbit)}}(\beta,z) + \tilde{\mathcal{X}}_{\mathpzc{h},r_{1}}^{\mathrm{(orbit)}}(\beta,z) + \tilde{\mathcal{X}}_{\mathpzc{h},r_{2}}^{\mathrm{(orbit)}}(\beta,z)+ \tilde{\mathcal{X}}_{\mathpzc{h},r_{3}}^{\mathrm{(orbit)}}(\beta,z),\quad \textrm{with:}\nonumber\\
\label{chip}
\tilde{\mathscr{X}}_{\mathpzc{h}}^{\mathrm{(orbit)}}(\beta,z) := - \mathpzc{h}^{2} \frac{\varkappa_{\mathrm{o}}}{8 \pi^{2}} \frac{1}{\beta} \frac{i}{2\pi} \int_{\mathscr{C}_{\beta}} \mathrm{d}\xi\, \mathfrak{f}(\beta,z;\xi)
\int_{\Omega_{\mathpzc{h}}} \mathrm{d}\mathbf{x} \int_{\mathbb{R}^{3}} \mathrm{d}\bold{z} \frac{\mathrm{e}^{- 2\sqrt{-2\left(\xi - V(\mathpzc{h}\mathbf{x})\right)} \vert \mathbf{x} - \mathbf{z}\vert}}{\vert \mathbf{x} - \mathbf{z}\vert^{2}} \bold{a}^{2}(\mathbf{x}-\mathbf{z}).
\end{gather}

Now, we need the following lemma whose proof is placed in Appendix, see Sec. \ref{append22}.

\begin{lema}
\label{retro1}
For any $0<\alpha<1$, $0 < \theta \leq 1$, $\beta>0$ and $z>0$, it holds:
\begin{gather*}
\tilde{\mathcal{X}}_{\mathpzc{h},r_{1}}^{\mathrm{(orbit)}}(\beta,z) + \tilde{\mathcal{X}}_{\mathpzc{h},r_{2}}^{\mathrm{(orbit)}}(\beta,z) = \mathcal{O}\left(\mathpzc{h}^{-1 + \theta(1-\alpha)}\right),\\
 \tilde{\mathcal{X}}_{\mathpzc{h},r_{3}}^{\mathrm{(orbit)}}(\beta,z) = \mathcal{O}\left(\mathpzc{h}^{N}\right),\quad \forall N>0.
\end{gather*}
\end{lema}

It remains to prove that \eqref{chip} is nothing but the leading term in \eqref{semisuo}. By performing some change of variables, and by using this explicit calculation:
\begin{equation*}
\int_{\mathbb{R}^{3}} \mathrm{d}\mathbf{z}\,  \frac{\mathrm{e}^{- 2 \mathpzc{h}^{-1}\sqrt{-2\left(\xi - V(\mathbf{x})\right)} \vert \mathbf{x} - \mathbf{z}\vert}}{\vert \mathbf{x} - \mathbf{z}\vert^{2}} \left[(z_{1} - x_{1})^{2} + (z_{2} - x_{2})^{2}\right] =  2\pi \frac{4}{3} \frac{\mathpzc{h}^{3}}{8\sqrt{2} \left(-\left(\xi - V(\mathbf{x})\right)\right)^{\frac{3}{2}}},
\end{equation*}
then it follows that \eqref{chip} can be rewritten under the same conditions as:
\begin{equation*}
\tilde{\mathscr{X}}_{\mathpzc{h}}^{\mathrm{(orbit)}}(\beta,z) = - \frac{1}{\mathpzc{h}} \frac{\varkappa_{\mathrm{o}}}{96\sqrt{2} \pi} \frac{1}{\beta}   \frac{i}{2\pi} \int_{\Omega} \mathrm{d}\mathbf{x} \int_{\tilde{\mathscr{C}}_{\beta}} \mathrm{d}\xi\, \ln(1 + z \mathrm{e}^{-\beta V(\mathbf{x})} \mathrm{e}^{-\beta \xi}) \left(-\xi\right)^{-\frac{3}{2}},
\end{equation*}
where $\tilde{\mathscr{C}}_{\beta}$ denotes the contour defined as in \eqref{Gamma} but with $\tilde{\delta} := -1$ instead of the $\delta$ in \eqref{delta}. Next, by performing an integration by parts w.r.t. $\xi$, followed by the change of variable $t:= \beta \xi$:
\begin{equation}
\label{Xzero}
\tilde{\mathscr{X}}_{\mathpzc{h}}^{\mathrm{(orbit)}}(\beta,z) = -\frac{1}{\mathpzc{h}} \frac{\varkappa_{\mathrm{o}}}{48\sqrt{2} \pi} \frac{1}{\sqrt{\beta}}   \frac{i}{2\pi}  \int_{\Omega} \mathrm{d}\mathbf{x}\, \bigg(\frac{i}{2\pi}\bigg) \int_{\hat{\mathscr{C}}_{1}} \mathrm{d}t\, \frac{z \mathrm{e}^{-\beta V(\mathbf{x})}}{\mathrm{e}^{t} + z\mathrm{e}^{-\beta V(\mathbf{x})}} (-t)^{-\frac{1}{2}},
\end{equation}
where $\hat{\mathscr{C}}_{1}$ denotes the contour defined as in \eqref{Gamma} with $\beta=1$ and $\hat{\delta} = -\beta$ instead of the $\delta$ in \eqref{delta}. It remains to use the following rewriting of the Fermi-Dirac function, see \cite[Eqs. (A.3)-(A.5)]{McDS}:
\begin{equation}
\label{rel}
\forall u \geq 0,\quad f_{\nu}(u) = - \frac{\Gamma(1-\nu)}{2i\pi} \int_{\mathrm{Hl}} \mathrm{d}t\, \frac{u}{\mathrm{e}^{t} + u}(-t)^{\nu-1} \quad \textrm{with $\nu>0$ and $\nu\notin \mathbb{N}^{*}$},
\end{equation}
where $\mathrm{Hl}$ stands for a Hankel-type contour. Gathering \eqref{Xzero} and \eqref{rel} together, one arrives at:
\begin{equation*}
\tilde{\mathscr{X}}_{\mathpzc{h}}^{\mathrm{(orbit)}}(\beta,z) = - \frac{1}{\mathpzc{h}} \frac{\varkappa_{\mathrm{o}}}{48\sqrt{2} \Gamma(\frac{1}{2}) \pi} \frac{1}{\sqrt{\beta}} \int_{\Omega} \mathrm{d}\mathbf{x}\, f_{\frac{1}{2}}\left(z \mathrm{e}^{-\beta V(\bold{x})}\right),\quad \varkappa_{\mathrm{o}} = \left(\frac{q}{c}\right)^{2} \frac{4}{\vert \Omega\vert}. \tag*{\qed}
\end{equation*}

\begin{itemize}
\item \textbf{Proof of \eqref{semisup}}.
\end{itemize}

Set $\varkappa_{\mathrm{s}} :=
(\frac{q}{c})^{2} \frac{g^{2}}{4 \vert \Omega\vert}$. In view of \eqref{susp5}, denote  $\forall 0<\alpha<1$, $\forall \beta>0$, $\forall z>0$ and $\forall\mathpzc{h} \in (0,\mathpzc{h}_{0}]$:
\begin{multline}
\label{prlarefs}
\tilde{\mathcal{X}}_{\mathpzc{h}}^{\mathrm{(spin)}}(\beta,z) :=  \mathpzc{h}^{2} \frac{\varkappa_{\mathrm{s}}}{\beta} \frac{i}{2\pi} \int_{\mathscr{C}_{\beta}} \mathrm{d}\xi\, \mathfrak{f}(\beta,z;\xi)
\int_{\Omega_{\mathpzc{h}}} \mathrm{d}\mathbf{z}_{0} \int_{\mathbb{R}^{3}} \mathrm{d}\mathbf{z}_{1} \int_{\mathbb{R}^{3}} \mathrm{d}\mathbf{z}_{2} \sum_{\boldsymbol{\gamma}_{1} \in \mathscr{E}} \left(\tilde{H}_{\mathpzc{h},\boldsymbol{\gamma}_{1}}^{\mathrm{(cste)}} - \xi\right)^{-1}(\mathbf{z}_{0},\mathbf{z}_{1}) \\
\times \tau_{\mathpzc{h},\boldsymbol{\gamma}_{1}}(\mathbf{z}_{1}) \sum_{\boldsymbol{\gamma}_{2} \in \mathscr{E}} \left(\tilde{H}_{\mathpzc{h},\boldsymbol{\gamma}_{2}}^{\mathrm{(cste)}} - \xi\right)^{-1}(\mathbf{z}_{1},\mathbf{z}_{2}) \tau_{\mathpzc{h},\boldsymbol{\gamma}_{2}}(\mathbf{z}_{2}) \sum_{\boldsymbol{\gamma}_{3} \in \mathscr{E}} \left(\tilde{H}_{\mathpzc{h},\boldsymbol{\gamma}_{3}}^{\mathrm{(cste)}} - \xi\right)^{-1}(\mathbf{z}_{2},\mathbf{z}_{0}) \tau_{\mathpzc{h},\boldsymbol{\gamma}_{3}}(\mathbf{z}_{0}).
\end{multline}
Splitting the integrals w.r.t. $\mathbf{z}_{l}$, $l=1,2$ into two parts, introduce under the conditions of \eqref{prlarefs}:
\begin{multline*}
\tilde{\mathcal{X}}_{\mathpzc{h},r_{1}}^{\mathrm{(spin)}}(\beta,z) := \tilde{\mathcal{X}}_{\mathpzc{h}}^{\mathrm{(spin)}}(\beta,z)
- \mathpzc{h}^{2} \frac{\varkappa_{\mathrm{s}}}{\beta} \frac{i}{2\pi} \int_{\mathscr{C}_{\beta}} \mathrm{d}\xi\, \mathfrak{f}(\beta,z;\xi) \times \\
\int_{\Omega_{\mathpzc{h}}} \mathrm{d}\mathbf{z}_{0} \dotsb \int_{\Omega_{\mathpzc{h}}} \mathrm{d}\bold{z}_{2}\, \prod_{l=0}^{2} \sum_{\boldsymbol{\gamma}_{l+1} \in \mathscr{E}} \left(\tilde{H}_{\mathpzc{h},\boldsymbol{\gamma}_{l+1}}^{\mathrm{(cste)}} - \xi\right)^{-1}(\mathbf{z}_{l},\mathbf{z}_{l+1}) \tau_{\mathpzc{h},\boldsymbol{\gamma}_{l+1}}(\mathbf{z}_{l+1}),\quad \mathbf{z}_{3}:=\mathbf{z}_{0}.
\end{multline*}
Replacing the three above kernels with their explicit expression in \eqref{Greenf}, then by using \eqref{astuc}, let us introduce under the conditions of \eqref{prlarefs}:
\begin{multline*}
\tilde{\mathcal{X}}_{\mathpzc{h},r_{2}}^{\mathrm{(spin)}}(\beta,z) := \tilde{\mathcal{X}}_{\mathpzc{h}}^{\mathrm{(spin)}}(\beta,z) - \tilde{\mathcal{X}}_{\mathpzc{h},r_{1}}^{\mathrm{(spin)}}(\beta,z) \\
 - \mathpzc{h}^{2} \frac{\varkappa_{\mathrm{s}}}{ (2\pi)^{3}} \frac{1}{\beta} \frac{i}{2\pi} \int_{\mathscr{C}_{\beta}} \mathrm{d}\xi\, \mathfrak{f}(\beta,z;\xi)
\int_{\Omega_{\mathpzc{h}}} \mathrm{d}\mathbf{z}_{0} \dotsb  \int_{\Omega_{\mathpzc{h}}} \mathrm{d}\bold{z}_{2} \prod_{l=0}^{2} \frac{\mathrm{e}^{- \sqrt{-2\left(\xi - V(\mathpzc{h}\mathbf{z}_{0})\right)} \vert \mathbf{z}_{l} - \mathbf{z}_{l+1}\vert}}{\vert \mathbf{z}_{l} - \mathbf{z}_{l+1}\vert},\quad \mathbf{z}_{3}=\mathbf{z}_{0}.
\end{multline*}
Finally, introduce under the conditions of \eqref{prlarefs}:
\begin{multline*}
\tilde{\mathcal{X}}_{\mathpzc{h},r_{3}}^{\mathrm{(spin)}}(\beta,z) := \tilde{\mathcal{X}}_{\mathpzc{h}}^{\mathrm{(spin)}}(\beta,z) - \tilde{\mathcal{X}}_{\mathpzc{h},r_{1}}^{\mathrm{(spin)}}(\beta,z) - \tilde{\mathcal{X}}_{\mathpzc{h},r_{2}}^{\mathrm{(spin)}}(\beta,z) \\
 - \mathpzc{h}^{2} \frac{\varkappa_{\mathrm{s}}}{8 \pi^{3}} \frac{1}{\beta} \frac{i}{2\pi} \int_{\mathscr{C}_{\beta}} \mathrm{d}\xi\, \mathfrak{f}(\beta,z;\xi)
\int_{\Omega_{\mathpzc{h}}} \mathrm{d}\mathbf{z}_{0}\int_{\mathbb{R}^{3}} \mathrm{d}\mathbf{z}_{1}  \int_{\mathbb{R}^{3}} \mathrm{d}\mathbf{z}_{2} \prod_{l=0}^{2} \frac{\mathrm{e}^{- \sqrt{-2\left(\xi - V(\mathpzc{h}\mathbf{z}_{0})\right)} \vert \mathbf{z}_{l} - \mathbf{z}_{l+1}\vert}}{\vert \mathbf{z}_{l} - \mathbf{z}_{l+1}\vert},\quad \mathbf{z}_{3}=\mathbf{z}_{0}.
\end{multline*}
From the foregoing, one arrives $\forall 0<\alpha<1$, $\forall \beta>0$, $\forall z>0$ and $\forall\mathpzc{h} \in (0,\mathpzc{h}_{0}]$ at the decomposition:
\begin{equation*}
\tilde{\mathcal{X}}_{\mathpzc{h}}^{\mathrm{(spin)}}(\beta,z) = \tilde{\mathscr{X}}_{\mathpzc{h}}^{\mathrm{(spin)}}(\beta,z) + \tilde{\mathcal{X}}_{\mathpzc{h},r_{1}}^{\mathrm{(spin)}}(\beta,z) + \tilde{\mathcal{X}}_{\mathpzc{h},r_{2}}^{\mathrm{(spin)}}(\beta,z)+ \tilde{\mathcal{X}}_{\mathpzc{h},r_{3}}^{\mathrm{(spin)}}(\beta,z), \quad \textrm{with:}
\end{equation*}
\begin{multline}
\label{chis}
\tilde{\mathscr{X}}_{\mathpzc{h}}^{\mathrm{(spin)}}(\beta,z) :=  \mathpzc{h}^{2} \frac{\varkappa_{\mathrm{s}}}{8 \pi^{3}} \frac{1}{\beta} \frac{i}{2\pi} \int_{\mathscr{C}_{\beta}} \mathrm{d}\xi\, \mathfrak{f}(\beta,z;\xi) \\
\times \int_{\Omega_{\mathpzc{h}}} \mathrm{d}\mathbf{z}_{0} \int_{\mathbb{R}^{3}} \mathrm{d}\bold{z}_{1} \int_{\mathbb{R}^{3}} \mathrm{d}\mathbf{z}_{2}\, \prod_{l=0}^{2} \frac{\mathrm{e}^{- \sqrt{-2\left(\xi - V(\mathpzc{h}\mathbf{z}_{0})\right)} \vert \mathbf{z}_{l} - \mathbf{z}_{l+1}\vert}}{\vert \mathbf{z}_{l} - \mathbf{z}_{l+1}\vert},\quad \mathbf{z}_{3} = \mathbf{z}_{0}.
\end{multline}

Now, we need the following lemma whose proof is placed in Appendix, see Sec. \ref{append22}.

\begin{lema}
\label{retro2}
For any $0<\alpha<1$, $0 < \theta \leq 1$, $\beta>0$ and $z>0$, it holds:
\begin{gather*}
\tilde{\mathcal{X}}_{\mathpzc{h},r_{1}}^{\mathrm{(spin)}}(\beta,z) + \tilde{\mathcal{X}}_{\mathpzc{h},r_{2}}^{\mathrm{(spin)}}(\beta,z)  = \mathcal{O}\left(\mathpzc{h}^{-1 + \theta(1-\alpha)}\right),\\ \tilde{\mathcal{X}}_{\mathpzc{h},r_{3}}^{\mathrm{(spin)}}(\beta,z) = \mathcal{O}\left(\mathpzc{h}^{N}\right),\quad \forall N > 0.
\end{gather*}
\end{lema}

It remains to prove that \eqref{chis} is nothing but the leading term in \eqref{semisup}. By performing some change of variables, then by using that (below, we set $\mathbf{z}_{3}=\mathbf{z}_{0}$):
\begin{equation*}
\int_{\mathbb{R}^{3}} \mathrm{d}\mathbf{z}_{1} \int_{\mathbb{R}^{3}} \mathbf{z}_{2}\,  \prod_{l=0}^{2} \frac{\mathrm{e}^{- 2 \mathpzc{h}^{-1}\sqrt{-2\left(\xi - V(\mathbf{z}_{0})\right)} \vert \mathbf{z}_{l} - \mathbf{z}_{l+1}\vert}}{\vert \mathbf{z}_{l} - \mathbf{z}_{l+1}\vert}  =  \sqrt{\pi} (2\pi)^{\frac{3}{2}} \frac{\mathpzc{h}^{3}}{4 \left(-\left(\xi - V(\mathbf{z}_{0})\right)\right)^{\frac{3}{2}}},
\end{equation*}
it follows that \eqref{chis} can be rewritten under the same conditions as:
\begin{equation*}
\tilde{\mathscr{X}}_{\mathpzc{h}}^{\mathrm{(spin)}}(\beta,z) = \frac{1}{\mathpzc{h}} \frac{\sqrt{\pi}\varkappa_{\mathrm{s}}}{4 (2\pi)^{\frac{3}{2}}} \frac{1}{\beta}   \frac{i}{2\pi} \int_{\Omega} \mathrm{d}\mathbf{x} \int_{\tilde{\mathscr{C}}_{\beta}} \mathrm{d}\xi\, \ln(1 + z \mathrm{e}^{-\beta V(\mathbf{x})} \mathrm{e}^{-\beta \xi}) \left(-\xi\right)^{-\frac{3}{2}}.
\end{equation*}
Next, by performing an integration by parts w.r.t. $\xi$, followed by the change of variable $t:= \beta \xi$:
\begin{equation}
\label{Xzeros}
\tilde{\mathscr{X}}_{\mathpzc{h}}^{\mathrm{(spin)}}(\beta,z) = \frac{1}{\mathpzc{h}} \frac{\sqrt{\pi}\varkappa_{\mathrm{s}}}{2 (2\pi)^{\frac{3}{2}}} \frac{1}{\sqrt{\beta}}   \frac{i}{2\pi}  \int_{\Omega} \mathrm{d}\mathbf{x}\, \bigg(\frac{i}{2\pi}\bigg) \int_{\hat{\mathscr{C}}_{1}} \mathrm{d}t\, \frac{z \mathrm{e}^{-\beta V(\mathbf{x})}}{\mathrm{e}^{t} + z\mathrm{e}^{-\beta V(\mathbf{x})}} (-t)^{-\frac{1}{2}},
\end{equation}
where $\hat{\mathscr{C}}_{1}$ is the contour as in \eqref{Xzero}. Gathering \eqref{Xzeros} and \eqref{rel} together, one arrives finally at:
\begin{equation*}
\tilde{\mathscr{X}}_{\mathpzc{h}}^{\mathrm{(spin)}}(\beta,z) =  \frac{1}{\mathpzc{h}} \frac{\sqrt{\pi}\varkappa_{\mathrm{s}}}{2 \Gamma(\frac{1}{2}) (2\pi)^{\frac{3}{2}}} \frac{1}{\sqrt{\beta}} \int_{\Omega} \mathrm{d}\mathbf{x}\, f_{\frac{1}{2}}\left(z \mathrm{e}^{-\beta V(\bold{x})}\right),\quad \varkappa_{\mathrm{s}} = \left(\frac{q}{c}\right)^{2} \frac{g^{2}}{4 \vert \Omega\vert}. \tag*{\qed}
\end{equation*}

\section{Appendix.}
\label{append}

Throughout this section, $(\mathfrak{I}_{2}(L^{2}(\mathbb{R}^{3})), \Vert \cdot\, \Vert_{\mathfrak{I}_{2}})$ and $(\mathfrak{I}_{1}(L^{2}(\mathbb{R}^{3})), \Vert \cdot\,\Vert_{\mathfrak{I}_{1}})$ denote the Banach space of Hilbert-Schmidt (H-S) and trace-class operators on $L^{2}(\mathbb{R}^{3})$ respectively.

\subsection{Proof of Lemmas \ref{isola0}-\ref{isola1}.}
\label{append21}

\indent When using the estimates from Lemmas \ref{lem01} and \ref{lem1}, we set
$\eta = \min\{ 1,\frac{\pi}{2\beta}\}>0$, see \eqref{Gamma}.\\

\noindent \textbf{Proof of Lemma \ref{isola0}}. Let $0<\alpha<1$ and $\beta>0$. We start with $\mathrm{(i)}$. Since the integral kernels of $(\tilde{H}_{\mathpzc{h}} -\xi)^{-1}$ and $\mathcal{W}_{\mathpzc{h}}(\xi)$ obey \eqref{trures} and  \eqref{kerT} respectively, then each one is locally H-S. Ergo,
$(\tilde{H}_{\mathpzc{h}} -\xi)^{-1} \mathcal{W}_{\mathpzc{h}}(\xi)$ is locally trace-class on $L^{2}(\mathbb{R}^{3})$. From \eqref{trures} and  \eqref{kerT} again, $(\tilde{H}_{\mathpzc{h}} -\xi)^{-1}\mathcal{W}_{\mathpzc{h}}(\xi)$ has a jointly continuous integral kernel on $\mathbb{R}^{6}$, see  \cite[Lem. A.1]{BS}. From the foregoing, it follows:
\begin{equation*}
\mathrm{Tr}_{L^{2}(\mathbb{R}^{3})}\left\{\chi_{\Omega_{\mathpzc{h}}} \left(\tilde{H}_{\mathpzc{h}} - \xi\right)^{-1} \mathcal{W}_{\mathpzc{h}}(\xi)\chi_{\Omega_{\mathpzc{h}}}\right\} =  \int_{\Omega_{\mathpzc{h}}} \mathrm{d}\mathbf{x}\, \left(\left(\tilde{H}_{\mathpzc{h}} - \xi\right)^{-1} \mathcal{W}_{\mathpzc{h}}(\xi)\right)(\mathbf{x},\mathbf{x}).
\end{equation*}
Under the conditions of Lemma \ref{isola0},  there exists a $\vartheta>0$ and a polynomial $p(\cdot\,)$ s.t. $\forall \mathpzc{h} \in (0,\mathpzc{h}_{0}]$:
\begin{equation}
\label{kerresT}
\forall (\mathbf{x},\mathbf{y}) \in \mathbb{R}^{6}, \quad \left\vert\left(\left(\tilde{H}_{\mathpzc{h}} - \xi\right)^{-1} \mathcal{W}_{\mathpzc{h}}(\xi)\right)(\mathbf{x},\mathbf{y})\right\vert \leq p(\vert \xi \vert) \mathrm{e}^{- \frac{\vartheta}{1 + \vert \xi\vert} \mathpzc{h}^{- \alpha}} \mathrm{e}^{- \frac{\vartheta}{1 + \vert \xi\vert} \vert \mathbf{x} - \mathbf{y}\vert}.
\end{equation}
This leads to:
\begin{equation*}
\forall \mathpzc{h} \in (0,\mathpzc{h}_{0}],\, \forall \xi \in \mathscr{C}_{\beta},\quad \left \vert \mathrm{Tr}_{L^{2}(\mathbb{R}^{3})}\left\{\chi_{\Omega_{\mathpzc{h}}} \left(\tilde{H}_{\mathpzc{h}} - \xi\right)^{-1} \mathcal{W}_{\mathpzc{h}}(\xi)\chi_{\Omega_{\mathpzc{h}}}\right\}\right\vert \leq p(\vert \xi \vert) \vert \Omega_{\mathpzc{h}} \vert \mathrm{e}^{- \frac{\vartheta}{1 + \vert \xi\vert} \mathpzc{h}^{- \alpha}},
\end{equation*}
for another $\mathpzc{h}$-independent $p(\cdot\,)$. It remains to use \eqref{keyes} to get rid of the factor $\vert \Omega_{\mathpzc{h}}\vert = \mathpzc{h}^{-3}$ and use the following inequality: $\forall M>0$ there exists a constant $C_{M}>0$ s.t. $\forall \varkappa\geq 0$ and $\forall t \geq 0$:
\begin{equation}
\label{attendu}
\mathrm{e}^{-\varkappa t} \leq C_{M} (1 + \varkappa t)^{-M}.
\end{equation}
Next, we turn to $\mathrm{(ii)}$. From \eqref{I2n1} along with \eqref{accroissfi}, then under the conditions of Lemma \ref{isola0}, there exists a polynomial $p(\cdot\,)$ s.t. $\forall \theta \in (0,1]$, $\forall \mathpzc{h} \in (0,\mathpzc{h}_{0}]$ and $\forall \xi \in \mathscr{C}_{\beta}$:
\begin{equation*}
 \left\Vert \sum_{\boldsymbol{\gamma} \in \mathscr{E}} \hat{\tau}_{\mathpzc{h},\boldsymbol{\gamma}} \left(\tilde{H}_{\mathpzc{h},\boldsymbol{\gamma}}^{\mathrm{(ref)}} - \xi\right)^{-1} \left\{\hat{\hat{\tau}}_{\mathpzc{h},\boldsymbol{\gamma}} \left(V(\mathpzc{h}^{1-\alpha} \boldsymbol{\gamma}) - V(\mathpzc{h}\cdot\,)\right)\right\} \left(\tilde{H}_{\mathpzc{h},\boldsymbol{\gamma}}^{\mathrm{(cste)}} - \xi\right)^{-1} \tau_{\mathpzc{h},\boldsymbol{\gamma}} \right\Vert_{\mathfrak{I}_{1}} \leq p(\vert \xi\vert) \mathpzc{h}^{-3 + \theta(1-\alpha)}.
\end{equation*}
Here we used that $\mathrm{Card}(\mathscr{E}) = \mathcal{O}( \mathpzc{h}^{3 \alpha - 3})$ to get rid of the sum. Such an estimate still holds true when sandwiching the sum with two indicator functions $\chi_{\Omega_{\mathpzc{h}}}$. Finally, \eqref{rftgh2} follows from:
\begin{equation*}
\left\vert \mathrm{Tr}_{L^{2}(\mathbb{R}^{3})}\left\{\chi_{\Omega_{\mathpzc{h}}} \mathscr{W}_{\mathpzc{h}}(\xi) \chi_{\Omega_{\mathpzc{h}}}\right\}\right\vert \leq \left\Vert \chi_{\Omega_{\mathpzc{h}}} \mathscr{W}_{\mathpzc{h}}(\xi) \chi_{\Omega_{\mathpzc{h}}}\right\Vert_{\mathfrak{I}_{1}}. \tag*{\qed}
\end{equation*}

\noindent \textbf{Proof of Lemma \ref{isola1}}. Let $0<\alpha<1$. From \eqref{Sepsi} and by using the Dunford functional calculus in \cite[Sec. VI.3]{DS}, $\forall \beta>0$, $\forall z >0$ and $\forall \mathpzc{h} \in (0,\mathpzc{h}_{0}]$ it takes place in the bounded operators sense:
\begin{equation}
\label{fermdirc}
\frac{i}{2\pi} \int_{\mathscr{C}_{\beta}} \mathrm{d}\xi\, \mathfrak{f}_{FD}(\beta,z;\xi) \mathscr{R}_{\mathpzc{h}}(\xi) = \sum_{\boldsymbol{\gamma} \in \mathscr{E}} \hat{\tau}_{\mathpzc{h},\boldsymbol{\gamma}} \mathfrak{f}_{FD}\left(\beta,z;\tilde{H}_{\mathpzc{h},\boldsymbol{\gamma}}^{\mathrm{(cste)}}\right) \tau_{\mathpzc{h},\boldsymbol{\gamma}}.
\end{equation}
Next, we need to write down an expression for the kernel of $\mathfrak{f}_{FD}(\beta,z;\tilde{H}_{\mathpzc{h},\boldsymbol{\gamma}}^{\mathrm{(cste)}})$. Since $\tilde{H}_{\mathpzc{h},\boldsymbol{\gamma}}^{\mathrm{(cste)}}$ has a constant potential, see \eqref{HconstV}, then performing a Fourier transform leads in the kernels sense to:
\begin{equation*}
\forall (\mathbf{x},\mathbf{y}) \in \mathbb{R}^{6},\quad \left(\mathfrak{f}_{FD}\left(\beta,z; \tilde{H}_{\mathpzc{h},\boldsymbol{\gamma}}^{\mathrm{(cste)}}\right)\right)(\mathbf{x},\mathbf{y}) = \frac{1}{(2\pi)^{3}} \int_{\mathbb{R}^{3}} \mathrm{d}\mathbf{k}\, \frac{z \mathrm{e}^{-\beta \left(\frac{1}{2}\mathbf{k}^{2} + V(\mathpzc{h}^{1-\alpha} \boldsymbol{\gamma})\right)}}{1+ z\mathrm{e}^{-\beta \left(\frac{1}{2}\mathbf{k}^{2}  + V(\mathpzc{h}^{1-\alpha} \boldsymbol{\gamma})\right)}} \mathrm{e}^{i \mathbf{k} \cdot (\mathbf{x} - \mathbf{y})}.
\end{equation*}
By setting $\mathbf{y}=\mathbf{x}$, the diagonal part is nothing but a constant (in accordance with the fact that the operator $\tilde{H}_{\mathpzc{h},\boldsymbol{\gamma}}^{\mathrm{(cste)}}$ commutes with the real translations). Now, denote for any $\beta>0$ and $z>0$:
\begin{equation}
\label{grandphi}
\forall \mathbf{w} \in \mathbb{R}^{3},\quad \Phi(\beta,z;\mathbf{w}) := \frac{1}{(2\pi)^{3}} \int_{\mathbb{R}^{3}} \mathrm{d}\mathbf{k}\, \frac{z \mathrm{e}^{-\beta \left(\frac{1}{2}\mathbf{k}^{2} + V(\mathbf{w})\right)}}{1 + z\mathrm{e}^{-\beta \left(\frac{1}{2}\mathbf{k}^{2} + V(\mathbf{w})\right)}}.
\end{equation}
In view of \eqref{fermdirc} and \eqref{grandphi}, the l.h.s. of \eqref{rftgh3} can be rewritten $\forall \beta>0$, $\forall z>0$ and $\forall \mathpzc{h} \in (0,\mathpzc{h}_{0}]$ as:
\begin{multline}
\label{groPh}
\frac{2}{\vert \Omega \vert} \mathrm{Tr}_{L^{2}(\mathbb{R}^{3})}\left\{\chi_{\Omega_{\mathpzc{h}}} \left(\frac{i}{2\pi} \int_{\mathscr{C}_{\beta}} \mathrm{d}\xi\, \mathfrak{f}_{FD}(\beta,z;\xi) \mathscr{R}_{\mathpzc{h}}(\xi)\right)\chi_{\Omega_{\mathpzc{h}}}\right\} = \\\frac{2}{\vert \Omega\vert} \frac{1}{\mathpzc{h}^{3}} \int_{\Omega} \mathrm{d}\mathbf{x}\, \Phi(\beta,z;\mathbf{x})
 + \frac{2}{\vert \Omega\vert} \frac{1}{\mathpzc{h}^{3}} \int_{\Omega} \mathrm{d}\mathbf{x}\, \sum_{\boldsymbol{\gamma} \in \mathscr{E}} \left\{\Phi\left(\beta,z;\mathpzc{h}^{1-\alpha} \boldsymbol{\gamma}\right) - \Phi(\beta,z;\mathbf{x})\right\} \tau_{\mathpzc{h},\boldsymbol{\gamma}}\left(\frac{\mathbf{x}}{\mathpzc{h}}\right),
\end{multline}
where we used \eqref{partiden} in the first term of the r.h.s. of \eqref{groPh}. It remains to prove that the second term behaves like $\mathcal{O}(z\mathpzc{h}^{-3 + \theta(1-\alpha)})$. To do that, we use that the map $\mathbb{R}^{3} \owns \mathbf{w} \mapsto \Phi(\beta,z;\mathbf{w})$ is globally H\"older continuous with exponent $\theta \in (0,1]$ as a result of \eqref{deftheta}. Indeed:
\begin{equation*}
\forall (\mathbf{w},\mathbf{w}_{0}) \in \mathbb{R}^{6},\quad \left\vert \Phi(\beta,z;\mathbf{w})-\Phi(\beta,z;\mathbf{w}_{0})\right\vert \leq  \beta z \mathrm{e}^{\beta \Vert V \Vert_{\infty}} \left(\int_{\mathbb{R}^{3}} \mathrm{d}\mathbf{k}\, \mathrm{e}^{-\frac{1}{2} \beta \mathbf{k}^{2}}\right)\left\vert V(\mathbf{w}) - V(\mathbf{w}_{0})\right\vert.
\end{equation*}
From \eqref{accroissfi}, for any $\beta>0$ there exists a $c=c(\beta)>0$ s.t. $\forall z>0$, $\forall \mathpzc{h} \in (0,\mathpzc{h}_{0}]$ and $\forall \boldsymbol{\gamma} \in \mathscr{E}$:
\begin{equation*}
\forall \mathbf{x} \in \Omega \,\,\textrm{s.t.}\,\, \mathbf{x}\mathpzc{h}^{-1} \in \mathrm{Supp}\left(\tau_{\mathpzc{h},\boldsymbol{\gamma}}\right),\quad \left\vert \Phi\left(\beta,z;\mathpzc{h}^{1-\alpha}\boldsymbol{\gamma}\right) - \Phi\left(\beta,z;\mathbf{x}\right) \right\vert \leq c z  \mathpzc{h}^{\theta(1 - \alpha)}.
\end{equation*}
The behavior of the remainder in \eqref{rftgh3} follows from the above estimate along with \eqref{partiden}.
\qed

\subsection{Proof of Lemma \ref{derPstar} and Proposition \ref{weakequi}.}
\label{append25}

\noindent \textit{\textbf{Proof of Lemma \ref{derPstar}}}. Let $\beta>0$, $\rho^{\mathpzc{s}}>0$ and $\mathpzc{h}>0$ be fixed. Remind that $(z,b) \mapsto P_{\mathpzc{h}}^{\mathpzc{s}}(\beta,z,b)$ is jointly smooth on $(0,\infty) \times \mathbb{R}$, see Sec. \ref{mainre}. In view of \eqref{Tleg}, it remains to prove that $b \mapsto \overline{z}_{\mathpzc{h}}^{\mathpzc{s}}(\beta,\rho^{\mathpzc{s}},b)$ is a $\mathcal{C}^{\infty}$-function. Pick $b_{0} \in \mathbb{R}$, and consider the equation $\rho_{\mathpzc{h}}^{\mathpzc{s}}(\beta,\overline{z}_{\mathpzc{h}}^{\mathpzc{s}}(\beta,\rho^{\mathpzc{s}},b_{0}) ,b_{0}) - \rho^{\mathpzc{s}} = 0$.
Due to \eqref{densitylim}, $\rho_{\mathpzc{h}}^{\mathpzc{s}}(\beta,\cdot\,,\cdot\,)-\rho^{\mathpzc{s}}$ is jointly smooth in $(z,b) \in (0,\infty)\times\mathbb{R}$. Besides, since $z \mapsto \rho_{\mathpzc{h}}^{\mathpzc{s}}(\beta,z,b)$ is strictly increasing and does not vanish on $(0,\infty)$, then $(\partial_{z} \rho_{\mathpzc{h}}^{\mathpzc{s}})(\beta,\overline{z}_{\mathpzc{h}}^{\mathpzc{s}}(\beta,\rho^{\mathpzc{s}},b_{0}) ,b_{0})>0$. Therefore, $b \mapsto \overline{z}_{\mathpzc{h}}^{\mathpzc{s}}(\beta,\rho^{\mathpzc{s}},b)$ is smooth in a neighborhood $I(b_{0})$ of $b_{0}$ by the implicit function theorem, and:
\begin{equation}
\label{derz0}
\forall b \in I(b_{0}), \quad \frac{\partial \overline{z}_{\mathpzc{h}}^{\mathpzc{s}}}{\partial b}(\beta,\rho^{\mathpzc{s}},b) = -\frac{\partial \rho_{\mathpzc{h}}^{\mathpzc{s}}}{\partial b}\left(\beta,\overline{z}_{\mathpzc{h}}^{\mathpzc{s}}(\beta,\rho^{\mathpzc{s}},b),b\right) \left(\frac{\partial \rho_{\mathpzc{h}}^{\mathpzc{s}}}{\partial z}\left(\beta,\overline{z}_{\mathpzc{h}}^{\mathpzc{s}}(\beta,\rho^{\mathpzc{s}},b),b\right)\right)^{-1}.
\end{equation}
This proves the first statement. By using that $(\partial_{z} P_{\mathpzc{h}}^{\mathpzc{s}})(\beta,\overline{z}_{\mathpzc{h}}^{\mathpzc{s}}(\beta,\rho^{\mathpzc{s}},b),b) = (\beta \overline{z}_{\mathpzc{h}}^{\mathpzc{s}}(\beta,\rho^{\mathpzc{s}},b))^{-1}\rho^{\mathpzc{s}}$, then:
\begin{gather*}
\frac{\partial (P_{\mathpzc{h}}^{\mathpzc{s}})^{*}}{\partial b}(\beta,\rho^{\mathpzc{s}},b) =  - \frac{\partial P_{\mathpzc{h}}^{\mathpzc{s}}}{\partial b}\left(\beta, \overline{z}_{\mathpzc{h}}^{\mathpzc{s}}(\beta,\rho^{\mathpzc{s}},b),b\right),\\
\frac{\partial^{2} (P_{\mathpzc{h}}^{\mathpzc{s}})^{*}}{\partial b^{2}}(\beta,\rho^{\mathpzc{s}},b) = - \frac{\partial^{2} P_{\mathpzc{h}}^{\mathpzc{s}}}{\partial b^{2}}\left(\beta, \overline{z}_{\mathpzc{h}}^{\mathpzc{s}}(\beta,\rho^{\mathpzc{s}},b),b\right) - \frac{\partial \overline{z}_{\mathpzc{h}}^{\mathpzc{s}}}{\partial b}(\beta,\rho^{\mathpzc{s}},b)\frac{\partial^{2} P_{\mathpzc{h}}^{\mathpzc{s}}}{\partial z \partial b}\left(\beta, \overline{z}_{\mathpzc{h}}^{\mathpzc{s}}(\beta,\rho^{\mathpzc{s}},b),b\right).
\end{gather*}
It remains to use that $b \mapsto \rho_{\mathpzc{h}}^{\mathpzc{s}}(\beta,z,b)$ is an even function as point-wise limit of the sequence of even functions $\{\rho_{\mathpzc{h},L}^{\mathpzc{s}}(\beta,z,\cdot\,)\}_{L >0}$. Due to \eqref{derz0}, this implies that $(\partial_{b} \overline{z}_{\mathpzc{h}}^{\mathpzc{s}})(\beta,\rho^{\mathpzc{s}},0) = 0$.\qed


\begin{remark}
\label{canorem} From the above proof, we can see that in non-vanishing magnetic field:
\begin{equation*}
-\left(\frac{q}{c}\right)^{2} \frac{\partial^{2} (P_{\mathpzc{h}}^{\mathpzc{s}})^{*}}{\partial b^{2}}(\beta,\rho^{\mathpzc{s}},b) =  \mathcal{X}_{\mathpzc{h}}^{\mathpzc{s}\,\mathrm{(GC)}}(\beta,\rho^{\mathpzc{s}},b) + \left(\frac{q}{c}\right)^{2} \frac{\partial \overline{z}_{\mathpzc{h}}^{\mathpzc{s}}}{\partial b}(\beta,\rho^{\mathpzc{s}},b)\frac{\partial^{2} P_{\mathpzc{h}}^{\mathpzc{s}}}{\partial z \partial b}\left(\beta, \overline{z}_{\mathpzc{h}}^{\mathpzc{s}}(\beta,\rho^{\mathpzc{s}},b),b\right).
\end{equation*}
\end{remark}

\noindent \textit{\textbf{Proof of Proposition \ref{weakequi}}.} We start by proving \eqref{enlib}. For any $L>0$, $\beta>0$, $\rho^{\mathpzc{s}}>0$, $b \in \mathbb{R}$ and $\mathpzc{h}>0$, define the Legendre-transform of the finite-volume grand-canonical pressure in \eqref{pressurefv} as:
\begin{equation}
\label{LTPL}
\begin{split}
(P_{\mathpzc{h},L}^{\mathpzc{s}})^{*}(\beta,\rho^{\mathpzc{s}},b) :&= \sup_{\mu \in \mathbb{R}}\left(\rho^{\mathpzc{s}} \mu - P_{\mathpzc{h},L}^{\mathpzc{s}}\left(\beta,\mathrm{e}^{\beta \mu},b\right)\right) \\
&= \frac{\rho^{\mathpzc{s}}}{\beta} \log\left(\overline{z}_{\mathpzc{h},L}^{\mathpzc{s}}(\beta,\rho^{\mathpzc{s}},b)\right) - P_{\mathpzc{h},L}^{\mathpzc{s}}\left(\beta, \overline{z}_{\mathpzc{h},L}^{\mathpzc{s}}(\beta,\rho^{\mathpzc{s}},b),b\right),
\end{split}
\end{equation}
where $\overline{z}_{\mathpzc{h},L}^{\mathpzc{s}}(\beta,\rho^{\mathpzc{s}},b)>0$ is the unique solution of the equation $\rho_{\mathpzc{h},L}^{\mathpzc{s}}(\beta,z,b) = \rho^{\mathpzc{s}}$, see \eqref{knock}.
In view of \eqref{Tleg}, one has the following point-wise convergence:
\begin{equation}
\label{limTL}
\lim_{L \uparrow \infty} (P_{\mathpzc{h},L}^{\mathpzc{s}})^{*}(\beta,\rho^{\mathpzc{s}},b) = (P_{\mathpzc{h}}^{\mathpzc{s}})^{*}(\beta,\rho^{\mathpzc{s}},b),\quad \mathpzc{h}>0,\,\beta>0,\,\rho^{\mathpzc{s}}>0,\,b\in \mathbb{R}.
\end{equation}
This comes from the convergence in \eqref{bulkpres} which is compact in $z$, along with, see e.g. \cite{Sa}:
\begin{equation}
\label{limzL}
\lim_{L \uparrow \infty} \overline{z}_{\mathpzc{h},L}^{\mathpzc{s}}(\beta,\rho^{\mathpzc{s}},b) = \overline{z}_{\mathpzc{h}}^{\mathpzc{s}}(\beta,\rho^{\mathpzc{s}},b).
\end{equation}
The starting-point consists in expressing the free energy density in \eqref{Helm} in terms of the Legendre transform in \eqref{LTPL} from \eqref{goore}. The contour appearing in \eqref{goore} has to be included in the holomorphic domain $\mathbb{C}\setminus (-\infty,-\mathrm{e}^{\beta E_{\mathpzc{h}}^{\mathpzc{s}}(b)}]$ of $z \mapsto \widehat{P}_{\mathpzc{h},L}^{\mathpzc{s}}(\beta,z,b)$ while surrounding the origin. Under the assumptions of Proposition \ref{weakequi}, we claim that it can be chosen for $L$ large enough as:
\begin{equation}
\label{cotour}
\mathscr{C} := \left\{\overline{z}_{\mathpzc{h},L}^{\mathpzc{s}}(\beta,\rho^{\mathpzc{s}},b) \mathrm{e}^{i \phi},\, \phi \in [-\pi,\pi]\right\}.
\end{equation}
Indeed, if $\rho \in (0,\rho_{\mathpzc{h}}^{\mathpzc{s}}(\beta, \mathrm{e}^{\beta E_{\mathpzc{h}}^{\mathpzc{s}}(b)},b))$, then
$0< \overline{z}_{\mathpzc{h}}^{\mathpzc{s}}(\beta,\rho^{\mathpzc{s}},b) < \mathrm{e}^{\beta E_{\mathpzc{h}}^{\mathpzc{s}}(b)}$ since $z\mapsto \rho_{h}^{\mathpzc{s}}(\beta,z,b)$ is strictly increasing. Now by using \eqref{limzL}, then one has for $L$ sufficiently large:
\begin{equation*}
\overline{z}_{\mathpzc{h},L}^{\mathpzc{s}}(\beta,\rho^{\mathpzc{s}},b) \leq \frac{\overline{z}_{\mathpzc{h}}^{\mathpzc{s}}(\beta,\rho^{\mathpzc{s}},b) + \mathrm{e}^{\beta E_{\mathpzc{h}}^{\mathpzc{s}}(b)}}{2} < \mathrm{e}^{\beta E_{\mathpzc{h}}^{\mathpzc{s}}(b)}.
\end{equation*}
We emphasize that the restriction set on $\rho^{\mathpzc{s}}$ in Proposition \ref{weakequi} allows us to consider a contour of type \eqref{cotour}. This feature is important for the following. Performing a change of variable in \eqref{goore}, and expressing $\overline{z}_{\mathpzc{h},L}^{\mathpzc{s}}(\beta,\rho^{\mathpzc{s}},b)$ in terms of $(P_{\mathpzc{h},L}^{\mathpzc{s}})^{*}(\beta,\rho^{\mathpzc{s}},b)$ from \eqref{LTPL}, we get for $L$ large enough:
\begin{gather}
\label{ademo}
\mathcal{F}_{\mathpzc{h},L}^{\mathpzc{s}}(\beta,\rho^{\mathpzc{s}},b) = (P_{\mathpzc{h},L}^{\mathpzc{s}})^{*}(\beta,\rho^{\mathpzc{s}},b) + \frac{\ln(N_{L}^{\mathpzc{s}})}{2 \beta \vert \Lambda_{L}\vert} - \frac{1}{\beta \vert \Lambda_{L}\vert} \ln(\mathcal{A}_{\mathpzc{h},L}^{\mathpzc{s}}(\beta,\rho^{\mathpzc{s}},b)), \quad N_{L}^{\mathpzc{s}}:= \rho^{\mathpzc{s}} \vert \Lambda_{L}\vert,\\
\label{calA}
\mathcal{A}_{\mathpzc{h},L}^{\mathpzc{s}}(\beta,\rho^{\mathpzc{s}},b) := \frac{\sqrt{N_{L}^{\mathpzc{s}}}}{2\pi} \int_{-\pi}^{\pi} \mathrm{d}\phi\, \mathrm{e}^{\beta\frac{N_{L}^{\mathpzc{s}}}{\rho^{\mathpzc{s}}}\left(\widehat{P}_{\mathpzc{h},L}^{\mathpzc{s}} \left(\beta,\overline{z}_{\mathpzc{h},L}^{\mathpzc{s}}(\beta,\rho^{\mathpzc{s}},b) \mathrm{e}^{i\phi},b\right) - P_{\mathpzc{h},L}^{\mathpzc{s}}\left(\beta,\overline{z}_{\mathpzc{h},L}^{\mathpzc{s}}(\beta,\rho^{\mathpzc{s}},b) ,b\right)\right)} \mathrm{e}^{-i N_{L}^{\mathpzc{s}} \phi}.
\end{gather}
\eqref{calA} corresponds to \cite[Eq. (3.3)]{C1}, in which the case of a spin-0 Bose gas interacting only with a constant magnetic field has been treated (for densities smaller than the critical density). Since the contour in \eqref{cotour} is of the same type as \cite[Eq. (3.1)]{C1}, it is enough to mimic the proof of
\cite[Thm. 2]{C1}.
Let us explain the strategy. It consists in taking a well-chosen parametrization of the contour \eqref{cotour} in a neighborhood of $\phi=0$. To do so, set $\phi_{L}^{\mathpzc{s}} := (N_{L}^{\mathpzc{s}})^{\kappa}$, with $\kappa \in (-\frac{1}{2},-\frac{1}{3})$. Then, the contour can be decomposed into two parts: one corresponding to $\vert \phi\vert \geq \phi_{L}^{\mathpzc{s}}$ and the other one to $\vert \phi\vert\leq \phi_{L}^{\mathpzc{s}}$. The rest of the strategy consists in showing that:
\begin{itemize}
\item[(1)] the contribution to $\mathcal{A}_{\mathpzc{h},L}^{\mathpzc{s}}$ coming from the region $\vert \phi\vert \geq \phi_{L}^{\mathpzc{s}}$ is exponentially small in $L$.
\item[(2)] the contribution to $\mathcal{A}_{\mathpzc{h},L}^{\mathpzc{s}}$ coming from the region $\vert \phi\vert \leq \phi_{L}^{\mathpzc{s}}$ has a strictly positive limit.
\end{itemize}

We collect in the following lemma, stated without proof, all the results needed to mimic the proof of (1) and (2) in \cite[pp. 7--10]{C1}. The same results are also used in the proof of \cite[Thm. 2]{C1}, then their proof given in \cite{C1} can be easily adapted to our situation. For simplicity's sake, we set:
\begin{gather*}
P_{\mathpzc{h},L}^{\mathpzc{s}}\left(\overline{z}_{\mathpzc{h},L}^{\mathpzc{s}}\right)= P_{\mathpzc{h},L}^{\mathpzc{s}}\left(\beta,\overline{z}_{\mathpzc{h},L}^{\mathpzc{s}}(\beta,\rho^{\mathpzc{s}},b) ,b\right),\quad \widehat{P}_{\mathpzc{h},L}^{\mathpzc{s}}\left(\overline{z}_{\mathpzc{h},L}^{\mathpzc{s}} \mathrm{e}^{i\phi}\right)= \widehat{P}_{\mathpzc{h},L}^{\mathpzc{s}} \left(\beta,\overline{z}_{\mathpzc{h},L}^{\mathpzc{s}}(\beta,\rho^{\mathpzc{s}},b) \mathrm{e}^{i\phi},b\right),\\
\Re \widehat{P}_{\mathpzc{h},L}^{\mathpzc{s}}\left(\overline{z}_{\mathpzc{h},L}^{\mathpzc{s}} \mathrm{e}^{i\phi}\right) := \Re \widehat{P}_{\mathpzc{h},L}^{\mathpzc{s}}\left(\beta,\overline{z}_{\mathpzc{h},L}^{\mathpzc{s}}(\beta, \rho^{\mathpzc{s}},b) \mathrm{e}^{i\phi},b\right),\,\,
\Im \widehat{P}_{\mathpzc{h},L}^{\mathpzc{s}}\left(\overline{z}_{\mathpzc{h},L}^{\mathpzc{s}}\mathrm{e}^{i\phi} \right) := \Im \widehat{P}_{\mathpzc{h},L}^{\mathpzc{s}}\left(\beta,\overline{z}_{\mathpzc{h},L}^{\mathpzc{s}}(\beta, \rho^{\mathpzc{s}} ,b)\mathrm{e}^{i\phi},b\right),
\end{gather*}
and similar notation hold for the corresponding bulk quantities (obtained by dropping the '$L$').

\begin{lema}
\label{resimp}
For any $\mathpzc{h}>0$, $\beta>0$, $\rho^{\mathpzc{s}}>0$ and $b\in \mathbb{R}$: \\
$\mathrm{(i)}$.
\begin{equation*}
\lim_{L \uparrow \infty} \frac{\partial^{2} \Re \widehat{P}_{\mathpzc{h},L}^{\mathpzc{s}}}{\partial \phi^{2}} \left(\overline{z}_{\mathpzc{h},L}^{\mathpzc{s}}\right) =  \frac{\partial^{2} \Re \widehat{P}_{\mathpzc{h}}^{\mathpzc{s}}}{\partial \phi^{2}} \left(\overline{z}_{\mathpzc{h}}^{\mathpzc{s}}\right),
\end{equation*}
where $\widehat{P}_{\mathpzc{h}}^{\mathpzc{s}}$ denotes the analytic continuation of the bulk pressure $P_{\mathpzc{h}}^{\mathpzc{s}}(\beta,\cdot\,,b)$ to $\mathbb{C}\setminus (-\infty,-\mathrm{e}^{\beta E_{\mathpzc{h}}^{\mathpzc{s}}(b)}]$.\\
$\mathrm{(ii)}$.
\begin{equation*}
\frac{\partial^{j} \Re \widehat{P}_{\mathpzc{h},L}^{\mathpzc{s}}}{\partial \phi^{j}} \left(\overline{z}_{\mathpzc{h},L}^{\mathpzc{s}}\right)\left\{\begin{array}{ll}
= P_{\mathpzc{h},L}^{\mathpzc{s}}\left(\overline{z}_{\mathpzc{h},L}^{\mathpzc{s}}\right),\,\, &\textrm{if $j=0$},\\
= 0,\,\, &\textrm{if $j=1$},\\
<0,\,\, &\textrm{if $j=2$},
\end{array}\right.;\quad
\frac{\partial^{j} \Im \widehat{P}_{\mathpzc{h},L}^{\mathpzc{s}}}{\partial \phi^{j}} \left(\overline{z}_{\mathpzc{h},L}^{\mathpzc{s}}\right) = \left\{\begin{array}{ll}
0,\,\, &\textrm{if $j=0$},\\
\frac{\rho^{\mathpzc{s}}}{\beta},\,\, &\textrm{if $j=1$},\\
0,\,\, &\textrm{if $j=2$}
\end{array}\right..
\end{equation*}
$\mathrm{(iii)}$. $[-\pi,\pi] \owns \phi \mapsto \Re \widehat{P}_{\mathpzc{h},L}^{\mathpzc{s}}(\overline{z}_{\mathpzc{h},L}^{\mathpzc{s}}\mathrm{e}^{i\phi})$ is increasing on $[-\pi,0]$, decreasing on $[0,\pi]$.\\
$\mathrm{(iv)}$. There exist two constants $C_{l}= C_{l}(\beta,\rho^{\mathpzc{s}},b,\mathpzc{h})>0$, $l=1,2$ s.t. for $L\geq 1$ sufficiently large:
\begin{gather}
\label{firelm}
\frac{\partial^{2} \Re \widehat{P}_{\mathpzc{h},L}^{\mathpzc{s}}}{\partial \phi^{2}} \left(\overline{z}_{\mathpzc{h},L}^{\mathpzc{s}}\right)
 \leq - C_{1} < 0,\\
\label{secelm}
\forall \phi \in [-\pi,\pi],\quad \left\vert \frac{\partial^{3} \widehat{P}_{\mathpzc{h},L}^{\mathpzc{s}}}{\partial \phi^{3}} \left(\overline{z}_{\mathpzc{h},L}^{\mathpzc{s}}\mathrm{e}^{i\phi}\right)\right\vert \leq C_{2}.
\end{gather}
\end{lema}

Let us sketch the proof of $(1)$. Consider the following contribution:
\begin{equation}
\label{calAr}
\mathcal{A}_{\mathpzc{h},L}^{\mathpzc{s}\,\mathrm{(r)}}(\beta,\rho^{\mathpzc{s}},b) :=  \frac{\sqrt{N_{L}^{\mathpzc{s}}}}{2\pi} \int_{\phi_{L}^{\mathpzc{s}}}^{\pi} \mathrm{d}\phi\, \mathrm{e}^{\beta \frac{N_{L}^{\mathpzc{s}}}{\rho^{\mathpzc{s}}} \left(\widehat{P}_{\mathpzc{h},L}^{\mathpzc{s}}\left(\overline{z}_{\mathpzc{h},L}^{\mathpzc{s}} \mathrm{e}^{i\phi}\right) - P_{\mathpzc{h},L}^{\mathpzc{s}}\left(\overline{z}_{\mathpzc{h},L}^{\mathpzc{s}}\right)\right)} \mathrm{e}^{-i N_{L}^{\mathpzc{s}} \phi}.
\end{equation}
By Lemma \ref{resimp} $\mathrm{(iii)}$, $\phi \mapsto \Re \widehat{P}_{\mathpzc{h},L}^{\mathpzc{s}}(\overline{z}_{\mathpzc{h},L}^{\mathpzc{s}}\mathrm{e}^{i\phi})$ decreases on $[0,\pi]$. By expanding $\Re \widehat{P}_{\mathpzc{h},L}^{\mathpzc{s}}(\overline{z}_{\mathpzc{h},L}^{\mathpzc{s}} \mathrm{e}^{i\phi_{L}^{\mathpzc{s}}})$ by the Taylor's formula with integral remainder, then Lemma \ref{resimp} $\mathrm{(ii)}$ leads for $L$ sufficiently large to:
\begin{equation}
\label{why}
\left\vert \mathcal{A}_{\mathpzc{h},L}^{\mathpzc{s}\,\mathrm{(r)}}(\beta,\rho^{\mathpzc{s}},b)\right\vert \leq \frac{\sqrt{N_{L}^{\mathpzc{s}}}}{2\pi} \left(\pi - \phi_{L}^{\mathpzc{s}}\right) \mathrm{e}^{\beta\frac{N_{L}^{\mathpzc{s}}}{2 \rho^{\mathpzc{s}}} (\phi_{L}^{\mathpzc{s}})^{2} \frac{\partial^{2} \Re \widehat{P}_{\mathpzc{h},L}^{\mathpzc{s}}}{\partial \phi^{2}} \left(\overline{z}_{\mathpzc{h},L}^{\mathpzc{s}}\right)} \mathrm{e}^{\beta\frac{N_{L}^{\mathpzc{s}}}{2 \rho^{\mathpzc{s}}} \int_{0}^{\phi_{L}^{\mathpzc{s}}} \mathrm{d}u\, \left(\phi_{L}^{\mathpzc{s}} - u\right)^{2} \frac{\partial^{3} \Re \widehat{P}_{\mathpzc{h},L}^{\mathpzc{s}}}{\partial \phi^{3}} \left(\overline{z}_{\mathpzc{h},L}^{\mathpzc{s}}\mathrm{e}^{iu}\right)}.
\end{equation}
It remains to use \eqref{firelm} and \eqref{secelm} which lead respectively for $L$ sufficiently large to:
\begin{gather}
\label{gtred1}
\mathrm{e}^{\beta\frac{N_{L}^{\mathpzc{s}}}{2\rho^{\mathpzc{s}}} (\phi_{L}^{\mathpzc{s}})^{2} \frac{\partial^{2} \Re \widehat{P}_{\mathpzc{h},L}^{\mathpzc{s}}}{\partial \phi^{2}}  \left(\overline{z}_{\mathpzc{h},L}^{\mathpzc{s}}\right)} \leq \mathrm{e}^{- \frac{\beta}{2\rho^{\mathpzc{s}}} C_{1} (N_{L}^{\mathpzc{s}})^{2\kappa +1}},\\
\label{gtred2}
\mathrm{e}^{\beta \frac{N_{L}^{\mathpzc{s}}}{2\rho^{\mathpzc{s}}} \int_{0}^{\phi_{L}^{\mathpzc{s}}} \mathrm{d}u\, \left(\phi_{L}^{\mathpzc{s}} - u\right)^{2} \frac{\partial^{3}\Re \widehat{P}_{\mathpzc{h},L}^{\mathpzc{s}}}{\partial \phi^{3}} \left(\overline{z}_{\mathpzc{h},L}^{\mathpzc{s}}\mathrm{e}^{iu}\right)} \leq \mathrm{e}^{\frac{\beta}{6\rho^{\mathpzc{s}}} C_{2} (N_{L}^{\mathpzc{s}})^{3\kappa +1}}.
\end{gather}
The other contribution (as in \eqref{calAr} but with the integral over $[-\pi,-\phi_{L}^{\mathpzc{s}}]$) can be treated similarly. \\ Next, let us sketch the proof of (2). By expanding   $\widehat{P}_{\mathpzc{h},L}^{\mathpzc{s}}(\overline{z}_{\mathpzc{h},L}^{\mathpzc{s}}\mathrm{e}^{i\phi})$ by the Taylor's formula with integral remainder, followed by Lemma \ref{resimp} $\mathrm{(ii)}$, then one has for $L$ sufficiently large:
\begin{equation}
\label{calAmp}
\mathcal{A}_{\mathpzc{h},L}^{\mathpzc{s}\,\mathrm{(mp)}}(\beta,\rho^{\mathpzc{s}},b)  := \frac{\sqrt{N_{L}^{\mathpzc{s}}}}{2\pi} \int_{-(N_{L}^{\mathpzc{s}})^{\kappa}}^{(N_{L}^{\mathpzc{s}})^{\kappa}} \mathrm{d}\phi\, \mathrm{e}^{\beta\frac{N_{L}^{\mathpzc{s}}}{2\rho^{\mathpzc{s}}}\phi^{2}  \frac{\partial^{2} \Re \widehat{P}_{\mathpzc{h},L}^{\mathpzc{s}}}{\partial \phi^{2}} \left(\overline{z}_{\mathpzc{h},L}^{\mathpzc{s}}\right)} \mathrm{e}^{\beta\frac{N_{L}^{\mathpzc{s}}}{2\rho^{\mathpzc{s}}} \int_{0}^{\phi} \mathrm{d}u\, \left(\phi - u\right)^{2} \frac{\partial^{3} \widehat{P}_{\mathpzc{h},L}^{\mathpzc{s}}}{\partial \phi^{3}} \left(\overline{z}_{\mathpzc{h},L}^{\mathpzc{s}}\mathrm{e}^{iu}\right)}.
\end{equation}
Performing the change of variable $t= \sqrt{N_{L}^{\mathpzc{s}}} \phi$, one has for $L$  sufficiently large:
\begin{equation}
\label{calAmprex}
\mathcal{A}_{\mathpzc{h},L}^{\mathpzc{s}\,\mathrm{(mp)}}(\beta,\rho^{\mathpzc{s}},b)  = \frac{1}{2\pi} \int_{-(N_{L}^{\mathpzc{s}})^{\kappa + \frac{1}{2}}}^{(N_{L}^{\mathpzc{s}})^{\kappa + \frac{1}{2}}} \mathrm{d}t\, \mathrm{e}^{\frac{\beta}{2 \rho^{\mathpzc{s}}} \frac{\partial^{2} \Re \widehat{P}_{\mathpzc{h},L}^{\mathpzc{s}}}{\partial \phi^{2}} \left(\overline{z}_{\mathpzc{h},L}^{\mathpzc{s}}\right) t^{2}} \mathrm{e}^{\frac{\beta}{2 \rho^{\mathpzc{s}}} \int_{0}^{\frac{t}{\sqrt{N_{L}^{\mathpzc{s}}}}} \mathrm{d}u\, \left(t - \sqrt{N_{L}^{\mathpzc{s}}}u\right)^{2} \frac{\partial^{3} \widehat{P}_{\mathpzc{h},L}^{\mathpzc{s}}}{\partial \phi^{3}} \left(\overline{z}_{\mathpzc{h},L}^{\mathpzc{s}}\mathrm{e}^{iu}\right)}.
\end{equation}
It remains to use \eqref{secelm}, and \eqref{firelm} along with Lemma \ref{resimp} $\mathrm{(i)}$, which lead for $L$ large enough to:
\begin{gather*}
\int_{0}^{\frac{t}{\sqrt{N_{L}^{\mathpzc{s}}}}} \mathrm{d}u\, \left(t - \sqrt{N_{L}^{\mathpzc{s}}}u\right)^{2} \frac{\partial^{3} \widehat{P}_{\mathpzc{h},L}^{\mathpzc{s}}}{\partial \phi^{3}} \left(\overline{z}_{\mathpzc{h},L}^{\mathpzc{s}}\mathrm{e}^{iu}\right) = \mathcal{O}\left((N_{L}^{\mathpzc{s}})^{3\kappa + 1}\right),\\
\lim_{L \uparrow \infty} \int_{-(N_{L}^{\mathpzc{s}})^{\kappa + \frac{1}{2}}}^{(N_{L}^{\mathpzc{s}})^{\kappa + \frac{1}{2}}} \mathrm{d}t\, \mathrm{e}^{\frac{\beta}{2 \rho^{\mathpzc{s}}} \frac{\partial^{2} \Re \widehat{P}_{\mathpzc{h},L}^{\mathpzc{s}}}{\partial \phi^{2}} \left(\overline{z}_{\mathpzc{h},L}^{\mathpzc{s}}\right) t^{2}} = \sqrt{- \frac{2\pi \rho^{\mathpzc{s}}}{\beta} \left(\frac{\partial^{2} \Re \widehat{P}_{\mathpzc{h}}^{\mathpzc{s}}}{\partial \phi^{2}}\left(\overline{z}_{\mathpzc{h}}^{\mathpzc{s}}\right)\right)^{-1}} > 0,
\end{gather*}
where the last identity is obtained from the Lebesgue dominated convergence theorem. From the foregoing, one concludes that for any $\beta>0$, $\rho^{\mathpzc{s}}>0$, $b\in \mathbb{R}$ and $\mathpzc{h}>0$:
\begin{equation}
\label{limAl}
\mathcal{A}_{\mathpzc{h}}^{\mathpzc{s}}(\beta,\rho^{\mathpzc{s}},b) := \lim_{L \uparrow \infty} \mathcal{A}_{\mathpzc{h},L}^{\mathpzc{s}}(\beta,\rho^{\mathpzc{s}},b) =  \sqrt{- \frac{\rho^{\mathpzc{s}}}{2\pi \beta} \left(\frac{\partial^{2} \Re \widehat{P}_{\mathpzc{h}}^{\mathpzc{s}}}{\partial \phi^{2}}\left(\overline{z}_{\mathpzc{h}}^{\mathpzc{s}}\right)\right)^{-1}} > 0.
\end{equation}
In view of \eqref{ademo}, then \eqref{enlib} follows from  \eqref{limAl} along with \eqref{limTL}.\\
\indent Let us turn to the proof of \eqref{succano}. We only give the main arguments. In view of \eqref{ademo}, one has for any $\beta>0$, $\rho^{\mathpzc{s}}>0$, $b\in \mathbb{R}$, $\mathpzc{h}>0$ and for $L$ sufficiently large:
\begin{gather}
\label{finhft}
\frac{\partial \mathcal{F}_{\mathpzc{h},L}^{\mathpzc{s}}}{\partial b}(\beta,\rho^{\mathpzc{s}},b) - \frac{\partial (P_{\mathpzc{h},L}^{\mathpzc{s}})^{*}}{\partial b}(\beta,\rho^{\mathpzc{s}},b) = -\frac{1}{\beta \vert \Lambda_{L}\vert} \frac{\frac{\partial \mathcal{A}_{\mathpzc{h},L}^{\mathpzc{s}}}{\partial b}(\beta,\rho^{\mathpzc{s}},b)}{\mathcal{A}_{\mathpzc{h},L}^{\mathpzc{s}}(\beta,\rho^{\mathpzc{s}},b)},\\
\label{sinhft}
\frac{\partial^{2} \mathcal{F}_{\mathpzc{h},L}^{\mathpzc{s}}}{\partial b^{2}}(\beta,\rho^{\mathpzc{s}},b) - \frac{\partial^{2} (P_{\mathpzc{h},L}^{\mathpzc{s}})^{*}}{\partial b^{2}}(\beta,\rho^{\mathpzc{s}},b) = - \frac{1}{\beta \vert \Lambda_{L}\vert}\frac{ \frac{\partial^{2} \mathcal{A}_{\mathpzc{h},L}^{\mathpzc{s}}}{\partial b^{2}}(\beta,\rho^{\mathpzc{s}},b)}{\mathcal{A}_{\mathpzc{h},L}^{\mathpzc{s}}(\beta,\rho^{\mathpzc{s}},b)} + \frac{1}{\beta \vert \Lambda_{L}\vert} \frac{\left(\frac{\partial \mathcal{A}_{\mathpzc{h},L}^{\mathpzc{s}}}{\partial b}(\beta,\rho^{\mathpzc{s}},b)\right)^{2}}{\left(\mathcal{A}_{\mathpzc{h},L}^{\mathpzc{s}}(\beta,\rho^{\mathpzc{s}} ,b)\right)^{2}}.
\end{gather}
Let us prove that there exists a constant $C=C(\beta,\rho^{\mathpzc{s}},b,\mathpzc{h})>0$ s.t. for $L$ sufficiently large:
\begin{equation}
\label{onmor}
\left\vert \frac{\partial^{l} \mathcal{F}_{\mathpzc{h},L}^{\mathpzc{s}}}{\partial b^{l}}(\beta,\rho^{\mathpzc{s}},b) - \frac{\partial^{l} (P_{\mathpzc{h},L}^{\mathpzc{s}})^{*}}{\partial b^{l}}(\beta,\rho^{\mathpzc{s}},b)\right\vert \leq \frac{C}{\vert \Lambda_{L}\vert}, \quad l=1,2.
\end{equation}
From \eqref{finhft}-\eqref{sinhft} and due to \eqref{limAl}, this means that one has to prove for $L$ large enough that:
\begin{equation}
\label{up}
\left\vert \frac{\partial^{l} \mathcal{A}_{\mathpzc{h},L}^{\mathpzc{s}}}{\partial b^{l}}(\beta,\rho^{\mathpzc{s}},b) \right\vert \leq C,\quad l=1,2.
\end{equation}

To do that, we need the following lemma stated without proof. The same results are also used in the proof of \cite[Thm. 2]{C1}. Below, we use the shorthand notation introduced above Lemma \ref{resimp}.

\begin{lema}
\label{resimp2}
For any $\mathpzc{h}>0$, $\beta>0$, $\rho^{\mathpzc{s}}>0$ and $b\in \mathbb{R}$:\\
$\mathrm{(i)}$. There exist two constants $C_{l}>0$, $l=1,2$ s.t. for $L$ sufficiently large:
\begin{equation*}
\left\vert \frac{\partial^{l}}{\partial b^{l}} \left[\widehat{P}_{\mathpzc{h},L}^{\mathpzc{s}}\left(\overline{z}_{\mathpzc{h},L}^{\mathpzc{s}} \mathrm{e}^{i\phi}\right) - P_{\mathpzc{h},L}^{\mathpzc{s}}\left(\overline{z}_{\mathpzc{h},L}^{\mathpzc{s}}\right)\right] \right\vert \leq C_{l},\quad l=1,2.
\end{equation*}
$\mathrm{(ii)}$. There exist four constants $C_{l}^{'}, C_{l}^{''}>0$, $l=1,2$ s.t. for $L$ sufficiently large:
\begin{gather*}
\left\vert \frac{\partial^{l}}{\partial b^{l}} \left(\frac{\partial^{2} \widehat{P}_{\mathpzc{h},L}^{\mathpzc{s}}}{\partial \phi^{2}}\left(\overline{z}_{\mathpzc{h},L}^{\mathpzc{s}} \mathrm{e}^{i\phi}\right)\right)\right\vert\leq C_{l}^{'},\quad l=1,2,\\
\left\vert \int_{0}^{\frac{t}{\sqrt{N_{L}^{\mathpzc{s}}}}} \mathrm{d}u\, \left(t- \sqrt{N_{L}^{\mathpzc{s}}} u\right)^{2} \frac{\partial^{l}}{\partial b^{l}} \left(\frac{\partial^{3} \widehat{P}_{\mathpzc{h},L}^{\mathpzc{s}}}{\partial \phi^{3}}\left(\overline{z}_{\mathpzc{h},L}^{\mathpzc{s}} \mathrm{e}^{iu}\right) \right) \right\vert\leq C_{l}^{''} \frac{\vert t\vert^{3}}{\sqrt{N_{L}^{\mathpzc{s}}}},\quad l=1,2.
\end{gather*}
\end{lema}

We now prove \eqref{up}. From \eqref{calAr} and by using Lemma \ref{resimp2} $\mathrm{(i)}$, then for $L$ sufficiently large:
\begin{equation*}
\left\vert \frac{\partial^{l} \mathcal{A}_{\mathpzc{h},L}^{\mathpzc{s}\,\mathrm{(r)}}}{\partial b^{l}}(\beta,\rho^{\mathpzc{s}},b)\right\vert \leq \frac{(N_{L}^{\mathpzc{s}})^{\frac{3}{2}}}{2\pi} \frac{\beta}{\rho^{\mathpzc{s}}}\left(C_{l} + (l-1)C_{1}^{2} \frac{\beta}{\rho^{\mathpzc{s}}} N_{L}^{\mathpzc{s}}\right) \int_{\phi_{L}^{\mathpzc{s}}}^{\pi} \mathrm{d}\phi\, \mathrm{e}^{\beta \frac{N_{L}^{\mathpzc{s}}}{\rho^{\mathpzc{s}}}\left(\Re \widehat{P}_{\mathpzc{h},L}^{\mathpzc{s}}\left(\overline{z}_{\mathpzc{h},L}^{\mathpzc{s}} \mathrm{e}^{i \phi}\right) - P_{\mathpzc{h},L}^{\mathpzc{s}}\left(\overline{z}_{\mathpzc{h},L}^{\mathpzc{s}}\right)\right)}.
\end{equation*}
It remains to use the arguments leading to \eqref{why}, and the estimates in \eqref{gtred1}-\eqref{gtred2} to conclude that the contributions as in \eqref{calAr} are exponentially small in $L$. We continue with the contribution in \eqref{calAmp}. From \eqref{calAmprex} followed by Lemma \ref{resimp2} $\mathrm{(ii)}$, then for $L$ sufficiently large:
\begin{equation*}
\left\vert \frac{\partial^{l} \mathcal{A}_{\mathpzc{h},L}^{\mathpzc{s}\,\mathrm{(mp)}}}{\partial b^{l}}(\beta,\rho^{\mathpzc{s}},b)\right\vert \leq \frac{1}{2\pi}  \int_{-(N_{L}^{\mathpzc{s}})^{\kappa + \frac{1}{2}}}^{(N_{L}^{\mathpzc{s}})^{\kappa + \frac{1}{2}}} \mathrm{d}t\, \sum_{k=1}^{l} \left(\frac{\beta}{2\rho^{\mathpzc{s}}}\right)^{\frac{l}{k}} \left(C_{k}^{'} t^{2} + C_{k}^{''} \frac{\vert t\vert^{3}}{\sqrt{N_{L}^{\mathpzc{s}}}}\right)^{\frac{l}{k}}  \mathrm{e}^{ \frac{\beta}{2\rho^{\mathpzc{s}}} \frac{\partial^{2} \Re \widehat{P}_{\mathpzc{h},L}^{\mathpzc{s}}}{\partial \phi^{2}}\left(\overline{z}_{\mathpzc{h},L}^{\mathpzc{s}}\right) t^{2}} \mathrm{e}^{\mathcal{O}\left((N_{L}^{\mathpzc{s}})^{3\kappa+1}\right)}.
\end{equation*}
It remains to use that $\vert t \vert^{3} \leq C(\rho^{\mathpzc{s}}) t^{2}$ on $[-(N_{L}^{\mathpzc{s}})^{\kappa + \frac{1}{2}}, (N_{L}^{\mathpzc{s}})^{\kappa + \frac{1}{2}}]$ for $L$ large enough, along with:
\begin{equation*}
\lim_{L \uparrow \infty} \int_{-\infty}^{+\infty} \mathrm{d}t\, t^{2m}  \mathrm{e}^{ \frac{\beta}{2\rho^{\mathpzc{s}}} \frac{\partial^{2} \Re \widehat{P}_{\mathpzc{h},L}^{\mathpzc{s}}}{\partial \phi^{2}}\left(\overline{z}_{\mathpzc{h},L}^{\mathpzc{s}}\right) t^{2}} =  \Gamma\left(\frac{1}{2} + m\right) \left(\frac{\beta}{2\rho^{\mathpzc{s}}}\right)^{-\frac{1}{2}-m} \left(- \frac{\partial^{2} \Re \widehat{P}_{\mathpzc{h}}^{\mathpzc{s}}}{\partial \phi^{2}}\left(\overline{z}_{\mathpzc{h}}^{\mathpzc{s}}\right)\right)^{-\frac{1}{2}-m}>0,
\end{equation*}
for any $m \in \mathbb{N}^{*}$. Gathering the above results together, then \eqref{up} follows. The proof of \eqref{succano} follows from \eqref{onmor} together with the fact that $\lim_{L \uparrow \infty} (\partial_{b}^{l} (P_{\mathpzc{h},L}^{\mathpzc{s}})^{*})(\beta,\rho^{\mathpzc{s}},b) = (\partial_{b}^{l} (P_{\mathpzc{h}}^{\mathpzc{s}})^{*})(\beta,\rho^{\mathpzc{s}},b)$, $l=1,2$. The last statement comes from \eqref{limTL},  \eqref{limzL} and \eqref{Tleg}.\qed

\subsection{Proof of Lemmas \ref{isola11}-\ref{isola13}.}
\label{append23}

\indent When using the estimates from Lemmas \ref{lem01}-\ref{lem1}, we set
$\eta = \min\{ 1,\frac{\pi}{2\beta}\}>0$, see \eqref{Gamma}.\\

\noindent \textbf{Proof of Lemma \ref{isola11}.} We start with \eqref{trace01}. Denote
$\mathcal{Y}_{\mathpzc{h},0}(\xi) := \mathrm{Tr}_{L^{2}(\mathbb{R}^{3})}\{\chi_{\Omega_{\mathpzc{h}}} (\tilde{H}_{\mathpzc{h}} - \xi)^{-3} \chi_{\Omega_{\mathpzc{h}}}\}$.
By replacing $(\tilde{H}_{\mathpzc{h}} - \xi)^{-1}$ with the r.h.s. of \eqref{approxres} in $\mathcal{Y}_{\mathpzc{h},0}(\xi)$, then:
\begin{equation*}
\mathcal{Y}_{\mathpzc{h},0}(\xi) = \mathrm{Tr}_{L^{2}(\mathbb{R}^{3})}\left\{\chi_{\Omega_{\mathpzc{h}}} \mathcal{R}_{\mathpzc{h}}(\xi)\mathcal{R}_{\mathpzc{h}}(\xi)\mathcal{R}_{\mathpzc{h}}(\xi) \chi_{\Omega_{\mathpzc{h}}}\right\} + \mathcal{Q}_{\mathpzc{h},0}(\xi),
\end{equation*}
where $\mathcal{Q}_{\mathpzc{h},0}(\xi)$ consists of seven terms. Let us show that, under the conditions of Lemma \ref{isola11}, the quantity $\vert \mathcal{Q}_{\mathpzc{h},0}(\xi) \vert$ obeys an estimate of type \eqref{trace01}. To do so, take a generical term from $\mathcal{Q}_{\mathpzc{h},0}(\xi)$:
\begin{equation*}
q_{\mathpzc{h},0}(\xi) := - \mathrm{Tr}_{L^{2}(\mathbb{R}^{3})}\left\{\chi_{\Omega_{\mathpzc{h}}}\mathcal{R}_{\mathpzc{h}}(\xi)  \mathcal{R}_{\mathpzc{h}}(\xi)\left(\tilde{H}_{\mathpzc{h}} - \xi\right)^{-1} \mathcal{W}_{\mathpzc{h}}(\xi)\chi_{\Omega_{\mathpzc{h}}}\right\}.
\end{equation*}
Let $0<\alpha<1$ be fixed. From \eqref{kerSep} and \eqref{kerT}, one has under the conditions of Lemma \ref{lem1}:
\begin{equation}
\label{intermo}
\left \Vert \chi_{\Omega_{\mathpzc{h}}} \mathcal{R}_{\mathpzc{h}}(\xi)\right\Vert_{\mathfrak{I}_{2}} \leq p(\vert \xi\vert) \sqrt{\vert \Omega_{\mathpzc{h}}\vert}, \quad
\left \Vert \mathcal{W}_{\mathpzc{h}}(\xi) \chi_{\Omega_{\mathpzc{h}}} \right\Vert_{\mathfrak{I}_{2}} \leq p(\vert \xi \vert) \sqrt{\vert \Omega_{\mathpzc{h}}\vert} \mathrm{e}^{-\vartheta_{\xi} \mathpzc{h}^{-\alpha}},
\end{equation}
for another $\mathpzc{h}$-independent constant $\vartheta>0$ and polynomial $p(\cdot\,)$. From \eqref{intermo} along with \eqref{esfgv}, one concludes that there exists another $\vartheta>0$ and  polynomial $p(\cdot\,)$ s.t. $\forall \mathpzc{h} \in (0,\mathpzc{h}_{0}]$ and $\forall \xi \in \mathscr{C}_{\beta}$, $\vert q_{\mathpzc{h},0}(\xi)\vert \leq p(\vert \xi\vert) \mathrm{e}^{-\vartheta_{\xi} \mathpzc{h}^{-\alpha}}$. Here we used \eqref{keyes} to get rid of the factor $\vert \Omega_{\mathpzc{h}}\vert$. Finally, it remains to use \eqref{attendu}. The other terms coming from $\mathcal{Q}_{\mathpzc{h},0}(\xi)$ can be treated by using similar arguments.\\
Next, we turn to \eqref{trace1}. Denote $\mathcal{Y}_{\mathpzc{h},1}(\xi):= \mathrm{Tr}_{L^{2}(\mathbb{R}^{3})} \{\chi_{\Omega_{\mathpzc{h}}}(\tilde{H}_{\mathpzc{h}} - \xi)^{-1} \tilde{T}_{\mathpzc{h},1}(\xi)\tilde{T}_{\mathpzc{h},1}(\xi)\chi_{\Omega_{\mathpzc{h}}}\}$ and
$\mathcal{Y}_{\mathpzc{h},2}(\xi):= \mathrm{Tr}_{L^{2}(\mathbb{R}^{3})} \{\chi_{\Omega_{\mathpzc{h}}}(\tilde{H}_{\mathpzc{h}} - \xi)^{-1} \tilde{T}_{\mathpzc{h},2}(\xi)\chi_{\Omega_{\mathpzc{h}}}\}$. By replacing $(\tilde{H}_{\mathpzc{h}} - \xi)^{-1}$ with the r.h.s. of \eqref{approxres}, one gets:
\begin{gather*}
\mathcal{Y}_{\mathpzc{h},1}(\xi) = \int_{\Omega_{\mathpzc{h}}} \mathrm{d}\mathbf{x}\, \left(\mathcal{R}_{\mathpzc{h}}(\xi) \mathcal{T}_{\mathpzc{h},1}(\xi) \mathcal{T}_{\mathpzc{h},1}(\xi)\right)(\mathbf{x},\mathbf{x}) + \mathcal{Q}_{\mathpzc{h},1}(\xi),\\
\mathcal{Y}_{\mathpzc{h},2}(\xi) = \int_{\Omega_{\mathpzc{h}}} \mathrm{d}\mathbf{x}\, \left(\mathcal{R}_{\mathpzc{h}}(\xi) \mathcal{T}_{\mathpzc{h},2}(\xi)\right)(\mathbf{x},\mathbf{x}) + \mathcal{Q}_{\mathpzc{h},2}(\xi),
\end{gather*}
where $\mathcal{Q}_{\mathpzc{h},1}(\xi)$ and $\mathcal{Q}_{\mathpzc{h},2}(\xi)$ consist of seven and three terms respectively. Let $0<\alpha<1$ be fixed. Let us show that, under the conditions of Lemma \ref{isola11}, the quantities $\vert \mathcal{Q}_{\mathpzc{h},j}(\xi)\vert$, $j=1,2$ obey an estimate of type \eqref{trace1}. To do so, take one generical term from $\mathcal{Q}_{\mathpzc{h},1}(\xi)$ and one from $\mathcal{Q}_{\mathpzc{h},2}(\xi)$:
\begin{multline*}
q_{\mathpzc{h},1}(\xi) := \int_{\Omega_{\mathpzc{h}}} \mathrm{d}\mathbf{x} \int_{\mathbb{R}^{3}} \mathrm{d}\mathbf{z}_{1} \int_{\mathbb{R}^{3}} \mathrm{d}\mathbf{z}_{2}\, \left(\mathcal{R}_{\mathpzc{h}}(\xi)\right)(\mathbf{x},\mathbf{z}_{1}) \mathbf{a}(\mathbf{z}_{1} - \mathbf{z}_{2}) \cdot \nabla_{\mathbf{z}_{1}} \left(\mathcal{R}_{\mathpzc{h}}(\xi)\right)(\mathbf{z}_{1},\mathbf{z}_{2}) \times \\
\times \mathbf{a}(\mathbf{z}_{2} - \mathbf{x}) \cdot \nabla_{\mathbf{z}_{2}} \left(\left(\tilde{H}_{\mathpzc{h}} - \xi\right)^{-1} \mathcal{W}_{\mathpzc{h}}(\xi)\right)(\mathbf{z}_{2},\mathbf{x}),
\end{multline*}
\begin{equation*}
q_{\mathpzc{h},2}(\xi) := -\int_{\Omega_{\mathpzc{h}}} \mathrm{d}\mathbf{x} \int_{\mathbb{R}^{3}} \mathrm{d}\mathbf{z}\, \left(\mathcal{R}_{\mathpzc{h}}(\xi)\right)(\mathbf{x},\mathbf{z}) \frac{1}{2} \mathbf{a}^{2}(\mathbf{z} - \mathbf{x}) \left(\left(\tilde{H}_{\mathpzc{h}} - \xi\right)^{-1} \mathcal{W}_{\mathpzc{h}}(\xi)\right)(\mathbf{z},\mathbf{x}).
\end{equation*}
From \eqref{kerSep} and \eqref{kerresT}, there exists a constant $\vartheta>0$ and a polynomial $p(\cdot\,)$ s.t.
\begin{equation*}
\forall \mathpzc{h} \in (0,\mathpzc{h}_{0}],\,\forall \xi \in \mathscr{C}_{\beta},\quad \left\vert q_{\mathpzc{h},2}(\xi)\right\vert \leq  p(\vert \xi \vert) \mathrm{e}^{- \vartheta_{\xi} \mathpzc{h}^{- \alpha}}.
\end{equation*}
Here, we used the bound $\vert \mathbf{a}(\mathbf{z}-\mathbf{x})\vert \leq \vert \mathbf{z}-\mathbf{x}\vert$, the estimate \eqref{keyes} to get rid of a factor $\vert \mathbf{z}-\mathbf{x}\vert$, followed by \cite[Lem. A.2]{BS}. To control the quantity $\vert q_{\mathpzc{h},1}(\xi)\vert$, we need the following. From \eqref{trures} and \eqref{kerT}, there exists another $\vartheta>0$ and polynomial $p(\cdot\,)$ s.t. $\forall \mathpzc{h} \in (0,\mathpzc{h}_{0}]$ and $\forall \xi \in \mathscr{C}_{\beta}$:
\begin{equation}
\label{et1}
\forall(\bold{x},\bold{y}) \in \mathbb{R}^{6}\setminus D,\quad \left\vert \int_{\mathbb{R}^{3}} \mathrm{d}\mathbf{z}\, \nabla_{\mathbf{x}} \left(\tilde{H}_{\mathpzc{h}} - \xi\right)^{-1}(\mathbf{x},\mathbf{z}) \left(\mathcal{W}_{\mathpzc{h}}(\xi)\right)(\mathbf{z},\mathbf{y})\right\vert \leq p(\vert \xi\vert) \mathrm{e}^{- \vartheta_{\xi} \mathpzc{h}^{-\alpha}} \frac{\mathrm{e}^{- \vartheta_{\xi} \vert \mathbf{x} - \mathbf{y}\vert}}{\vert \mathbf{x}-\mathbf{y}\vert}.
\end{equation}
It follows from \eqref{kerSep}, \eqref{kerSepder}, \eqref{et1} combined with \cite[Lem. A.2]{BS}:
\begin{equation*}
\forall \mathpzc{h} \in (0,\mathpzc{h}_{0}],\,\forall \xi \in \mathscr{C}_{\beta},\quad \vert q_{\mathpzc{h},1}(\xi) \vert \leq  p(\vert \xi \vert) \mathrm{e}^{- \vartheta_{\xi} \mathpzc{h}^{- \alpha}},
\end{equation*}
for another constant $\vartheta>0$ and polynomial $p(\cdot\,)$ both $\mathpzc{h}$-independent. Note that, when estimating $\vert q_{\mathpzc{h},j}(\xi) \vert$, we got rid of the factor $\vert \Omega_{\mathpzc{h}}\vert= \mathpzc{h}^{-3}$ via \eqref{keyes}. Then, an estimate of type \eqref{trace1} follows from \eqref{attendu}. The other terms coming from $\mathcal{Q}_{\mathpzc{h},j}(\xi)$, $j=1,2$ can be treated by using similar arguments. \qed\\

\noindent \textbf{Proof of Lemma \ref{isola12}}. Start with \eqref{trace02}. Denote
$\mathscr{Y}_{\mathpzc{h},0}(\xi) := \mathrm{Tr}_{L^{2}(\mathbb{R}^{3})}\{\chi_{\Omega_{\mathpzc{h}}} \mathcal{R}_{\mathpzc{h}}(\xi)\mathcal{R}_{\mathpzc{h}}(\xi)\mathcal{R}_{\mathpzc{h}}(\xi)  \chi_{\Omega_{\mathpzc{h}}}\}$.
By replacing $\mathcal{R}_{\mathpzc{h}}(\xi)$ with the r.h.s. of \eqref{idd} in $\mathscr{Y}_{\mathpzc{h},0}(\xi)$, then:
\begin{equation*}
\mathscr{Y}_{\mathpzc{h},0}(\xi) = \mathrm{Tr}_{L^{2}(\mathbb{R}^{3})}\left\{\chi_{\Omega_{\mathpzc{h}}} \mathscr{R}_{\mathpzc{h}}(\xi)\mathscr{R}_{\mathpzc{h}}(\xi)\mathscr{R}_{\mathpzc{h}}(\xi)  \chi_{\Omega_{\mathpzc{h}}}\right\} + \mathscr{Q}_{\mathpzc{h},0}(\xi),
\end{equation*}
where $\mathscr{Q}_{\mathpzc{h},0}(\xi)$ consists of seven terms. Let us show that, under the conditions of Lemma \ref{isola12}, the quantity $\vert \mathscr{Q}_{\mathpzc{h},0}(\xi) \vert$ obeys an estimate of type \eqref{trace02}. To do so, take a generical term from $\mathscr{Q}_{\mathpzc{h},0}(\xi)$:
\begin{equation*}
\mathfrak{q}_{\mathpzc{h},0}(\xi) := \mathrm{Tr}_{L^{2}(\mathbb{R}^{3})}\left\{\chi_{\Omega_{\mathpzc{h}}} \mathscr{R}_{\mathpzc{h}}(\xi)\mathscr{R}_{\mathpzc{h}}(\xi)\mathscr{W}_{\mathpzc{h}}(\xi)  \chi_{\Omega_{\mathpzc{h}}}\right\}.
\end{equation*}
Fix $\alpha \in (0,1)$ and $\theta \in (0,1]$. From \eqref{kerSep} and \eqref{kerresT2}, one has under the conditions of Lemma \ref{lem1}:
\begin{equation}
\label{esresI2}
\left \Vert \chi_{\Omega_{\mathpzc{h}}} \mathscr{R}_{\mathpzc{h}}(\xi)\right\Vert_{\mathfrak{I}_{2}} \leq p(\vert \xi\vert) \sqrt{\vert \Omega_{\mathpzc{h}}\vert},\quad
\left \Vert \mathscr{W}_{\mathpzc{h}}(\xi) \chi_{\Omega_{\mathpzc{h}}} \right\Vert_{\mathfrak{I}_{2}} \leq p(\vert \xi \vert) \sqrt{\vert \Omega_{\mathpzc{h}}\vert} \mathpzc{h}^{\theta(1-\alpha)},
\end{equation}
for another $\mathpzc{h}$-independent polynomial $p(\cdot\,)$. From \eqref{esresI2} along with \eqref{esfgv}, one concludes that there exists another polynomial $p(\cdot\,)$ s.t. $\forall \mathpzc{h} \in (0,\mathpzc{h}_{0}]$ and $\forall \xi \in \mathscr{C}_{\beta}$, $\vert \mathfrak{q}_{\mathpzc{h},0}(\xi)\vert \leq p(\vert \xi\vert) \mathpzc{h}^{-3 + \theta(1-\alpha)}$. The other terms coming from $\mathscr{Q}_{\mathpzc{h},0}(\xi)$ can be treated by using similar arguments.\\
Next, we turn to \eqref{trace2}. Denote: $\mathscr{Y}_{\mathpzc{h},1}(\xi):= \mathrm{Tr}_{L^{2}(\mathbb{R}^{3})} \{\chi_{\Omega_{\mathpzc{h}}} \mathcal{R}_{\mathpzc{h}}(\xi) \mathcal{T}_{\mathpzc{h},1}(\xi)\mathcal{T}_{\mathpzc{h},1}(\xi)\chi_{\Omega_{\mathpzc{h}}}\}$ and
$\mathscr{Y}_{\mathpzc{h},2}(\xi):= \mathrm{Tr}_{L^{2}(\mathbb{R}^{3})}\{\chi_{\Omega_{\mathpzc{h}}} \mathcal{R}_{\mathpzc{h}}(\xi) \mathcal{T}_{\mathpzc{h},2}(\xi)\chi_{\Omega_{\mathpzc{h}}}\}$. By replacing $\mathcal{R}_{\mathpzc{h}}(\xi)$ with the r.h.s. of \eqref{idd} in $\mathscr{Y}_{\mathpzc{h},j}(\xi)$, $j=1,2$:
\begin{gather*}
\mathscr{Y}_{\mathpzc{h},1}(\xi) = \int_{\Omega_{\mathpzc{h}}} \mathrm{d}\mathbf{x}\, \left(\mathscr{R}_{\mathpzc{h}}(\xi) \mathscr{T}_{\mathpzc{h},1}(\xi) \mathscr{T}_{\mathpzc{h},1}(\xi)\right)(\mathbf{x},\mathbf{x}) + \mathscr{Q}_{\mathpzc{h},1}(\xi),\\
\mathscr{Y}_{\mathpzc{h},2}(\xi) = \int_{\Omega_{\mathpzc{h}}} \mathrm{d}\mathbf{x}\, \left(\mathscr{R}_{\mathpzc{h}}(\xi) \mathscr{T}_{\mathpzc{h},2}(\xi)\right)(\mathbf{x},\mathbf{x}) + \mathscr{Q}_{\mathpzc{h},2}(\xi),
\end{gather*}
where $\mathscr{Q}_{\mathpzc{h},1}(\xi)$ and $\mathscr{Q}_{\mathpzc{h},2}(\xi)$ consist of seven and three terms respectively. Fix $\alpha \in (0,1)$ and $\theta \in (0,1]$. Show that, under the conditions of Lemma \ref{isola12}, the quantities $\vert \mathscr{Q}_{\mathpzc{h},j}(\xi)\vert$, $j=1,2$ obey an estimate of type \eqref{trace2}. To do so, take one generical term from $\mathscr{Q}_{\mathpzc{h},1}(\xi)$ and one from $\mathscr{Q}_{\mathpzc{h},2}(\xi)$:
\begin{multline*}
\mathfrak{q}_{\mathpzc{h},1}(\xi) := - \int_{\Omega_{\mathpzc{h}}} \mathrm{d}\mathbf{x} \int_{\mathbb{R}^{3}} \mathrm{d}\mathbf{z}_{1} \int_{\mathbb{R}^{3}} \mathrm{d}\mathbf{z}_{2}\, \left(\mathscr{R}_{\mathpzc{h}}(\xi)\right)(\mathbf{x},\mathbf{z}_{1}) \mathbf{a}(\mathbf{z}_{1} - \mathbf{z}_{2}) \cdot \nabla_{\mathbf{z}_{1}} \left(\mathscr{R}_{\mathpzc{h}}(\xi)\right)(\mathbf{z}_{1},\mathbf{z}_{2}) \times \\
\times \mathbf{a}(\mathbf{z}_{2} - \mathbf{x}) \cdot \nabla_{\mathbf{z}_{2}} \left(\mathscr{W}_{\mathpzc{h}}(\xi)\right)(\mathbf{z}_{2},\mathbf{x}),
\end{multline*}
\begin{equation*}
\mathfrak{q}_{\mathpzc{h},2}(\xi) := \int_{\Omega_{\mathpzc{h}}} \mathrm{d}\mathbf{x} \int_{\mathbb{R}^{3}} \mathrm{d}\mathbf{z}\, \left(\mathscr{R}_{\mathpzc{h}}(\xi)\right)(\mathbf{x},\mathbf{z}) \frac{1}{2} \mathbf{a}^{2}(\mathbf{z} - \mathbf{x}) \left(\mathscr{W}_{\mathpzc{h}}(\xi)\right)(\mathbf{z},\mathbf{x}).
\end{equation*}
From \eqref{kerresT2} and by using that $\vert \mathbf{a}(\mathbf{x} - \mathbf{y})\vert \leq \vert \mathbf{x} - \mathbf{y}\vert$, then under the conditions of Lemma \ref{lem1}:
\begin{equation}
\label{interm1}
\forall (\mathbf{x},\mathbf{y})\in \mathbb{R}^{6},\quad \left\vert \mathbf{a}^{2}(\mathbf{x} - \mathbf{y}) \left(\mathscr{W}_{\mathpzc{h}}(\xi)\right)(\mathbf{x},\mathbf{y})\right\vert \leq p(\vert \xi\vert) \mathpzc{h}^{\theta(1-\alpha)} \mathrm{e}^{-\vartheta_{\xi} \vert \mathbf{x} - \mathbf{y}\vert},
\end{equation}
for another $\vartheta>0$ and polynomial $p(\cdot\,)$ both $\mathpzc{h}$-independent. Here we used \eqref{keyes}. From \eqref{interm1}, \eqref{kerSep} along with \cite[Lem. A.2]{BS}, then there exists another constant $\vartheta>0$ and polynomial $p(\cdot\,)$ s.t.
\begin{equation*}
\forall \mathpzc{h} \in (0,\mathpzc{h}_{0}],\,\forall \xi \in \mathscr{C}_{\beta},\quad \left\vert \mathfrak{q}_{\mathpzc{h},2}(\xi)\right\vert \leq  p(\vert \xi \vert) \mathpzc{h}^{-3 + \theta(1-\alpha)}.
\end{equation*}
To control the quantity $\vert \mathfrak{q}_{\mathpzc{h},1}(\xi)\vert$, we need the following estimate. From \eqref{Teps}, by using \eqref{otyp}-\eqref{otypder} together with \eqref{accroissfi}, then one can prove that under the conditions of Lemma \ref{lem1}:
\begin{equation}
\label{interm2}
\forall (\mathbf{x},\mathbf{y})\in \mathbb{R}^{6},\quad \left\vert \mathbf{a}(\mathbf{x} - \mathbf{y}) \cdot \nabla_{\mathbf{x}} \left(\mathscr{W}_{\mathpzc{h}}(\xi)\right)(\mathbf{x},\mathbf{y})\right\vert \leq p(\vert \xi\vert) \mathpzc{h}^{\theta(1-\alpha)} \mathrm{e}^{-\vartheta_{\xi} \vert \mathbf{x} - \mathbf{y}\vert},
\end{equation}
for another $\vartheta>0$ and polynomial $p(\cdot\,)$ both $\mathpzc{h}$-independent. Here we used \eqref{keyes} again. From \eqref{trures}, \eqref{kerSepder} and \eqref{interm2} combined with \cite[Lem. A.2]{BS}:
\begin{equation*}
\forall \mathpzc{h} \in (0,\mathpzc{h}_{0}],\,\forall \xi \in \mathscr{C}_{\beta},\quad \vert \mathfrak{q}_{\mathpzc{h},1}(\xi) \vert \leq  p(\vert \xi \vert) \mathpzc{h}^{-3 + \theta(1-\alpha)},
\end{equation*}
for another constant $\vartheta>0$ and polynomial $p(\cdot\,)$ both $\mathpzc{h}$-independent. The other terms coming from $\mathscr{Q}_{\mathpzc{h},j}(\xi)$, $j=1,2$ can be treated by using similar arguments. \qed \\

\noindent \textbf{Proof of Lemma \ref{isola13}}. We start with \eqref{moust1}. Note that \eqref{Sepsi} can be rewritten $\forall \mathpzc{h} \in (0,\mathpzc{h}_{0}]$ as:
\begin{gather}
\label{ldevp}
\mathscr{R}_{\mathpzc{h}}(\xi) = \sum_{\boldsymbol{\gamma} \in \mathscr{E}} \left(\tilde{H}_{\mathpzc{h},\boldsymbol{\gamma}}^{\mathrm{(cste)}} - \xi\right)^{-1} \tau_{\mathpzc{h},\boldsymbol{\gamma}} + \sum_{\boldsymbol{\gamma} \in \mathscr{E}} \mathcal{S}_{\mathpzc{h},\boldsymbol{\gamma}}^{\mathrm{(cste)}}(\xi), \quad \textrm{with:}\\
\label{calSa}
\mathcal{S}_{\mathpzc{h},\boldsymbol{\gamma}}^{\mathrm{(cste)}}(\xi) := \left(\tilde{H}_{\mathpzc{h},\boldsymbol{\gamma}}^{\mathrm{(cste)}} - \xi\right)^{-1}  \left[\tilde{H}_{\mathpzc{h},\boldsymbol{\gamma}}^{\mathrm{(cste)}},\hat{\tau}_{\mathpzc{h},\boldsymbol{\gamma}} \right] \left(\tilde{H}_{\mathpzc{h},\boldsymbol{\gamma}}^{\mathrm{(cste)}} - \xi\right)^{-1} \tau_{\mathpzc{h},\boldsymbol{\gamma}},
\end{gather}
resulting from the commutation of $\hat{\tau}_{\mathpzc{h},\boldsymbol{\gamma}}$ with $(\tilde{H}_{\mathpzc{h},\boldsymbol{\gamma}}^{\mathrm{(cste)}} - \xi)^{-1}$. Then, it follows that:
\begin{equation*}
\mathrm{Tr}_{L^{2}(\mathbb{R}^{3})}\left\{\chi_{\Omega_{\mathpzc{h}}} \left(\mathscr{R}_{\mathpzc{h}}(\xi)\right)^{3} \chi_{\Omega_{\mathpzc{h}}}\right\} =
\mathrm{Tr}_{L^{2}(\mathbb{R}^{3})}\left\{\chi_{\Omega_{\mathpzc{h}}}\left(\prod_{l=1}^{3} \sum_{\boldsymbol{\gamma}_{l} \in \mathscr{E}} \left(\tilde{H}_{\mathpzc{h},\boldsymbol{\gamma}_{l}}^{\mathrm{(cste)}} - \xi\right)^{-1} \tau_{\mathpzc{h},\boldsymbol{\gamma}_{l}}\right)\chi_{\Omega_{\mathpzc{h}}}\right\} + \mathfrak{Q}_{\mathpzc{h},0}(\xi),
\end{equation*}
where $\mathfrak{Q}_{\mathpzc{h},0}(\xi)$ consists of seven terms. Let us show that, under the conditions of Lemma \ref{isola13}, the quantity $\vert \mathfrak{Q}_{\mathpzc{h},0}(\xi)\vert$ obeys an estimate of type \eqref{moust1}. To do so, take a generical term from $\mathfrak{Q}_{\mathpzc{h},0}(\xi)$:
\begin{equation*}
\textswab{q}_{\mathpzc{h},0}(\xi) := \mathrm{Tr}_{L^{2}(\mathbb{R}^{3})}\left\{\chi_{\Omega_{\mathpzc{h}}}\left(\prod_{l=1}^{2} \sum_{\boldsymbol{\gamma}_{l} \in \mathscr{E}} \left(\tilde{H}_{\mathpzc{h},\boldsymbol{\gamma}_{l}}^{\mathrm{(cste)}} - \xi\right)^{-1} \tau_{\mathpzc{h},\boldsymbol{\gamma}_{l}}\right) \left(\sum_{\boldsymbol{\gamma} \in \mathscr{E}} \mathcal{S}_{\mathpzc{h},\boldsymbol{\gamma}}^{\mathrm{(cste)}}(\xi)\right)\chi_{\Omega_{\mathpzc{h}}}\right\}.
\end{equation*}
Firstly, by using \eqref{otyp} followed by \eqref{sumim}, then under the conditions of Lemma \ref{lem01}:
\begin{equation}
\label{txonI2}
\left \Vert \chi_{\Omega_{\mathpzc{h}}}\left(\sum_{\boldsymbol{\gamma} \in \mathscr{E}} \left(\tilde{H}_{\mathpzc{h},\boldsymbol{\gamma}}^{\mathrm{(cste)}} - \xi\right)^{-1} \tau_{\mathpzc{h},\boldsymbol{\gamma}}\right) \right\Vert_{\mathfrak{I}_{2}} \leq p(\vert \xi\vert) \sqrt{\vert \Omega_{\mathpzc{h}}\vert},
\end{equation}
for another $\mathpzc{h}$-independent polynomial $p(\cdot\,)$. Next, use that $[\tilde{H}_{\mathpzc{h},\boldsymbol{\gamma}}^{\mathrm{(cste)}},\hat{\tau}_{\mathpzc{h},\boldsymbol{\gamma}}] = -\frac{1}{2} (\Delta \hat{\tau}_{\mathpzc{h},\boldsymbol{\gamma}}) - (\nabla \hat{\tau}_{\mathpzc{h},\boldsymbol{\gamma}})\cdot \nabla$. Then, from \eqref{bigl} (which is unchanged when replacing $(\tilde{H}_{\mathpzc{h},\boldsymbol{\gamma}}^{(\mathrm{ref})} - \xi)^{-1}$ with $(\tilde{H}_{\mathpzc{h},\boldsymbol{\gamma}}^{\mathrm{(cste)}} - \xi)^{-1}$) and \eqref{otyp} combined with \cite[Lem. A.2]{BS}, one has on $\mathbb{R}^{6}$ under the conditions of Lemma \ref{lem1}:
\begin{equation*}
\sum_{\boldsymbol{\gamma} \in \mathscr{E}} \left\vert \left(  \left(\tilde{H}_{\mathpzc{h},\boldsymbol{\gamma}}^{\mathrm{(cste)}} - \xi\right)^{-1}  \left[\tilde{H}_{\mathpzc{h},\boldsymbol{\gamma}}^{\mathrm{(cste)}},\hat{\tau}_{\mathpzc{h},\boldsymbol{\gamma}} \right] \left(\tilde{H}_{\mathpzc{h},\boldsymbol{\gamma}}^{\mathrm{(cste)}} - \xi\right)^{-1} \tau_{\mathpzc{h},\boldsymbol{\gamma}}\right)(\mathbf{x},\mathbf{y})\right\vert \leq p(\vert \xi\vert) \mathrm{e}^{-\vartheta_{\xi} \mathpzc{h}^{-\alpha}} \mathrm{e}^{-\vartheta_{\xi}\vert \mathbf{x}-\mathbf{y}\vert},
\end{equation*}
for another constant $\vartheta>0$ and polynomial $p(\cdot\,)$ both $\mathpzc{h}$-independent. Therefore, it follows that:
\begin{equation}
\label{rappe}
\left\Vert\left(\sum_{\boldsymbol{\gamma} \in \mathscr{E}} \mathcal{S}_{\mathpzc{h},\boldsymbol{\gamma}}^{\mathrm{(cste)}}(\xi)\right) \chi_{\Omega_{\mathpzc{h}}} \right\Vert_{\mathfrak{I}_{2}} \leq p(\vert \xi\vert) \sqrt{\vert \Omega_{\mathpzc{h}}\vert} \mathrm{e}^{-\vartheta_{\xi} \mathpzc{h}^{-\alpha}},
\end{equation}
for another $\vartheta>0$ and polynomial $p(\cdot\,)$ both $\mathpzc{h}$-independent. Gathering \eqref{txonI2} and \eqref{rappe} together and using \eqref{esfgv}, then there exists another constant $\vartheta>0$ and polynomial $p(\cdot\,)$ s.t. $\forall \mathpzc{h} \in (0,\mathpzc{h}_{0}]$ and $\forall \xi \in \mathscr{C}_{\beta}$, $\vert \textswab{q}_{\mathpzc{h},0}(\xi) \vert \leq  p(\vert \xi \vert)\mathrm{e}^{-\vartheta_{\xi} \mathpzc{h}^{-\alpha}}$. Finally, it remains to use \eqref{attendu} to get an estimate of type \eqref{moust1}. The other terms coming from $\mathfrak{Q}_{\mathpzc{h},0}(\xi)$ can be treated by using similar arguments.\\
\indent Afterwards, let us turn to \eqref{moust}. Introduce $\forall 0<\alpha<1$, $\forall \mathpzc{h} \in (0,\mathpzc{h}_{0}]$ and $\forall \boldsymbol{\gamma} \in \mathscr{E}$ the operators $\tilde{T}_{\mathpzc{h},\boldsymbol{\gamma};0}^{\mathrm{(cste)}}(\xi)$ and $\tilde{T}_{\mathpzc{h},\boldsymbol{\gamma};1}^{\mathrm{(cste)}}(\xi)$ on $L^{2}(\mathbb{R}^{3})$ generated via their kernel  respectively defined as:
\begin{align}
\label{Tgam1hat}
\forall(\mathbf{x},\mathbf{y}) \in \mathbb{R}^{6}\setminus D,\quad \tilde{T}_{\mathpzc{h},\boldsymbol{\gamma};0}^{\mathrm{(cste)}}(\mathbf{x},\mathbf{y};\xi) &:= \mathbf{a}(\mathbf{x} - \mathbf{y}) \left(\tilde{H}_{\mathpzc{h},\boldsymbol{\gamma}}^{\mathrm{(cste)}} - \xi\right)^{-1}(\mathbf{x},\mathbf{y}),\\
\tilde{T}_{\mathpzc{h},\boldsymbol{\gamma};1}^{\mathrm{(cste)}}(\mathbf{x},\mathbf{y};\xi) &:= \mathbf{a}(\mathbf{x} - \mathbf{y}) \cdot (i \nabla_{\mathbf{x}}) \left(\tilde{H}_{\mathpzc{h},\boldsymbol{\gamma}}^{\mathrm{(cste)}} - \xi\right)^{-1}(\mathbf{x},\mathbf{y}) \nonumber.
\end{align}
Note an important thing. From the explicit expression in \eqref{Greenf}, one gets by direct calculations:
\begin{equation*}
\label{annul}
\forall(\mathbf{x},\mathbf{y}) \in \mathbb{R}^{6}\setminus D,\quad \mathbf{a}(\mathbf{x}-\mathbf{y}) \cdot (\nabla_{\mathbf{x}}) \left(\tilde{H}_{\mathpzc{h},\boldsymbol{\gamma}}^{\mathrm{(cste)}} - \xi\right)^{-1}(\mathbf{x},\mathbf{y}) = 0.
\end{equation*}
In view of \eqref{Sepsi}, the kernel $\mathscr{T}_{\mathpzc{h},1}(\cdot\,,\cdot\,;\xi)$ defined in \eqref{Wep1} can be therefore rewritten on $\mathbb{R}^{6} \setminus D$ as:
\begin{equation*}
\mathscr{T}_{\mathpzc{h},1}(\mathbf{x},\mathbf{y};\xi) = i \sum_{\boldsymbol{\gamma} \in \mathscr{E}} \left(\nabla  \hat{\tau}_{\mathpzc{h},\boldsymbol{\gamma}}\right)(\mathbf{x}) \tilde{T}_{\mathpzc{h},\boldsymbol{\gamma};0}^{\mathrm{(cste)}}(\mathbf{x},\mathbf{y};\xi) \tau_{\mathpzc{h},\boldsymbol{\gamma}}(\mathbf{y}).
\end{equation*}
As a result, the following identity holds:
\begin{multline*}
\mathrm{Tr}_{L^{2}(\mathbb{R}^{3})}\left\{\chi_{\Omega_{\mathpzc{h}}} \mathscr{R}_{\mathpzc{h}}(\xi) \mathscr{T}_{\mathpzc{h},1}(\xi) \mathscr{T}_{\mathpzc{h},1}(\xi)\chi_{\Omega_{\mathpzc{h}}}\right\} = \\
- \mathrm{Tr}_{L^{2}(\mathbb{R}^{3})}\left\{ \chi_{\Omega_{\mathpzc{h}}} \mathscr{R}_{\mathpzc{h}}(\xi) \left(\prod_{l=1}^{2} \sum_{\boldsymbol{\gamma}_{l} \in \mathscr{E}} (\nabla \hat{\tau}_{\mathpzc{h},\boldsymbol{\gamma}_{l}}) \tilde{T}_{\mathpzc{h},\boldsymbol{\gamma}_{l};0}^{\mathrm{(cste)}}(\xi) \tau_{\mathpzc{h},\boldsymbol{\gamma}_{l}}\right)\chi_{\Omega_{\mathpzc{h}}}\right\}.
\end{multline*}
Next, from \eqref{support} together with \eqref{otyp}, then under the conditions of Lemma \ref{lem01}:
\begin{equation}
\label{surest}
\left\Vert \left(\sum_{\boldsymbol{\gamma} \in \mathscr{E}} \left(\nabla \hat{\tau}_{\mathpzc{h},\boldsymbol{\gamma}}\right) \tilde{T}_{\mathpzc{h},\boldsymbol{\gamma};0}^{\mathrm{(cste)}}(\xi) \tau_{\mathpzc{h},\boldsymbol{\gamma}}\right)\chi_{\Omega_{\mathpzc{h}}} \right\Vert_{\mathfrak{I}_{2}} \leq p(\vert \xi\vert) \sqrt{\vert \Omega_{\mathpzc{h}}\vert} \mathrm{e}^{- \vartheta_{\xi} \mathpzc{h}^{-\alpha}},
\end{equation}
for another $\vartheta>0$ and polynomial $p(\cdot\,)$ both $\mathpzc{h}$-independent. From \eqref{surest}, \eqref{esresI2}, \eqref{esfgv} and \eqref{keyes}, one concludes that there exists another $\vartheta>0$ and polynomial $p(\cdot\,)$ s.t. $\forall \mathpzc{h} \in (0,\mathpzc{h}_{0}]$ and $\forall \xi \in \mathscr{C}_{\beta}$, $\Vert \chi_{\Omega_{\mathpzc{h}}} \mathscr{R}_{\mathpzc{h}}(\xi) \mathscr{T}_{\mathpzc{h},1}(\xi) \mathscr{T}_{\mathpzc{h},1}(\xi) \chi_{\Omega_{\mathpzc{h}}} \Vert_{\mathfrak{I}_{1}} \leq  p(\vert \xi \vert)\mathrm{e}^{-\vartheta_{\xi} \mathpzc{h}^{-\alpha}}$. It remains to use \eqref{attendu}, what leads to \eqref{moust}.\\
\indent Finally, let us turn to \eqref{etenc}. Remind that the operator $\mathscr{T}_{\mathpzc{h},2}(\xi)$ is generated via its kernel defined in \eqref{Wep2}. By replacing $\mathscr{R}_{\mathpzc{h}}(\xi)$ with the r.h.s. of \eqref{ldevp}, one has:
\begin{multline*}
\mathrm{Tr}_{L^{2}(\mathbb{R}^{3})}\left\{\chi_{\Omega_{\mathpzc{h}}}\mathscr{R}_{\mathpzc{h}}(\xi) \mathscr{T}_{\mathpzc{h},2}(\xi)\chi_{\Omega_{\mathpzc{h}}}\right\} = \\ \mathrm{Tr}_{L^{2}(\mathbb{R}^{3})}\left\{\chi_{\Omega_{\mathpzc{h}}}\left(
\sum_{\boldsymbol{\gamma}_{1} \in \mathscr{E}}  \left(\tilde{H}_{\mathpzc{h},\boldsymbol{\gamma}_{1}}^{\mathrm{(cste)}} - \xi\right)^{-1} \tau_{\mathpzc{h},\boldsymbol{\gamma}_{1}} \sum_{\boldsymbol{\gamma}_{2} \in \mathscr{E}} \tilde{T}_{\mathpzc{h},\boldsymbol{\gamma}_{2};2}^{\mathrm{(cste)}}(\xi) \tau_{\mathpzc{h},\boldsymbol{\gamma}_{2}}\right)\chi_{\Omega_{\mathpzc{h}}}\right\} + \mathfrak{Q}_{\mathpzc{h},2}(\xi),
\end{multline*}
where $\tilde{T}_{\mathpzc{h},\boldsymbol{\gamma};2}^{\mathrm{(cste)}}(\xi)$ is the operator generated by the kernel in \eqref{Tgam2hat} and $\mathfrak{Q}_{\mathpzc{h},2}(\xi)$ consists of three terms. Let us show that, under the conditions of Lemma \ref{isola13}, the quantity $\vert \mathfrak{Q}_{\mathpzc{h},2}(\xi)\vert$ obeys an estimate of type \eqref{etenc}. To do so, take a generical term from $\mathfrak{Q}_{\mathpzc{h},2}(\xi)$:
\begin{equation*}
\textswab{q}_{\mathpzc{h},2}(\xi) := \mathrm{Tr}_{L^{2}(\mathbb{R}^{3})}\left\{\chi_{\Omega_{\mathpzc{h}}} \left(\sum_{\boldsymbol{\gamma}_{1} \in \mathscr{E}} \left(\tilde{H}_{\mathpzc{h},\boldsymbol{\gamma}_{1}}^{\mathrm{(cste)}} - \xi\right)^{-1} \tau_{\mathpzc{h},\boldsymbol{\gamma}_{1}}\right) \left(\sum_{\boldsymbol{\gamma}_{2} \in \mathscr{E}} \mathscr{V}_{\mathpzc{h},\boldsymbol{\gamma}_{2}}^{\mathrm{(cste)}}(\xi) \tau_{\mathpzc{h},\boldsymbol{\gamma}_{2}}\right)\chi_{\Omega_{\mathpzc{h}}}\right\},
\end{equation*}
where:
\begin{equation*}
\begin{split}
\mathscr{V}_{\mathpzc{h},\boldsymbol{\gamma}}^{\mathrm{(cste)}}(\xi) := &\tilde{T}_{\mathpzc{h},\boldsymbol{\gamma};2}^{\mathrm{(cste)}}(\xi)
 \left[\tilde{H}_{\mathpzc{h},\boldsymbol{\gamma}}^{\mathrm{(cste)}},
\hat{\tau}_{\mathpzc{h},\boldsymbol{\gamma}}\right]
\left(\tilde{H}_{\mathpzc{h},\boldsymbol{\gamma}}^{\mathrm{(cste)}} - \xi\right)^{-1}  + \left(\tilde{H}_{\mathpzc{h},\boldsymbol{\gamma}}^{\mathrm{(cste)}} - \xi\right)^{-1} \left[\tilde{H}_{\mathpzc{h},\boldsymbol{\gamma}}^{\mathrm{(cste)}}, \hat{\tau}_{\mathpzc{h},\boldsymbol{\gamma}}\right] \tilde{T}_{\mathpzc{h},\boldsymbol{\gamma};2}^{\mathrm{(cste)}}(\xi) \\
&+ \tilde{T}_{\mathpzc{h},\boldsymbol{\gamma};0}^{\mathrm{(cste)}}(\xi) \left[\tilde{H}_{\mathpzc{h},\boldsymbol{\gamma}}^{\mathrm{(cste)}}, \hat{\tau}_{\mathpzc{h},\boldsymbol{\gamma}}\right] \tilde{T}_{\mathpzc{h},\boldsymbol{\gamma};0}^{\mathrm{(cste)}}(\xi).
\end{split}
\end{equation*}
Here, $\tilde{T}_{\mathpzc{h},\boldsymbol{\gamma};0}^{\mathrm{(cste)}}(\xi)$ is the operator generated by the kernel in \eqref{Tgam1hat}. To arrive at such an identity, we used that $\mathbf{a}^{2}(\mathbf{x}-\mathbf{y}) = \{\mathbf{a}(\mathbf{x}-\mathbf{z}) + \mathbf{a}(\mathbf{z}-\mathbf{y})\}^{2}$ $\forall (\mathbf{x},\mathbf{y},\mathbf{z}) \in \mathbb{R}^{9}$. From \eqref{Wep2} and \eqref{Tgam1hat}, then under the conditions of Lemma \ref{lem01}, the kernel of the operators $\tilde{T}_{\mathpzc{h},\boldsymbol{\gamma};k}^{\mathrm{(cste)}}(\xi)$, $k\in \{0,2\}$ obey an estimate of type \eqref{otyp}. Ergo, the estimate on the H-S norm in \eqref{rappe} still holds true when replacing in \eqref{calSa} one or the two resolvents with $\tilde{T}_{\mathpzc{h},\boldsymbol{\gamma};k}^{\mathrm{(cste)}}(\xi)$, $k\in \{0,2\}$. From this and \eqref{txonI2}, then there exists another $\vartheta>0$ and polynomial $p(\cdot\,)$ s.t. $\forall \mathpzc{h} \in (0,\mathpzc{h}_{0}]$ and $\forall \xi \in \mathscr{C}_{\beta}$, $\vert \textswab{q}_{\mathpzc{h},2}(\xi) \vert \leq  p(\vert \xi \vert)\mathrm{e}^{-\vartheta_{\xi} \mathpzc{h}^{-\alpha}}$. It remains to use \eqref{attendu}. The other terms coming from $\mathfrak{Q}_{\mathpzc{h},2}(\xi)$ can be treated similarly. \qed

\begin{remark}
\label{bite}
The methods we use to estimate the traces in the proof of Lemmas \ref{isola11}-\ref{isola12}-\ref{isola13} heavily rely on the presence of the characteristic functions $\chi_{\Omega_{\mathpzc{h}}}$. But in fact, the $\chi_{\Omega_{\mathpzc{h}}}$'s can be both removed from the traces since the operators sandwiched between the $\chi_{\Omega_{\mathpzc{h}}}$'s are trace-class on $L^{2}(\mathbb{R}^{3})$. Indeed, these operators can always be written as a product of bounded operators of type:
\begin{equation}
\label{cc}
\mathpzc{K}_{i}(\xi) \mathpzc{O}(\xi) \mathpzc{K}_{j}(\xi)\,\,\,\,\textrm{or}\,\,\,\, \mathpzc{O}(\xi) \mathpzc{K}_{i}(\xi) \mathpzc{K}_{j}(\xi)\,\,\,\,\textrm{or}\,\,\,\, \mathpzc{K}_{i}(\xi)\mathpzc{K}_{j}(\xi)\mathpzc{O}(\xi),\quad i,j \in \{1,2\}^{2},
\end{equation}
where, under the conditions of Lemma \ref{lem01}:\\
$\mathrm{(i)}$. The kernel $O(\cdot\,,\cdot\,;\xi)$ of $\mathpzc{O}(\xi)$ obeys an estimate of type \eqref{otyp} or \eqref{otypder};\\
$\mathrm{(ii)}$. $\mathpzc{K}_{1}(\xi)$ has the form:
\begin{equation*}
\mathpzc{K}_{1}(\xi) = \sum_{\boldsymbol{\gamma} \in \mathscr{E}} \hat{\tau}_{\mathpzc{h},\boldsymbol{\gamma}} K_{1,\boldsymbol{\gamma}}(\xi) \tau_{\mathpzc{h},\boldsymbol{\gamma}},
\end{equation*}
and the kernel $K_{1,\boldsymbol{\gamma}}(\cdot\,,\cdot\,;\xi)$ of $K_{1,\boldsymbol{\gamma}}(\xi)$ obeys an estimate of type \eqref{otyp};\\
$\mathrm{(iii)}$. $\mathpzc{K}_{2}(\xi)$ has one of the two following forms:
\begin{equation*}
(1)\,\,\,\sum_{\boldsymbol{\gamma} \in \mathscr{E}} (\nabla \hat{\tau}_{\mathpzc{h},\boldsymbol{\gamma}}) K_{2,\boldsymbol{\gamma}}(\xi) \tau_{\mathpzc{h},\boldsymbol{\gamma}},\,\,\,\,or\,\,\,\, (2)\,\,\,
\sum_{\boldsymbol{\gamma} \in \mathscr{E}}  (\Delta \hat{\tau}_{\mathpzc{h},\boldsymbol{\gamma}}) K_{2,\boldsymbol{\gamma}}(\xi) \tau_{\mathpzc{h},\boldsymbol{\gamma}},
\end{equation*}
and the kernel $K_{2,\boldsymbol{\gamma}}(\cdot\,,\cdot\,;\xi)$ of $K_{2,\boldsymbol{\gamma}}(\xi)$ obeys an estimate of type \eqref{otyp} or \eqref{otypder}.\\
Consider for instance $\mathpzc{K}_{1}(\xi) \mathpzc{O}(\xi) \mathpzc{K}_{2}(\xi)$ with $\mathpzc{K}_{2}(\xi)$ having the form $(1)$. Then, one has:
\begin{equation*}
\left\vert \mathrm{Tr}_{L^{2}(\mathbb{R}^{3})}\left\{\mathpzc{K}_{1}(\xi) \mathpzc{O}(\xi) \mathpzc{K}_{2}(\xi)\right\}\right\vert \leq \sum_{\boldsymbol{\gamma}_{1} \in \mathscr{E}} \sum_{\boldsymbol{\gamma}_{2} \in \mathscr{E}} \left\Vert \hat{\tau}_{\mathpzc{h},\boldsymbol{\gamma}_{1}} K_{1,\boldsymbol{\gamma}_{1}}(\xi) \tau_{\mathpzc{h},\boldsymbol{\gamma}_{1}} \mathpzc{O}(\xi) (\nabla \hat{\tau}_{\mathpzc{h},\boldsymbol{\gamma}_{2}}) K_{2,\boldsymbol{\gamma}_{2}}(\xi) \tau_{\mathpzc{h},\boldsymbol{\gamma}_{2}}\right\Vert_{\mathfrak{I}_{1}}.
\end{equation*}
Note that the double sum only contains $\mathcal{O}(\mathpzc{h}^{3\alpha - 3})$ non-zero terms since when keeping $\boldsymbol{\gamma}_{1}$ fixed, only a finite number ($\mathpzc{h}$-independent) of $\boldsymbol{\gamma}_{2}$'s have an overlapping support. By mimicking the arguments leading to the estimates on H-S norms in Remark \ref{HSn}, then under the conditions of Lemma \ref{lem01}:
\begin{equation*}
\left\Vert \hat{\tau}_{\mathpzc{h},\boldsymbol{\gamma}} K_{1,\boldsymbol{\gamma}}(\xi) \tau_{\mathpzc{h},\boldsymbol{\gamma}} \right\Vert_{\mathfrak{I}_{2}} \leq p(\vert \xi\vert) \mathpzc{h}^{-\frac{3}{2}\alpha},\quad \left\Vert (\nabla \hat{\tau}_{\mathpzc{h},\boldsymbol{\gamma}}) K_{2,\boldsymbol{\gamma}}(\xi) \tau_{\mathpzc{h},\boldsymbol{\gamma}}\right\Vert_{\mathfrak{I}_{2}} \leq p(\vert \xi\vert) \mathrm{e}^{-\vartheta_{\xi} \mathpzc{h}^{-\alpha}} \mathpzc{h}^{-\frac{3}{2} \alpha},
\end{equation*}
for another constant $\vartheta>0$ and polynomial $p(\cdot\,)$ both $\mathpzc{h}$-independent.
It follows that:
\begin{equation*}
\left\vert \mathrm{Tr}_{L^{2}(\mathbb{R}^{3})}\left\{\mathpzc{K}_{1}(\xi) \mathpzc{O}(\xi) \mathpzc{K}_{2}(\xi)\right\}\right\vert \leq p(\vert \xi\vert) \mathpzc{h}^{-3}  \mathrm{e}^{-\vartheta_{\xi} \mathpzc{h}^{-\alpha}},
\end{equation*}
for another $\mathpzc{h}$-independent polynomial $p(\cdot\,)$. All the terms in \eqref{cc} can be treated similarly.
\end{remark}

\subsection{Proof of Lemmas \ref{retro1}-\ref{retro2}.}
\label{append22}

\indent When using the estimates from Lemmas \ref{lem01}-\ref{lem1}, we set
$\eta = \min\{ 1,\frac{\pi}{2\beta}\}>0$, see \eqref{Gamma}.\\

\noindent \textit{\textbf{Proof of Lemma \ref{retro1}}}. Under the conditions of \eqref{prlaref}, the contribution $\tilde{\mathcal{X}}_{\mathpzc{h},r_{1}}^{\mathrm{(orbit)}}(\beta,z)$ reads as:
\begin{multline}
\label{chir1}
\tilde{\mathcal{X}}_{\mathpzc{h},r_{1}}^{\mathrm{(orbit)}}(\beta,z) := -\mathpzc{h}^{2} \frac{\varkappa_{\mathrm{o}}}{\beta} \frac{i}{2\pi} \int_{\mathscr{C}_{\beta}} \mathrm{d}\xi\, \mathfrak{f}(\beta,z;\xi) \\
\times \int_{\Omega_{\mathpzc{h}}} \mathrm{d}\mathbf{x} \int_{\mathbb{R}^{3}\setminus\Omega_{\mathpzc{h}}} \mathrm{d}\bold{z} \sum_{\boldsymbol{\gamma}_{1} \in \mathscr{E}} \left(\tilde{H}_{\mathpzc{h},\boldsymbol{\gamma}_{1}}^{\mathrm{(cste)}} - \xi\right)^{-1}(\mathbf{x},\mathbf{z}) \tau_{\mathpzc{h},\boldsymbol{\gamma}_{1}}(\mathbf{z})  \sum_{\boldsymbol{\gamma}_{2} \in \mathscr{E}} \tilde{T}_{\mathpzc{h},\boldsymbol{\gamma}_{2};2}^{\mathrm{(cste)}}(\mathbf{z},\mathbf{x};\xi) \tau_{\mathpzc{h},\boldsymbol{\gamma}_{2}}(\mathbf{x}).
\end{multline}
The integral w.r.t. $\bold{z}$ can be reduced to the integral over the set $\mathpzc{S}_{\mathpzc{h}}:= (\cup_{\boldsymbol{\gamma} \in \mathscr{E}} \mathrm{Supp}(\tau_{\mathpzc{h},\boldsymbol{\gamma}})) \setminus \Omega_{\mathpzc{h}}$ whose Lebesgue-measure is of order $\mathcal{O}(\mathpzc{h}^{-2-\alpha})$.
From \eqref{otyp}, under the conditions of Lemma \ref{lem01}:
\begin{equation}
\label{rdg}
\forall(\bold{x},\bold{y}) \in \mathbb{R}^{6}\setminus D,\quad \max\left\{\left\vert \left(\tilde{H}_{\mathpzc{h},\boldsymbol{\gamma}}^{\mathrm{(cste)}} - \xi\right)^{-1}(\mathbf{x},\mathbf{y})\right\vert, \left\vert \tilde{T}_{\mathpzc{h},\boldsymbol{\gamma};2}^{\mathrm{(cste)}}(\mathbf{x},\mathbf{y};\xi)\right\vert\right\} \leq p(\vert \xi\vert) \frac{\mathrm{e}^{-\vartheta_{\xi} \vert \mathbf{x} - \mathbf{y}\vert}}{\vert \mathbf{x} - \mathbf{y}\vert},
\end{equation}
for another $\vartheta>0$ and polynomial $p(\cdot\,)$ both independent of $\mathpzc{h}$. By using \eqref{rdg}, \eqref{sumim} combined with \cite[Lem. A.2]{BS}, then there exists another polynomial $p(\cdot\,)$ s.t. $\forall \xi \in \mathscr{C}_{\beta}$ and $\forall \mathpzc{h} \in (0,\mathpzc{h}_{0}]$:
\begin{equation*}
\int_{\mathpzc{S}_{\mathpzc{h}}} \mathrm{d}\mathbf{z}\int_{\mathbb{R}^{3}} \mathrm{d}\mathbf{x} \left\vert \sum_{\boldsymbol{\gamma}_{1} \in \mathscr{E}} \left(\tilde{H}_{\mathpzc{h},\boldsymbol{\gamma}_{1}}^{\mathrm{(cste)}} - \xi\right)^{-1}(\mathbf{x},\mathbf{z}) \tau_{\mathpzc{h},\boldsymbol{\gamma}_{1}}(\mathbf{z})  \sum_{\boldsymbol{\gamma}_{2} \in \mathscr{E}} \tilde{T}_{\mathpzc{h},\boldsymbol{\gamma}_{2};2}^{\mathrm{(cste)}}(\mathbf{z},\mathbf{x};\xi) \tau_{\mathpzc{h},\boldsymbol{\gamma}_{2}}(\mathbf{x})\right\vert \leq p(\vert \xi\vert) \mathpzc{h}^{-2-\alpha}.
\end{equation*}
Therefore, the integrals w.r.t. $\mathbf{x}$ and $\mathbf{z}$ in \eqref{chir1} can be commuted by the Tonelli's theorem. By using the above estimate along with \eqref{estimo2}, then there exists a constant $C=C(\beta)>0$ s.t.
\begin{equation*}
\forall \mathpzc{h} \in (0,\mathpzc{h}_{0}],\quad \left\vert \tilde{\mathcal{X}}_{\mathpzc{h},r_{1}}^{\mathrm{(orbit)}}(\beta,z) \right\vert \leq C z \mathpzc{h}^{-\alpha}.
\end{equation*}
Next, let us turn to $\tilde{\mathcal{X}}_{\mathpzc{h},r_{2}}^{\mathrm{(orbit)}}(\beta,z)$. It is made up of three terms, whose a generical term reads as:
\begin{equation}
\label{crisp}
\tilde{\mathpzc{X}}_{\mathpzc{h},r_{2}}^{\mathrm{(orbit)}}(\beta,z) := - \mathpzc{h}^{2} \frac{\varkappa_{\mathrm{o}}}{2} \frac{1}{\beta} \frac{i}{2\pi} \int_{\mathscr{C}_{\beta}} \mathrm{d}\xi\, \mathfrak{f}(\beta,z;\xi)  \int_{\Omega_{\mathpzc{h}}} \mathrm{d}\mathbf{x} \, \mathcal{K}_{\mathpzc{h},r_{2}}(\mathbf{x};\xi),\\
\end{equation}
where, $\forall \xi \in \mathscr{C}_{\beta}$, $\forall \mathpzc{h} \in (0,\mathpzc{h}_{0}]$ and $\forall \bold{x} \in \Omega_{\mathpzc{h}}$:
\begin{multline}
\label{calK}
\mathcal{K}_{\mathpzc{h},r_{2}}(\mathbf{x};\xi) :=
\frac{1}{(2\pi)^{2}} \int_{\Omega_{\mathpzc{h}}} \mathrm{d}\mathbf{z}\, \frac{\mathrm{e}^{- \sqrt{-2\left(\xi - V(\mathpzc{h}\mathbf{x})\right)}\vert \mathbf{x} - \mathbf{z}\vert}}{\vert \mathbf{x} - \mathbf{z}\vert} \mathbf{a}^{2}(\mathbf{z}-\mathbf{x}) \\
 \times \sum_{\boldsymbol{\gamma} \in \mathscr{E}} \left\{\frac{\mathrm{e}^{-\varsigma_{\mathpzc{h},\boldsymbol{\gamma}}(\xi) \vert \mathbf{z} - \mathbf{x}\vert}}{\vert \mathbf{z} - \mathbf{x}\vert} -  \frac{\mathrm{e}^{-\sqrt{-2\left(\xi - V(\mathpzc{h}\mathbf{x})\right)} \vert \mathbf{z} - \mathbf{x}\vert}}{\vert \mathbf{z} - \mathbf{x}\vert}\right\} \tau_{\mathpzc{h},\boldsymbol{\gamma}}(\mathbf{x}).
\end{multline}
Let $\mathbf{x}_{0} \in \Omega_{\mathpzc{h}}$ kept fixed. Introduce in $L^{2}(\mathbb{R}^{3})$ the self-adjoint realization of the operator $\hat{H}_{\mathpzc{h},\mathbf{x}_{0}}^{\mathrm{(cste)}} := -\frac{1}{2} \Delta + V(\mathpzc{h} \mathbf{x}_{0})$ defined originally on $\mathcal{C}_{0}^{\infty}(\mathbb{R}^{3})$. Its Green function is  explicitly known and it is given by \eqref{Greenf} but with $V(\mathpzc{h}\mathbf{x}_{0})$ instead of $V(\mathpzc{h}^{1-\alpha} \boldsymbol{\gamma})$. Then, by using a resolvent identity in \eqref{calK}:
\begin{multline}
\label{follw}
\mathcal{K}_{\mathpzc{h},r_{2}}(\mathbf{x}_{0};\xi) = \int_{\Omega_{\mathpzc{h}}} \mathrm{d}\mathbf{z}\, \left(\hat{H}_{\mathpzc{h},\mathbf{x}_{0}}^{\mathrm{(cste)}} - \xi\right)^{-1}(\mathbf{x}_{0},\mathbf{z}) \mathbf{a}^{2}(\mathbf{z}-\mathbf{x}_{0}) \times \\
\times \sum_{\boldsymbol{\gamma} \in \mathscr{E}} \left\{\left(\tilde{H}_{\mathpzc{h},\boldsymbol{\gamma}}^{\mathrm{(cste)}} - \xi\right)^{-1}\left[V(\mathpzc{h} \mathbf{x}_{0}) - V\left(\mathpzc{h}^{1-\alpha} \boldsymbol{\gamma}\right)\right] \left(\hat{H}_{\mathpzc{h},\mathbf{x}_{0}}^{\mathrm{(cste)}} - \xi\right)^{-1}\right\}(\mathbf{z},\mathbf{x}_{0}) \tau_{\mathpzc{h},\boldsymbol{\gamma}}(\mathbf{x}_{0}).
\end{multline}
Now by using \eqref{accroissfi}, the estimate \eqref{otyp} combined with \cite[Lem A.2]{BS}, then under the conditions of Lemma \ref{lem01}, there exists a polynomial $p(\cdot\,)$ independent of $\bold{x}_{0}$ s.t. $\forall \xi \in \mathscr{C}_{\beta}$ and $\forall \mathpzc{h} \in (0,\mathpzc{h}_{0}]$:
\begin{equation}
\label{esGcal}
\left\vert \mathcal{K}_{\mathpzc{h},r_{2}}(\mathbf{x}_{0};\xi) \right\vert \leq p(\vert \xi \vert) \mathpzc{h}^{\theta(1 - \alpha)}.
\end{equation}
In view of \eqref{crisp}, we conclude from \eqref{esGcal} and \eqref{estimo2} that there exists another $C=C(\beta)>0$ s.t.
\begin{equation*}
\forall \mathpzc{h} \in (0,\mathpzc{h}_{0}],\quad \left\vert \tilde{\mathpzc{X}}_{\mathpzc{h},r_{2}}^{\mathrm{(orbit)}}(\beta,z) \right\vert \leq C z \mathpzc{h}^{-1 + \theta(1-\alpha)}.
\end{equation*}
The two other terms coming from $\tilde{\mathcal{X}}_{\mathpzc{h},r_{2}}^{\mathrm{(orbit)}}(\beta,z)$ can be treated by similar arguments. Finally, let us turn to the contribution $\tilde{\mathcal{X}}_{\mathpzc{h},r_{3}}^{\mathrm{(orbit)}}(\beta,z)$ which reads under the conditions of \eqref{prlaref} as:
\begin{equation}
\label{pzcK'}
\tilde{\mathcal{X}}_{\mathpzc{h},r_{3}}^{\mathrm{(orbit)}}(\beta,z) := \frac{\mathpzc{h}^{2}}{\mathpzc{h}^{6}} \frac{\varkappa_{\mathrm{o}}}{2\beta} \frac{i}{2\pi} \int_{\mathscr{C}_{\beta}} \mathrm{d}\xi\, \mathfrak{f}(\beta,z;\xi) \int_{\Omega} \mathrm{d}\mathbf{x}\, \mathcal{K}_{\mathpzc{h},r_{3}}\left(\frac{\mathbf{x}}{\mathpzc{h}};\xi\right),\\
\end{equation}
with, $\forall \xi \in \mathscr{C}_{\beta}$, $\forall \mathpzc{h} \in (0,\mathpzc{h}_{0}]$ and $\forall \bold{x} \in \Omega$:
\begin{equation*}
\mathcal{K}_{\mathpzc{h},r_{3}}\left(\frac{\mathbf{x}}{\mathpzc{h}};\xi\right) :=
\frac{1}{(2\pi)^{2}} \int_{\mathbb{R}^{3}\setminus \Omega} \mathrm{d}\mathbf{z}\, \frac{\mathrm{e}^{- \sqrt{-2\left(\xi - V(\mathbf{x})\right)}\left\vert \frac{\mathbf{x}}{\mathpzc{h}} - \frac{\mathbf{z}}{\mathpzc{h}}\right\vert}}{\left\vert \frac{\mathbf{x}}{\mathpzc{h}} - \frac{\mathbf{z}}{\mathpzc{h}}\right\vert^{2}} \mathbf{a}^{2}\left(\frac{\mathbf{z}}{\mathpzc{h}}- \frac{\mathbf{x}}{\mathpzc{h}}\right).
\end{equation*}
By using that $\vert \bold{a}(\bold{x}-\bold{y})\vert \leq \vert \bold{x} - \bold{y}\vert$, then the above integrand obeys an estimate of type \eqref{otyp}. Fix $\mathbf{x}_{0} \in \Omega$. Switching to the spherical coordinates, then under the conditions of Lemma \ref{lem01}, there exists a constant $\vartheta>0$ and a polynomial $p(\cdot\,)$ independent of $\mathbf{x}_{0}$ s.t. $\forall \xi \in \mathcal{C}_{\beta}$ and $\forall \mathpzc{h} \in (0,\mathpzc{h}_{0}]$:
\begin{equation}
\label{hestcalK}
\left\vert \mathcal{K}_{\mathpzc{h},r_{3}}\left(\frac{\mathbf{x}_{0}}{\mathpzc{h}};\xi\right) \right\vert \leq p(\vert \xi\vert) \int_{\frac{1}{2}}^{\infty} \mathrm{d}r\, r^{2} \mathrm{e}^{-\mathpzc{h}^{-1} \frac{\vartheta}{1 + \vert \xi\vert} r}
\leq p(\vert \xi\vert) \mathrm{e}^{-\frac{\mathpzc{h}^{-1}}{2} \frac{\vartheta}{1 + \vert \xi\vert}} \int_{0}^{\infty} \mathrm{d}r\, r^{2} \mathrm{e}^{-\frac{\mathpzc{h}^{-1}}{2} \frac{\vartheta}{1 + \vert \xi\vert} r}.
\end{equation}
Performing the integration w.r.t. the $r$-variable, one arrives at the estimate:
\begin{equation}
\label{cruess}
\forall \xi \in \mathcal{C}_{\beta},\, \forall \mathpzc{h} \in (0,\mathpzc{h}_{0}],\quad \left\vert \mathcal{K}_{\mathpzc{h},r_{3}}\left(\frac{\mathbf{x}_{0}}{\mathpzc{h}};\xi\right) \right\vert \leq p(\vert \xi\vert) \mathpzc{h}^{3} \mathrm{e}^{-\frac{\mathpzc{h}^{-1}}{2} \frac{\vartheta}{1 + \vert \xi\vert}},
\end{equation}
for another polynomial $p(\cdot\,)$ independent of $\mathpzc{h}, \mathbf{x}_{0}$. In view of \eqref{pzcK'}, from \eqref{cruess}, \eqref{attendu} with $N \geq 1$ and \eqref{estimo2}, then one concludes that $\forall M > 0$ there exists a constant $C_{M}= C_{M}(\beta) >0$ s.t.
\begin{equation*}
\forall \mathpzc{h} \in (0,\mathpzc{h}_{0}],\quad \left\vert \tilde{\mathcal{X}}_{\mathpzc{h},r_{3}}^{\mathrm{(orbit)}}(\beta,z) \right\vert \leq C_{M} z \mathpzc{h}^{M}. \tag*{\qed}
\end{equation*}

\noindent \textit{\textbf{Proof of Lemma \ref{retro2}}}. We start with the contribution $\tilde{\mathcal{X}}_{\mathpzc{h},r_{1}}^{\mathrm{(spin)}}(\beta,z)$ which is made up of two terms. Under the conditions of \eqref{prlarefs}, one of them  reads as (hereafter we set $\mathbf{z}_{3}:=\mathbf{z}_{0}$):
\begin{multline}
\label{chis1}
\tilde{\mathpzc{X}}_{\mathpzc{h},r_{1}}^{\mathrm{(spin)}}(\beta,z) :=\mathpzc{h}^{2} \frac{\varkappa_{\mathrm{s}}}{\beta} \frac{i}{2\pi} \int_{\mathscr{C}_{\beta}} \mathrm{d}\xi\, \mathfrak{f}(\beta,z;\xi) \\
\times
\int_{\Omega_{\mathpzc{h}}} \mathrm{d}\mathbf{z}_{0} \int_{\mathbb{R}^{3} \setminus \Omega_{\mathpzc{h}}} \mathrm{d}\mathbf{z}_{1} \int_{\mathbb{R}^{3}} \mathrm{d}\bold{z}_{2}\, \prod_{l=0}^{2} \sum_{\boldsymbol{\gamma}_{l+1} \in \mathscr{E}} \left(\tilde{H}_{\mathpzc{h},\boldsymbol{\gamma}_{l+1}}^{\mathrm{(cste)}} - \xi\right)^{-1}(\mathbf{z}_{l},\mathbf{z}_{l+1}) \tau_{\mathpzc{h},\boldsymbol{\gamma}_{l+1}}(\mathbf{z}_{l+1}).
\end{multline}
Reducing the integral w.r.t. $\mathbf{z}$ to the integral over $\mathpzc{S}_{\mathpzc{h}}:= (\cup_{\boldsymbol{\gamma} \in \mathscr{E}} \mathrm{Supp}(\tau_{\mathpzc{h},\boldsymbol{\gamma}})) \setminus \Omega_{\mathpzc{h}}$, from \eqref{otyp}, \eqref{sumim} combined with
\cite[Lem. A.2]{BS}, then there exists a polynomial $p(\cdot\,)$ s.t. $\forall \xi \in \mathscr{C}_{\beta}$ and $\forall \mathpzc{h} \in (0,\mathpzc{h}_{0}]$:
\begin{equation*}
\int_{\mathpzc{S}_{\mathpzc{h}}} \mathrm{d}\mathbf{z}_{1} \int_{\mathbb{R}^{3}} \mathrm{d}\mathbf{z}_{0} \int_{\mathbb{R}^{3}} \mathrm{d}\mathbf{z}_{2} \left\vert \prod_{l=0}^{2} \sum_{\boldsymbol{\gamma}_{l+1} \in \mathscr{E}} \left(\tilde{H}_{\mathpzc{h},\boldsymbol{\gamma}_{l+1}}^{\mathrm{(cste)}} - \xi\right)^{-1}(\mathbf{z}_{l},\mathbf{z}_{l+1}) \tau_{\mathpzc{h},\boldsymbol{\gamma}_{l+1}}(\mathbf{z}_{l+1})\right\vert \leq p(\vert \xi\vert) \mathpzc{h}^{-2-\alpha}.
\end{equation*}
Therefore, the integrals w.r.t. $\mathbf{z}_{0}$ and $\mathbf{z}_{1}$ in \eqref{chis1} can be commuted by the Tonelli's theorem, and by using the estimate just above along with \eqref{estimo2}, then there exists a constant $C=C(\beta)>0$ s.t.
\begin{equation*}
\forall \mathpzc{h} \in (0,\mathpzc{h}_{0}],\quad \left\vert \tilde{\mathpzc{X}}_{\mathpzc{h},r_{1}}^{\mathrm{(spin)}}(\beta,z) \right\vert \leq C z \mathpzc{h}^{-\alpha}.
\end{equation*}
The other term coming from $\tilde{\mathcal{X}}_{\mathpzc{h},r_{1}}^{\mathrm{(spin)}}(\beta,z)$ can be treated by the same method. Next, let us turn to the contribution  $\tilde{\mathcal{X}}_{\mathpzc{h},r_{2}}^{\mathrm{(spin)}}(\beta,z)$. It is made up of seven terms, whose a generical term reads as:
\begin{equation}
\label{crisp2}
\tilde{\mathpzc{X}}_{\mathpzc{h},r_{2}}^{\mathrm{(spin)}}(\beta,z) := \mathpzc{h}^{2} \frac{\varkappa_{\mathrm{s}}}{\beta} \frac{i}{2\pi} \int_{\mathscr{C}_{\beta}} \mathrm{d}\xi\, \mathfrak{f}(\beta,z;\xi)  \int_{\Omega_{\mathpzc{h}}} \mathrm{d}\mathbf{x} \, \mathcal{J}_{\mathpzc{h},r_{2}}(\mathbf{x};\xi),\\
\end{equation}
where, $\forall \xi \in \mathscr{C}_{\beta}$, $\forall \mathpzc{h} \in (0,\mathpzc{h}_{0}]$ and $\forall \bold{x} \in \Omega_{\mathpzc{h}}$:
\begin{multline*}
\mathcal{J}_{\mathpzc{h},r_{2}}(\mathbf{x};\xi) :=
\frac{1}{(2\pi)^{3}} \int_{\Omega_{\mathpzc{h}}} \mathrm{d}\mathbf{z}_{1} \int_{\Omega_{\mathpzc{h}}} \mathrm{d}\mathbf{z}_{2}\, \frac{\mathrm{e}^{- \sqrt{-2\left(\xi - V(\mathpzc{h}\mathbf{x})\right)}\vert \mathbf{x} - \mathbf{z}_{1}\vert}}{\vert \mathbf{x} - \mathbf{z}_{1}\vert} \\
 \times \sum_{\boldsymbol{\gamma} \in \mathscr{E}} \left\{\frac{\mathrm{e}^{-\varsigma_{\mathpzc{h},\boldsymbol{\gamma}}(\xi) \vert \mathbf{z}_{1} - \mathbf{z}_{2}\vert}}{\vert \mathbf{z}_{1} - \mathbf{z}_{2}\vert} -  \frac{\mathrm{e}^{-\sqrt{-2\left(\xi - V(\mathpzc{h}\mathbf{x})\right)} \vert \mathbf{z}_{1} - \mathbf{z}_{2}\vert}}{\vert \mathbf{z}_{1} - \mathbf{z}_{2}\vert}\right\} \tau_{\mathpzc{h},\boldsymbol{\gamma}}(\mathbf{z}_{2}) \frac{\mathrm{e}^{- \sqrt{-2\left(\xi - V(\mathpzc{h}\mathbf{x})\right)}\vert \mathbf{z}_{2} - \mathbf{x}\vert}}{\vert \mathbf{z}_{2} - \mathbf{x}\vert}.
\end{multline*}
Let $\mathbf{x}_{0} \in \Omega_{\mathpzc{h}}$ kept fixed. By following the same arguments as the ones leading to \eqref{follw} from \eqref{calK} in the proof of Lemma \ref{retro1}, then $\forall \xi \in \mathscr{C}_{\beta}$ and $\forall \mathpzc{h} \in (0,\mathpzc{h}_{0}]$, one arrives at:
\begin{multline*}
\mathcal{J}_{\mathpzc{h},r_{2}}(\mathbf{x}_{0};\xi) = \int_{\Omega_{\mathpzc{h}}} \mathrm{d}\mathbf{z}_{1} \int_{\Omega_{\mathpzc{h}}} \mathrm{d}\mathbf{z}_{2} \, \left(\hat{H}_{\mathpzc{h},\mathbf{x}_{0}}^{\mathrm{(cste)}} - \xi\right)^{-1}(\mathbf{x}_{0},\mathbf{z}_{1})  \sum_{\boldsymbol{\gamma} \in \mathscr{E}} \left\{\left(\tilde{H}_{\mathpzc{h},\boldsymbol{\gamma}}^{\mathrm{(cste)}} - \xi\right)^{-1} \right. \\
\left. \times \left[V(\mathpzc{h} \mathbf{x}_{0}) - V\left(\mathpzc{h}^{1-\alpha} \boldsymbol{\gamma}\right)\right] \left(\hat{H}_{\mathpzc{h},\mathbf{x}_{0}}^{\mathrm{(cste)}} - \xi\right)^{-1}\right\}(\mathbf{z}_{1},\mathbf{z}_{2}) \tau_{\mathpzc{h},\boldsymbol{\gamma}}(\mathbf{z}_{2})
 \left(\hat{H}_{\mathpzc{h},\mathbf{x}_{0}}^{\mathrm{(cste)}} - \xi\right)^{-1}(\mathbf{z}_{2},\mathbf{x}_{0}).
\end{multline*}
Now by using \eqref{accroissfi}, the estimate \eqref{otyp} combined with \cite[Lem A.2]{BS}, then under the conditions of Lemma \ref{lem01}, there exists a polynomial $p(\cdot\,)$ independent of $\mathbf{x}_{0}$ s.t. $\forall \xi \in \mathscr{C}_{\beta}$ and $\forall \mathpzc{h} \in (0,\mathpzc{h}_{0}]$:
\begin{equation}
\label{esGcal2}
\left\vert \mathcal{J}_{\mathpzc{h},r_{2}}(\mathbf{x}_{0};\xi) \right\vert \leq p(\vert \xi \vert) \mathpzc{h}^{\theta(1 - \alpha)}.
\end{equation}
In view of \eqref{crisp2}, we conclude from \eqref{esGcal2} and \eqref{estimo2} that there exists another $C=C(\beta)>0$ s.t.
\begin{equation*}
\forall \mathpzc{h} \in (0,\mathpzc{h}_{0}],\quad \left\vert \tilde{\mathpzc{X}}_{\mathpzc{h},r_{2}}^{\mathrm{(spin)}}(\beta,z) \right\vert \leq C z \mathpzc{h}^{-1 + \theta(1-\alpha)}.
\end{equation*}
The six other terms coming from   $\tilde{\mathcal{X}}_{\mathpzc{h},r_{2}}^{\mathrm{(spin)}}(\beta,z)$ can be treated by the same method. Finally, let us turn to the contribution $\tilde{\mathcal{X}}_{\mathpzc{h},r_{3}}^{\mathrm{(spin)}}(\beta,z)$ which is made up of two terms. Take one of them:
\begin{equation}
\label{pzcJ'}
\tilde{\mathpzc{X}}_{\mathpzc{h},r_{3}}^{\mathrm{(spin)}}(\beta,z) := -\frac{\mathpzc{h}^{2}}{\mathpzc{h}^{9}} \frac{\varkappa_{\mathrm{s}}}{\beta} \frac{i}{2\pi} \int_{\mathscr{C}_{\beta}} \mathrm{d}\xi\, \mathfrak{f}(\beta,z;\xi) \int_{\Omega} \mathrm{d}\mathbf{x}\, \mathcal{J}_{\mathpzc{h},r_{3}}\left(\frac{\mathbf{x}}{\mathpzc{h}};\xi\right),\\
\end{equation}
with, $\forall \xi \in \mathscr{C}_{\beta}$, $\forall \mathpzc{h} \in (0,\mathpzc{h}_{0}]$ and $\forall \bold{x} \in \Omega$:
\begin{equation*}
\mathcal{J}_{\mathpzc{h},r_{3}}\left(\frac{\mathbf{x}}{\mathpzc{h}};\xi\right) :=
\frac{1}{(2\pi)^{3}} \int_{\mathbb{R}^{3}\setminus \Omega} \mathrm{d}\mathbf{z}_{1} \int_{\Omega} \mathrm{d}\mathbf{z}_{2}\, \frac{\mathrm{e}^{- \sqrt{-\left(\xi - V(\mathbf{x})\right)}\left\vert \frac{\mathbf{x}}{\mathpzc{h}} - \frac{\mathbf{z}_{1}}{\mathpzc{h}}\right\vert}}{\left\vert \frac{\mathbf{x}}{\mathpzc{h}} - \frac{\mathbf{z}_{1}}{\mathpzc{h}}\right\vert} \frac{\mathrm{e}^{- \sqrt{-\left(\xi - V(\mathbf{x})\right)}\left(\left\vert \frac{\mathbf{z}_{1}}{\mathpzc{h}} - \frac{\mathbf{z}_{2}}{\mathpzc{h}}\right\vert + \left\vert \frac{\mathbf{z}_{2}}{\mathpzc{h}} - \frac{\mathbf{x}}{\mathpzc{h}}\right\vert\right)}}{\left\vert \frac{\mathbf{z}_{1}}{\mathpzc{h}} - \frac{\mathbf{z}_{2}}{\mathpzc{h}}\right\vert \left\vert \frac{\mathbf{z}_{2}}{\mathpzc{h}} - \frac{\mathbf{x}}{\mathpzc{h}}\right\vert}.
\end{equation*}
Let $\mathbf{x}_{0} \in \Omega$ kept fixed. Under the conditions of Lemma \ref{lem01} and from \cite[Lem. A.2]{BS}, then there exists a constant $\vartheta>0$ and a polynomial $p(\cdot\,)$ independent of $\mathbf{x}_{0}$ s.t. $\forall \xi \in \mathcal{C}_{\beta}$ and $\forall \mathpzc{h} \in (0,\mathpzc{h}_{0}]$:
\begin{equation*}
\int_{\Omega} \mathrm{d}\mathbf{z}_{2}\, \frac{\mathrm{e}^{- \sqrt{-\left(\xi - V(\mathbf{x}_{0})\right)}\left(\left\vert \frac{\mathbf{z}_{1}}{\mathpzc{h}} - \frac{\mathbf{z}_{2}}{\mathpzc{h}}\right\vert + \left\vert \frac{\mathbf{z}_{2}}{\mathpzc{h}} - \frac{\mathbf{x}_{0}}{\mathpzc{h}}\right\vert\right)}}{\left\vert \frac{\mathbf{z}_{1}}{\mathpzc{h}} - \frac{\mathbf{z}_{2}}{\mathpzc{h}}\right\vert \left\vert \frac{\mathbf{z}_{2}}{\mathpzc{h}} - \frac{\mathbf{x}_{0}}{\mathpzc{h}}\right\vert} \leq p(\vert \xi\vert) \mathpzc{h}^{3} \mathrm{e}^{- \frac{\vartheta}{1+\vert \xi\vert}\left\vert \frac{\mathbf{z}_{1}}{\mathpzc{h}} - \frac{\mathbf{x}_{0}}{\mathpzc{h}}\right\vert}.
\end{equation*}
By using the same arguments as the ones leading to \eqref{hestcalK}, one has under the same conditions:
\begin{equation}
\label{cruess2}
\forall \xi \in \mathcal{C}_{\beta},\,\forall \mathpzc{h} \in (0,\mathpzc{h}_{0}],\quad \left\vert \mathcal{J}_{\mathpzc{h},r_{3}}\left(\frac{\mathbf{x}_{0}}{\mathpzc{h}};\xi\right) \right\vert \leq p(\vert \xi\vert) \mathpzc{h}^{6} \mathrm{e}^{-\frac{\mathpzc{h}^{-1}}{2} \frac{\vartheta}{1 + \vert \xi\vert}},
\end{equation}
for another $\vartheta>0$ and polynomial $p(\cdot\,)$ independent of $\mathpzc{h},\mathbf{x}_{0}$. In view of \eqref{pzcJ'}, from \eqref{cruess2}, \eqref{attendu} with $N \geq1$  and \eqref{estimo2}, then one concludes that $\forall M >0$ there exists a $C_{M}= C_{M}(\beta) >0$ s.t.
\begin{equation*}
\forall \mathpzc{h} \in (0,\mathpzc{h}_{0}],\quad \left\vert \tilde{\mathpzc{X}}_{\mathpzc{h},r_{3}}^{\mathrm{(spin)}}(\beta,z) \right\vert \leq C_{M} z \mathpzc{h}^{M}.
\end{equation*}
The other term coming from $\tilde{\mathcal{X}}_{\mathpzc{h},r_{3}}^{\mathrm{(spin)}}(\beta,z)$ can be treated by similar arguments. \qed

\section{Acknowledgments.}

A part of this work was done while the author was a member of the Simion Stoilow Institute of Mathematics of the
Romanian Academy, Bucharest.
B.S. thanks Horia D. Cornean and Tony C. Dorlas for many fruitful and stimulating discussions.

\end{document}